\documentclass[10pt,conference]{IEEEtran}

\IEEEoverridecommandlockouts
\usepackage{cite}
\usepackage{amsmath,amssymb,amsfonts}

\usepackage[]{algorithm}
\newfloat{algorithm}{t}{lop}
\usepackage[noend]{algpseudocode}
\newcommand{\AlgoComment}[1]{\Comment{{\small\color{blue}#1}}}
\usepackage{balance}
\usepackage{textcomp}
\usepackage{xspace}
\usepackage{xcolor}
\definecolor{annotatecolor}{rgb}{0.59,0,0.09}
\usepackage{graphicx}
\usepackage[normalem]{ulem} %
\usepackage{enumitem}

\usepackage{hyperref}
\hypersetup{linkcolor=black,citecolor=black,anchorcolor=black,filecolor=black,menucolor=black,runcolor=black,urlcolor=black,hidelinks}
\usepackage{breakurl}

\usepackage{booktabs}
\usepackage{multirow}
\usepackage{makecell}
\usepackage{ragged2e}

\usepackage{caption}
\usepackage{subcaption}
\usepackage{subfloat}

\usepackage{listings}

\definecolor{gray}{RGB}{211,211,211}
\newcommand{\jbasicstyle}{\small\ttfamily} %

\newcommand{\jnumberstyle}{\scriptsize}

\lstdefinelanguage{pseudo}
{
  morekeywords={},
  keywordstyle=\bfseries,
  lineskip=-0.1em,
  numbers=left, %
  numberstyle=\jnumberstyle,
  numbersep=4pt,
  basicstyle=\jbasicstyle,
  breaklines=true,
  breakautoindent=true,
  tabsize=2,
  columns=fullflexible,
  morecomment=*[l][\textsl]{//},
  mathescape=true,
  xleftmargin=10pt,
}

\lstdefinelanguage{todo-comment}
{
  morekeywords={},
  keywordstyle=\bfseries,
  lineskip=-0.1em,
  numbers=none,
  basicstyle=\scriptsize\ttfamily,
  breaklines=true,
  breakautoindent=true,
  tabsize=2,
  columns=fullflexible,
  morecomment=*[l][\textsl]{//},
  mathescape=true,
  xleftmargin=0pt,
}

\definecolor{keywordcolor}{rgb}{0,0,1}      %
\definecolor{modifiercolor}{rgb}{0.5,0,0.5} %
\definecolor{datatypecolor}{rgb}{0.82,0.16,0.46} %
\definecolor{methodcolor}{rgb}{0.25,0.5,0.35} %
\definecolor{byzantine}{rgb}{0.74, 0.2, 0.64}  %
\definecolor{cadetblue}{rgb}{0.37, 0.62, 0.63}  %
\definecolor{cadet}{rgb}{0.0, 0.42, 0.24}
\definecolor{brown(web)}{rgb}{0.65, 0.16, 0.16}  %
\definecolor{bluegray}{rgb}{0.2, 0.2, 0.6}

\lstdefinelanguage{java-pretty}
{
  language=java,
  numbers=left,
  basicstyle=\scriptsize\ttfamily,
  numberstyle=\scriptsize,
  breaklines=true,
  columns=fullflexible,
  xleftmargin=14pt,
  tabsize=2,
  showstringspaces=false,
  deletekeywords={public, private, protected, static, final, class, interface, abstract, implements, extends, if, else, while, do, for, switch, case, default, break, continue, return, int, long, double, float, boolean, char, void, String,this},
  morekeywords=[1]{if, else, while, do, for, switch, case, default, break, continue, return},
  keywordstyle=[1]\color{byzantine}\bfseries,
  morekeywords=[2]{public, private, protected, static, final, class, interface, abstract, implements, extends},
  keywordstyle=[2]\color{bluegray}\bfseries,
  morekeywords=[3]{int, long, double, float, boolean, char, void, String, @Override, @Test},
  keywordstyle=[3]\color{cadet}\bfseries,
  morekeywords=[4]{class, interface, extends, implements, new, super, throw, throws, try, catch, finally},
  keywordstyle=[4]\color{methodcolor},
  morecomment=[l]{//},
  commentstyle=\color{cadet},
  stringstyle=\color{brown(web)},
}

\usepackage{tikz}
\usetikzlibrary{calc}
\usetikzlibrary{shapes.geometric}
\usetikzlibrary{decorations.pathreplacing}
\usetikzlibrary{positioning}

\usepackage{datetime} %

\definecolor{darkgreen}{rgb}{0.0, 0.5, 0.0}

\newcommand{\XSpace}[1]{}
\newcommand{\XComment}[1]{}
\newcommand{\Fix}[1]{\textcolor{red}{#1}}
\newcommand{\MR}[1]{\textcolor{black}{#1}}

\newcommand{\DefMacro}[2]{\expandafter\newcommand\csname rmk-#1\endcsname{#2}}
\newcommand{\UseMacro}[1]{\csname rmk-#1\endcsname}

\newcommand{\MyPara}[1]{\noindent\textbf{#1}.}
\newcommand{\MyParaOnly}[1]{\noindent\textbf{#1}}
\newcommand{\reducedstrut}{\vrule width 0pt height .9\ht\strutbox depth .9\dp\strutbox\relax}
\newcommand{\InputWithSpace}[1]{\bgroup\def\arraystretch{1.1}\input{#1}\egroup}
\newcommand{\Code}[1]{{\ifmmode{\mathtt{#1}}\else$\mathtt{#1}$\fi}}
\newcommand{\CodeIn}[1]{{\ifmmode{\mathtt{#1}}\else$\mathtt{#1}$\fi}}
\newcommand{\ColorBack}[1]{%
  \begingroup
  \setlength{\fboxsep}{0pt}%
  \colorbox{purple!20}{\reducedstrut#1\/}%
  \endgroup
}
\newcommand{\ColorBackX}[2]{%
  \begingroup
  \setlength{\fboxsep}{0pt}%
  \colorbox{#1}{\reducedstrut#2\/}%
  \endgroup
}

\newcolumntype{R}[1]{>{\RaggedLeft\arraybackslash}p{#1}}
\newcolumntype{L}[1]{>{\RaggedRight\arraybackslash}p{#1}}

\newcommand{\ltrue}{\top} %
\newcommand{\lfalse}{\bot} %

\newcommand{\lAnd}[1]{\bigwedge_{\substack{#1}}} %

\newcommand{\Tool}{\textsc{exLong}\xspace}

\newcommand{\CAT}{CAT-LM\xspace}
\newcommand{\GPTThree}{GPT3.5\xspace}
\newcommand{\GPTFour}{GPT-4o\xspace}
\newcommand{\exlongGPTFour}{\Tool--GPT-4o\xspace}

\newcommand{\Java}{Java\xspace}
\newcommand{\Python}{Python\xspace}
\newcommand{\TeCo}{TeCo\xspace}
\newcommand{\Title}{exLong: Generating Exceptional Behavior Tests with Large Language Models}

\newcommand{\ExampleException}{\CodeIn{IllegalStateException}\xspace}

\newcommand{\eg}{e.g.\xspace}
\newcommand{\ie}{i.e.\xspace}
\newcommand{\userView}{developer-oriented use case\xspace}
\newcommand{\UserView}{Developer-oriented use case\xspace}
\newcommand{\XUserView}{Developer-Oriented Use Case\xspace}
\newcommand{\machineView}{machine-oriented use case\xspace}
\newcommand{\umViews}{developer- and machine-oriented use cases\xspace}
\newcommand{\MachineView}{Machine-oriented use case\xspace}
\newcommand{\XMachineView}{Machine-Oriented Use Case\xspace}
\newcommand{\EBTs}{EBTs\xspace}
\newcommand{\EBT}{EBT\xspace}
\newcommand{\etest}{exceptional-behavior test\xspace}

\newcommand{\nEBTs}{non-EBTs\xspace}
\newcommand{\nEBT}{non-EBT\xspace}

\newcommand{\MUT}{MUT\xspace}
\newcommand{\methodundertest}{method under test\xspace}
\newcommand{\LLM}{LLM\xspace}
\newcommand{\LLMs}{LLMs\xspace}
\newcommand{\stacktrace}{stack trace\xspace}
\newcommand{\stacktraces}{stack traces\xspace}
\newcommand{\Stacktrace}{Stack trace\xspace}

\newcommand{\guardexp}{guard expression\xspace}
\newcommand{\guardexps}{guard expressions\xspace}

\newcommand{\testfile}{destination test file\xspace}
\newcommand{\Testfile}{Destination test file\xspace}
\newcommand{\ts}{target throw statement\xspace}

\newcommand{\tss}{target throw statements\xspace}

\newcommand{\pretrained}{pretrained\xspace}
\newcommand{\gt}{ground-truth\xspace}
\newcommand{\CodeLlama}{CodeLlama\xspace}
\newcommand{\finetune}{fine-tune\xspace}
\newcommand{\finetuned}{fine-tuned\xspace}
\newcommand{\finetuning}{fine-tuning\xspace}
\newcommand{\train}{training\xspace}
\newcommand{\val}{validation\xspace}
\newcommand{\eval}{evaluation\xspace}
\newcommand{\developerwritten}{developer-written\xspace}
\newcommand{\repo}{repository\xspace}
\newcommand{\repos}{repositories\xspace}
\newcommand{\instdata}{SFT data\xspace}

\newcommand{\Randoop}{Randoop\xspace}
\newcommand{\EvoSuite}{EvoSuite\xspace}

\newcommand{\etype}{etype\xspace}
\newcommand{\aTest}{\ensuremath{\mathtt{t}}\xspace}
\newcommand{\aETest}{\ensuremath{\aTest_{\mathtt{eb}}}\xspace}
\newcommand{\aNETest}{\ensuremath{\aTest_{\mathtt{neb}}}\xspace}

\newcommand{\aTestSet}{\ensuremath{\mathtt{T}}\xspace}
\newcommand{\aETestSet}{\ensuremath{\aTestSet_{\mathtt{eb}}}\xspace}
\newcommand{\aNETestSet}{\ensuremath{\aTestSet_{\mathtt{neb}}}\xspace}
\newcommand{\aNETestSetRel}{\ensuremath{\hat{\aTestSet}_{\mathtt{neb}}}\xspace}

\newcommand{\aThrowStmt}{\ensuremath{\mathtt{s}}\xspace}
\newcommand{\aThrowStmtTarget}{\aThrowStmt}
\newcommand{\aMethod}{\ensuremath{\mathtt{m}}\xspace}
\newcommand{\aMUT}{\ensuremath{\mathtt{m_{ut}}}\xspace}

\newcommand{\aStackTrace}{\ensuremath{\mathtt{r}}\xspace}

\newcommand{\aGuardExp}{\ensuremath{\mathtt{g}}\xspace}
\newcommand{\aTestFile}{\ensuremath{\mathtt{d}}\xspace}
\newcommand{\aPrompt}{\ensuremath{\mathtt{p}}\xspace}
\newcommand{\aSUT}{\ensuremath{\mathtt{SUT}}\xspace}

\newcommand{\aCorpus}{\ensuremath{\mathcal{C}}\xspace}
\newcommand{\aCorpusItem}{\ensuremath{\mathtt{c}}\xspace}
\newcommand{\aNode}{\ensuremath{\mathtt{n}}\xspace}
\newcommand{\aNodes}{\ensuremath{\mathtt{N}}\xspace}
\newcommand{\aCond}{\ensuremath{\mathtt{e}}\xspace}
\newcommand{\aCondSet}{\ensuremath{\mathcal{E}}\xspace}
\newcommand{\aArgMap}{\ensuremath{\mathtt{argmap}}\xspace}
\newcommand{\aCurrent}{\ensuremath{\mathtt{current}}\xspace}
\newcommand{\aParent}{\ensuremath{\mathtt{parent}}\xspace}

\newcommand{\aStackTraceTuple}{\ensuremath{\mathtt{q}}\xspace}
\newcommand{\aStackTraceTupleSet}{\ensuremath{\mathcal{Q}}\xspace}
\newcommand{\aThrows}{\ensuremath{\mathtt{E}}\xspace}
\newcommand{\sTrace}{\ensuremath{STrace}\xspace}

\newcommand{\aThrowsCovered}{\ensuremath{\aThrows^{\ltrue}}\xspace}
\newcommand{\aThrowsUncovered}{\ensuremath{\aThrows^{\lfalse}}\xspace}
\newcommand*\circled[1]{\tikz[baseline=(char.base)]{
    \node[shape=circle, draw=black, minimum size=0.1, inner sep=0.8pt, fill=white, thick, text=black] (char) {#1};}}

\DefMacro{TCap-rq1-eval-dataset-stats}{Statistics of the evaluation dataset for \userView.\xspace\#\MUT is the number of unique method under test; \#Exception Types is the number of unique exception types tested by the \EBTs; \#Throw Statements is the number of unique throw statements covered by the \EBTs.\vspace{-5pt}}
\DefMacro{TCap-analysis-based-tools}{Randoop and EvoSuite generated tests statistics for the
  30 projects we studied. We report number of tests generated by each tool and the time taken for generation per project. 0 means the tool fails to execute on that project. We report the number of throw statements that are within the public method and number of throw statements we used for evaluating tools for each project.\vspace{-5pt}}
\DefMacro{TCap-dataset}{Statistics of the collected dataset. \#MUTs is the number of unique method under test; \#Exception Types is the number of unique exception types tested by the \EBTs.\vspace{-5pt}}
\DefMacro{TCap-eval-dataset-stats}{Statistics of the evaluation dataset for \machineView. \#Throw Statements is the number of \tss we extracted from the repository according to Section~\ref{sec:tech:eval:usecases}.\xspace\#Exception Types is the number of unique exception types thrown by the \tss.
  \vspace{-5pt}}
\DefMacro{TCap-exlong-runtime}{Runtime cost of \Tool for each project.}
\DefMacro{TCap-Tool-cmp-table}{Throw statements coverage rate for \Tool, \Randoop and \EvoSuite.\vspace{-5pt}}%
\DefMacro{TCap-models-model-size}{Results for \Tool-base 13B and \Tool 13B models; \Tool-base 13B does not have access to \nEBTs, \stacktrace and \guardexp.\vspace{-5pt}}
\DefMacro{TCap-models-user-view-with-name}{Results on \userView with \gt \EBT's name in the prompt.\vspace{-5pt}}
\DefMacro{TCap-models-user-view-no-name}{Results on \userView without \gt \EBT's name in the prompt.\vspace{-5pt}}
\DefMacro{TCap-models-exlong-ablation}{Ablations on different context of \Tool.\vspace{-5pt}}
\DefMacro{TCap-models-netest-diversity}{Comparison between using different \nEBTs to sample and using the same \nEBT but sampling multiple times. Note that we add the target \EBT's name to the prompt and we only report results on examples that have more than one candidate \nEBT.\vspace{-5pt}}
\DefMacro{TCap-mut2e-pattern}{Number of instances for each \EBT pattern~\cite{MarcilioFuria21HowJavaProgrammersTestExceptionalBehavior} in the collected corpus.\vspace{-5pt}}
\DefMacro{TCap-models-13b-and-7b}{Comparison between 7B and 13B results.\vspace{-5pt}}
\DefMacro{TCap-models-gpt4o-with-name}{\MR{Comparison between \GPTFour-few-shot and \exlongGPTFour.}}
\DefMacro{TH-num-project}{\#project\xspace}
\DefMacro{TH-num-module}{\#module\xspace}
\DefMacro{TH-num-test}{\#\aTest\xspace}
\DefMacro{TH-num-etest-cover-tc}{\# ET-tc \xspace}
\DefMacro{TH-num-etest-undeclared-exception}{\# ET udex\xspace}
\DefMacro{TH-num-etest-unk-coverage}{\# tests nomut\xspace}
\DefMacro{TH-num-etest}{\#\aETest\xspace}
\DefMacro{TH-num-throws}{\#\aThrows\xspace}
\DefMacro{TH-num-covered}{\#\aThrowsCovered\xspace}
\DefMacro{TH-num-self-thrown}{\#SelfThrown\xspace}
\DefMacro{TH-coverage}{coverage[\%]\xspace}
\DefMacro{TH-num-ne-e-pair}{\#$(\aNETest,\aETest)$\xspace}
\DefMacro{TH-num-uncovered-w-ne}{\#\aThrowsUncovered w/ \aNETest\xspace}
\DefMacro{TH-source-m}{Manual\xspace}
\DefMacro{TH-source-r}{\Randoop\xspace}
\DefMacro{TH-source-u}{Union\xspace}

\DefMacro{TH-num-etest-not-runnable}{\#no-run\xspace}
\DefMacro{TH-num-etest-covered-throw-clause}{\# check\xspace}
\DefMacro{TH-num-etest-unchecked-exception}{\# uncheck \xspace}
\DefMacro{TH-num-etest-unmatched-exception}{\# unma \xspace}
\DefMacro{TH-num-etest-no-mut}{\#no mut\xspace}
\DefMacro{TH-num-etest-no-throw-stmt}{\# no thr \xspace}
\DefMacro{TH-num-w-condition}{\# with conditions\xspace}

\DefMacro{THead-subset-projects}{Subset Projects}
\DefMacro{THead-all-projects}{All Projects}

\DefMacro{THead-ebts}{\#\EBTs\xspace}
\DefMacro{THead-count-mut}{\#\MUT\xspace}
\DefMacro{THead-rq1-eval}{Developer-Oriented}
\DefMacro{THead-models}{Models\xspace}
\DefMacro{THead-coverage-max}{ThrowCov\%}
\DefMacro{THead-bleu-avg}{BLEU\xspace}
\DefMacro{THead-bleu-max}{BLEU\xspace}
\DefMacro{THead-code-bleu-avg}{CodeBLEU\xspace}
\DefMacro{THead-code-bleu-max}{CodeBLEU\xspace}
\DefMacro{THead-edit-sim-avg}{EditSim\xspace}
\DefMacro{THead-edit-sim-max}{EditSim\xspace}
\DefMacro{THead-xmatch-top1}{xMatch\xspace}
\DefMacro{THead-compilable-avg}{Compilable\%\xspace}
\DefMacro{THead-compilable-max}{Compilable\%\xspace}
\DefMacro{THead-match-avg}{Matched-E\%\xspace}
\DefMacro{THead-match-max}{Matched-E\%\xspace}
\DefMacro{THead-runnable-avg}{Runnable\%\xspace}
\DefMacro{THead-runnable-max}{Runnable\%\xspace}
\DefMacro{THead-runnable-overall-max}{Runnable\%\xspace}
\DefMacro{THead-train}{Train\xspace}
\DefMacro{THead-test}{Eval\xspace}
\DefMacro{THead-valid}{Valid\xspace}
\DefMacro{THead-project-count}{\#Projects\xspace}
\DefMacro{THead-count-project}{\#Projects\xspace}
\DefMacro{THead-with-stack-trace-count}{With \sTrace\xspace}
\DefMacro{THead-max-stack-trace-len}{Max. len \sTrace\xspace}
\DefMacro{THead-median-stack-trace-len}{Med. len \sTrace\xspace}
\DefMacro{THead-min-stack-trace-len}{Min. len \sTrace\xspace}
\DefMacro{THead-median-stack-trace-count}{Med. \# \sTrace\xspace}
\DefMacro{THead-max-stack-trace-count}{Max. \# \sTrace\xspace}
\DefMacro{THead-with-netest-count}{With \aNETest\xspace}
\DefMacro{THead-max-netest-count}{Max. \# \aNETest\xspace}
\DefMacro{THead-median-netest-count}{Med. \# \aNETest\xspace}
\DefMacro{THead-exception-count}{\# \etype\xspace}
\DefMacro{THead-count-eval-throw-stmt-data}{\# \aETest \xspace}
\DefMacro{THead-count-projects}{\#Projects}
\DefMacro{THead-count-module}{\#Modules}
\DefMacro{THead-count-etype}{\#Ex. Type\xspace}
\DefMacro{THead-count-ts}{\#ThrowSt}
\DefMacro{THead-rq1}{User View}
\DefMacro{THead-rq2}{Machine View}

\DefMacro{THead-count-static-analysis-helps}{\# Analysis find\xspace}
\DefMacro{THead-count-methods-with-ts}{\# M. w. ts}
\DefMacro{THead-count-methods-with-ts-public}{\# Public M}

\DefMacro{THead-count-ts-public}{\# Pub. Ts.}
\DefMacro{THead-mean-ts-per-method}{Avg. \# TS}
\DefMacro{THead-coverage}{Exception coverage}
\DefMacro{THead-exception-coverage-evosuite}{\EvoSuite Coverage}
\DefMacro{THead-exception-coverage-randoop}{\Randoop Coverage}

\DefMacro{TCap-eval-dataset-ts-stats}{Stats on Methods from testset projects with Throw Statements. \# methods with throw statements, \# non-private and non-protected methods among them, \# total throw statements, Avg. \# throw statements per method.}
\DefMacro{THead-selected-253-conditionnestack2e-with-name-ft-lora-codellama-7b}{\Tool}
\DefMacro{THead-selected-253-conditionnestack2e-all-with-name-ft-lora-codellama-7b}{\Fix{\ToolMoreNebt w/ different \nEBTs}} %
\DefMacro{THead-conditionnestack2e-sample-with-name-ft-lora-codellama-7b}{\ToolMoreNebt w/ same \nEBT} %
\DefMacro{THead-diversity-conditionnestack2e-all-with-name-ft-lora-codellama-7b}{\ToolMoreNebt w/ different \nEBT}
\DefMacro{THead-selected-434-mut2e-no-name-ft-lora-codellama-7b}{MUT+FILE2E\xspace}
\DefMacro{THead-selected-434-mut2e-with-name-ft-lora-codellama-7b}{No \stacktrace, \guardexp, \nEBT}
\DefMacro{THead-mut2e-with-name-ft-lora-codellama-7b}{No \stacktrace, \guardexp, \nEBT}
\DefMacro{THead-selected-434-mut2e-with-name-ft-lora-codellama-13b}{\Tool-base 13B\xspace}
\DefMacro{THead-selected-434-mut2e-no-name-ft-lora-codellama-13b}{MUT+FILE2E(13B)\xspace}
\DefMacro{THead-selected-434-catlm-mut2e-with-name-catlm}{CAT-LM\xspace}
\DefMacro{THead-selected-434-catlm-mut2e-no-name-catlm}{CAT-LM\xspace}
\DefMacro{THead-catlm-ne2e-with-name-catlm}{CAT-LM\xspace}
\DefMacro{THead-catlm-ne2e-no-name-catlm}{CAT-LM\xspace}
\DefMacro{THead-selected-434-ne2e-with-name-ft-lora-codellama-7b}{No \stacktrace, \guardexp}
\DefMacro{THead-ne2e-with-name-ft-lora-codellama-7b}{No \stacktrace, \guardexp}
\DefMacro{THead-selected-434-ne2e-no-name-ft-lora-codellama-7b}{MUT+FILE+NE2E\xspace}
\DefMacro{THead-selected-434-ne2e-all-no-name-ft-lora-codellama-7b}{MUT+FILE+NE2E-sample5\xspace}
\DefMacro{THead-selected-434-ne2e-all-with-name-ft-lora-codellama-7b}{MUT+FILE+NE2E-sample5\xspace}
\DefMacro{THead-selected-434-nestack2e-noexlong-name-ft-lora-codellama-7b}{MUT+FILE+NE+STACK2E\xspace}
\DefMacro{THead-selected-434-nestack2e-with-name-ft-lora-codellama-7b}{MUT+FILE+NE+STACK2E\xspace}
\DefMacro{THead-selected-434-nestack2e-with-name-ft-lora-codellama-13b}{\Tool 13B\xspace}
\DefMacro{THead-conditionnestack2e-with-name-zero-shot-lora-codellama-7b}{\CodeLlama zero-shot}
\DefMacro{THead-conditionnestack2e-no-name-zero-shot-lora-codellama-7b}{\CodeLlama zero-shot}
\DefMacro{THead-selected-434-conditionnestack2e-no-name-zero-shot-lora-codellama-7b}{\CodeLlama zero-shot}
\DefMacro{THead-selected-434-conditionnestack2e-with-name-zero-shot-lora-codellama-7b}{\CodeLlama zero-shot}
\DefMacro{THead-mut2e-with-name-ft-lora-codellama-13b}{\Tool-base 13B\xspace}
\DefMacro{THead-conditionnestack2e-with-name-ft-lora-codellama-13b}{\Tool 13B\xspace}
\DefMacro{THead-conditionnestack2e-few-shot-no-name-gpt-4o}{\Tool(\GPTFour)\xspace}
\DefMacro{THead-conditionnestack2e-few-shot-with-name-gpt-4o}{\Tool--\GPTFour\xspace}
\DefMacro{THead-selected-434-nestack2e-no-name-ft-lora-codellama-13b}{MUT+FILE+NE+STACK2E(13B)\xspace}
\DefMacro{THead-selected-434-ne2e-few-shot-with-name-gpt-3.5-turbo-16k}{GPT3.5-few-shot\xspace}
\DefMacro{THead-ne2e-few-shot-with-name-gpt-4o}{\GPTFour-few-shot\xspace}
\DefMacro{THead-ne2e-few-shot-no-name-gpt-4o}{GPT4o-few-shot\xspace}
\DefMacro{THead-selected-434-ne2e-few-shot-no-name-gpt-3.5-turbo-16k}{GPT3.5-few-shot\xspace}
\DefMacro{THead-ne2e-few-shot-with-name-gpt-3.5-turbo-16k}{GPT3.5-few-shot\xspace}
\DefMacro{THead-ne2e-few-shot-no-name-gpt-3.5-turbo-16k}{GPT3.5-few-shot\xspace}
\DefMacro{THead-selected-434-conditionne2e-no-name-ft-lora-codellama-7b}{MUT+FILE+NE+CONDITION2E\xspace}
\DefMacro{THead-conditionne2e-with-name-ft-lora-codellama-7b}{No \stacktrace}
\DefMacro{THead-selected-434-conditionne2e-with-name-ft-lora-codellama-7b}{No \stacktrace}
\DefMacro{THead-selected-434-conditionnestack2e-no-name-ft-lora-codellama-7b}{\Tool\xspace}
\DefMacro{THead-selected-434-conditionnestack2e-with-name-ft-lora-codellama-7b}{\Tool\xspace}
\DefMacro{THead-selected-434-nestack2e-all-with-name-ft-lora-codellama-7b}{MUT+FILE+NE+STACK2E-sample5\xspace}
\DefMacro{THead-selected-434-nestack2e-all-no-name-ft-lora-codellama-7b}{MUT+FILE+NE+STACK2E-sample5\xspace}
\DefMacro{THead-selected-434-conditionnestack2e-all-with-name-ft-lora-codellama-7b}{\Tool-sample\xspace}
\DefMacro{THead-conditionnestack2e-all-with-name-ft-lora-codellama-7b}{\Tool-sample\xspace}
\DefMacro{THead-selected-434-conditionnestack2e-all-no-name-ft-lora-codellama-7b}{\Tool-sample\xspace}
\DefMacro{THead-conditionnestack2e-all-no-name-ft-lora-codellama-7b}{\Tool-sample\xspace}

\DefMacro{THead-mut2e-no-name-ft-lora-codellama-7b}{MUT+FILE2E\xspace}
\DefMacro{THead-ne2e-no-name-ft-lora-codellama-7b}{MUT+FILE+NE2E\xspace}
\DefMacro{THead-ne2e-all-with-name-ft-lora-codellama-7b}{MUT+FILE+NE2E-sample5\xspace}
\DefMacro{THead-ne2e-all-no-name-ft-lora-codellama-7b}{MUT+FILE+NE2E-sample5\xspace}
\DefMacro{THead-nestack2e-no-name-ft-lora-codellama-7b}{MUT+FILE+NE+STACK2E\xspace}
\DefMacro{THead-nestack2e-with-name-ft-lora-codellama-7b}{MUT+FILE+NE+STACK2E\xspace}
\DefMacro{THead-nestack2e-with-name-ft-lora-codellama-7b-real}{MUT+FILE+NE+STACK2E\xspace}
\DefMacro{THead-nestack2e-no-name-ft-lora-codellama-7b-real}{MUT+FILE+NE+STACK2E\xspace}
\DefMacro{THead-nestack2e-with-name-ft-lora-codellama-13b-real}{MUT+FILE+NE+STACK2E(13B)\xspace}
\DefMacro{THead-nestack2e-no-name-ft-lora-codellama-13b-real}{MUT+FILE+NE+STACK2E(13B)\xspace}
\DefMacro{THead-nestack2e-no-name-ft-lora-codellama-13b}{MUT+FILE+NE+STACKTRACE(13B)\xspace}
\DefMacro{THead-conditionne2e-no-name-ft-lora-codellama-7b}{MUT+FILE+NE+CONDITION2E\xspace}
\DefMacro{THead-conditionne2e-with-name-ft-lora-codellama-7b-real}{MUT+FILE+NE+CONDITION2E\xspace}
\DefMacro{THead-conditionne2e-no-name-ft-lora-codellama-7b-real}{MUT+FILE+NE+CONDITION2E\xspace}
\DefMacro{THead-conditionnestack2e-no-name-ft-lora-codellama-7b}{\Tool\xspace}
\DefMacro{THead-conditionnestack2e-with-name-ft-lora-codellama-7b}{\Tool\xspace}
\DefMacro{THead-conditionnestack2e-no-name-ft-lora-codellama-7b-real}{MUT+FILE+NE+CONDITION+STACK2E\xspace}
\DefMacro{THead-conditionnestack2e-with-name-ft-lora-codellama-7b-real}{MUT+FILE+NE+CONDITION+STACK2E\xspace}

\DefMacro{THead-mut2e-zero-shot-gpt-4-0613}{GPT4 zero-shot MUT+FILE2E\xspace}
\DefMacro{THead-mut2e-zero-shot-gpt3.5}{GPT3.5 zero-shot MUT+FILE2E\xspace}
\DefMacro{THead-mut2e-ft-axolotl-llama}{CodeLlama-7b MUT+FILE2E}
\DefMacro{THead-nestack2e-axolotl-codellama-7b}{CodeLlama-7b NESTACK2E}
\DefMacro{THead-cov2e-1-ft-axolotl-codellama-7b}{CodeLlama-7b NE2E}
\DefMacro{THead-nestack2e-ft-axolotl-llama}{CodeLlama-7b NESTACK2E\xspace}
\DefMacro{THead-nestack2e-no-stack-trace-axolotl-codellama-7b}{CodeLlama-7b (w.o. s.t.)}
\DefMacro{THead-nestack2e-with-stack-trace-axolotl-codellama-7b}{CodeLlama-7b (w. s.t.)}
\DefMacro{THead-stack2e-ft-axolotl-llama}{CodeLlama-7b STACK2E}
\DefMacro{THead-ne2e-ft-axolotl-llama}{CodeLlama-7b NE2E}
\DefMacro{THead-gpt4-nestack2e-cot-gpt-4-0613}{GPT4 COT (one-shot)\xspace}
\DefMacro{THead-gpt4-nestack2e-few-shot-gpt-4-0613}{GPT4 one-shot\xspace}

\DefMacro{THead-gt-nestack2e-few-shot-no-name-gpt-3.5-turbo-16k}{GPT3.5-gt-st-no-name\xspace}
\DefMacro{THead-gt-conditionne2e-few-shot-no-name-gpt-3.5-turbo-16k}{GPT3.5-gt-condition-no-name\xspace}
\DefMacro{THead-nestack2e-few-shot-no-name-gpt-3.5-turbo-16k}{GPT3.5-st-no-name\xspace}

\DefMacro{THead-analysis-time}{Analysis Time\xspace}
\DefMacro{THead-inference-time}{Inference Time\xspace}
\DefMacro{THead-total-time}{Total\xspace}

\DefMacro{THead-expect}{Test(expected)\xspace}
\DefMacro{THead-assertThrows}{assertThrow\xspace}
\DefMacro{THead-expectedRule}{expectedRule\xspace}
\DefMacro{THead-try{fail}catch}{tryFailCatch\xspace}

\DefMacro{THead-mut2e-stmt-otest-chatgpt-basic-top1}{ChatGPT-zero-shot\xspace}
\DefMacro{THead-mut2e-stmt-otest-SingleEvaluator-base-4197-bs10-last}{TeCo-basic\xspace}
\DefMacro{THead-mut2e-stmt-otest-chatgpt-one-shot-top1}{ChatGPT-one-shot\xspace}
\DefMacro{THead-mut2e-stmt-test-chatgpt-zero-shot-full-context-top1}{ChatGPT-zero-shot-full-context\xspace}
\DefMacro{THead-mean-stmts-toks}{len(stmts)\xspace}
\DefMacro{THead-num-estests}{\# \aETest}
\DefMacro{THead-num-projects}{\#project\xspace}
\DefMacro{THead-num-stmts}{\#stmts\xspace}

\DefMacro{THead-etest-count}{\#\EBTs\xspace}
\DefMacro{THead-test-count}{\#Tests\xspace}
\DefMacro{THead-mut-count}{\#MUTs\xspace}
\DefMacro{THead-total-test-count}{\# Tests\xspace}
\DefMacro{THead-module-count}{\#Modules\xspace}
\DefMacro{THead-etype-count}{\# Ex. Type\xspace}

\DefMacro{THead-num-tests}{\#Tests\xspace}
\DefMacro{THead-evosuite-num-tests}{\EvoSuite\xspace}
\DefMacro{THead-randoop-num-tests}{\Randoop\xspace}
\DefMacro{THead-time}{Time (s)\xspace}
\DefMacro{THead-evosuite-time}{\EvoSuite\xspace}
\DefMacro{THead-randoop-time}{\Randoop\xspace}
\DefMacro{THead-num-public-throw-stmts}{\makecell{\#Throw Statements\\in Public Methods}}
\DefMacro{THead-num-eval-throw-stmts}{\makecell{\#Throw Statements\\for Evaluation}}

\DefMacro{NumDiversitySubset}{253}
\DefMacro{TotalJavaProj}{4,767}
\DefMacro{UsedJavaProj}{}
\DefMacro{expectedPct}{65\%}
\DefMacro{tryfailPct}{24\%}
\DefMacro{realStracePct}{61\%}
\DefMacro{testSize}{842}
\DefMacro{tsEtests}{434}
\DefMacro{machineViewSize}{649}
\DefMacro{RandoopSuccProjs}{27}
\DefMacro{RandoopFailProjs}{three}
\DefMacro{EvosuiteAverageTime}{1752.6s}
\DefMacro{RandoopAverageTime}{6621.4s}
\DefMacro{exLongAverageTime}{501.3s}
\DefMacro{exLongAnalysisTime}{276.9s}
\DefMacro{exLongGenTime}{224.4s}
\DefMacro{Evosuite-all-projects-coverage}{20.95}
\DefMacro{Evosuite-subset-projects-coverage}{20.37}
\DefMacro{Randoop-all-projects-coverage}{18.95}
\DefMacro{Randoop-subset-projects-coverage}{21.87}

\DefMacro{res-tools-cmp-conditionnestack2e-all-no-name-ft-lora-codellama-7b-eval-rq2-compilable-avg}{66.22}
\DefMacro{res-tools-cmp-conditionnestack2e-all-no-name-ft-lora-codellama-7b-eval-rq2-compilable-max}{80.43}
\DefMacro{res-tools-cmp-conditionnestack2e-all-no-name-ft-lora-codellama-7b-eval-rq2-compilable-min}{51.62}
\DefMacro{res-tools-cmp-conditionnestack2e-all-no-name-ft-lora-codellama-7b-eval-rq2-coverage-avg}{25.28}
\DefMacro{res-tools-cmp-conditionnestack2e-all-no-name-ft-lora-codellama-7b-eval-rq2-coverage-max}{28.81}
\DefMacro{res-tools-cmp-conditionnestack2e-all-no-name-ft-lora-codellama-7b-eval-rq2-coverage-min}{19.88}
\DefMacro{res-tools-cmp-conditionnestack2e-all-no-name-ft-lora-codellama-7b-eval-rq2-match-avg}{99.43}
\DefMacro{res-tools-cmp-conditionnestack2e-all-no-name-ft-lora-codellama-7b-eval-rq2-match-max}{99.43}
\DefMacro{res-tools-cmp-conditionnestack2e-all-no-name-ft-lora-codellama-7b-eval-rq2-match-min}{99.43}
\DefMacro{res-tools-cmp-conditionnestack2e-all-no-name-ft-lora-codellama-7b-eval-rq2-runnable-avg}{56.50}
\DefMacro{res-tools-cmp-conditionnestack2e-all-no-name-ft-lora-codellama-7b-eval-rq2-runnable-max}{64.16}
\DefMacro{res-tools-cmp-conditionnestack2e-all-no-name-ft-lora-codellama-7b-eval-rq2-runnable-min}{47.78}
\DefMacro{res-tools-cmp-conditionnestack2e-all-no-name-ft-lora-codellama-7b-eval-rq2-runnable-overall}{51.31}
\DefMacro{res-tools-cmp-conditionnestack2e-all-no-name-ft-lora-codellama-7b-eval-rq2-runnable-overall-avg}{40.20}
\DefMacro{res-tools-cmp-conditionnestack2e-all-no-name-ft-lora-codellama-7b-eval-rq2-runnable-overall-max}{51.31}
\DefMacro{res-tools-cmp-conditionnestack2e-all-no-name-ft-lora-codellama-7b-eval-rq2-runnable-overall-min}{29.12}
\DefMacro{res-tools-cmp-conditionnestack2e-all-no-name-ft-lora-codellama-7b-eval-rq2-timeout-avg}{2.75}
\DefMacro{res-tools-cmp-conditionnestack2e-all-no-name-ft-lora-codellama-7b-eval-rq2-timeout-max}{3.28}
\DefMacro{res-tools-cmp-conditionnestack2e-all-no-name-ft-lora-codellama-7b-eval-rq2-timeout-min}{2.50}
\DefMacro{res-tools-cmp-conditionnestack2e-all-no-name-ft-lora-codellama-7b-eval-rq2-intersect-compilable-avg}{64.97}
\DefMacro{res-tools-cmp-conditionnestack2e-all-no-name-ft-lora-codellama-7b-eval-rq2-intersect-compilable-max}{80.37}
\DefMacro{res-tools-cmp-conditionnestack2e-all-no-name-ft-lora-codellama-7b-eval-rq2-intersect-compilable-min}{50.28}
\DefMacro{res-tools-cmp-conditionnestack2e-all-no-name-ft-lora-codellama-7b-eval-rq2-intersect-coverage-avg}{25.87}
\DefMacro{res-tools-cmp-conditionnestack2e-all-no-name-ft-lora-codellama-7b-eval-rq2-intersect-coverage-max}{29.72}
\DefMacro{res-tools-cmp-conditionnestack2e-all-no-name-ft-lora-codellama-7b-eval-rq2-intersect-coverage-min}{20.19}
\DefMacro{res-tools-cmp-conditionnestack2e-all-no-name-ft-lora-codellama-7b-eval-rq2-intersect-match-avg}{100.00}
\DefMacro{res-tools-cmp-conditionnestack2e-all-no-name-ft-lora-codellama-7b-eval-rq2-intersect-match-max}{100.00}
\DefMacro{res-tools-cmp-conditionnestack2e-all-no-name-ft-lora-codellama-7b-eval-rq2-intersect-match-min}{100.00}
\DefMacro{res-tools-cmp-conditionnestack2e-all-no-name-ft-lora-codellama-7b-eval-rq2-intersect-runnable-avg}{54.52}
\DefMacro{res-tools-cmp-conditionnestack2e-all-no-name-ft-lora-codellama-7b-eval-rq2-intersect-runnable-max}{61.86}
\DefMacro{res-tools-cmp-conditionnestack2e-all-no-name-ft-lora-codellama-7b-eval-rq2-intersect-runnable-min}{46.05}
\DefMacro{res-tools-cmp-conditionnestack2e-all-no-name-ft-lora-codellama-7b-eval-rq2-intersect-runnable-overall}{49.72}
\DefMacro{res-tools-cmp-conditionnestack2e-all-no-name-ft-lora-codellama-7b-eval-rq2-intersect-runnable-overall-avg}{38.42}
\DefMacro{res-tools-cmp-conditionnestack2e-all-no-name-ft-lora-codellama-7b-eval-rq2-intersect-runnable-overall-max}{49.72}
\DefMacro{res-tools-cmp-conditionnestack2e-all-no-name-ft-lora-codellama-7b-eval-rq2-intersect-runnable-overall-min}{27.85}
\DefMacro{res-tools-cmp-conditionnestack2e-all-no-name-ft-lora-codellama-7b-eval-rq2-intersect-timeout-avg}{3.32}
\DefMacro{res-tools-cmp-conditionnestack2e-all-no-name-ft-lora-codellama-7b-eval-rq2-intersect-timeout-max}{3.95}
\DefMacro{res-tools-cmp-conditionnestack2e-all-no-name-ft-lora-codellama-7b-eval-rq2-intersect-timeout-min}{3.02}

\DefMacro{Coreoz_Wisp-analysis-time}{113.57}
\DefMacro{Harium_keel-analysis-time}{92.09}
\DefMacro{JodaOrg_joda-beans-analysis-time}{265.02}
\DefMacro{OpenHFT_Chronicle-Map-analysis-time}{2,175.34}
\DefMacro{OpenHFT_Chronicle-Network-analysis-time}{188.57}
\DefMacro{OpenNMS_newts-analysis-time}{322.89}
\DefMacro{analogweb_core-analysis-time}{291.96}
\DefMacro{arquillian_arquillian-core-analysis-time}{533.09}
\DefMacro{bingoohuang_westcache-analysis-time}{241.82}
\DefMacro{craftercms_engine-analysis-time}{99.74}
\DefMacro{davidmoten_ppk-analysis-time}{41.26}
\DefMacro{entrusc_xdata-analysis-time}{33.06}
\DefMacro{globocom_GloboDNS-Client-analysis-time}{29.89}
\DefMacro{greenmail-mail-test_greenmail-analysis-time}{569.82}
\DefMacro{javadev_moneytostr-russian-analysis-time}{39.13}
\DefMacro{jpmml_jpmml-model-analysis-time}{315.25}
\DefMacro{kevinsawicki_http-request-analysis-time}{161.09}
\DefMacro{mguymon_model-citizen-analysis-time}{50.81}
\DefMacro{microfocus-idol_java-content-parameter-api-analysis-time}{197.76}
\DefMacro{mistraltechnologies_smog-analysis-time}{71.85}
\DefMacro{mp911de_logstash-gelf-analysis-time}{40.26}
\DefMacro{opencb_java-common-libs-analysis-time}{776.64}
\DefMacro{pinterest_secor-analysis-time}{180.08}
\DefMacro{ralscha_wampspring-analysis-time}{315.00}
\DefMacro{sbtourist_Journal.IO-analysis-time}{219.02}
\DefMacro{sharneng_gm4java-analysis-time}{59.76}
\DefMacro{spotify_apollo-analysis-time}{401.95}
\DefMacro{spotify_async-google-pubsub-client-analysis-time}{103.72}
\DefMacro{stackify_stackify-api-java-analysis-time}{159.50}
\DefMacro{statefulj_statefulj-analysis-time}{216.47}
\DefMacro{sum-analysis-time}{8,306.43}
\DefMacro{sum-inference-time}{6,732.38}
\DefMacro{sum-total-time}{15,038.81}
\DefMacro{mean-analysis-time}{276.88}
\DefMacro{mean-inference-time}{224.41}
\DefMacro{mean-total-time}{501.29}
\DefMacro{Coreoz_Wisp-inference-time}{141.53}
\DefMacro{Harium_keel-inference-time}{177.76}
\DefMacro{JodaOrg_joda-beans-inference-time}{543.73}
\DefMacro{OpenHFT_Chronicle-Map-inference-time}{182.58}
\DefMacro{OpenHFT_Chronicle-Network-inference-time}{189.28}
\DefMacro{OpenNMS_newts-inference-time}{144.79}
\DefMacro{analogweb_core-inference-time}{336.82}
\DefMacro{arquillian_arquillian-core-inference-time}{432.17}
\DefMacro{bingoohuang_westcache-inference-time}{140.65}
\DefMacro{craftercms_engine-inference-time}{137.45}
\DefMacro{davidmoten_ppk-inference-time}{213.74}
\DefMacro{entrusc_xdata-inference-time}{169.13}
\DefMacro{greenmail-mail-test_greenmail-inference-time}{147.92}
\DefMacro{globocom_GloboDNS-Client-inference-time}{789.90}
\DefMacro{javadev_moneytostr-russian-inference-time}{283.01}
\DefMacro{jpmml_jpmml-model-inference-time}{222.33}
\DefMacro{kevinsawicki_http-request-inference-time}{155.67}
\DefMacro{mguymon_model-citizen-inference-time}{137.72}
\DefMacro{microfocus-idol_java-content-parameter-api-inference-time}{235.88}
\DefMacro{mistraltechnologies_smog-inference-time}{163.10}
\DefMacro{mp911de_logstash-gelf-inference-time}{145.68}
\DefMacro{opencb_java-common-libs-inference-time}{140.25}
\DefMacro{pinterest_secor-inference-time}{157.49}
\DefMacro{ralscha_wampspring-inference-time}{184.25}
\DefMacro{sbtourist_Journal.IO-inference-time}{248.85}
\DefMacro{sharneng_gm4java-inference-time}{212.17}
\DefMacro{spotify_apollo-inference-time}{190.81}
\DefMacro{spotify_async-google-pubsub-client-inference-time}{141.57}
\DefMacro{stackify_stackify-api-java-inference-time}{172.89}
\DefMacro{statefulj_statefulj-inference-time}{193.28}
\DefMacro{Coreoz_Wisp-total-time}{255.10}
\DefMacro{Harium_keel-total-time}{269.85}
\DefMacro{JodaOrg_joda-beans-total-time}{808.74}
\DefMacro{OpenHFT_Chronicle-Map-total-time}{2,357.92}
\DefMacro{OpenHFT_Chronicle-Network-total-time}{377.85}
\DefMacro{OpenNMS_newts-total-time}{467.69}
\DefMacro{analogweb_core-total-time}{628.78}
\DefMacro{arquillian_arquillian-core-total-time}{965.26}
\DefMacro{bingoohuang_westcache-total-time}{382.48}
\DefMacro{craftercms_engine-total-time}{237.19}
\DefMacro{davidmoten_ppk-total-time}{255.00}
\DefMacro{entrusc_xdata-total-time}{202.19}
\DefMacro{globocom_GloboDNS-Client-total-time}{819.79}
\DefMacro{greenmail-mail-test_greenmail-total-time}{717.74}
\DefMacro{javadev_moneytostr-russian-total-time}{322.14}
\DefMacro{jpmml_jpmml-model-total-time}{537.58}
\DefMacro{kevinsawicki_http-request-total-time}{316.76}
\DefMacro{mguymon_model-citizen-total-time}{188.54}
\DefMacro{microfocus-idol_java-content-parameter-api-total-time}{433.64}
\DefMacro{mistraltechnologies_smog-total-time}{234.95}
\DefMacro{mp911de_logstash-gelf-total-time}{185.94}
\DefMacro{opencb_java-common-libs-total-time}{916.89}
\DefMacro{pinterest_secor-total-time}{337.57}
\DefMacro{ralscha_wampspring-total-time}{499.24}
\DefMacro{sbtourist_Journal.IO-total-time}{467.87}
\DefMacro{sharneng_gm4java-total-time}{271.93}
\DefMacro{spotify_apollo-total-time}{592.76}
\DefMacro{spotify_async-google-pubsub-client-total-time}{245.29}
\DefMacro{stackify_stackify-api-java-total-time}{332.39}
\DefMacro{statefulj_statefulj-total-time}{409.75}

\DefMacro{ds-rq1-eval-count-etype}{41}
\DefMacro{ds-rq1-eval-count-module}{42}
\DefMacro{ds-rq1-eval-count-mut}{267}
\DefMacro{ds-rq1-eval-count-project}{30}
\DefMacro{ds-rq1-eval-count-ts}{278}
\DefMacro{ds-rq1-eval-ebts}{434}

\DefMacro{res-CSN-stmt-mut2e-stmt-otest-chatgpt-basic-top1-bleu}{37.56}
\DefMacro{res-CSN-stmt-mut2e-stmt-otest-chatgpt-basic-top1-bleu-avg}{37.56}
\DefMacro{res-CSN-stmt-mut2e-stmt-otest-chatgpt-basic-top1-bleu-max}{37.56}
\DefMacro{res-CSN-stmt-mut2e-stmt-otest-chatgpt-basic-top1-bleu-min}{37.56}
\DefMacro{res-CSN-stmt-mut2e-stmt-otest-chatgpt-basic-top1-code-bleu}{33.86}
\DefMacro{res-CSN-stmt-mut2e-stmt-otest-chatgpt-basic-top1-code-bleu-avg}{33.86}
\DefMacro{res-CSN-stmt-mut2e-stmt-otest-chatgpt-basic-top1-code-bleu-max}{33.86}
\DefMacro{res-CSN-stmt-mut2e-stmt-otest-chatgpt-basic-top1-code-bleu-min}{33.86}
\DefMacro{res-CSN-stmt-mut2e-stmt-otest-chatgpt-basic-top1-edit-sim}{54.78}
\DefMacro{res-CSN-stmt-mut2e-stmt-otest-chatgpt-basic-top1-edit-sim-avg}{54.78}
\DefMacro{res-CSN-stmt-mut2e-stmt-otest-chatgpt-basic-top1-edit-sim-max}{54.78}
\DefMacro{res-CSN-stmt-mut2e-stmt-otest-chatgpt-basic-top1-edit-sim-min}{54.78}
\DefMacro{res-CSN-stmt-mut2e-stmt-otest-chatgpt-basic-top1-rouge-f}{57.43}
\DefMacro{res-CSN-stmt-mut2e-stmt-otest-chatgpt-basic-top1-rouge-f-avg}{57.43}
\DefMacro{res-CSN-stmt-mut2e-stmt-otest-chatgpt-basic-top1-rouge-f-max}{57.43}
\DefMacro{res-CSN-stmt-mut2e-stmt-otest-chatgpt-basic-top1-rouge-f-min}{57.43}
\DefMacro{res-CSN-stmt-mut2e-stmt-otest-chatgpt-basic-top1-rouge-p}{58.54}
\DefMacro{res-CSN-stmt-mut2e-stmt-otest-chatgpt-basic-top1-rouge-p-avg}{58.54}
\DefMacro{res-CSN-stmt-mut2e-stmt-otest-chatgpt-basic-top1-rouge-p-max}{58.54}
\DefMacro{res-CSN-stmt-mut2e-stmt-otest-chatgpt-basic-top1-rouge-p-min}{58.54}
\DefMacro{res-CSN-stmt-mut2e-stmt-otest-chatgpt-basic-top1-rouge-r}{58.37}
\DefMacro{res-CSN-stmt-mut2e-stmt-otest-chatgpt-basic-top1-rouge-r-avg}{58.37}
\DefMacro{res-CSN-stmt-mut2e-stmt-otest-chatgpt-basic-top1-rouge-r-max}{58.37}
\DefMacro{res-CSN-stmt-mut2e-stmt-otest-chatgpt-basic-top1-rouge-r-min}{58.37}
\DefMacro{res-CSN-stmt-mut2e-stmt-otest-chatgpt-basic-top1-xmatch}{17.28}
\DefMacro{res-CSN-stmt-mut2e-stmt-otest-chatgpt-basic-top1-xmatch-top1}{17.28}
\DefMacro{res-CSN-stmt-mut2e-stmt-otest-SingleEvaluator-base-4197-bs10-last-xmatch}{3.70}
\DefMacro{res-CSN-stmt-mut2e-stmt-otest-SingleEvaluator-base-4197-bs10-last-xmatch-top3}{3.70}
\DefMacro{res-CSN-stmt-mut2e-stmt-otest-SingleEvaluator-base-4197-bs10-last-xmatch-top5}{3.70}
\DefMacro{res-CSN-stmt-mut2e-stmt-otest-SingleEvaluator-base-4197-bs10-last-xmatch-top10}{4.94}
\DefMacro{res-CSN-stmt-mut2e-stmt-otest-SingleEvaluator-base-4197-bs10-last-xmatch-top100}{4.94}
\DefMacro{res-CSN-stmt-mut2e-stmt-otest-SingleEvaluator-base-4197-bs10-last-bleu}{26.20}
\DefMacro{res-CSN-stmt-mut2e-stmt-otest-SingleEvaluator-base-4197-bs10-last-bleu-max}{33.69}
\DefMacro{res-CSN-stmt-mut2e-stmt-otest-SingleEvaluator-base-4197-bs10-last-bleu-min}{14.59}
\DefMacro{res-CSN-stmt-mut2e-stmt-otest-SingleEvaluator-base-4197-bs10-last-bleu-avg}{20.55}
\DefMacro{res-CSN-stmt-mut2e-stmt-otest-SingleEvaluator-base-4197-bs10-last-code-bleu}{19.87}
\DefMacro{res-CSN-stmt-mut2e-stmt-otest-SingleEvaluator-base-4197-bs10-last-code-bleu-max}{28.95}
\DefMacro{res-CSN-stmt-mut2e-stmt-otest-SingleEvaluator-base-4197-bs10-last-code-bleu-min}{7.85}
\DefMacro{res-CSN-stmt-mut2e-stmt-otest-SingleEvaluator-base-4197-bs10-last-code-bleu-avg}{14.48}
\DefMacro{res-CSN-stmt-mut2e-stmt-otest-SingleEvaluator-base-4197-bs10-last-edit-sim}{48.96}
\DefMacro{res-CSN-stmt-mut2e-stmt-otest-SingleEvaluator-base-4197-bs10-last-edit-sim-max}{59.16}
\DefMacro{res-CSN-stmt-mut2e-stmt-otest-SingleEvaluator-base-4197-bs10-last-edit-sim-min}{36.38}
\DefMacro{res-CSN-stmt-mut2e-stmt-otest-SingleEvaluator-base-4197-bs10-last-edit-sim-avg}{45.17}
\DefMacro{res-CSN-stmt-mut2e-stmt-otest-SingleEvaluator-base-4197-bs10-last-rouge-p}{49.27}
\DefMacro{res-CSN-stmt-mut2e-stmt-otest-SingleEvaluator-base-4197-bs10-last-rouge-p-max}{60.41}
\DefMacro{res-CSN-stmt-mut2e-stmt-otest-SingleEvaluator-base-4197-bs10-last-rouge-p-min}{35.50}
\DefMacro{res-CSN-stmt-mut2e-stmt-otest-SingleEvaluator-base-4197-bs10-last-rouge-p-avg}{43.93}
\DefMacro{res-CSN-stmt-mut2e-stmt-otest-SingleEvaluator-base-4197-bs10-last-rouge-r}{48.10}
\DefMacro{res-CSN-stmt-mut2e-stmt-otest-SingleEvaluator-base-4197-bs10-last-rouge-r-max}{55.30}
\DefMacro{res-CSN-stmt-mut2e-stmt-otest-SingleEvaluator-base-4197-bs10-last-rouge-r-min}{39.40}
\DefMacro{res-CSN-stmt-mut2e-stmt-otest-SingleEvaluator-base-4197-bs10-last-rouge-r-avg}{45.58}
\DefMacro{res-CSN-stmt-mut2e-stmt-otest-SingleEvaluator-base-4197-bs10-last-rouge-f}{47.17}
\DefMacro{res-CSN-stmt-mut2e-stmt-otest-SingleEvaluator-base-4197-bs10-last-rouge-f-max}{55.61}
\DefMacro{res-CSN-stmt-mut2e-stmt-otest-SingleEvaluator-base-4197-bs10-last-rouge-f-min}{35.75}
\DefMacro{res-CSN-stmt-mut2e-stmt-otest-SingleEvaluator-base-4197-bs10-last-rouge-f-avg}{42.95}
\DefMacro{res-CSN-stmt-mut2e-stmt-otest-SingleEvaluator-base-4197-bs10-last-num-seq}{10.00}
\DefMacro{res-CSN-stmt-mut2e-stmt-test-chatgpt-zero-shot-full-context-top1-bleu}{41.05}
\DefMacro{res-CSN-stmt-mut2e-stmt-test-chatgpt-zero-shot-full-context-top1-bleu-avg}{41.05}
\DefMacro{res-CSN-stmt-mut2e-stmt-test-chatgpt-zero-shot-full-context-top1-bleu-max}{41.05}
\DefMacro{res-CSN-stmt-mut2e-stmt-test-chatgpt-zero-shot-full-context-top1-bleu-min}{41.05}
\DefMacro{res-CSN-stmt-mut2e-stmt-test-chatgpt-zero-shot-full-context-top1-code-bleu}{43.36}
\DefMacro{res-CSN-stmt-mut2e-stmt-test-chatgpt-zero-shot-full-context-top1-code-bleu-avg}{43.36}
\DefMacro{res-CSN-stmt-mut2e-stmt-test-chatgpt-zero-shot-full-context-top1-code-bleu-max}{43.36}
\DefMacro{res-CSN-stmt-mut2e-stmt-test-chatgpt-zero-shot-full-context-top1-code-bleu-min}{43.36}
\DefMacro{res-CSN-stmt-mut2e-stmt-test-chatgpt-zero-shot-full-context-top1-edit-sim}{58.14}
\DefMacro{res-CSN-stmt-mut2e-stmt-test-chatgpt-zero-shot-full-context-top1-edit-sim-avg}{58.14}
\DefMacro{res-CSN-stmt-mut2e-stmt-test-chatgpt-zero-shot-full-context-top1-edit-sim-max}{58.14}
\DefMacro{res-CSN-stmt-mut2e-stmt-test-chatgpt-zero-shot-full-context-top1-edit-sim-min}{58.14}
\DefMacro{res-CSN-stmt-mut2e-stmt-test-chatgpt-zero-shot-full-context-top1-rouge-f}{58.95}
\DefMacro{res-CSN-stmt-mut2e-stmt-test-chatgpt-zero-shot-full-context-top1-rouge-f-avg}{58.95}
\DefMacro{res-CSN-stmt-mut2e-stmt-test-chatgpt-zero-shot-full-context-top1-rouge-f-max}{58.95}
\DefMacro{res-CSN-stmt-mut2e-stmt-test-chatgpt-zero-shot-full-context-top1-rouge-f-min}{58.95}
\DefMacro{res-CSN-stmt-mut2e-stmt-test-chatgpt-zero-shot-full-context-top1-rouge-p}{58.37}
\DefMacro{res-CSN-stmt-mut2e-stmt-test-chatgpt-zero-shot-full-context-top1-rouge-p-avg}{58.37}
\DefMacro{res-CSN-stmt-mut2e-stmt-test-chatgpt-zero-shot-full-context-top1-rouge-p-max}{58.37}
\DefMacro{res-CSN-stmt-mut2e-stmt-test-chatgpt-zero-shot-full-context-top1-rouge-p-min}{58.37}
\DefMacro{res-CSN-stmt-mut2e-stmt-test-chatgpt-zero-shot-full-context-top1-rouge-r}{66.72}
\DefMacro{res-CSN-stmt-mut2e-stmt-test-chatgpt-zero-shot-full-context-top1-rouge-r-avg}{66.72}
\DefMacro{res-CSN-stmt-mut2e-stmt-test-chatgpt-zero-shot-full-context-top1-rouge-r-max}{66.72}
\DefMacro{res-CSN-stmt-mut2e-stmt-test-chatgpt-zero-shot-full-context-top1-rouge-r-min}{66.72}
\DefMacro{res-CSN-stmt-mut2e-stmt-test-chatgpt-zero-shot-full-context-top1-xmatch}{19.75}
\DefMacro{res-CSN-stmt-mut2e-stmt-test-chatgpt-zero-shot-full-context-top1-xmatch-top1}{19.75}

\DefMacro{ds-mut2e-stmt-eval-otest-mean-stmts-toks}{12.80}
\DefMacro{ds-mut2e-stmt-eval-otest-num-estests}{16}
\DefMacro{ds-mut2e-stmt-eval-otest-num-projects}{5}
\DefMacro{ds-mut2e-stmt-eval-otest-num-stmts}{81}

\DefMacro{res-nestack2e-no-stack-trace-axolotl-codellama-7b-bleu}{53.20}
\DefMacro{res-nestack2e-no-stack-trace-axolotl-codellama-7b-bleu-avg}{53.20}
\DefMacro{res-nestack2e-no-stack-trace-axolotl-codellama-7b-bleu-max}{53.20}
\DefMacro{res-nestack2e-no-stack-trace-axolotl-codellama-7b-bleu-min}{53.20}
\DefMacro{res-nestack2e-no-stack-trace-axolotl-codellama-7b-code-bleu}{60.33}
\DefMacro{res-nestack2e-no-stack-trace-axolotl-codellama-7b-code-bleu-avg}{60.33}
\DefMacro{res-nestack2e-no-stack-trace-axolotl-codellama-7b-code-bleu-max}{60.33}
\DefMacro{res-nestack2e-no-stack-trace-axolotl-codellama-7b-code-bleu-min}{60.33}
\DefMacro{res-nestack2e-no-stack-trace-axolotl-codellama-7b-edit-sim}{78.78}
\DefMacro{res-nestack2e-no-stack-trace-axolotl-codellama-7b-edit-sim-avg}{78.78}
\DefMacro{res-nestack2e-no-stack-trace-axolotl-codellama-7b-edit-sim-max}{78.78}
\DefMacro{res-nestack2e-no-stack-trace-axolotl-codellama-7b-edit-sim-min}{78.78}
\DefMacro{res-nestack2e-no-stack-trace-axolotl-codellama-7b-rouge-f}{73.21}
\DefMacro{res-nestack2e-no-stack-trace-axolotl-codellama-7b-rouge-f-avg}{73.21}
\DefMacro{res-nestack2e-no-stack-trace-axolotl-codellama-7b-rouge-f-max}{73.21}
\DefMacro{res-nestack2e-no-stack-trace-axolotl-codellama-7b-rouge-f-min}{73.21}
\DefMacro{res-nestack2e-no-stack-trace-axolotl-codellama-7b-rouge-p}{74.17}
\DefMacro{res-nestack2e-no-stack-trace-axolotl-codellama-7b-rouge-p-avg}{74.17}
\DefMacro{res-nestack2e-no-stack-trace-axolotl-codellama-7b-rouge-p-max}{74.17}
\DefMacro{res-nestack2e-no-stack-trace-axolotl-codellama-7b-rouge-p-min}{74.17}
\DefMacro{res-nestack2e-no-stack-trace-axolotl-codellama-7b-rouge-r}{75.17}
\DefMacro{res-nestack2e-no-stack-trace-axolotl-codellama-7b-rouge-r-avg}{75.17}
\DefMacro{res-nestack2e-no-stack-trace-axolotl-codellama-7b-rouge-r-max}{75.17}
\DefMacro{res-nestack2e-no-stack-trace-axolotl-codellama-7b-rouge-r-min}{75.17}
\DefMacro{res-nestack2e-no-stack-trace-axolotl-codellama-7b-xmatch}{11.85}
\DefMacro{res-nestack2e-no-stack-trace-axolotl-codellama-7b-xmatch-top1}{11.85}
\DefMacro{res-nestack2e-no-stack-trace-axolotl-codellama-7b-compilable-avg}{75.72}
\DefMacro{res-nestack2e-no-stack-trace-axolotl-codellama-7b-compilable-max}{75.72}
\DefMacro{res-nestack2e-no-stack-trace-axolotl-codellama-7b-compilable-min}{75.72}
\DefMacro{res-nestack2e-no-stack-trace-axolotl-codellama-7b-match-avg}{98.99}
\DefMacro{res-nestack2e-no-stack-trace-axolotl-codellama-7b-match-max}{98.99}
\DefMacro{res-nestack2e-no-stack-trace-axolotl-codellama-7b-match-min}{98.99}
\DefMacro{res-nestack2e-no-stack-trace-axolotl-codellama-7b-runnable-avg}{72.96}
\DefMacro{res-nestack2e-no-stack-trace-axolotl-codellama-7b-runnable-max}{72.96}
\DefMacro{res-nestack2e-no-stack-trace-axolotl-codellama-7b-runnable-min}{72.96}
\DefMacro{res-nestack2e-no-stack-trace-axolotl-codellama-7b-timeout-avg}{0.26}
\DefMacro{res-nestack2e-no-stack-trace-axolotl-codellama-7b-timeout-max}{0.26}
\DefMacro{res-nestack2e-no-stack-trace-axolotl-codellama-7b-timeout-min}{0.26}
\DefMacro{res-nestack2e-no-stack-trace-axolotl-codellama-7b-runnable-overall}{54.68}
\DefMacro{res-nestack2e-with-stack-trace-axolotl-codellama-7b-bleu}{53.48}
\DefMacro{res-nestack2e-with-stack-trace-axolotl-codellama-7b-bleu-avg}{53.48}
\DefMacro{res-nestack2e-with-stack-trace-axolotl-codellama-7b-bleu-max}{53.48}
\DefMacro{res-nestack2e-with-stack-trace-axolotl-codellama-7b-bleu-min}{53.48}
\DefMacro{res-nestack2e-with-stack-trace-axolotl-codellama-7b-code-bleu}{60.59}
\DefMacro{res-nestack2e-with-stack-trace-axolotl-codellama-7b-code-bleu-avg}{60.59}
\DefMacro{res-nestack2e-with-stack-trace-axolotl-codellama-7b-code-bleu-max}{60.59}
\DefMacro{res-nestack2e-with-stack-trace-axolotl-codellama-7b-code-bleu-min}{60.59}
\DefMacro{res-nestack2e-with-stack-trace-axolotl-codellama-7b-edit-sim}{79.24}
\DefMacro{res-nestack2e-with-stack-trace-axolotl-codellama-7b-edit-sim-avg}{79.24}
\DefMacro{res-nestack2e-with-stack-trace-axolotl-codellama-7b-edit-sim-max}{79.24}
\DefMacro{res-nestack2e-with-stack-trace-axolotl-codellama-7b-edit-sim-min}{79.24}
\DefMacro{res-nestack2e-with-stack-trace-axolotl-codellama-7b-rouge-f}{73.53}
\DefMacro{res-nestack2e-with-stack-trace-axolotl-codellama-7b-rouge-f-avg}{73.53}
\DefMacro{res-nestack2e-with-stack-trace-axolotl-codellama-7b-rouge-f-max}{73.53}
\DefMacro{res-nestack2e-with-stack-trace-axolotl-codellama-7b-rouge-f-min}{73.53}
\DefMacro{res-nestack2e-with-stack-trace-axolotl-codellama-7b-rouge-p}{74.62}
\DefMacro{res-nestack2e-with-stack-trace-axolotl-codellama-7b-rouge-p-avg}{74.62}
\DefMacro{res-nestack2e-with-stack-trace-axolotl-codellama-7b-rouge-p-max}{74.62}
\DefMacro{res-nestack2e-with-stack-trace-axolotl-codellama-7b-rouge-p-min}{74.62}
\DefMacro{res-nestack2e-with-stack-trace-axolotl-codellama-7b-rouge-r}{75.10}
\DefMacro{res-nestack2e-with-stack-trace-axolotl-codellama-7b-rouge-r-avg}{75.10}
\DefMacro{res-nestack2e-with-stack-trace-axolotl-codellama-7b-rouge-r-max}{75.10}
\DefMacro{res-nestack2e-with-stack-trace-axolotl-codellama-7b-rouge-r-min}{75.10}
\DefMacro{res-nestack2e-with-stack-trace-axolotl-codellama-7b-xmatch}{11.28}
\DefMacro{res-nestack2e-with-stack-trace-axolotl-codellama-7b-xmatch-top1}{11.28}
\DefMacro{res-nestack2e-with-stack-trace-axolotl-codellama-7b-compilable-avg}{77.25}
\DefMacro{res-nestack2e-with-stack-trace-axolotl-codellama-7b-compilable-max}{77.25}
\DefMacro{res-nestack2e-with-stack-trace-axolotl-codellama-7b-compilable-min}{77.25}
\DefMacro{res-nestack2e-with-stack-trace-axolotl-codellama-7b-match-avg}{99.50}
\DefMacro{res-nestack2e-with-stack-trace-axolotl-codellama-7b-match-max}{99.50}
\DefMacro{res-nestack2e-with-stack-trace-axolotl-codellama-7b-match-min}{99.50}
\DefMacro{res-nestack2e-with-stack-trace-axolotl-codellama-7b-runnable-avg}{74.13}
\DefMacro{res-nestack2e-with-stack-trace-axolotl-codellama-7b-runnable-max}{74.13}
\DefMacro{res-nestack2e-with-stack-trace-axolotl-codellama-7b-runnable-min}{74.13}
\DefMacro{res-nestack2e-with-stack-trace-axolotl-codellama-7b-timeout-avg}{0.00}
\DefMacro{res-nestack2e-with-stack-trace-axolotl-codellama-7b-timeout-max}{0.00}
\DefMacro{res-nestack2e-with-stack-trace-axolotl-codellama-7b-timeout-min}{0.00}
\DefMacro{res-nestack2e-with-stack-trace-axolotl-codellama-7b-runnable-overall}{56.98}

\DefMacro{res-gpt4-nestack2e-cot-gpt-4-0613-bleu}{33.39}
\DefMacro{res-gpt4-nestack2e-cot-gpt-4-0613-bleu-avg}{33.39}
\DefMacro{res-gpt4-nestack2e-cot-gpt-4-0613-bleu-max}{33.39}
\DefMacro{res-gpt4-nestack2e-cot-gpt-4-0613-bleu-min}{33.39}
\DefMacro{res-gpt4-nestack2e-cot-gpt-4-0613-code-bleu}{42.28}
\DefMacro{res-gpt4-nestack2e-cot-gpt-4-0613-code-bleu-avg}{42.28}
\DefMacro{res-gpt4-nestack2e-cot-gpt-4-0613-code-bleu-max}{42.28}
\DefMacro{res-gpt4-nestack2e-cot-gpt-4-0613-code-bleu-min}{42.28}
\DefMacro{res-gpt4-nestack2e-cot-gpt-4-0613-edit-sim}{64.10}
\DefMacro{res-gpt4-nestack2e-cot-gpt-4-0613-edit-sim-avg}{64.10}
\DefMacro{res-gpt4-nestack2e-cot-gpt-4-0613-edit-sim-max}{64.10}
\DefMacro{res-gpt4-nestack2e-cot-gpt-4-0613-edit-sim-min}{64.10}
\DefMacro{res-gpt4-nestack2e-cot-gpt-4-0613-rouge-f}{55.19}
\DefMacro{res-gpt4-nestack2e-cot-gpt-4-0613-rouge-f-avg}{55.19}
\DefMacro{res-gpt4-nestack2e-cot-gpt-4-0613-rouge-f-max}{55.19}
\DefMacro{res-gpt4-nestack2e-cot-gpt-4-0613-rouge-f-min}{55.19}
\DefMacro{res-gpt4-nestack2e-cot-gpt-4-0613-rouge-p}{51.35}
\DefMacro{res-gpt4-nestack2e-cot-gpt-4-0613-rouge-p-avg}{51.35}
\DefMacro{res-gpt4-nestack2e-cot-gpt-4-0613-rouge-p-max}{51.35}
\DefMacro{res-gpt4-nestack2e-cot-gpt-4-0613-rouge-p-min}{51.35}
\DefMacro{res-gpt4-nestack2e-cot-gpt-4-0613-rouge-r}{62.50}
\DefMacro{res-gpt4-nestack2e-cot-gpt-4-0613-rouge-r-avg}{62.50}
\DefMacro{res-gpt4-nestack2e-cot-gpt-4-0613-rouge-r-max}{62.50}
\DefMacro{res-gpt4-nestack2e-cot-gpt-4-0613-rouge-r-min}{62.50}
\DefMacro{res-gpt4-nestack2e-cot-gpt-4-0613-xmatch}{4.00}
\DefMacro{res-gpt4-nestack2e-cot-gpt-4-0613-xmatch-top1}{4.00}
\DefMacro{res-gpt4-nestack2e-cot-gpt-4-0613-compilable-avg}{66.00}
\DefMacro{res-gpt4-nestack2e-cot-gpt-4-0613-compilable-max}{66.00}
\DefMacro{res-gpt4-nestack2e-cot-gpt-4-0613-compilable-min}{66.00}
\DefMacro{res-gpt4-nestack2e-cot-gpt-4-0613-match-avg}{100.00}
\DefMacro{res-gpt4-nestack2e-cot-gpt-4-0613-match-max}{100.00}
\DefMacro{res-gpt4-nestack2e-cot-gpt-4-0613-match-min}{100.00}
\DefMacro{res-gpt4-nestack2e-cot-gpt-4-0613-runnable-avg}{93.94}
\DefMacro{res-gpt4-nestack2e-cot-gpt-4-0613-runnable-max}{93.94}
\DefMacro{res-gpt4-nestack2e-cot-gpt-4-0613-runnable-min}{93.94}
\DefMacro{res-gpt4-nestack2e-cot-gpt-4-0613-runnable-overall}{62.00}
\DefMacro{res-gpt4-nestack2e-cot-gpt-4-0613-timeout-avg}{0.00}
\DefMacro{res-gpt4-nestack2e-cot-gpt-4-0613-timeout-max}{0.00}
\DefMacro{res-gpt4-nestack2e-cot-gpt-4-0613-timeout-min}{0.00}
\DefMacro{res-gpt4-nestack2e-few-shot-gpt-4-0613-bleu}{33.80}
\DefMacro{res-gpt4-nestack2e-few-shot-gpt-4-0613-bleu-avg}{33.80}
\DefMacro{res-gpt4-nestack2e-few-shot-gpt-4-0613-bleu-max}{33.80}
\DefMacro{res-gpt4-nestack2e-few-shot-gpt-4-0613-bleu-min}{33.80}
\DefMacro{res-gpt4-nestack2e-few-shot-gpt-4-0613-code-bleu}{44.29}
\DefMacro{res-gpt4-nestack2e-few-shot-gpt-4-0613-code-bleu-avg}{44.29}
\DefMacro{res-gpt4-nestack2e-few-shot-gpt-4-0613-code-bleu-max}{44.29}
\DefMacro{res-gpt4-nestack2e-few-shot-gpt-4-0613-code-bleu-min}{44.29}
\DefMacro{res-gpt4-nestack2e-few-shot-gpt-4-0613-edit-sim}{64.22}
\DefMacro{res-gpt4-nestack2e-few-shot-gpt-4-0613-edit-sim-avg}{64.22}
\DefMacro{res-gpt4-nestack2e-few-shot-gpt-4-0613-edit-sim-max}{64.22}
\DefMacro{res-gpt4-nestack2e-few-shot-gpt-4-0613-edit-sim-min}{64.22}
\DefMacro{res-gpt4-nestack2e-few-shot-gpt-4-0613-rouge-f}{55.28}
\DefMacro{res-gpt4-nestack2e-few-shot-gpt-4-0613-rouge-f-avg}{55.28}
\DefMacro{res-gpt4-nestack2e-few-shot-gpt-4-0613-rouge-f-max}{55.28}
\DefMacro{res-gpt4-nestack2e-few-shot-gpt-4-0613-rouge-f-min}{55.28}
\DefMacro{res-gpt4-nestack2e-few-shot-gpt-4-0613-rouge-p}{51.77}
\DefMacro{res-gpt4-nestack2e-few-shot-gpt-4-0613-rouge-p-avg}{51.77}
\DefMacro{res-gpt4-nestack2e-few-shot-gpt-4-0613-rouge-p-max}{51.77}
\DefMacro{res-gpt4-nestack2e-few-shot-gpt-4-0613-rouge-p-min}{51.77}
\DefMacro{res-gpt4-nestack2e-few-shot-gpt-4-0613-rouge-r}{62.01}
\DefMacro{res-gpt4-nestack2e-few-shot-gpt-4-0613-rouge-r-avg}{62.01}
\DefMacro{res-gpt4-nestack2e-few-shot-gpt-4-0613-rouge-r-max}{62.01}
\DefMacro{res-gpt4-nestack2e-few-shot-gpt-4-0613-rouge-r-min}{62.01}
\DefMacro{res-gpt4-nestack2e-few-shot-gpt-4-0613-xmatch}{4.00}
\DefMacro{res-gpt4-nestack2e-few-shot-gpt-4-0613-xmatch-top1}{4.00}
\DefMacro{res-gpt4-nestack2e-few-shot-gpt-4-0613-compilable-avg}{68.00}
\DefMacro{res-gpt4-nestack2e-few-shot-gpt-4-0613-compilable-max}{68.00}
\DefMacro{res-gpt4-nestack2e-few-shot-gpt-4-0613-compilable-min}{68.00}
\DefMacro{res-gpt4-nestack2e-few-shot-gpt-4-0613-match-avg}{100.00}
\DefMacro{res-gpt4-nestack2e-few-shot-gpt-4-0613-match-max}{100.00}
\DefMacro{res-gpt4-nestack2e-few-shot-gpt-4-0613-match-min}{100.00}
\DefMacro{res-gpt4-nestack2e-few-shot-gpt-4-0613-runnable-avg}{94.12}
\DefMacro{res-gpt4-nestack2e-few-shot-gpt-4-0613-runnable-max}{94.12}
\DefMacro{res-gpt4-nestack2e-few-shot-gpt-4-0613-runnable-min}{94.12}
\DefMacro{res-gpt4-nestack2e-few-shot-gpt-4-0613-runnable-overall}{64.00}
\DefMacro{res-gpt4-nestack2e-few-shot-gpt-4-0613-timeout-avg}{0.00}
\DefMacro{res-gpt4-nestack2e-few-shot-gpt-4-0613-timeout-max}{0.00}
\DefMacro{res-gpt4-nestack2e-few-shot-gpt-4-0613-timeout-min}{0.00}

\DefMacro{ds-mut2e-avg-stack-trace-len}{2.35}
\DefMacro{ds-mut2e-etest-count}{12,574}
\DefMacro{ds-mut2e-exception-count}{821}
\DefMacro{ds-mut2e-max-stack-trace-len}{16}
\DefMacro{ds-mut2e-median-stack-trace-len}{2.00}
\DefMacro{ds-mut2e-min-stack-trace-len}{1}
\DefMacro{ds-mut2e-module-count}{699}
\DefMacro{ds-mut2e-mut-count}{6,250}
\DefMacro{ds-mut2e-no-stack-trace-count}{0}
\DefMacro{ds-mut2e-project-count}{562}
\DefMacro{ds-mut2e-stack-trace-len>6}{634}
\DefMacro{ds-mut2e-total-test-count}{111,230}

\DefMacro{ds-mut2e-@Test(expected)}{8,234}
\DefMacro{ds-mut2e-assertThrows}{44}
\DefMacro{ds-mut2e-expectedException.expect()}{1,252}
\DefMacro{ds-mut2e-try{fail}catch}{3,044}

\DefMacro{ds-test-etest-count}{842}
\DefMacro{ds-test-etype-count}{65}
\DefMacro{ds-test-module-count}{50}
\DefMacro{ds-test-mut-count}{464}
\DefMacro{ds-test-project-count}{32}
\DefMacro{ds-test-test-count}{5,902}
\DefMacro{ds-train-etest-count}{11,182}
\DefMacro{ds-train-etype-count}{725}
\DefMacro{ds-train-module-count}{610}
\DefMacro{ds-train-mut-count}{5,508}
\DefMacro{ds-train-project-count}{501}
\DefMacro{ds-train-test-count}{100,030}
\DefMacro{ds-valid-etest-count}{550}
\DefMacro{ds-valid-etype-count}{66}
\DefMacro{ds-valid-module-count}{39}
\DefMacro{ds-valid-mut-count}{279}
\DefMacro{ds-valid-project-count}{29}
\DefMacro{ds-valid-test-count}{5,298}

\DefMacro{ds-eval-count-eval-throw-stmt-data}{487}
\DefMacro{ds-eval-count-projects}{30}
\DefMacro{ds-eval-count-st-more-than-one}{1}
\DefMacro{ds-eval-count-static-analysis-helps}{434}
\DefMacro{ds-eval-avg-condition-count}{41.17}
\DefMacro{ds-eval-avg-netest-count}{7.12}
\DefMacro{ds-eval-avg-stack-trace-count}{48.74}
\DefMacro{ds-eval-avg-stack-trace-len}{13.01}
\DefMacro{ds-eval-max-netest-count}{70}
\DefMacro{ds-eval-max-stack-trace-count}{1,820}
\DefMacro{ds-eval-max-stack-trace-len}{23}
\DefMacro{ds-eval-median-netest-count}{2.00}
\DefMacro{ds-eval-median-stack-trace-count}{1.00}
\DefMacro{ds-eval-median-stack-trace-len}{14}
\DefMacro{ds-eval-min-stack-trace-len}{1}
\DefMacro{ds-eval-no-condition-count}{383}
\DefMacro{ds-eval-no-netest-count}{102}
\DefMacro{ds-eval-no-stack-trace-count}{328}
\DefMacro{ds-eval-with-netest-count}{740}
\DefMacro{ds-eval-with-stack-trace-count}{514}

\DefMacro{res-ne2e-no-name-few-shot-gpt-3.5-turbo-16k-bleu}{31.17}
\DefMacro{res-ne2e-no-name-few-shot-gpt-3.5-turbo-16k-bleu-avg}{31.17}
\DefMacro{res-ne2e-no-name-few-shot-gpt-3.5-turbo-16k-bleu-max}{31.17}
\DefMacro{res-ne2e-no-name-few-shot-gpt-3.5-turbo-16k-bleu-min}{31.17}
\DefMacro{res-ne2e-no-name-few-shot-gpt-3.5-turbo-16k-code-bleu}{41.02}
\DefMacro{res-ne2e-no-name-few-shot-gpt-3.5-turbo-16k-code-bleu-avg}{41.02}
\DefMacro{res-ne2e-no-name-few-shot-gpt-3.5-turbo-16k-code-bleu-max}{41.02}
\DefMacro{res-ne2e-no-name-few-shot-gpt-3.5-turbo-16k-code-bleu-min}{41.02}
\DefMacro{res-ne2e-no-name-few-shot-gpt-3.5-turbo-16k-edit-sim}{65.48}
\DefMacro{res-ne2e-no-name-few-shot-gpt-3.5-turbo-16k-edit-sim-avg}{65.48}
\DefMacro{res-ne2e-no-name-few-shot-gpt-3.5-turbo-16k-edit-sim-max}{65.48}
\DefMacro{res-ne2e-no-name-few-shot-gpt-3.5-turbo-16k-edit-sim-min}{65.48}
\DefMacro{res-ne2e-no-name-few-shot-gpt-3.5-turbo-16k-rouge-f}{58.83}
\DefMacro{res-ne2e-no-name-few-shot-gpt-3.5-turbo-16k-rouge-f-avg}{58.83}
\DefMacro{res-ne2e-no-name-few-shot-gpt-3.5-turbo-16k-rouge-f-max}{58.83}
\DefMacro{res-ne2e-no-name-few-shot-gpt-3.5-turbo-16k-rouge-f-min}{58.83}
\DefMacro{res-ne2e-no-name-few-shot-gpt-3.5-turbo-16k-rouge-p}{58.70}
\DefMacro{res-ne2e-no-name-few-shot-gpt-3.5-turbo-16k-rouge-p-avg}{58.70}
\DefMacro{res-ne2e-no-name-few-shot-gpt-3.5-turbo-16k-rouge-p-max}{58.70}
\DefMacro{res-ne2e-no-name-few-shot-gpt-3.5-turbo-16k-rouge-p-min}{58.70}
\DefMacro{res-ne2e-no-name-few-shot-gpt-3.5-turbo-16k-rouge-r}{62.52}
\DefMacro{res-ne2e-no-name-few-shot-gpt-3.5-turbo-16k-rouge-r-avg}{62.52}
\DefMacro{res-ne2e-no-name-few-shot-gpt-3.5-turbo-16k-rouge-r-max}{62.52}
\DefMacro{res-ne2e-no-name-few-shot-gpt-3.5-turbo-16k-rouge-r-min}{62.52}
\DefMacro{res-ne2e-no-name-few-shot-gpt-3.5-turbo-16k-xmatch}{0.24}
\DefMacro{res-ne2e-no-name-few-shot-gpt-3.5-turbo-16k-xmatch-top1}{0.24}
\DefMacro{res-ne2e-no-name-few-shot-gpt-3.5-turbo-16k-compilable-avg}{53.21}
\DefMacro{res-ne2e-no-name-few-shot-gpt-3.5-turbo-16k-compilable-max}{53.21}
\DefMacro{res-ne2e-no-name-few-shot-gpt-3.5-turbo-16k-compilable-min}{53.21}
\DefMacro{res-ne2e-no-name-few-shot-gpt-3.5-turbo-16k-match-avg}{98.21}
\DefMacro{res-ne2e-no-name-few-shot-gpt-3.5-turbo-16k-match-max}{98.21}
\DefMacro{res-ne2e-no-name-few-shot-gpt-3.5-turbo-16k-match-min}{98.21}
\DefMacro{res-ne2e-no-name-few-shot-gpt-3.5-turbo-16k-runnable-avg}{65.23}
\DefMacro{res-ne2e-no-name-few-shot-gpt-3.5-turbo-16k-runnable-max}{65.23}
\DefMacro{res-ne2e-no-name-few-shot-gpt-3.5-turbo-16k-runnable-min}{65.23}
\DefMacro{res-ne2e-no-name-few-shot-gpt-3.5-turbo-16k-runnable-overall}{34.09}
\DefMacro{res-ne2e-no-name-few-shot-gpt-3.5-turbo-16k-timeout-avg}{0.23}
\DefMacro{res-ne2e-no-name-few-shot-gpt-3.5-turbo-16k-timeout-max}{0.23}
\DefMacro{res-ne2e-no-name-few-shot-gpt-3.5-turbo-16k-timeout-min}{0.23}
\DefMacro{res-mut2e-no-name-ft-lora-codellama-7b-bleu}{38.08}
\DefMacro{res-mut2e-no-name-ft-lora-codellama-7b-bleu-avg}{38.08}
\DefMacro{res-mut2e-no-name-ft-lora-codellama-7b-bleu-max}{38.08}
\DefMacro{res-mut2e-no-name-ft-lora-codellama-7b-bleu-min}{38.08}
\DefMacro{res-mut2e-no-name-ft-lora-codellama-7b-code-bleu}{46.34}
\DefMacro{res-mut2e-no-name-ft-lora-codellama-7b-code-bleu-avg}{46.34}
\DefMacro{res-mut2e-no-name-ft-lora-codellama-7b-code-bleu-max}{46.34}
\DefMacro{res-mut2e-no-name-ft-lora-codellama-7b-code-bleu-min}{46.34}
\DefMacro{res-mut2e-no-name-ft-lora-codellama-7b-edit-sim}{73.02}
\DefMacro{res-mut2e-no-name-ft-lora-codellama-7b-edit-sim-avg}{73.02}
\DefMacro{res-mut2e-no-name-ft-lora-codellama-7b-edit-sim-max}{73.02}
\DefMacro{res-mut2e-no-name-ft-lora-codellama-7b-edit-sim-min}{73.02}
\DefMacro{res-mut2e-no-name-ft-lora-codellama-7b-rouge-f}{66.15}
\DefMacro{res-mut2e-no-name-ft-lora-codellama-7b-rouge-f-avg}{66.15}
\DefMacro{res-mut2e-no-name-ft-lora-codellama-7b-rouge-f-max}{66.15}
\DefMacro{res-mut2e-no-name-ft-lora-codellama-7b-rouge-f-min}{66.15}
\DefMacro{res-mut2e-no-name-ft-lora-codellama-7b-rouge-p}{68.72}
\DefMacro{res-mut2e-no-name-ft-lora-codellama-7b-rouge-p-avg}{68.72}
\DefMacro{res-mut2e-no-name-ft-lora-codellama-7b-rouge-p-max}{68.72}
\DefMacro{res-mut2e-no-name-ft-lora-codellama-7b-rouge-p-min}{68.72}
\DefMacro{res-mut2e-no-name-ft-lora-codellama-7b-rouge-r}{65.84}
\DefMacro{res-mut2e-no-name-ft-lora-codellama-7b-rouge-r-avg}{65.84}
\DefMacro{res-mut2e-no-name-ft-lora-codellama-7b-rouge-r-max}{65.84}
\DefMacro{res-mut2e-no-name-ft-lora-codellama-7b-rouge-r-min}{65.84}
\DefMacro{res-mut2e-no-name-ft-lora-codellama-7b-xmatch}{0.24}
\DefMacro{res-mut2e-no-name-ft-lora-codellama-7b-xmatch-top1}{0.24}
\DefMacro{res-mut2e-no-name-ft-lora-codellama-7b-compilable-avg}{64.85}
\DefMacro{res-mut2e-no-name-ft-lora-codellama-7b-compilable-max}{64.85}
\DefMacro{res-mut2e-no-name-ft-lora-codellama-7b-compilable-min}{64.85}
\DefMacro{res-mut2e-no-name-ft-lora-codellama-7b-match-avg}{100.00}
\DefMacro{res-mut2e-no-name-ft-lora-codellama-7b-match-max}{100.00}
\DefMacro{res-mut2e-no-name-ft-lora-codellama-7b-match-min}{100.00}
\DefMacro{res-mut2e-no-name-ft-lora-codellama-7b-runnable-avg}{73.63}
\DefMacro{res-mut2e-no-name-ft-lora-codellama-7b-runnable-max}{73.63}
\DefMacro{res-mut2e-no-name-ft-lora-codellama-7b-runnable-min}{73.63}
\DefMacro{res-mut2e-no-name-ft-lora-codellama-7b-runnable-overall}{47.74}
\DefMacro{res-mut2e-no-name-ft-lora-codellama-7b-timeout-avg}{0.37}
\DefMacro{res-mut2e-no-name-ft-lora-codellama-7b-timeout-max}{0.37}
\DefMacro{res-mut2e-no-name-ft-lora-codellama-7b-timeout-min}{0.37}
\DefMacro{res-ne2e-no-name-ft-lora-codellama-7b-bleu}{43.32}
\DefMacro{res-ne2e-no-name-ft-lora-codellama-7b-bleu-avg}{43.32}
\DefMacro{res-ne2e-no-name-ft-lora-codellama-7b-bleu-max}{43.32}
\DefMacro{res-ne2e-no-name-ft-lora-codellama-7b-bleu-min}{43.32}
\DefMacro{res-ne2e-no-name-ft-lora-codellama-7b-code-bleu}{51.19}
\DefMacro{res-ne2e-no-name-ft-lora-codellama-7b-code-bleu-avg}{51.19}
\DefMacro{res-ne2e-no-name-ft-lora-codellama-7b-code-bleu-max}{51.19}
\DefMacro{res-ne2e-no-name-ft-lora-codellama-7b-code-bleu-min}{51.19}
\DefMacro{res-ne2e-no-name-ft-lora-codellama-7b-edit-sim}{77.48}
\DefMacro{res-ne2e-no-name-ft-lora-codellama-7b-edit-sim-avg}{77.48}
\DefMacro{res-ne2e-no-name-ft-lora-codellama-7b-edit-sim-max}{77.48}
\DefMacro{res-ne2e-no-name-ft-lora-codellama-7b-edit-sim-min}{77.48}
\DefMacro{res-ne2e-no-name-ft-lora-codellama-7b-rouge-f}{71.02}
\DefMacro{res-ne2e-no-name-ft-lora-codellama-7b-rouge-f-avg}{71.02}
\DefMacro{res-ne2e-no-name-ft-lora-codellama-7b-rouge-f-max}{71.02}
\DefMacro{res-ne2e-no-name-ft-lora-codellama-7b-rouge-f-min}{71.02}
\DefMacro{res-ne2e-no-name-ft-lora-codellama-7b-rouge-p}{72.97}
\DefMacro{res-ne2e-no-name-ft-lora-codellama-7b-rouge-p-avg}{72.97}
\DefMacro{res-ne2e-no-name-ft-lora-codellama-7b-rouge-p-max}{72.97}
\DefMacro{res-ne2e-no-name-ft-lora-codellama-7b-rouge-p-min}{72.97}
\DefMacro{res-ne2e-no-name-ft-lora-codellama-7b-rouge-r}{71.08}
\DefMacro{res-ne2e-no-name-ft-lora-codellama-7b-rouge-r-avg}{71.08}
\DefMacro{res-ne2e-no-name-ft-lora-codellama-7b-rouge-r-max}{71.08}
\DefMacro{res-ne2e-no-name-ft-lora-codellama-7b-rouge-r-min}{71.08}
\DefMacro{res-ne2e-no-name-ft-lora-codellama-7b-xmatch}{0.71}
\DefMacro{res-ne2e-no-name-ft-lora-codellama-7b-xmatch-top1}{0.71}
\DefMacro{res-ne2e-no-name-ft-lora-codellama-7b-compilable-avg}{83.14}
\DefMacro{res-ne2e-no-name-ft-lora-codellama-7b-compilable-max}{83.14}
\DefMacro{res-ne2e-no-name-ft-lora-codellama-7b-compilable-min}{83.14}
\DefMacro{res-ne2e-no-name-ft-lora-codellama-7b-match-avg}{100.00}
\DefMacro{res-ne2e-no-name-ft-lora-codellama-7b-match-max}{100.00}
\DefMacro{res-ne2e-no-name-ft-lora-codellama-7b-match-min}{100.00}
\DefMacro{res-ne2e-no-name-ft-lora-codellama-7b-runnable-avg}{75.29}
\DefMacro{res-ne2e-no-name-ft-lora-codellama-7b-runnable-max}{75.29}
\DefMacro{res-ne2e-no-name-ft-lora-codellama-7b-runnable-min}{75.29}
\DefMacro{res-ne2e-no-name-ft-lora-codellama-7b-runnable-overall}{62.59}
\DefMacro{res-ne2e-no-name-ft-lora-codellama-7b-timeout-avg}{0.29}
\DefMacro{res-ne2e-no-name-ft-lora-codellama-7b-timeout-max}{0.29}
\DefMacro{res-ne2e-no-name-ft-lora-codellama-7b-timeout-min}{0.29}
\DefMacro{res-nestack2e-no-name-ft-lora-codellama-7b-real-bleu}{42.13}
\DefMacro{res-nestack2e-no-name-ft-lora-codellama-7b-real-bleu-avg}{42.09}
\DefMacro{res-nestack2e-no-name-ft-lora-codellama-7b-real-bleu-max}{43.56}
\DefMacro{res-nestack2e-no-name-ft-lora-codellama-7b-real-bleu-min}{40.74}
\DefMacro{res-nestack2e-no-name-ft-lora-codellama-7b-real-code-bleu}{50.54}
\DefMacro{res-nestack2e-no-name-ft-lora-codellama-7b-real-code-bleu-avg}{50.46}
\DefMacro{res-nestack2e-no-name-ft-lora-codellama-7b-real-code-bleu-max}{51.85}
\DefMacro{res-nestack2e-no-name-ft-lora-codellama-7b-real-code-bleu-min}{49.19}
\DefMacro{res-nestack2e-no-name-ft-lora-codellama-7b-real-edit-sim}{76.16}
\DefMacro{res-nestack2e-no-name-ft-lora-codellama-7b-real-edit-sim-avg}{76.07}
\DefMacro{res-nestack2e-no-name-ft-lora-codellama-7b-real-edit-sim-max}{77.64}
\DefMacro{res-nestack2e-no-name-ft-lora-codellama-7b-real-edit-sim-min}{74.46}
\DefMacro{res-nestack2e-no-name-ft-lora-codellama-7b-real-rouge-f}{69.94}
\DefMacro{res-nestack2e-no-name-ft-lora-codellama-7b-real-rouge-f-avg}{69.90}
\DefMacro{res-nestack2e-no-name-ft-lora-codellama-7b-real-rouge-f-max}{71.39}
\DefMacro{res-nestack2e-no-name-ft-lora-codellama-7b-real-rouge-f-min}{68.44}
\DefMacro{res-nestack2e-no-name-ft-lora-codellama-7b-real-rouge-p}{71.53}
\DefMacro{res-nestack2e-no-name-ft-lora-codellama-7b-real-rouge-p-avg}{71.46}
\DefMacro{res-nestack2e-no-name-ft-lora-codellama-7b-real-rouge-p-max}{73.29}
\DefMacro{res-nestack2e-no-name-ft-lora-codellama-7b-real-rouge-p-min}{69.43}
\DefMacro{res-nestack2e-no-name-ft-lora-codellama-7b-real-rouge-r}{70.66}
\DefMacro{res-nestack2e-no-name-ft-lora-codellama-7b-real-rouge-r-avg}{70.69}
\DefMacro{res-nestack2e-no-name-ft-lora-codellama-7b-real-rouge-r-max}{71.90}
\DefMacro{res-nestack2e-no-name-ft-lora-codellama-7b-real-rouge-r-min}{69.63}
\DefMacro{res-nestack2e-no-name-ft-lora-codellama-7b-real-xmatch}{0.95}
\DefMacro{res-nestack2e-no-name-ft-lora-codellama-7b-real-xmatch-top1}{0.95}
\DefMacro{res-nestack2e-no-name-ft-lora-codellama-7b-real-compilable-avg}{77.22}
\DefMacro{res-nestack2e-no-name-ft-lora-codellama-7b-real-compilable-max}{81.47}
\DefMacro{res-nestack2e-no-name-ft-lora-codellama-7b-real-compilable-min}{72.80}
\DefMacro{res-nestack2e-no-name-ft-lora-codellama-7b-real-match-avg}{100.00}
\DefMacro{res-nestack2e-no-name-ft-lora-codellama-7b-real-match-max}{100.00}
\DefMacro{res-nestack2e-no-name-ft-lora-codellama-7b-real-match-min}{100.00}
\DefMacro{res-nestack2e-no-name-ft-lora-codellama-7b-real-runnable-avg}{73.05}
\DefMacro{res-nestack2e-no-name-ft-lora-codellama-7b-real-runnable-max}{78.28}
\DefMacro{res-nestack2e-no-name-ft-lora-codellama-7b-real-runnable-min}{67.35}
\DefMacro{res-nestack2e-no-name-ft-lora-codellama-7b-real-runnable-overall}{63.78}
\DefMacro{res-nestack2e-no-name-ft-lora-codellama-7b-real-runnable-overall-avg}{56.97}
\DefMacro{res-nestack2e-no-name-ft-lora-codellama-7b-real-runnable-overall-max}{63.78}
\DefMacro{res-nestack2e-no-name-ft-lora-codellama-7b-real-runnable-overall-min}{50.36}
\DefMacro{res-nestack2e-no-name-ft-lora-codellama-7b-real-timeout-avg}{0.15}
\DefMacro{res-nestack2e-no-name-ft-lora-codellama-7b-real-timeout-max}{0.15}
\DefMacro{res-nestack2e-no-name-ft-lora-codellama-7b-real-timeout-min}{0.15}
\DefMacro{res-conditionne2e-no-name-ft-lora-codellama-7b-real-bleu}{42.17}
\DefMacro{res-conditionne2e-no-name-ft-lora-codellama-7b-real-bleu-avg}{42.48}
\DefMacro{res-conditionne2e-no-name-ft-lora-codellama-7b-real-bleu-max}{44.36}
\DefMacro{res-conditionne2e-no-name-ft-lora-codellama-7b-real-bleu-min}{40.77}
\DefMacro{res-conditionne2e-no-name-ft-lora-codellama-7b-real-code-bleu}{51.08}
\DefMacro{res-conditionne2e-no-name-ft-lora-codellama-7b-real-code-bleu-avg}{51.21}
\DefMacro{res-conditionne2e-no-name-ft-lora-codellama-7b-real-code-bleu-max}{53.06}
\DefMacro{res-conditionne2e-no-name-ft-lora-codellama-7b-real-code-bleu-min}{49.46}
\DefMacro{res-conditionne2e-no-name-ft-lora-codellama-7b-real-edit-sim}{76.21}
\DefMacro{res-conditionne2e-no-name-ft-lora-codellama-7b-real-edit-sim-avg}{76.43}
\DefMacro{res-conditionne2e-no-name-ft-lora-codellama-7b-real-edit-sim-max}{78.17}
\DefMacro{res-conditionne2e-no-name-ft-lora-codellama-7b-real-edit-sim-min}{74.43}
\DefMacro{res-conditionne2e-no-name-ft-lora-codellama-7b-real-rouge-f}{69.76}
\DefMacro{res-conditionne2e-no-name-ft-lora-codellama-7b-real-rouge-f-avg}{69.99}
\DefMacro{res-conditionne2e-no-name-ft-lora-codellama-7b-real-rouge-f-max}{71.76}
\DefMacro{res-conditionne2e-no-name-ft-lora-codellama-7b-real-rouge-f-min}{68.09}
\DefMacro{res-conditionne2e-no-name-ft-lora-codellama-7b-real-rouge-p}{70.79}
\DefMacro{res-conditionne2e-no-name-ft-lora-codellama-7b-real-rouge-p-avg}{71.04}
\DefMacro{res-conditionne2e-no-name-ft-lora-codellama-7b-real-rouge-p-max}{73.16}
\DefMacro{res-conditionne2e-no-name-ft-lora-codellama-7b-real-rouge-p-min}{68.68}
\DefMacro{res-conditionne2e-no-name-ft-lora-codellama-7b-real-rouge-r}{71.13}
\DefMacro{res-conditionne2e-no-name-ft-lora-codellama-7b-real-rouge-r-avg}{71.25}
\DefMacro{res-conditionne2e-no-name-ft-lora-codellama-7b-real-rouge-r-max}{72.84}
\DefMacro{res-conditionne2e-no-name-ft-lora-codellama-7b-real-rouge-r-min}{69.76}
\DefMacro{res-conditionne2e-no-name-ft-lora-codellama-7b-real-xmatch}{1.19}
\DefMacro{res-conditionne2e-no-name-ft-lora-codellama-7b-real-xmatch-top1}{1.19}
\DefMacro{res-conditionne2e-no-name-ft-lora-codellama-7b-real-compilable-avg}{76.75}
\DefMacro{res-conditionne2e-no-name-ft-lora-codellama-7b-real-compilable-max}{81.83}
\DefMacro{res-conditionne2e-no-name-ft-lora-codellama-7b-real-compilable-min}{71.62}
\DefMacro{res-conditionne2e-no-name-ft-lora-codellama-7b-real-match-avg}{100.00}
\DefMacro{res-conditionne2e-no-name-ft-lora-codellama-7b-real-match-max}{100.00}
\DefMacro{res-conditionne2e-no-name-ft-lora-codellama-7b-real-match-min}{100.00}
\DefMacro{res-conditionne2e-no-name-ft-lora-codellama-7b-real-runnable-avg}{66.90}
\DefMacro{res-conditionne2e-no-name-ft-lora-codellama-7b-real-runnable-max}{72.13}
\DefMacro{res-conditionne2e-no-name-ft-lora-codellama-7b-real-runnable-min}{61.54}
\DefMacro{res-conditionne2e-no-name-ft-lora-codellama-7b-real-runnable-overall}{59.03}
\DefMacro{res-conditionne2e-no-name-ft-lora-codellama-7b-real-timeout-avg}{0.33}
\DefMacro{res-conditionne2e-no-name-ft-lora-codellama-7b-real-timeout-max}{0.44}
\DefMacro{res-conditionne2e-no-name-ft-lora-codellama-7b-real-timeout-min}{0.29}
\DefMacro{res-conditionnestack2e-no-name-ft-lora-codellama-7b-real-bleu}{42.56}
\DefMacro{res-conditionnestack2e-no-name-ft-lora-codellama-7b-real-bleu-avg}{42.53}
\DefMacro{res-conditionnestack2e-no-name-ft-lora-codellama-7b-real-bleu-max}{45.22}
\DefMacro{res-conditionnestack2e-no-name-ft-lora-codellama-7b-real-bleu-min}{40.07}
\DefMacro{res-conditionnestack2e-no-name-ft-lora-codellama-7b-real-code-bleu}{51.00}
\DefMacro{res-conditionnestack2e-no-name-ft-lora-codellama-7b-real-code-bleu-avg}{50.89}
\DefMacro{res-conditionnestack2e-no-name-ft-lora-codellama-7b-real-code-bleu-max}{53.55}
\DefMacro{res-conditionnestack2e-no-name-ft-lora-codellama-7b-real-code-bleu-min}{48.24}
\DefMacro{res-conditionnestack2e-no-name-ft-lora-codellama-7b-real-edit-sim}{75.85}
\DefMacro{res-conditionnestack2e-no-name-ft-lora-codellama-7b-real-edit-sim-avg}{75.78}
\DefMacro{res-conditionnestack2e-no-name-ft-lora-codellama-7b-real-edit-sim-max}{78.25}
\DefMacro{res-conditionnestack2e-no-name-ft-lora-codellama-7b-real-edit-sim-min}{73.09}
\DefMacro{res-conditionnestack2e-no-name-ft-lora-codellama-7b-real-rouge-f}{69.43}
\DefMacro{res-conditionnestack2e-no-name-ft-lora-codellama-7b-real-rouge-f-avg}{69.39}
\DefMacro{res-conditionnestack2e-no-name-ft-lora-codellama-7b-real-rouge-f-max}{72.10}
\DefMacro{res-conditionnestack2e-no-name-ft-lora-codellama-7b-real-rouge-f-min}{66.47}
\DefMacro{res-conditionnestack2e-no-name-ft-lora-codellama-7b-real-rouge-p}{70.58}
\DefMacro{res-conditionnestack2e-no-name-ft-lora-codellama-7b-real-rouge-p-avg}{70.57}
\DefMacro{res-conditionnestack2e-no-name-ft-lora-codellama-7b-real-rouge-p-max}{73.74}
\DefMacro{res-conditionnestack2e-no-name-ft-lora-codellama-7b-real-rouge-p-min}{67.10}
\DefMacro{res-conditionnestack2e-no-name-ft-lora-codellama-7b-real-rouge-r}{70.52}
\DefMacro{res-conditionnestack2e-no-name-ft-lora-codellama-7b-real-rouge-r-avg}{70.47}
\DefMacro{res-conditionnestack2e-no-name-ft-lora-codellama-7b-real-rouge-r-max}{72.91}
\DefMacro{res-conditionnestack2e-no-name-ft-lora-codellama-7b-real-rouge-r-min}{67.98}
\DefMacro{res-conditionnestack2e-no-name-ft-lora-codellama-7b-real-xmatch}{0.95}
\DefMacro{res-conditionnestack2e-no-name-ft-lora-codellama-7b-real-xmatch-top1}{0.95}
\DefMacro{res-conditionnestack2e-no-name-ft-lora-codellama-7b-real-compilable-avg}{76.53}
\DefMacro{res-conditionnestack2e-no-name-ft-lora-codellama-7b-real-compilable-max}{82.42}
\DefMacro{res-conditionnestack2e-no-name-ft-lora-codellama-7b-real-compilable-min}{70.90}
\DefMacro{res-conditionnestack2e-no-name-ft-lora-codellama-7b-real-match-avg}{100.00}
\DefMacro{res-conditionnestack2e-no-name-ft-lora-codellama-7b-real-match-max}{100.00}
\DefMacro{res-conditionnestack2e-no-name-ft-lora-codellama-7b-real-match-min}{100.00}
\DefMacro{res-conditionnestack2e-no-name-ft-lora-codellama-7b-real-runnable-avg}{74.39}
\DefMacro{res-conditionnestack2e-no-name-ft-lora-codellama-7b-real-runnable-max}{79.39}
\DefMacro{res-conditionnestack2e-no-name-ft-lora-codellama-7b-real-runnable-min}{69.31}
\DefMacro{res-conditionnestack2e-no-name-ft-lora-codellama-7b-real-runnable-overall}{65.44}
\DefMacro{res-conditionnestack2e-no-name-ft-lora-codellama-7b-real-timeout-avg}{0.32}
\DefMacro{res-conditionnestack2e-no-name-ft-lora-codellama-7b-real-timeout-max}{0.43}
\DefMacro{res-conditionnestack2e-no-name-ft-lora-codellama-7b-real-timeout-min}{0.29}

\DefMacro{res-ne2e-with-name-few-shot-gpt-3.5-turbo-16k-bleu}{50.14}
\DefMacro{res-ne2e-with-name-few-shot-gpt-3.5-turbo-16k-bleu-avg}{50.14}
\DefMacro{res-ne2e-with-name-few-shot-gpt-3.5-turbo-16k-bleu-max}{50.14}
\DefMacro{res-ne2e-with-name-few-shot-gpt-3.5-turbo-16k-bleu-min}{50.14}
\DefMacro{res-ne2e-with-name-few-shot-gpt-3.5-turbo-16k-code-bleu}{57.51}
\DefMacro{res-ne2e-with-name-few-shot-gpt-3.5-turbo-16k-code-bleu-avg}{57.51}
\DefMacro{res-ne2e-with-name-few-shot-gpt-3.5-turbo-16k-code-bleu-max}{57.51}
\DefMacro{res-ne2e-with-name-few-shot-gpt-3.5-turbo-16k-code-bleu-min}{57.51}
\DefMacro{res-ne2e-with-name-few-shot-gpt-3.5-turbo-16k-edit-sim}{76.23}
\DefMacro{res-ne2e-with-name-few-shot-gpt-3.5-turbo-16k-edit-sim-avg}{76.23}
\DefMacro{res-ne2e-with-name-few-shot-gpt-3.5-turbo-16k-edit-sim-max}{76.23}
\DefMacro{res-ne2e-with-name-few-shot-gpt-3.5-turbo-16k-edit-sim-min}{76.23}
\DefMacro{res-ne2e-with-name-few-shot-gpt-3.5-turbo-16k-rouge-f}{69.44}
\DefMacro{res-ne2e-with-name-few-shot-gpt-3.5-turbo-16k-rouge-f-avg}{69.44}
\DefMacro{res-ne2e-with-name-few-shot-gpt-3.5-turbo-16k-rouge-f-max}{69.44}
\DefMacro{res-ne2e-with-name-few-shot-gpt-3.5-turbo-16k-rouge-f-min}{69.44}
\DefMacro{res-ne2e-with-name-few-shot-gpt-3.5-turbo-16k-rouge-p}{68.65}
\DefMacro{res-ne2e-with-name-few-shot-gpt-3.5-turbo-16k-rouge-p-avg}{68.65}
\DefMacro{res-ne2e-with-name-few-shot-gpt-3.5-turbo-16k-rouge-p-max}{68.65}
\DefMacro{res-ne2e-with-name-few-shot-gpt-3.5-turbo-16k-rouge-p-min}{68.65}
\DefMacro{res-ne2e-with-name-few-shot-gpt-3.5-turbo-16k-rouge-r}{73.30}
\DefMacro{res-ne2e-with-name-few-shot-gpt-3.5-turbo-16k-rouge-r-avg}{73.30}
\DefMacro{res-ne2e-with-name-few-shot-gpt-3.5-turbo-16k-rouge-r-max}{73.30}
\DefMacro{res-ne2e-with-name-few-shot-gpt-3.5-turbo-16k-rouge-r-min}{73.30}
\DefMacro{res-ne2e-with-name-few-shot-gpt-3.5-turbo-16k-xmatch}{11.05}
\DefMacro{res-ne2e-with-name-few-shot-gpt-3.5-turbo-16k-xmatch-top1}{11.05}
\DefMacro{res-ne2e-with-name-few-shot-gpt-3.5-turbo-16k-compilable-avg}{52.26}
\DefMacro{res-ne2e-with-name-few-shot-gpt-3.5-turbo-16k-compilable-max}{52.26}
\DefMacro{res-ne2e-with-name-few-shot-gpt-3.5-turbo-16k-compilable-min}{52.26}
\DefMacro{res-ne2e-with-name-few-shot-gpt-3.5-turbo-16k-match-avg}{98.41}
\DefMacro{res-ne2e-with-name-few-shot-gpt-3.5-turbo-16k-match-max}{98.41}
\DefMacro{res-ne2e-with-name-few-shot-gpt-3.5-turbo-16k-match-min}{98.41}
\DefMacro{res-ne2e-with-name-few-shot-gpt-3.5-turbo-16k-runnable-avg}{83.14}
\DefMacro{res-ne2e-with-name-few-shot-gpt-3.5-turbo-16k-runnable-max}{83.14}
\DefMacro{res-ne2e-with-name-few-shot-gpt-3.5-turbo-16k-runnable-min}{83.14}
\DefMacro{res-ne2e-with-name-few-shot-gpt-3.5-turbo-16k-runnable-overall}{42.76}
\DefMacro{res-ne2e-with-name-few-shot-gpt-3.5-turbo-16k-timeout-avg}{0.00}
\DefMacro{res-ne2e-with-name-few-shot-gpt-3.5-turbo-16k-timeout-max}{0.00}
\DefMacro{res-ne2e-with-name-few-shot-gpt-3.5-turbo-16k-timeout-min}{0.00}
\DefMacro{res-mut2e-with-name-ft-lora-codellama-7b-bleu}{57.17}
\DefMacro{res-mut2e-with-name-ft-lora-codellama-7b-bleu-avg}{57.17}
\DefMacro{res-mut2e-with-name-ft-lora-codellama-7b-bleu-max}{57.17}
\DefMacro{res-mut2e-with-name-ft-lora-codellama-7b-bleu-min}{57.17}
\DefMacro{res-mut2e-with-name-ft-lora-codellama-7b-code-bleu}{61.62}
\DefMacro{res-mut2e-with-name-ft-lora-codellama-7b-code-bleu-avg}{61.62}
\DefMacro{res-mut2e-with-name-ft-lora-codellama-7b-code-bleu-max}{61.62}
\DefMacro{res-mut2e-with-name-ft-lora-codellama-7b-code-bleu-min}{61.62}
\DefMacro{res-mut2e-with-name-ft-lora-codellama-7b-edit-sim}{82.64}
\DefMacro{res-mut2e-with-name-ft-lora-codellama-7b-edit-sim-avg}{82.64}
\DefMacro{res-mut2e-with-name-ft-lora-codellama-7b-edit-sim-max}{82.64}
\DefMacro{res-mut2e-with-name-ft-lora-codellama-7b-edit-sim-min}{82.64}
\DefMacro{res-mut2e-with-name-ft-lora-codellama-7b-rouge-f}{74.75}
\DefMacro{res-mut2e-with-name-ft-lora-codellama-7b-rouge-f-avg}{74.75}
\DefMacro{res-mut2e-with-name-ft-lora-codellama-7b-rouge-f-max}{74.75}
\DefMacro{res-mut2e-with-name-ft-lora-codellama-7b-rouge-f-min}{74.75}
\DefMacro{res-mut2e-with-name-ft-lora-codellama-7b-rouge-p}{77.17}
\DefMacro{res-mut2e-with-name-ft-lora-codellama-7b-rouge-p-avg}{77.17}
\DefMacro{res-mut2e-with-name-ft-lora-codellama-7b-rouge-p-max}{77.17}
\DefMacro{res-mut2e-with-name-ft-lora-codellama-7b-rouge-p-min}{77.17}
\DefMacro{res-mut2e-with-name-ft-lora-codellama-7b-rouge-r}{74.68}
\DefMacro{res-mut2e-with-name-ft-lora-codellama-7b-rouge-r-avg}{74.68}
\DefMacro{res-mut2e-with-name-ft-lora-codellama-7b-rouge-r-max}{74.68}
\DefMacro{res-mut2e-with-name-ft-lora-codellama-7b-rouge-r-min}{74.68}
\DefMacro{res-mut2e-with-name-ft-lora-codellama-7b-xmatch}{9.62}
\DefMacro{res-mut2e-with-name-ft-lora-codellama-7b-xmatch-top1}{9.62}
\DefMacro{res-mut2e-with-name-ft-lora-codellama-7b-compilable-avg}{62.95}
\DefMacro{res-mut2e-with-name-ft-lora-codellama-7b-compilable-max}{62.95}
\DefMacro{res-mut2e-with-name-ft-lora-codellama-7b-compilable-min}{62.95}
\DefMacro{res-mut2e-with-name-ft-lora-codellama-7b-match-avg}{100.00}
\DefMacro{res-mut2e-with-name-ft-lora-codellama-7b-match-max}{100.00}
\DefMacro{res-mut2e-with-name-ft-lora-codellama-7b-match-min}{100.00}
\DefMacro{res-mut2e-with-name-ft-lora-codellama-7b-runnable-avg}{82.08}
\DefMacro{res-mut2e-with-name-ft-lora-codellama-7b-runnable-max}{82.08}
\DefMacro{res-mut2e-with-name-ft-lora-codellama-7b-runnable-min}{82.08}
\DefMacro{res-mut2e-with-name-ft-lora-codellama-7b-runnable-overall}{51.66}
\DefMacro{res-mut2e-with-name-ft-lora-codellama-7b-timeout-avg}{0.38}
\DefMacro{res-mut2e-with-name-ft-lora-codellama-7b-timeout-max}{0.38}
\DefMacro{res-mut2e-with-name-ft-lora-codellama-7b-timeout-min}{0.38}
\DefMacro{res-ne2e-with-name-ft-lora-codellama-7b-bleu}{61.81}
\DefMacro{res-ne2e-with-name-ft-lora-codellama-7b-bleu-avg}{61.81}
\DefMacro{res-ne2e-with-name-ft-lora-codellama-7b-bleu-max}{61.81}
\DefMacro{res-ne2e-with-name-ft-lora-codellama-7b-bleu-min}{61.81}
\DefMacro{res-ne2e-with-name-ft-lora-codellama-7b-code-bleu}{66.27}
\DefMacro{res-ne2e-with-name-ft-lora-codellama-7b-code-bleu-avg}{66.27}
\DefMacro{res-ne2e-with-name-ft-lora-codellama-7b-code-bleu-max}{66.27}
\DefMacro{res-ne2e-with-name-ft-lora-codellama-7b-code-bleu-min}{66.27}
\DefMacro{res-ne2e-with-name-ft-lora-codellama-7b-edit-sim}{84.86}
\DefMacro{res-ne2e-with-name-ft-lora-codellama-7b-edit-sim-avg}{84.86}
\DefMacro{res-ne2e-with-name-ft-lora-codellama-7b-edit-sim-max}{84.86}
\DefMacro{res-ne2e-with-name-ft-lora-codellama-7b-edit-sim-min}{84.86}
\DefMacro{res-ne2e-with-name-ft-lora-codellama-7b-rouge-f}{79.03}
\DefMacro{res-ne2e-with-name-ft-lora-codellama-7b-rouge-f-avg}{79.03}
\DefMacro{res-ne2e-with-name-ft-lora-codellama-7b-rouge-f-max}{79.03}
\DefMacro{res-ne2e-with-name-ft-lora-codellama-7b-rouge-f-min}{79.03}
\DefMacro{res-ne2e-with-name-ft-lora-codellama-7b-rouge-p}{80.17}
\DefMacro{res-ne2e-with-name-ft-lora-codellama-7b-rouge-p-avg}{80.17}
\DefMacro{res-ne2e-with-name-ft-lora-codellama-7b-rouge-p-max}{80.17}
\DefMacro{res-ne2e-with-name-ft-lora-codellama-7b-rouge-p-min}{80.17}
\DefMacro{res-ne2e-with-name-ft-lora-codellama-7b-rouge-r}{80.00}
\DefMacro{res-ne2e-with-name-ft-lora-codellama-7b-rouge-r-avg}{80.00}
\DefMacro{res-ne2e-with-name-ft-lora-codellama-7b-rouge-r-max}{80.00}
\DefMacro{res-ne2e-with-name-ft-lora-codellama-7b-rouge-r-min}{80.00}
\DefMacro{res-ne2e-with-name-ft-lora-codellama-7b-xmatch}{14.61}
\DefMacro{res-ne2e-with-name-ft-lora-codellama-7b-xmatch-top1}{14.61}
\DefMacro{res-ne2e-with-name-ft-lora-codellama-7b-compilable-avg}{79.93}
\DefMacro{res-ne2e-with-name-ft-lora-codellama-7b-compilable-max}{79.93}
\DefMacro{res-ne2e-with-name-ft-lora-codellama-7b-compilable-min}{79.93}
\DefMacro{res-ne2e-with-name-ft-lora-codellama-7b-match-avg}{100.00}
\DefMacro{res-ne2e-with-name-ft-lora-codellama-7b-match-max}{100.00}
\DefMacro{res-ne2e-with-name-ft-lora-codellama-7b-match-min}{100.00}
\DefMacro{res-ne2e-with-name-ft-lora-codellama-7b-runnable-avg}{76.97}
\DefMacro{res-ne2e-with-name-ft-lora-codellama-7b-runnable-max}{76.97}
\DefMacro{res-ne2e-with-name-ft-lora-codellama-7b-runnable-min}{76.97}
\DefMacro{res-ne2e-with-name-ft-lora-codellama-7b-runnable-overall}{61.52}
\DefMacro{res-ne2e-with-name-ft-lora-codellama-7b-timeout-avg}{0.45}
\DefMacro{res-ne2e-with-name-ft-lora-codellama-7b-timeout-max}{0.45}
\DefMacro{res-ne2e-with-name-ft-lora-codellama-7b-timeout-min}{0.45}
\DefMacro{res-ne2e-all-with-name-ft-lora-codellama-7b-bleu}{61.54}
\DefMacro{res-ne2e-all-with-name-ft-lora-codellama-7b-bleu-avg}{61.95}
\DefMacro{res-ne2e-all-with-name-ft-lora-codellama-7b-bleu-max}{66.44}
\DefMacro{res-ne2e-all-with-name-ft-lora-codellama-7b-bleu-min}{57.69}
\DefMacro{res-ne2e-all-with-name-ft-lora-codellama-7b-code-bleu}{65.90}
\DefMacro{res-ne2e-all-with-name-ft-lora-codellama-7b-code-bleu-avg}{66.25}
\DefMacro{res-ne2e-all-with-name-ft-lora-codellama-7b-code-bleu-max}{70.30}
\DefMacro{res-ne2e-all-with-name-ft-lora-codellama-7b-code-bleu-min}{62.72}
\DefMacro{res-ne2e-all-with-name-ft-lora-codellama-7b-edit-sim}{84.65}
\DefMacro{res-ne2e-all-with-name-ft-lora-codellama-7b-edit-sim-avg}{84.81}
\DefMacro{res-ne2e-all-with-name-ft-lora-codellama-7b-edit-sim-max}{87.66}
\DefMacro{res-ne2e-all-with-name-ft-lora-codellama-7b-edit-sim-min}{81.84}
\DefMacro{res-ne2e-all-with-name-ft-lora-codellama-7b-rouge-f}{79.02}
\DefMacro{res-ne2e-all-with-name-ft-lora-codellama-7b-rouge-f-avg}{79.33}
\DefMacro{res-ne2e-all-with-name-ft-lora-codellama-7b-rouge-f-max}{82.46}
\DefMacro{res-ne2e-all-with-name-ft-lora-codellama-7b-rouge-f-min}{76.15}
\DefMacro{res-ne2e-all-with-name-ft-lora-codellama-7b-rouge-p}{80.23}
\DefMacro{res-ne2e-all-with-name-ft-lora-codellama-7b-rouge-p-avg}{80.66}
\DefMacro{res-ne2e-all-with-name-ft-lora-codellama-7b-rouge-p-max}{84.28}
\DefMacro{res-ne2e-all-with-name-ft-lora-codellama-7b-rouge-p-min}{76.52}
\DefMacro{res-ne2e-all-with-name-ft-lora-codellama-7b-rouge-r}{80.01}
\DefMacro{res-ne2e-all-with-name-ft-lora-codellama-7b-rouge-r-avg}{80.19}
\DefMacro{res-ne2e-all-with-name-ft-lora-codellama-7b-rouge-r-max}{82.98}
\DefMacro{res-ne2e-all-with-name-ft-lora-codellama-7b-rouge-r-min}{77.72}
\DefMacro{res-ne2e-all-with-name-ft-lora-codellama-7b-xmatch}{14.01}
\DefMacro{res-ne2e-all-with-name-ft-lora-codellama-7b-xmatch-avg}{14.70}
\DefMacro{res-ne2e-all-with-name-ft-lora-codellama-7b-xmatch-max}{18.05}
\DefMacro{res-ne2e-all-with-name-ft-lora-codellama-7b-xmatch-min}{12.35}
\DefMacro{res-ne2e-all-with-name-ft-lora-codellama-7b-xmatch-top1}{14.01}
\DefMacro{res-ne2e-all-with-name-ft-lora-codellama-7b-compilable-avg}{80.80}
\DefMacro{res-ne2e-all-with-name-ft-lora-codellama-7b-compilable-max}{87.29}
\DefMacro{res-ne2e-all-with-name-ft-lora-codellama-7b-compilable-min}{73.52}
\DefMacro{res-ne2e-all-with-name-ft-lora-codellama-7b-match-avg}{100.00}
\DefMacro{res-ne2e-all-with-name-ft-lora-codellama-7b-match-max}{100.00}
\DefMacro{res-ne2e-all-with-name-ft-lora-codellama-7b-match-min}{100.00}
\DefMacro{res-ne2e-all-with-name-ft-lora-codellama-7b-runnable-avg}{74.87}
\DefMacro{res-ne2e-all-with-name-ft-lora-codellama-7b-runnable-max}{79.59}
\DefMacro{res-ne2e-all-with-name-ft-lora-codellama-7b-runnable-min}{70.07}
\DefMacro{res-ne2e-all-with-name-ft-lora-codellama-7b-runnable-overall}{69.48}
\DefMacro{res-ne2e-all-with-name-ft-lora-codellama-7b-runnable-overall-avg}{61.74}
\DefMacro{res-ne2e-all-with-name-ft-lora-codellama-7b-runnable-overall-max}{69.48}
\DefMacro{res-ne2e-all-with-name-ft-lora-codellama-7b-runnable-overall-min}{53.80}
\DefMacro{res-ne2e-all-with-name-ft-lora-codellama-7b-timeout-avg}{0.41}
\DefMacro{res-ne2e-all-with-name-ft-lora-codellama-7b-timeout-max}{0.41}
\DefMacro{res-ne2e-all-with-name-ft-lora-codellama-7b-timeout-min}{0.41}
\DefMacro{res-nestack2e-with-name-ft-lora-codellama-7b-real-bleu}{61.17}
\DefMacro{res-nestack2e-with-name-ft-lora-codellama-7b-real-bleu-avg}{61.27}
\DefMacro{res-nestack2e-with-name-ft-lora-codellama-7b-real-bleu-max}{63.40}
\DefMacro{res-nestack2e-with-name-ft-lora-codellama-7b-real-bleu-min}{59.34}
\DefMacro{res-nestack2e-with-name-ft-lora-codellama-7b-real-code-bleu}{65.66}
\DefMacro{res-nestack2e-with-name-ft-lora-codellama-7b-real-code-bleu-avg}{65.70}
\DefMacro{res-nestack2e-with-name-ft-lora-codellama-7b-real-code-bleu-max}{67.51}
\DefMacro{res-nestack2e-with-name-ft-lora-codellama-7b-real-code-bleu-min}{64.11}
\DefMacro{res-nestack2e-with-name-ft-lora-codellama-7b-real-edit-sim}{84.62}
\DefMacro{res-nestack2e-with-name-ft-lora-codellama-7b-real-edit-sim-avg}{84.75}
\DefMacro{res-nestack2e-with-name-ft-lora-codellama-7b-real-edit-sim-max}{86.11}
\DefMacro{res-nestack2e-with-name-ft-lora-codellama-7b-real-edit-sim-min}{83.29}
\DefMacro{res-nestack2e-with-name-ft-lora-codellama-7b-real-rouge-f}{78.67}
\DefMacro{res-nestack2e-with-name-ft-lora-codellama-7b-real-rouge-f-avg}{78.77}
\DefMacro{res-nestack2e-with-name-ft-lora-codellama-7b-real-rouge-f-max}{80.21}
\DefMacro{res-nestack2e-with-name-ft-lora-codellama-7b-real-rouge-f-min}{77.24}
\DefMacro{res-nestack2e-with-name-ft-lora-codellama-7b-real-rouge-p}{79.92}
\DefMacro{res-nestack2e-with-name-ft-lora-codellama-7b-real-rouge-p-avg}{79.94}
\DefMacro{res-nestack2e-with-name-ft-lora-codellama-7b-real-rouge-p-max}{81.61}
\DefMacro{res-nestack2e-with-name-ft-lora-codellama-7b-real-rouge-p-min}{78.08}
\DefMacro{res-nestack2e-with-name-ft-lora-codellama-7b-real-rouge-r}{79.60}
\DefMacro{res-nestack2e-with-name-ft-lora-codellama-7b-real-rouge-r-avg}{79.72}
\DefMacro{res-nestack2e-with-name-ft-lora-codellama-7b-real-rouge-r-max}{80.98}
\DefMacro{res-nestack2e-with-name-ft-lora-codellama-7b-real-rouge-r-min}{78.51}
\DefMacro{res-nestack2e-with-name-ft-lora-codellama-7b-real-xmatch}{13.30}
\DefMacro{res-nestack2e-with-name-ft-lora-codellama-7b-real-xmatch-top1}{13.30}
\DefMacro{res-nestack2e-with-name-ft-lora-codellama-7b-real-compilable-avg}{77.92}
\DefMacro{res-nestack2e-with-name-ft-lora-codellama-7b-real-compilable-max}{80.76}
\DefMacro{res-nestack2e-with-name-ft-lora-codellama-7b-real-compilable-min}{73.75}
\DefMacro{res-nestack2e-with-name-ft-lora-codellama-7b-real-match-avg}{100.00}
\DefMacro{res-nestack2e-with-name-ft-lora-codellama-7b-real-match-max}{100.00}
\DefMacro{res-nestack2e-with-name-ft-lora-codellama-7b-real-match-min}{100.00}
\DefMacro{res-nestack2e-with-name-ft-lora-codellama-7b-real-runnable-avg}{80.85}
\DefMacro{res-nestack2e-with-name-ft-lora-codellama-7b-real-runnable-max}{83.24}
\DefMacro{res-nestack2e-with-name-ft-lora-codellama-7b-real-runnable-min}{77.79}
\DefMacro{res-nestack2e-with-name-ft-lora-codellama-7b-real-runnable-overall}{67.22}
\DefMacro{res-nestack2e-with-name-ft-lora-codellama-7b-real-timeout-avg}{0.44}
\DefMacro{res-nestack2e-with-name-ft-lora-codellama-7b-real-timeout-max}{0.44}
\DefMacro{res-nestack2e-with-name-ft-lora-codellama-7b-real-timeout-min}{0.44}
\DefMacro{res-conditionne2e-with-name-ft-lora-codellama-7b-real-bleu}{62.31}
\DefMacro{res-conditionne2e-with-name-ft-lora-codellama-7b-real-bleu-avg}{62.41}
\DefMacro{res-conditionne2e-with-name-ft-lora-codellama-7b-real-bleu-max}{64.20}
\DefMacro{res-conditionne2e-with-name-ft-lora-codellama-7b-real-bleu-min}{60.61}
\DefMacro{res-conditionne2e-with-name-ft-lora-codellama-7b-real-code-bleu}{66.34}
\DefMacro{res-conditionne2e-with-name-ft-lora-codellama-7b-real-code-bleu-avg}{66.37}
\DefMacro{res-conditionne2e-with-name-ft-lora-codellama-7b-real-code-bleu-max}{68.01}
\DefMacro{res-conditionne2e-with-name-ft-lora-codellama-7b-real-code-bleu-min}{64.81}
\DefMacro{res-conditionne2e-with-name-ft-lora-codellama-7b-real-edit-sim}{85.35}
\DefMacro{res-conditionne2e-with-name-ft-lora-codellama-7b-real-edit-sim-avg}{85.36}
\DefMacro{res-conditionne2e-with-name-ft-lora-codellama-7b-real-edit-sim-max}{86.59}
\DefMacro{res-conditionne2e-with-name-ft-lora-codellama-7b-real-edit-sim-min}{83.90}
\DefMacro{res-conditionne2e-with-name-ft-lora-codellama-7b-real-rouge-f}{79.27}
\DefMacro{res-conditionne2e-with-name-ft-lora-codellama-7b-real-rouge-f-avg}{79.38}
\DefMacro{res-conditionne2e-with-name-ft-lora-codellama-7b-real-rouge-f-max}{80.73}
\DefMacro{res-conditionne2e-with-name-ft-lora-codellama-7b-real-rouge-f-min}{77.97}
\DefMacro{res-conditionne2e-with-name-ft-lora-codellama-7b-real-rouge-p}{80.45}
\DefMacro{res-conditionne2e-with-name-ft-lora-codellama-7b-real-rouge-p-avg}{80.65}
\DefMacro{res-conditionne2e-with-name-ft-lora-codellama-7b-real-rouge-p-max}{82.24}
\DefMacro{res-conditionne2e-with-name-ft-lora-codellama-7b-real-rouge-p-min}{78.90}
\DefMacro{res-conditionne2e-with-name-ft-lora-codellama-7b-real-rouge-r}{79.95}
\DefMacro{res-conditionne2e-with-name-ft-lora-codellama-7b-real-rouge-r-avg}{79.98}
\DefMacro{res-conditionne2e-with-name-ft-lora-codellama-7b-real-rouge-r-max}{81.22}
\DefMacro{res-conditionne2e-with-name-ft-lora-codellama-7b-real-rouge-r-min}{78.79}
\DefMacro{res-conditionne2e-with-name-ft-lora-codellama-7b-real-xmatch}{13.90}
\DefMacro{res-conditionne2e-with-name-ft-lora-codellama-7b-real-xmatch-top1}{13.90}
\DefMacro{res-conditionne2e-with-name-ft-lora-codellama-7b-real-compilable-avg}{79.03}
\DefMacro{res-conditionne2e-with-name-ft-lora-codellama-7b-real-compilable-max}{82.66}
\DefMacro{res-conditionne2e-with-name-ft-lora-codellama-7b-real-compilable-min}{74.70}
\DefMacro{res-conditionne2e-with-name-ft-lora-codellama-7b-real-match-avg}{100.00}
\DefMacro{res-conditionne2e-with-name-ft-lora-codellama-7b-real-match-max}{100.00}
\DefMacro{res-conditionne2e-with-name-ft-lora-codellama-7b-real-match-min}{100.00}
\DefMacro{res-conditionne2e-with-name-ft-lora-codellama-7b-real-runnable-avg}{77.76}
\DefMacro{res-conditionne2e-with-name-ft-lora-codellama-7b-real-runnable-max}{80.89}
\DefMacro{res-conditionne2e-with-name-ft-lora-codellama-7b-real-runnable-min}{74.57}
\DefMacro{res-conditionne2e-with-name-ft-lora-codellama-7b-real-runnable-overall}{66.86}
\DefMacro{res-conditionne2e-with-name-ft-lora-codellama-7b-real-timeout-avg}{0.43}
\DefMacro{res-conditionne2e-with-name-ft-lora-codellama-7b-real-timeout-max}{0.43}
\DefMacro{res-conditionne2e-with-name-ft-lora-codellama-7b-real-timeout-min}{0.43}
\DefMacro{res-conditionnestack2e-with-name-ft-lora-codellama-7b-real-bleu}{61.49}
\DefMacro{res-conditionnestack2e-with-name-ft-lora-codellama-7b-real-bleu-avg}{61.48}
\DefMacro{res-conditionnestack2e-with-name-ft-lora-codellama-7b-real-bleu-max}{64.57}
\DefMacro{res-conditionnestack2e-with-name-ft-lora-codellama-7b-real-bleu-min}{58.54}
\DefMacro{res-conditionnestack2e-with-name-ft-lora-codellama-7b-real-code-bleu}{65.97}
\DefMacro{res-conditionnestack2e-with-name-ft-lora-codellama-7b-real-code-bleu-avg}{65.83}
\DefMacro{res-conditionnestack2e-with-name-ft-lora-codellama-7b-real-code-bleu-max}{68.61}
\DefMacro{res-conditionnestack2e-with-name-ft-lora-codellama-7b-real-code-bleu-min}{63.13}
\DefMacro{res-conditionnestack2e-with-name-ft-lora-codellama-7b-real-edit-sim}{84.35}
\DefMacro{res-conditionnestack2e-with-name-ft-lora-codellama-7b-real-edit-sim-avg}{84.15}
\DefMacro{res-conditionnestack2e-with-name-ft-lora-codellama-7b-real-edit-sim-max}{86.43}
\DefMacro{res-conditionnestack2e-with-name-ft-lora-codellama-7b-real-edit-sim-min}{81.66}
\DefMacro{res-conditionnestack2e-with-name-ft-lora-codellama-7b-real-rouge-f}{78.42}
\DefMacro{res-conditionnestack2e-with-name-ft-lora-codellama-7b-real-rouge-f-avg}{78.32}
\DefMacro{res-conditionnestack2e-with-name-ft-lora-codellama-7b-real-rouge-f-max}{80.87}
\DefMacro{res-conditionnestack2e-with-name-ft-lora-codellama-7b-real-rouge-f-min}{75.64}
\DefMacro{res-conditionnestack2e-with-name-ft-lora-codellama-7b-real-rouge-p}{79.38}
\DefMacro{res-conditionnestack2e-with-name-ft-lora-codellama-7b-real-rouge-p-avg}{79.33}
\DefMacro{res-conditionnestack2e-with-name-ft-lora-codellama-7b-real-rouge-p-max}{82.30}
\DefMacro{res-conditionnestack2e-with-name-ft-lora-codellama-7b-real-rouge-p-min}{76.09}
\DefMacro{res-conditionnestack2e-with-name-ft-lora-codellama-7b-real-rouge-r}{79.46}
\DefMacro{res-conditionnestack2e-with-name-ft-lora-codellama-7b-real-rouge-r-avg}{79.34}
\DefMacro{res-conditionnestack2e-with-name-ft-lora-codellama-7b-real-rouge-r-max}{81.71}
\DefMacro{res-conditionnestack2e-with-name-ft-lora-codellama-7b-real-rouge-r-min}{76.94}
\DefMacro{res-conditionnestack2e-with-name-ft-lora-codellama-7b-real-xmatch}{14.61}
\DefMacro{res-conditionnestack2e-with-name-ft-lora-codellama-7b-real-xmatch-top1}{14.61}
\DefMacro{res-conditionnestack2e-with-name-ft-lora-codellama-7b-real-compilable-avg}{76.12}
\DefMacro{res-conditionnestack2e-with-name-ft-lora-codellama-7b-real-compilable-max}{80.76}
\DefMacro{res-conditionnestack2e-with-name-ft-lora-codellama-7b-real-compilable-min}{71.85}
\DefMacro{res-conditionnestack2e-with-name-ft-lora-codellama-7b-real-match-avg}{100.00}
\DefMacro{res-conditionnestack2e-with-name-ft-lora-codellama-7b-real-match-max}{100.00}
\DefMacro{res-conditionnestack2e-with-name-ft-lora-codellama-7b-real-match-min}{100.00}
\DefMacro{res-conditionnestack2e-with-name-ft-lora-codellama-7b-real-runnable-avg}{78.79}
\DefMacro{res-conditionnestack2e-with-name-ft-lora-codellama-7b-real-runnable-max}{82.21}
\DefMacro{res-conditionnestack2e-with-name-ft-lora-codellama-7b-real-runnable-min}{75.74}
\DefMacro{res-conditionnestack2e-with-name-ft-lora-codellama-7b-real-runnable-overall}{66.39}
\DefMacro{res-conditionnestack2e-with-name-ft-lora-codellama-7b-real-timeout-avg}{0.29}
\DefMacro{res-conditionnestack2e-with-name-ft-lora-codellama-7b-real-timeout-max}{0.29}
\DefMacro{res-conditionnestack2e-with-name-ft-lora-codellama-7b-real-timeout-min}{0.29}

\DefMacro{res-nestack2e-no-name-ft-lora-codellama-7b-bleu}{45.12}
\DefMacro{res-nestack2e-no-name-ft-lora-codellama-7b-bleu-avg}{45.12}
\DefMacro{res-nestack2e-no-name-ft-lora-codellama-7b-bleu-max}{45.12}
\DefMacro{res-nestack2e-no-name-ft-lora-codellama-7b-bleu-min}{45.12}
\DefMacro{res-nestack2e-no-name-ft-lora-codellama-7b-code-bleu}{53.70}
\DefMacro{res-nestack2e-no-name-ft-lora-codellama-7b-code-bleu-avg}{53.70}
\DefMacro{res-nestack2e-no-name-ft-lora-codellama-7b-code-bleu-max}{53.70}
\DefMacro{res-nestack2e-no-name-ft-lora-codellama-7b-code-bleu-min}{53.70}
\DefMacro{res-nestack2e-no-name-ft-lora-codellama-7b-edit-sim}{79.03}
\DefMacro{res-nestack2e-no-name-ft-lora-codellama-7b-edit-sim-avg}{79.03}
\DefMacro{res-nestack2e-no-name-ft-lora-codellama-7b-edit-sim-max}{79.03}
\DefMacro{res-nestack2e-no-name-ft-lora-codellama-7b-edit-sim-min}{79.03}
\DefMacro{res-nestack2e-no-name-ft-lora-codellama-7b-rouge-f}{73.07}
\DefMacro{res-nestack2e-no-name-ft-lora-codellama-7b-rouge-f-avg}{73.07}
\DefMacro{res-nestack2e-no-name-ft-lora-codellama-7b-rouge-f-max}{73.07}
\DefMacro{res-nestack2e-no-name-ft-lora-codellama-7b-rouge-f-min}{73.07}
\DefMacro{res-nestack2e-no-name-ft-lora-codellama-7b-rouge-p}{74.74}
\DefMacro{res-nestack2e-no-name-ft-lora-codellama-7b-rouge-p-avg}{74.74}
\DefMacro{res-nestack2e-no-name-ft-lora-codellama-7b-rouge-p-max}{74.74}
\DefMacro{res-nestack2e-no-name-ft-lora-codellama-7b-rouge-p-min}{74.74}
\DefMacro{res-nestack2e-no-name-ft-lora-codellama-7b-rouge-r}{73.27}
\DefMacro{res-nestack2e-no-name-ft-lora-codellama-7b-rouge-r-avg}{73.27}
\DefMacro{res-nestack2e-no-name-ft-lora-codellama-7b-rouge-r-max}{73.27}
\DefMacro{res-nestack2e-no-name-ft-lora-codellama-7b-rouge-r-min}{73.27}
\DefMacro{res-nestack2e-no-name-ft-lora-codellama-7b-xmatch}{1.31}
\DefMacro{res-nestack2e-no-name-ft-lora-codellama-7b-xmatch-top1}{1.31}
\DefMacro{res-nestack2e-no-name-ft-lora-codellama-7b-compilable-avg}{81.24}
\DefMacro{res-nestack2e-no-name-ft-lora-codellama-7b-compilable-max}{81.24}
\DefMacro{res-nestack2e-no-name-ft-lora-codellama-7b-compilable-min}{81.24}
\DefMacro{res-nestack2e-no-name-ft-lora-codellama-7b-match-avg}{100.00}
\DefMacro{res-nestack2e-no-name-ft-lora-codellama-7b-match-max}{100.00}
\DefMacro{res-nestack2e-no-name-ft-lora-codellama-7b-match-min}{100.00}
\DefMacro{res-nestack2e-no-name-ft-lora-codellama-7b-runnable-avg}{82.60}
\DefMacro{res-nestack2e-no-name-ft-lora-codellama-7b-runnable-max}{82.60}
\DefMacro{res-nestack2e-no-name-ft-lora-codellama-7b-runnable-min}{82.60}
\DefMacro{res-nestack2e-no-name-ft-lora-codellama-7b-runnable-overall}{67.10}
\DefMacro{res-nestack2e-no-name-ft-lora-codellama-7b-timeout-avg}{0.15}
\DefMacro{res-nestack2e-no-name-ft-lora-codellama-7b-timeout-max}{0.15}
\DefMacro{res-nestack2e-no-name-ft-lora-codellama-7b-timeout-min}{0.15}
\DefMacro{res-nestack2e-no-name-ft-lora-codellama-13b-bleu}{45.82}
\DefMacro{res-nestack2e-no-name-ft-lora-codellama-13b-bleu-avg}{45.82}
\DefMacro{res-nestack2e-no-name-ft-lora-codellama-13b-bleu-max}{45.82}
\DefMacro{res-nestack2e-no-name-ft-lora-codellama-13b-bleu-min}{45.82}
\DefMacro{res-nestack2e-no-name-ft-lora-codellama-13b-code-bleu}{54.60}
\DefMacro{res-nestack2e-no-name-ft-lora-codellama-13b-code-bleu-avg}{54.60}
\DefMacro{res-nestack2e-no-name-ft-lora-codellama-13b-code-bleu-max}{54.60}
\DefMacro{res-nestack2e-no-name-ft-lora-codellama-13b-code-bleu-min}{54.60}
\DefMacro{res-nestack2e-no-name-ft-lora-codellama-13b-edit-sim}{79.44}
\DefMacro{res-nestack2e-no-name-ft-lora-codellama-13b-edit-sim-avg}{79.44}
\DefMacro{res-nestack2e-no-name-ft-lora-codellama-13b-edit-sim-max}{79.44}
\DefMacro{res-nestack2e-no-name-ft-lora-codellama-13b-edit-sim-min}{79.44}
\DefMacro{res-nestack2e-no-name-ft-lora-codellama-13b-rouge-f}{73.37}
\DefMacro{res-nestack2e-no-name-ft-lora-codellama-13b-rouge-f-avg}{73.37}
\DefMacro{res-nestack2e-no-name-ft-lora-codellama-13b-rouge-f-max}{73.37}
\DefMacro{res-nestack2e-no-name-ft-lora-codellama-13b-rouge-f-min}{73.37}
\DefMacro{res-nestack2e-no-name-ft-lora-codellama-13b-rouge-p}{74.76}
\DefMacro{res-nestack2e-no-name-ft-lora-codellama-13b-rouge-p-avg}{74.76}
\DefMacro{res-nestack2e-no-name-ft-lora-codellama-13b-rouge-p-max}{74.76}
\DefMacro{res-nestack2e-no-name-ft-lora-codellama-13b-rouge-p-min}{74.76}
\DefMacro{res-nestack2e-no-name-ft-lora-codellama-13b-rouge-r}{73.85}
\DefMacro{res-nestack2e-no-name-ft-lora-codellama-13b-rouge-r-avg}{73.85}
\DefMacro{res-nestack2e-no-name-ft-lora-codellama-13b-rouge-r-max}{73.85}
\DefMacro{res-nestack2e-no-name-ft-lora-codellama-13b-rouge-r-min}{73.85}
\DefMacro{res-nestack2e-no-name-ft-lora-codellama-13b-xmatch}{1.43}
\DefMacro{res-nestack2e-no-name-ft-lora-codellama-13b-xmatch-top1}{1.43}
\DefMacro{res-nestack2e-no-name-ft-lora-codellama-13b-compilable-avg}{81.24}
\DefMacro{res-nestack2e-no-name-ft-lora-codellama-13b-compilable-max}{81.24}
\DefMacro{res-nestack2e-no-name-ft-lora-codellama-13b-compilable-min}{81.24}
\DefMacro{res-nestack2e-no-name-ft-lora-codellama-13b-match-avg}{100.00}
\DefMacro{res-nestack2e-no-name-ft-lora-codellama-13b-match-max}{100.00}
\DefMacro{res-nestack2e-no-name-ft-lora-codellama-13b-match-min}{100.00}
\DefMacro{res-nestack2e-no-name-ft-lora-codellama-13b-runnable-avg}{85.53}
\DefMacro{res-nestack2e-no-name-ft-lora-codellama-13b-runnable-max}{85.53}
\DefMacro{res-nestack2e-no-name-ft-lora-codellama-13b-runnable-min}{85.53}
\DefMacro{res-nestack2e-no-name-ft-lora-codellama-13b-runnable-overall}{69.48}
\DefMacro{res-nestack2e-no-name-ft-lora-codellama-13b-runnable-overall-avg}{69.48}
\DefMacro{res-nestack2e-no-name-ft-lora-codellama-13b-runnable-overall-max}{69.48}
\DefMacro{res-nestack2e-no-name-ft-lora-codellama-13b-runnable-overall-min}{69.48}
\DefMacro{res-nestack2e-no-name-ft-lora-codellama-13b-timeout-avg}{0.00}
\DefMacro{res-nestack2e-no-name-ft-lora-codellama-13b-timeout-max}{0.00}
\DefMacro{res-nestack2e-no-name-ft-lora-codellama-13b-timeout-min}{0.00}
\DefMacro{res-conditionne2e-no-name-ft-lora-codellama-7b-bleu}{45.46}
\DefMacro{res-conditionne2e-no-name-ft-lora-codellama-7b-bleu-avg}{45.46}
\DefMacro{res-conditionne2e-no-name-ft-lora-codellama-7b-bleu-max}{45.46}
\DefMacro{res-conditionne2e-no-name-ft-lora-codellama-7b-bleu-min}{45.46}
\DefMacro{res-conditionne2e-no-name-ft-lora-codellama-7b-code-bleu}{53.49}
\DefMacro{res-conditionne2e-no-name-ft-lora-codellama-7b-code-bleu-avg}{53.49}
\DefMacro{res-conditionne2e-no-name-ft-lora-codellama-7b-code-bleu-max}{53.49}
\DefMacro{res-conditionne2e-no-name-ft-lora-codellama-7b-code-bleu-min}{53.49}
\DefMacro{res-conditionne2e-no-name-ft-lora-codellama-7b-edit-sim}{79.40}
\DefMacro{res-conditionne2e-no-name-ft-lora-codellama-7b-edit-sim-avg}{79.40}
\DefMacro{res-conditionne2e-no-name-ft-lora-codellama-7b-edit-sim-max}{79.40}
\DefMacro{res-conditionne2e-no-name-ft-lora-codellama-7b-edit-sim-min}{79.40}
\DefMacro{res-conditionne2e-no-name-ft-lora-codellama-7b-rouge-f}{73.33}
\DefMacro{res-conditionne2e-no-name-ft-lora-codellama-7b-rouge-f-avg}{73.33}
\DefMacro{res-conditionne2e-no-name-ft-lora-codellama-7b-rouge-f-max}{73.33}
\DefMacro{res-conditionne2e-no-name-ft-lora-codellama-7b-rouge-f-min}{73.33}
\DefMacro{res-conditionne2e-no-name-ft-lora-codellama-7b-rouge-p}{75.07}
\DefMacro{res-conditionne2e-no-name-ft-lora-codellama-7b-rouge-p-avg}{75.07}
\DefMacro{res-conditionne2e-no-name-ft-lora-codellama-7b-rouge-p-max}{75.07}
\DefMacro{res-conditionne2e-no-name-ft-lora-codellama-7b-rouge-p-min}{75.07}
\DefMacro{res-conditionne2e-no-name-ft-lora-codellama-7b-rouge-r}{73.36}
\DefMacro{res-conditionne2e-no-name-ft-lora-codellama-7b-rouge-r-avg}{73.36}
\DefMacro{res-conditionne2e-no-name-ft-lora-codellama-7b-rouge-r-max}{73.36}
\DefMacro{res-conditionne2e-no-name-ft-lora-codellama-7b-rouge-r-min}{73.36}
\DefMacro{res-conditionne2e-no-name-ft-lora-codellama-7b-xmatch}{1.07}
\DefMacro{res-conditionne2e-no-name-ft-lora-codellama-7b-xmatch-top1}{1.07}
\DefMacro{res-conditionne2e-no-name-ft-lora-codellama-7b-compilable-avg}{81.00}
\DefMacro{res-conditionne2e-no-name-ft-lora-codellama-7b-compilable-max}{81.00}
\DefMacro{res-conditionne2e-no-name-ft-lora-codellama-7b-compilable-min}{81.00}
\DefMacro{res-conditionne2e-no-name-ft-lora-codellama-7b-match-avg}{100.00}
\DefMacro{res-conditionne2e-no-name-ft-lora-codellama-7b-match-max}{100.00}
\DefMacro{res-conditionne2e-no-name-ft-lora-codellama-7b-match-min}{100.00}
\DefMacro{res-conditionne2e-no-name-ft-lora-codellama-7b-runnable-avg}{80.65}
\DefMacro{res-conditionne2e-no-name-ft-lora-codellama-7b-runnable-max}{80.65}
\DefMacro{res-conditionne2e-no-name-ft-lora-codellama-7b-runnable-min}{80.65}
\DefMacro{res-conditionne2e-no-name-ft-lora-codellama-7b-runnable-overall}{65.32}
\DefMacro{res-conditionne2e-no-name-ft-lora-codellama-7b-timeout-avg}{0.00}
\DefMacro{res-conditionne2e-no-name-ft-lora-codellama-7b-timeout-max}{0.00}
\DefMacro{res-conditionne2e-no-name-ft-lora-codellama-7b-timeout-min}{0.00}
\DefMacro{res-conditionnestack2e-no-name-ft-lora-codellama-7b-bleu}{45.63}
\DefMacro{res-conditionnestack2e-no-name-ft-lora-codellama-7b-bleu-avg}{45.63}
\DefMacro{res-conditionnestack2e-no-name-ft-lora-codellama-7b-bleu-max}{45.63}
\DefMacro{res-conditionnestack2e-no-name-ft-lora-codellama-7b-bleu-min}{45.63}
\DefMacro{res-conditionnestack2e-no-name-ft-lora-codellama-7b-code-bleu}{53.74}
\DefMacro{res-conditionnestack2e-no-name-ft-lora-codellama-7b-code-bleu-avg}{53.74}
\DefMacro{res-conditionnestack2e-no-name-ft-lora-codellama-7b-code-bleu-max}{53.74}
\DefMacro{res-conditionnestack2e-no-name-ft-lora-codellama-7b-code-bleu-min}{53.74}
\DefMacro{res-conditionnestack2e-no-name-ft-lora-codellama-7b-edit-sim}{79.53}
\DefMacro{res-conditionnestack2e-no-name-ft-lora-codellama-7b-edit-sim-avg}{79.53}
\DefMacro{res-conditionnestack2e-no-name-ft-lora-codellama-7b-edit-sim-max}{79.53}
\DefMacro{res-conditionnestack2e-no-name-ft-lora-codellama-7b-edit-sim-min}{79.53}
\DefMacro{res-conditionnestack2e-no-name-ft-lora-codellama-7b-rouge-f}{73.49}
\DefMacro{res-conditionnestack2e-no-name-ft-lora-codellama-7b-rouge-f-avg}{73.49}
\DefMacro{res-conditionnestack2e-no-name-ft-lora-codellama-7b-rouge-f-max}{73.49}
\DefMacro{res-conditionnestack2e-no-name-ft-lora-codellama-7b-rouge-f-min}{73.49}
\DefMacro{res-conditionnestack2e-no-name-ft-lora-codellama-7b-rouge-p}{75.53}
\DefMacro{res-conditionnestack2e-no-name-ft-lora-codellama-7b-rouge-p-avg}{75.53}
\DefMacro{res-conditionnestack2e-no-name-ft-lora-codellama-7b-rouge-p-max}{75.53}
\DefMacro{res-conditionnestack2e-no-name-ft-lora-codellama-7b-rouge-p-min}{75.53}
\DefMacro{res-conditionnestack2e-no-name-ft-lora-codellama-7b-rouge-r}{73.31}
\DefMacro{res-conditionnestack2e-no-name-ft-lora-codellama-7b-rouge-r-avg}{73.31}
\DefMacro{res-conditionnestack2e-no-name-ft-lora-codellama-7b-rouge-r-max}{73.31}
\DefMacro{res-conditionnestack2e-no-name-ft-lora-codellama-7b-rouge-r-min}{73.31}
\DefMacro{res-conditionnestack2e-no-name-ft-lora-codellama-7b-xmatch}{1.07}
\DefMacro{res-conditionnestack2e-no-name-ft-lora-codellama-7b-xmatch-top1}{1.07}
\DefMacro{res-conditionnestack2e-no-name-ft-lora-codellama-7b-compilable-avg}{81.00}
\DefMacro{res-conditionnestack2e-no-name-ft-lora-codellama-7b-compilable-max}{81.00}
\DefMacro{res-conditionnestack2e-no-name-ft-lora-codellama-7b-compilable-min}{81.00}
\DefMacro{res-conditionnestack2e-no-name-ft-lora-codellama-7b-match-avg}{100.00}
\DefMacro{res-conditionnestack2e-no-name-ft-lora-codellama-7b-match-max}{100.00}
\DefMacro{res-conditionnestack2e-no-name-ft-lora-codellama-7b-match-min}{100.00}
\DefMacro{res-conditionnestack2e-no-name-ft-lora-codellama-7b-runnable-avg}{85.92}
\DefMacro{res-conditionnestack2e-no-name-ft-lora-codellama-7b-runnable-max}{85.92}
\DefMacro{res-conditionnestack2e-no-name-ft-lora-codellama-7b-runnable-min}{85.92}
\DefMacro{res-conditionnestack2e-no-name-ft-lora-codellama-7b-runnable-overall}{69.60}
\DefMacro{res-conditionnestack2e-no-name-ft-lora-codellama-7b-timeout-avg}{0.29}
\DefMacro{res-conditionnestack2e-no-name-ft-lora-codellama-7b-timeout-max}{0.29}
\DefMacro{res-conditionnestack2e-no-name-ft-lora-codellama-7b-timeout-min}{0.29}

\DefMacro{res-nestack2e-with-name-ft-lora-codellama-7b-bleu}{63.19}
\DefMacro{res-nestack2e-with-name-ft-lora-codellama-7b-bleu-avg}{63.19}
\DefMacro{res-nestack2e-with-name-ft-lora-codellama-7b-bleu-max}{63.19}
\DefMacro{res-nestack2e-with-name-ft-lora-codellama-7b-bleu-min}{63.19}
\DefMacro{res-nestack2e-with-name-ft-lora-codellama-7b-code-bleu}{67.34}
\DefMacro{res-nestack2e-with-name-ft-lora-codellama-7b-code-bleu-avg}{67.34}
\DefMacro{res-nestack2e-with-name-ft-lora-codellama-7b-code-bleu-max}{67.34}
\DefMacro{res-nestack2e-with-name-ft-lora-codellama-7b-code-bleu-min}{67.34}
\DefMacro{res-nestack2e-with-name-ft-lora-codellama-7b-edit-sim}{85.39}
\DefMacro{res-nestack2e-with-name-ft-lora-codellama-7b-edit-sim-avg}{85.39}
\DefMacro{res-nestack2e-with-name-ft-lora-codellama-7b-edit-sim-max}{85.39}
\DefMacro{res-nestack2e-with-name-ft-lora-codellama-7b-edit-sim-min}{85.39}
\DefMacro{res-nestack2e-with-name-ft-lora-codellama-7b-rouge-f}{80.22}
\DefMacro{res-nestack2e-with-name-ft-lora-codellama-7b-rouge-f-avg}{80.22}
\DefMacro{res-nestack2e-with-name-ft-lora-codellama-7b-rouge-f-max}{80.22}
\DefMacro{res-nestack2e-with-name-ft-lora-codellama-7b-rouge-f-min}{80.22}
\DefMacro{res-nestack2e-with-name-ft-lora-codellama-7b-rouge-p}{81.65}
\DefMacro{res-nestack2e-with-name-ft-lora-codellama-7b-rouge-p-avg}{81.65}
\DefMacro{res-nestack2e-with-name-ft-lora-codellama-7b-rouge-p-max}{81.65}
\DefMacro{res-nestack2e-with-name-ft-lora-codellama-7b-rouge-p-min}{81.65}
\DefMacro{res-nestack2e-with-name-ft-lora-codellama-7b-rouge-r}{80.73}
\DefMacro{res-nestack2e-with-name-ft-lora-codellama-7b-rouge-r-avg}{80.73}
\DefMacro{res-nestack2e-with-name-ft-lora-codellama-7b-rouge-r-max}{80.73}
\DefMacro{res-nestack2e-with-name-ft-lora-codellama-7b-rouge-r-min}{80.73}
\DefMacro{res-nestack2e-with-name-ft-lora-codellama-7b-xmatch}{14.96}
\DefMacro{res-nestack2e-with-name-ft-lora-codellama-7b-xmatch-top1}{14.96}
\DefMacro{res-nestack2e-with-name-ft-lora-codellama-7b-compilable-avg}{80.17}
\DefMacro{res-nestack2e-with-name-ft-lora-codellama-7b-compilable-max}{80.17}
\DefMacro{res-nestack2e-with-name-ft-lora-codellama-7b-compilable-min}{80.17}
\DefMacro{res-nestack2e-with-name-ft-lora-codellama-7b-match-avg}{100.00}
\DefMacro{res-nestack2e-with-name-ft-lora-codellama-7b-match-max}{100.00}
\DefMacro{res-nestack2e-with-name-ft-lora-codellama-7b-match-min}{100.00}
\DefMacro{res-nestack2e-with-name-ft-lora-codellama-7b-runnable-avg}{83.26}
\DefMacro{res-nestack2e-with-name-ft-lora-codellama-7b-runnable-max}{83.26}
\DefMacro{res-nestack2e-with-name-ft-lora-codellama-7b-runnable-min}{83.26}
\DefMacro{res-nestack2e-with-name-ft-lora-codellama-7b-runnable-overall}{66.75}
\DefMacro{res-nestack2e-with-name-ft-lora-codellama-7b-timeout-avg}{0.30}
\DefMacro{res-nestack2e-with-name-ft-lora-codellama-7b-timeout-max}{0.30}
\DefMacro{res-nestack2e-with-name-ft-lora-codellama-7b-timeout-min}{0.30}
\DefMacro{res-conditionne2e-with-name-ft-lora-codellama-7b-bleu}{63.69}
\DefMacro{res-conditionne2e-with-name-ft-lora-codellama-7b-bleu-avg}{63.69}
\DefMacro{res-conditionne2e-with-name-ft-lora-codellama-7b-bleu-max}{63.69}
\DefMacro{res-conditionne2e-with-name-ft-lora-codellama-7b-bleu-min}{63.69}
\DefMacro{res-conditionne2e-with-name-ft-lora-codellama-7b-code-bleu}{67.49}
\DefMacro{res-conditionne2e-with-name-ft-lora-codellama-7b-code-bleu-avg}{67.49}
\DefMacro{res-conditionne2e-with-name-ft-lora-codellama-7b-code-bleu-max}{67.49}
\DefMacro{res-conditionne2e-with-name-ft-lora-codellama-7b-code-bleu-min}{67.49}
\DefMacro{res-conditionne2e-with-name-ft-lora-codellama-7b-edit-sim}{86.43}
\DefMacro{res-conditionne2e-with-name-ft-lora-codellama-7b-edit-sim-avg}{86.43}
\DefMacro{res-conditionne2e-with-name-ft-lora-codellama-7b-edit-sim-max}{86.43}
\DefMacro{res-conditionne2e-with-name-ft-lora-codellama-7b-edit-sim-min}{86.43}
\DefMacro{res-conditionne2e-with-name-ft-lora-codellama-7b-rouge-f}{80.60}
\DefMacro{res-conditionne2e-with-name-ft-lora-codellama-7b-rouge-f-avg}{80.60}
\DefMacro{res-conditionne2e-with-name-ft-lora-codellama-7b-rouge-f-max}{80.60}
\DefMacro{res-conditionne2e-with-name-ft-lora-codellama-7b-rouge-f-min}{80.60}
\DefMacro{res-conditionne2e-with-name-ft-lora-codellama-7b-rouge-p}{82.11}
\DefMacro{res-conditionne2e-with-name-ft-lora-codellama-7b-rouge-p-avg}{82.11}
\DefMacro{res-conditionne2e-with-name-ft-lora-codellama-7b-rouge-p-max}{82.11}
\DefMacro{res-conditionne2e-with-name-ft-lora-codellama-7b-rouge-p-min}{82.11}
\DefMacro{res-conditionne2e-with-name-ft-lora-codellama-7b-rouge-r}{80.82}
\DefMacro{res-conditionne2e-with-name-ft-lora-codellama-7b-rouge-r-avg}{80.82}
\DefMacro{res-conditionne2e-with-name-ft-lora-codellama-7b-rouge-r-max}{80.82}
\DefMacro{res-conditionne2e-with-name-ft-lora-codellama-7b-rouge-r-min}{80.82}
\DefMacro{res-conditionne2e-with-name-ft-lora-codellama-7b-xmatch}{14.13}
\DefMacro{res-conditionne2e-with-name-ft-lora-codellama-7b-xmatch-top1}{14.13}
\DefMacro{res-conditionne2e-with-name-ft-lora-codellama-7b-compilable-avg}{78.98}
\DefMacro{res-conditionne2e-with-name-ft-lora-codellama-7b-compilable-max}{78.98}
\DefMacro{res-conditionne2e-with-name-ft-lora-codellama-7b-compilable-min}{78.98}
\DefMacro{res-conditionne2e-with-name-ft-lora-codellama-7b-match-avg}{100.00}
\DefMacro{res-conditionne2e-with-name-ft-lora-codellama-7b-match-max}{100.00}
\DefMacro{res-conditionne2e-with-name-ft-lora-codellama-7b-match-min}{100.00}
\DefMacro{res-conditionne2e-with-name-ft-lora-codellama-7b-runnable-avg}{82.56}
\DefMacro{res-conditionne2e-with-name-ft-lora-codellama-7b-runnable-max}{82.56}
\DefMacro{res-conditionne2e-with-name-ft-lora-codellama-7b-runnable-min}{82.56}
\DefMacro{res-conditionne2e-with-name-ft-lora-codellama-7b-runnable-overall}{65.20}
\DefMacro{res-conditionne2e-with-name-ft-lora-codellama-7b-timeout-avg}{0.30}
\DefMacro{res-conditionne2e-with-name-ft-lora-codellama-7b-timeout-max}{0.30}
\DefMacro{res-conditionne2e-with-name-ft-lora-codellama-7b-timeout-min}{0.30}
\DefMacro{res-conditionnestack2e-with-name-ft-lora-codellama-7b-bleu}{62.41}
\DefMacro{res-conditionnestack2e-with-name-ft-lora-codellama-7b-bleu-avg}{62.41}
\DefMacro{res-conditionnestack2e-with-name-ft-lora-codellama-7b-bleu-max}{62.41}
\DefMacro{res-conditionnestack2e-with-name-ft-lora-codellama-7b-bleu-min}{62.41}
\DefMacro{res-conditionnestack2e-with-name-ft-lora-codellama-7b-code-bleu}{66.77}
\DefMacro{res-conditionnestack2e-with-name-ft-lora-codellama-7b-code-bleu-avg}{66.77}
\DefMacro{res-conditionnestack2e-with-name-ft-lora-codellama-7b-code-bleu-max}{66.77}
\DefMacro{res-conditionnestack2e-with-name-ft-lora-codellama-7b-code-bleu-min}{66.77}
\DefMacro{res-conditionnestack2e-with-name-ft-lora-codellama-7b-edit-sim}{85.37}
\DefMacro{res-conditionnestack2e-with-name-ft-lora-codellama-7b-edit-sim-avg}{85.37}
\DefMacro{res-conditionnestack2e-with-name-ft-lora-codellama-7b-edit-sim-max}{85.37}
\DefMacro{res-conditionnestack2e-with-name-ft-lora-codellama-7b-edit-sim-min}{85.37}
\DefMacro{res-conditionnestack2e-with-name-ft-lora-codellama-7b-rouge-f}{79.95}
\DefMacro{res-conditionnestack2e-with-name-ft-lora-codellama-7b-rouge-f-avg}{79.95}
\DefMacro{res-conditionnestack2e-with-name-ft-lora-codellama-7b-rouge-f-max}{79.95}
\DefMacro{res-conditionnestack2e-with-name-ft-lora-codellama-7b-rouge-f-min}{79.95}
\DefMacro{res-conditionnestack2e-with-name-ft-lora-codellama-7b-rouge-p}{80.99}
\DefMacro{res-conditionnestack2e-with-name-ft-lora-codellama-7b-rouge-p-avg}{80.99}
\DefMacro{res-conditionnestack2e-with-name-ft-lora-codellama-7b-rouge-p-max}{80.99}
\DefMacro{res-conditionnestack2e-with-name-ft-lora-codellama-7b-rouge-p-min}{80.99}
\DefMacro{res-conditionnestack2e-with-name-ft-lora-codellama-7b-rouge-r}{80.97}
\DefMacro{res-conditionnestack2e-with-name-ft-lora-codellama-7b-rouge-r-avg}{80.97}
\DefMacro{res-conditionnestack2e-with-name-ft-lora-codellama-7b-rouge-r-max}{80.97}
\DefMacro{res-conditionnestack2e-with-name-ft-lora-codellama-7b-rouge-r-min}{80.97}
\DefMacro{res-conditionnestack2e-with-name-ft-lora-codellama-7b-xmatch}{14.61}
\DefMacro{res-conditionnestack2e-with-name-ft-lora-codellama-7b-xmatch-top1}{14.61}
\DefMacro{res-conditionnestack2e-with-name-ft-lora-codellama-7b-compilable-avg}{79.22}
\DefMacro{res-conditionnestack2e-with-name-ft-lora-codellama-7b-compilable-max}{79.22}
\DefMacro{res-conditionnestack2e-with-name-ft-lora-codellama-7b-compilable-min}{79.22}
\DefMacro{res-conditionnestack2e-with-name-ft-lora-codellama-7b-match-avg}{100.00}
\DefMacro{res-conditionnestack2e-with-name-ft-lora-codellama-7b-match-max}{100.00}
\DefMacro{res-conditionnestack2e-with-name-ft-lora-codellama-7b-match-min}{100.00}
\DefMacro{res-conditionnestack2e-with-name-ft-lora-codellama-7b-runnable-avg}{84.41}
\DefMacro{res-conditionnestack2e-with-name-ft-lora-codellama-7b-runnable-max}{84.41}
\DefMacro{res-conditionnestack2e-with-name-ft-lora-codellama-7b-runnable-min}{84.41}
\DefMacro{res-conditionnestack2e-with-name-ft-lora-codellama-7b-runnable-overall}{66.86}
\DefMacro{res-conditionnestack2e-with-name-ft-lora-codellama-7b-timeout-avg}{0.30}
\DefMacro{res-conditionnestack2e-with-name-ft-lora-codellama-7b-timeout-max}{0.30}
\DefMacro{res-conditionnestack2e-with-name-ft-lora-codellama-7b-timeout-min}{0.30}

\DefMacro{ds-eval-count-methods-with-ts}{2,291}
\DefMacro{ds-eval-count-methods-with-ts-public}{1,723}
\DefMacro{ds-eval-count-public-ts}{2,580}
\DefMacro{ds-eval-count-ts}{3,419}
\DefMacro{ds-eval-mean-ts-per-method}{1.49}

\DefMacro{res-selected-434-ne2e-few-shot-no-name-gpt-3.5-turbo-16k-bleu}{31.12}
\DefMacro{res-selected-434-ne2e-few-shot-no-name-gpt-3.5-turbo-16k-bleu-avg}{31.12}
\DefMacro{res-selected-434-ne2e-few-shot-no-name-gpt-3.5-turbo-16k-bleu-max}{31.12}
\DefMacro{res-selected-434-ne2e-few-shot-no-name-gpt-3.5-turbo-16k-bleu-min}{31.12}
\DefMacro{res-selected-434-ne2e-few-shot-no-name-gpt-3.5-turbo-16k-code-bleu}{41.14}
\DefMacro{res-selected-434-ne2e-few-shot-no-name-gpt-3.5-turbo-16k-code-bleu-avg}{41.14}
\DefMacro{res-selected-434-ne2e-few-shot-no-name-gpt-3.5-turbo-16k-code-bleu-max}{41.14}
\DefMacro{res-selected-434-ne2e-few-shot-no-name-gpt-3.5-turbo-16k-code-bleu-min}{41.14}
\DefMacro{res-selected-434-ne2e-few-shot-no-name-gpt-3.5-turbo-16k-edit-sim}{64.34}
\DefMacro{res-selected-434-ne2e-few-shot-no-name-gpt-3.5-turbo-16k-edit-sim-avg}{64.34}
\DefMacro{res-selected-434-ne2e-few-shot-no-name-gpt-3.5-turbo-16k-edit-sim-max}{64.34}
\DefMacro{res-selected-434-ne2e-few-shot-no-name-gpt-3.5-turbo-16k-edit-sim-min}{64.34}
\DefMacro{res-selected-434-ne2e-few-shot-no-name-gpt-3.5-turbo-16k-rouge-f}{58.14}
\DefMacro{res-selected-434-ne2e-few-shot-no-name-gpt-3.5-turbo-16k-rouge-f-avg}{58.14}
\DefMacro{res-selected-434-ne2e-few-shot-no-name-gpt-3.5-turbo-16k-rouge-f-max}{58.14}
\DefMacro{res-selected-434-ne2e-few-shot-no-name-gpt-3.5-turbo-16k-rouge-f-min}{58.14}
\DefMacro{res-selected-434-ne2e-few-shot-no-name-gpt-3.5-turbo-16k-rouge-p}{58.47}
\DefMacro{res-selected-434-ne2e-few-shot-no-name-gpt-3.5-turbo-16k-rouge-p-avg}{58.47}
\DefMacro{res-selected-434-ne2e-few-shot-no-name-gpt-3.5-turbo-16k-rouge-p-max}{58.47}
\DefMacro{res-selected-434-ne2e-few-shot-no-name-gpt-3.5-turbo-16k-rouge-p-min}{58.47}
\DefMacro{res-selected-434-ne2e-few-shot-no-name-gpt-3.5-turbo-16k-rouge-r}{61.56}
\DefMacro{res-selected-434-ne2e-few-shot-no-name-gpt-3.5-turbo-16k-rouge-r-avg}{61.56}
\DefMacro{res-selected-434-ne2e-few-shot-no-name-gpt-3.5-turbo-16k-rouge-r-max}{61.56}
\DefMacro{res-selected-434-ne2e-few-shot-no-name-gpt-3.5-turbo-16k-rouge-r-min}{61.56}
\DefMacro{res-selected-434-ne2e-few-shot-no-name-gpt-3.5-turbo-16k-xmatch}{0.46}
\DefMacro{res-selected-434-ne2e-few-shot-no-name-gpt-3.5-turbo-16k-xmatch-top1}{0.46}
\DefMacro{res-selected-434-ne2e-few-shot-no-name-gpt-3.5-turbo-16k-compilable-avg}{56.45}
\DefMacro{res-selected-434-ne2e-few-shot-no-name-gpt-3.5-turbo-16k-compilable-max}{56.45}
\DefMacro{res-selected-434-ne2e-few-shot-no-name-gpt-3.5-turbo-16k-compilable-min}{56.45}
\DefMacro{res-selected-434-ne2e-few-shot-no-name-gpt-3.5-turbo-16k-coverage-avg}{23.27}
\DefMacro{res-selected-434-ne2e-few-shot-no-name-gpt-3.5-turbo-16k-coverage-max}{23.27}
\DefMacro{res-selected-434-ne2e-few-shot-no-name-gpt-3.5-turbo-16k-coverage-min}{23.27}
\DefMacro{res-selected-434-ne2e-few-shot-no-name-gpt-3.5-turbo-16k-match-avg}{97.96}
\DefMacro{res-selected-434-ne2e-few-shot-no-name-gpt-3.5-turbo-16k-match-max}{97.96}
\DefMacro{res-selected-434-ne2e-few-shot-no-name-gpt-3.5-turbo-16k-match-min}{97.96}
\DefMacro{res-selected-434-ne2e-few-shot-no-name-gpt-3.5-turbo-16k-runnable-avg}{65.00}
\DefMacro{res-selected-434-ne2e-few-shot-no-name-gpt-3.5-turbo-16k-runnable-max}{65.00}
\DefMacro{res-selected-434-ne2e-few-shot-no-name-gpt-3.5-turbo-16k-runnable-min}{65.00}
\DefMacro{res-selected-434-ne2e-few-shot-no-name-gpt-3.5-turbo-16k-runnable-overall}{35.94}
\DefMacro{res-selected-434-ne2e-few-shot-no-name-gpt-3.5-turbo-16k-runnable-overall-avg}{35.94}
\DefMacro{res-selected-434-ne2e-few-shot-no-name-gpt-3.5-turbo-16k-runnable-overall-max}{35.94}
\DefMacro{res-selected-434-ne2e-few-shot-no-name-gpt-3.5-turbo-16k-runnable-overall-min}{35.94}
\DefMacro{res-selected-434-ne2e-few-shot-no-name-gpt-3.5-turbo-16k-timeout-avg}{0.42}
\DefMacro{res-selected-434-ne2e-few-shot-no-name-gpt-3.5-turbo-16k-timeout-max}{0.42}
\DefMacro{res-selected-434-ne2e-few-shot-no-name-gpt-3.5-turbo-16k-timeout-min}{0.42}
\DefMacro{res-selected-434-mut2e-no-name-ft-lora-codellama-7b-bleu}{38.92}
\DefMacro{res-selected-434-mut2e-no-name-ft-lora-codellama-7b-bleu-avg}{38.92}
\DefMacro{res-selected-434-mut2e-no-name-ft-lora-codellama-7b-bleu-max}{38.92}
\DefMacro{res-selected-434-mut2e-no-name-ft-lora-codellama-7b-bleu-min}{38.92}
\DefMacro{res-selected-434-mut2e-no-name-ft-lora-codellama-7b-code-bleu}{47.70}
\DefMacro{res-selected-434-mut2e-no-name-ft-lora-codellama-7b-code-bleu-avg}{47.70}
\DefMacro{res-selected-434-mut2e-no-name-ft-lora-codellama-7b-code-bleu-max}{47.70}
\DefMacro{res-selected-434-mut2e-no-name-ft-lora-codellama-7b-code-bleu-min}{47.70}
\DefMacro{res-selected-434-mut2e-no-name-ft-lora-codellama-7b-edit-sim}{73.06}
\DefMacro{res-selected-434-mut2e-no-name-ft-lora-codellama-7b-edit-sim-avg}{73.06}
\DefMacro{res-selected-434-mut2e-no-name-ft-lora-codellama-7b-edit-sim-max}{73.06}
\DefMacro{res-selected-434-mut2e-no-name-ft-lora-codellama-7b-edit-sim-min}{73.06}
\DefMacro{res-selected-434-mut2e-no-name-ft-lora-codellama-7b-rouge-f}{66.41}
\DefMacro{res-selected-434-mut2e-no-name-ft-lora-codellama-7b-rouge-f-avg}{66.41}
\DefMacro{res-selected-434-mut2e-no-name-ft-lora-codellama-7b-rouge-f-max}{66.41}
\DefMacro{res-selected-434-mut2e-no-name-ft-lora-codellama-7b-rouge-f-min}{66.41}
\DefMacro{res-selected-434-mut2e-no-name-ft-lora-codellama-7b-rouge-p}{69.41}
\DefMacro{res-selected-434-mut2e-no-name-ft-lora-codellama-7b-rouge-p-avg}{69.41}
\DefMacro{res-selected-434-mut2e-no-name-ft-lora-codellama-7b-rouge-p-max}{69.41}
\DefMacro{res-selected-434-mut2e-no-name-ft-lora-codellama-7b-rouge-p-min}{69.41}
\DefMacro{res-selected-434-mut2e-no-name-ft-lora-codellama-7b-rouge-r}{65.64}
\DefMacro{res-selected-434-mut2e-no-name-ft-lora-codellama-7b-rouge-r-avg}{65.64}
\DefMacro{res-selected-434-mut2e-no-name-ft-lora-codellama-7b-rouge-r-max}{65.64}
\DefMacro{res-selected-434-mut2e-no-name-ft-lora-codellama-7b-rouge-r-min}{65.64}
\DefMacro{res-selected-434-mut2e-no-name-ft-lora-codellama-7b-xmatch}{0.46}
\DefMacro{res-selected-434-mut2e-no-name-ft-lora-codellama-7b-xmatch-top1}{0.46}
\DefMacro{res-selected-434-mut2e-no-name-ft-lora-codellama-7b-compilable-avg}{63.13}
\DefMacro{res-selected-434-mut2e-no-name-ft-lora-codellama-7b-compilable-max}{63.13}
\DefMacro{res-selected-434-mut2e-no-name-ft-lora-codellama-7b-compilable-min}{63.13}
\DefMacro{res-selected-434-mut2e-no-name-ft-lora-codellama-7b-coverage-avg}{32.72}
\DefMacro{res-selected-434-mut2e-no-name-ft-lora-codellama-7b-coverage-max}{32.72}
\DefMacro{res-selected-434-mut2e-no-name-ft-lora-codellama-7b-coverage-min}{32.72}
\DefMacro{res-selected-434-mut2e-no-name-ft-lora-codellama-7b-match-avg}{100.00}
\DefMacro{res-selected-434-mut2e-no-name-ft-lora-codellama-7b-match-max}{100.00}
\DefMacro{res-selected-434-mut2e-no-name-ft-lora-codellama-7b-match-min}{100.00}
\DefMacro{res-selected-434-mut2e-no-name-ft-lora-codellama-7b-runnable-avg}{74.45}
\DefMacro{res-selected-434-mut2e-no-name-ft-lora-codellama-7b-runnable-max}{74.45}
\DefMacro{res-selected-434-mut2e-no-name-ft-lora-codellama-7b-runnable-min}{74.45}
\DefMacro{res-selected-434-mut2e-no-name-ft-lora-codellama-7b-runnable-overall}{47.00}
\DefMacro{res-selected-434-mut2e-no-name-ft-lora-codellama-7b-runnable-overall-avg}{47.00}
\DefMacro{res-selected-434-mut2e-no-name-ft-lora-codellama-7b-runnable-overall-max}{47.00}
\DefMacro{res-selected-434-mut2e-no-name-ft-lora-codellama-7b-runnable-overall-min}{47.00}
\DefMacro{res-selected-434-mut2e-no-name-ft-lora-codellama-7b-timeout-avg}{0.73}
\DefMacro{res-selected-434-mut2e-no-name-ft-lora-codellama-7b-timeout-max}{0.73}
\DefMacro{res-selected-434-mut2e-no-name-ft-lora-codellama-7b-timeout-min}{0.73}
\DefMacro{res-selected-434-mut2e-no-name-ft-lora-codellama-13b-bleu}{39.00}
\DefMacro{res-selected-434-mut2e-no-name-ft-lora-codellama-13b-bleu-avg}{39.00}
\DefMacro{res-selected-434-mut2e-no-name-ft-lora-codellama-13b-bleu-max}{39.00}
\DefMacro{res-selected-434-mut2e-no-name-ft-lora-codellama-13b-bleu-min}{39.00}
\DefMacro{res-selected-434-mut2e-no-name-ft-lora-codellama-13b-code-bleu}{47.74}
\DefMacro{res-selected-434-mut2e-no-name-ft-lora-codellama-13b-code-bleu-avg}{47.74}
\DefMacro{res-selected-434-mut2e-no-name-ft-lora-codellama-13b-code-bleu-max}{47.74}
\DefMacro{res-selected-434-mut2e-no-name-ft-lora-codellama-13b-code-bleu-min}{47.74}
\DefMacro{res-selected-434-mut2e-no-name-ft-lora-codellama-13b-edit-sim}{73.49}
\DefMacro{res-selected-434-mut2e-no-name-ft-lora-codellama-13b-edit-sim-avg}{73.49}
\DefMacro{res-selected-434-mut2e-no-name-ft-lora-codellama-13b-edit-sim-max}{73.49}
\DefMacro{res-selected-434-mut2e-no-name-ft-lora-codellama-13b-edit-sim-min}{73.49}
\DefMacro{res-selected-434-mut2e-no-name-ft-lora-codellama-13b-rouge-f}{67.10}
\DefMacro{res-selected-434-mut2e-no-name-ft-lora-codellama-13b-rouge-f-avg}{67.10}
\DefMacro{res-selected-434-mut2e-no-name-ft-lora-codellama-13b-rouge-f-max}{67.10}
\DefMacro{res-selected-434-mut2e-no-name-ft-lora-codellama-13b-rouge-f-min}{67.10}
\DefMacro{res-selected-434-mut2e-no-name-ft-lora-codellama-13b-rouge-p}{70.29}
\DefMacro{res-selected-434-mut2e-no-name-ft-lora-codellama-13b-rouge-p-avg}{70.29}
\DefMacro{res-selected-434-mut2e-no-name-ft-lora-codellama-13b-rouge-p-max}{70.29}
\DefMacro{res-selected-434-mut2e-no-name-ft-lora-codellama-13b-rouge-p-min}{70.29}
\DefMacro{res-selected-434-mut2e-no-name-ft-lora-codellama-13b-rouge-r}{66.08}
\DefMacro{res-selected-434-mut2e-no-name-ft-lora-codellama-13b-rouge-r-avg}{66.08}
\DefMacro{res-selected-434-mut2e-no-name-ft-lora-codellama-13b-rouge-r-max}{66.08}
\DefMacro{res-selected-434-mut2e-no-name-ft-lora-codellama-13b-rouge-r-min}{66.08}
\DefMacro{res-selected-434-mut2e-no-name-ft-lora-codellama-13b-xmatch}{0.69}
\DefMacro{res-selected-434-mut2e-no-name-ft-lora-codellama-13b-xmatch-avg}{0.69}
\DefMacro{res-selected-434-mut2e-no-name-ft-lora-codellama-13b-xmatch-max}{0.69}
\DefMacro{res-selected-434-mut2e-no-name-ft-lora-codellama-13b-xmatch-min}{0.69}
\DefMacro{res-selected-434-mut2e-no-name-ft-lora-codellama-13b-xmatch-top1}{0.69}
\DefMacro{res-selected-434-mut2e-no-name-ft-lora-codellama-13b-compilable-avg}{66.36}
\DefMacro{res-selected-434-mut2e-no-name-ft-lora-codellama-13b-compilable-max}{66.36}
\DefMacro{res-selected-434-mut2e-no-name-ft-lora-codellama-13b-compilable-min}{66.36}
\DefMacro{res-selected-434-mut2e-no-name-ft-lora-codellama-13b-coverage-avg}{36.18}
\DefMacro{res-selected-434-mut2e-no-name-ft-lora-codellama-13b-coverage-max}{36.18}
\DefMacro{res-selected-434-mut2e-no-name-ft-lora-codellama-13b-coverage-min}{36.18}
\DefMacro{res-selected-434-mut2e-no-name-ft-lora-codellama-13b-match-avg}{100.00}
\DefMacro{res-selected-434-mut2e-no-name-ft-lora-codellama-13b-match-max}{100.00}
\DefMacro{res-selected-434-mut2e-no-name-ft-lora-codellama-13b-match-min}{100.00}
\DefMacro{res-selected-434-mut2e-no-name-ft-lora-codellama-13b-runnable-avg}{76.04}
\DefMacro{res-selected-434-mut2e-no-name-ft-lora-codellama-13b-runnable-max}{76.04}
\DefMacro{res-selected-434-mut2e-no-name-ft-lora-codellama-13b-runnable-min}{76.04}
\DefMacro{res-selected-434-mut2e-no-name-ft-lora-codellama-13b-runnable-overall}{50.46}
\DefMacro{res-selected-434-mut2e-no-name-ft-lora-codellama-13b-runnable-overall-avg}{50.46}
\DefMacro{res-selected-434-mut2e-no-name-ft-lora-codellama-13b-runnable-overall-max}{50.46}
\DefMacro{res-selected-434-mut2e-no-name-ft-lora-codellama-13b-runnable-overall-min}{50.46}
\DefMacro{res-selected-434-mut2e-no-name-ft-lora-codellama-13b-timeout-avg}{0.00}
\DefMacro{res-selected-434-mut2e-no-name-ft-lora-codellama-13b-timeout-max}{0.00}
\DefMacro{res-selected-434-mut2e-no-name-ft-lora-codellama-13b-timeout-min}{0.00}
\DefMacro{res-selected-434-ne2e-no-name-ft-lora-codellama-7b-bleu}{45.60}
\DefMacro{res-selected-434-ne2e-no-name-ft-lora-codellama-7b-bleu-avg}{45.60}
\DefMacro{res-selected-434-ne2e-no-name-ft-lora-codellama-7b-bleu-max}{45.60}
\DefMacro{res-selected-434-ne2e-no-name-ft-lora-codellama-7b-bleu-min}{45.60}
\DefMacro{res-selected-434-ne2e-no-name-ft-lora-codellama-7b-code-bleu}{53.56}
\DefMacro{res-selected-434-ne2e-no-name-ft-lora-codellama-7b-code-bleu-avg}{53.56}
\DefMacro{res-selected-434-ne2e-no-name-ft-lora-codellama-7b-code-bleu-max}{53.56}
\DefMacro{res-selected-434-ne2e-no-name-ft-lora-codellama-7b-code-bleu-min}{53.56}
\DefMacro{res-selected-434-ne2e-no-name-ft-lora-codellama-7b-edit-sim}{77.91}
\DefMacro{res-selected-434-ne2e-no-name-ft-lora-codellama-7b-edit-sim-avg}{77.91}
\DefMacro{res-selected-434-ne2e-no-name-ft-lora-codellama-7b-edit-sim-max}{77.91}
\DefMacro{res-selected-434-ne2e-no-name-ft-lora-codellama-7b-edit-sim-min}{77.91}
\DefMacro{res-selected-434-ne2e-no-name-ft-lora-codellama-7b-rouge-f}{72.38}
\DefMacro{res-selected-434-ne2e-no-name-ft-lora-codellama-7b-rouge-f-avg}{72.38}
\DefMacro{res-selected-434-ne2e-no-name-ft-lora-codellama-7b-rouge-f-max}{72.38}
\DefMacro{res-selected-434-ne2e-no-name-ft-lora-codellama-7b-rouge-f-min}{72.38}
\DefMacro{res-selected-434-ne2e-no-name-ft-lora-codellama-7b-rouge-p}{74.91}
\DefMacro{res-selected-434-ne2e-no-name-ft-lora-codellama-7b-rouge-p-avg}{74.91}
\DefMacro{res-selected-434-ne2e-no-name-ft-lora-codellama-7b-rouge-p-max}{74.91}
\DefMacro{res-selected-434-ne2e-no-name-ft-lora-codellama-7b-rouge-p-min}{74.91}
\DefMacro{res-selected-434-ne2e-no-name-ft-lora-codellama-7b-rouge-r}{72.05}
\DefMacro{res-selected-434-ne2e-no-name-ft-lora-codellama-7b-rouge-r-avg}{72.05}
\DefMacro{res-selected-434-ne2e-no-name-ft-lora-codellama-7b-rouge-r-max}{72.05}
\DefMacro{res-selected-434-ne2e-no-name-ft-lora-codellama-7b-rouge-r-min}{72.05}
\DefMacro{res-selected-434-ne2e-no-name-ft-lora-codellama-7b-xmatch}{1.38}
\DefMacro{res-selected-434-ne2e-no-name-ft-lora-codellama-7b-xmatch-top1}{1.38}
\DefMacro{res-selected-434-ne2e-no-name-ft-lora-codellama-7b-compilable-avg}{84.10}
\DefMacro{res-selected-434-ne2e-no-name-ft-lora-codellama-7b-compilable-max}{84.10}
\DefMacro{res-selected-434-ne2e-no-name-ft-lora-codellama-7b-compilable-min}{84.10}
\DefMacro{res-selected-434-ne2e-no-name-ft-lora-codellama-7b-coverage-avg}{40.09}
\DefMacro{res-selected-434-ne2e-no-name-ft-lora-codellama-7b-coverage-max}{40.09}
\DefMacro{res-selected-434-ne2e-no-name-ft-lora-codellama-7b-coverage-min}{40.09}
\DefMacro{res-selected-434-ne2e-no-name-ft-lora-codellama-7b-match-avg}{100.00}
\DefMacro{res-selected-434-ne2e-no-name-ft-lora-codellama-7b-match-max}{100.00}
\DefMacro{res-selected-434-ne2e-no-name-ft-lora-codellama-7b-match-min}{100.00}
\DefMacro{res-selected-434-ne2e-no-name-ft-lora-codellama-7b-runnable-avg}{71.78}
\DefMacro{res-selected-434-ne2e-no-name-ft-lora-codellama-7b-runnable-max}{71.78}
\DefMacro{res-selected-434-ne2e-no-name-ft-lora-codellama-7b-runnable-min}{71.78}
\DefMacro{res-selected-434-ne2e-no-name-ft-lora-codellama-7b-runnable-overall}{60.37}
\DefMacro{res-selected-434-ne2e-no-name-ft-lora-codellama-7b-runnable-overall-avg}{60.37}
\DefMacro{res-selected-434-ne2e-no-name-ft-lora-codellama-7b-runnable-overall-max}{60.37}
\DefMacro{res-selected-434-ne2e-no-name-ft-lora-codellama-7b-runnable-overall-min}{60.37}
\DefMacro{res-selected-434-ne2e-no-name-ft-lora-codellama-7b-timeout-avg}{0.27}
\DefMacro{res-selected-434-ne2e-no-name-ft-lora-codellama-7b-timeout-max}{0.27}
\DefMacro{res-selected-434-ne2e-no-name-ft-lora-codellama-7b-timeout-min}{0.27}
\DefMacro{res-selected-434-ne2e-all-no-name-ft-lora-codellama-7b-bleu}{44.43}
\DefMacro{res-selected-434-ne2e-all-no-name-ft-lora-codellama-7b-bleu-avg}{44.94}
\DefMacro{res-selected-434-ne2e-all-no-name-ft-lora-codellama-7b-bleu-max}{49.03}
\DefMacro{res-selected-434-ne2e-all-no-name-ft-lora-codellama-7b-bleu-min}{41.11}
\DefMacro{res-selected-434-ne2e-all-no-name-ft-lora-codellama-7b-code-bleu}{52.18}
\DefMacro{res-selected-434-ne2e-all-no-name-ft-lora-codellama-7b-code-bleu-avg}{52.51}
\DefMacro{res-selected-434-ne2e-all-no-name-ft-lora-codellama-7b-code-bleu-max}{57.41}
\DefMacro{res-selected-434-ne2e-all-no-name-ft-lora-codellama-7b-code-bleu-min}{48.62}
\DefMacro{res-selected-434-ne2e-all-no-name-ft-lora-codellama-7b-edit-sim}{77.25}
\DefMacro{res-selected-434-ne2e-all-no-name-ft-lora-codellama-7b-edit-sim-avg}{77.51}
\DefMacro{res-selected-434-ne2e-all-no-name-ft-lora-codellama-7b-edit-sim-max}{81.03}
\DefMacro{res-selected-434-ne2e-all-no-name-ft-lora-codellama-7b-edit-sim-min}{73.76}
\DefMacro{res-selected-434-ne2e-all-no-name-ft-lora-codellama-7b-rouge-f}{71.70}
\DefMacro{res-selected-434-ne2e-all-no-name-ft-lora-codellama-7b-rouge-f-avg}{72.03}
\DefMacro{res-selected-434-ne2e-all-no-name-ft-lora-codellama-7b-rouge-f-max}{75.56}
\DefMacro{res-selected-434-ne2e-all-no-name-ft-lora-codellama-7b-rouge-f-min}{68.55}
\DefMacro{res-selected-434-ne2e-all-no-name-ft-lora-codellama-7b-rouge-p}{74.61}
\DefMacro{res-selected-434-ne2e-all-no-name-ft-lora-codellama-7b-rouge-p-avg}{74.98}
\DefMacro{res-selected-434-ne2e-all-no-name-ft-lora-codellama-7b-rouge-p-max}{78.93}
\DefMacro{res-selected-434-ne2e-all-no-name-ft-lora-codellama-7b-rouge-p-min}{70.15}
\DefMacro{res-selected-434-ne2e-all-no-name-ft-lora-codellama-7b-rouge-r}{71.37}
\DefMacro{res-selected-434-ne2e-all-no-name-ft-lora-codellama-7b-rouge-r-avg}{71.59}
\DefMacro{res-selected-434-ne2e-all-no-name-ft-lora-codellama-7b-rouge-r-max}{75.08}
\DefMacro{res-selected-434-ne2e-all-no-name-ft-lora-codellama-7b-rouge-r-min}{68.78}
\DefMacro{res-selected-434-ne2e-all-no-name-ft-lora-codellama-7b-xmatch}{1.38}
\DefMacro{res-selected-434-ne2e-all-no-name-ft-lora-codellama-7b-xmatch-avg}{1.43}
\DefMacro{res-selected-434-ne2e-all-no-name-ft-lora-codellama-7b-xmatch-max}{1.84}
\DefMacro{res-selected-434-ne2e-all-no-name-ft-lora-codellama-7b-xmatch-min}{1.15}
\DefMacro{res-selected-434-ne2e-all-no-name-ft-lora-codellama-7b-xmatch-top1}{1.38}
\DefMacro{res-selected-434-ne2e-all-no-name-ft-lora-codellama-7b-compilable-avg}{84.16}
\DefMacro{res-selected-434-ne2e-all-no-name-ft-lora-codellama-7b-compilable-max}{90.32}
\DefMacro{res-selected-434-ne2e-all-no-name-ft-lora-codellama-7b-compilable-min}{76.50}
\DefMacro{res-selected-434-ne2e-all-no-name-ft-lora-codellama-7b-coverage-avg}{39.89}
\DefMacro{res-selected-434-ne2e-all-no-name-ft-lora-codellama-7b-coverage-max}{50.46}
\DefMacro{res-selected-434-ne2e-all-no-name-ft-lora-codellama-7b-coverage-min}{32.49}
\DefMacro{res-selected-434-ne2e-all-no-name-ft-lora-codellama-7b-match-avg}{100.00}
\DefMacro{res-selected-434-ne2e-all-no-name-ft-lora-codellama-7b-match-max}{100.00}
\DefMacro{res-selected-434-ne2e-all-no-name-ft-lora-codellama-7b-match-min}{100.00}
\DefMacro{res-selected-434-ne2e-all-no-name-ft-lora-codellama-7b-runnable-avg}{67.96}
\DefMacro{res-selected-434-ne2e-all-no-name-ft-lora-codellama-7b-runnable-max}{80.61}
\DefMacro{res-selected-434-ne2e-all-no-name-ft-lora-codellama-7b-runnable-min}{56.12}
\DefMacro{res-selected-434-ne2e-all-no-name-ft-lora-codellama-7b-runnable-overall}{72.81}
\DefMacro{res-selected-434-ne2e-all-no-name-ft-lora-codellama-7b-runnable-overall-avg}{57.85}
\DefMacro{res-selected-434-ne2e-all-no-name-ft-lora-codellama-7b-runnable-overall-max}{72.81}
\DefMacro{res-selected-434-ne2e-all-no-name-ft-lora-codellama-7b-runnable-overall-min}{44.70}
\DefMacro{res-selected-434-ne2e-all-no-name-ft-lora-codellama-7b-timeout-avg}{0.57}
\DefMacro{res-selected-434-ne2e-all-no-name-ft-lora-codellama-7b-timeout-max}{0.77}
\DefMacro{res-selected-434-ne2e-all-no-name-ft-lora-codellama-7b-timeout-min}{0.26}
\DefMacro{res-selected-434-nestack2e-no-name-ft-lora-codellama-7b-bleu}{47.17}
\DefMacro{res-selected-434-nestack2e-no-name-ft-lora-codellama-7b-bleu-avg}{47.17}
\DefMacro{res-selected-434-nestack2e-no-name-ft-lora-codellama-7b-bleu-max}{47.17}
\DefMacro{res-selected-434-nestack2e-no-name-ft-lora-codellama-7b-bleu-min}{47.17}
\DefMacro{res-selected-434-nestack2e-no-name-ft-lora-codellama-7b-code-bleu}{56.04}
\DefMacro{res-selected-434-nestack2e-no-name-ft-lora-codellama-7b-code-bleu-avg}{56.04}
\DefMacro{res-selected-434-nestack2e-no-name-ft-lora-codellama-7b-code-bleu-max}{56.04}
\DefMacro{res-selected-434-nestack2e-no-name-ft-lora-codellama-7b-code-bleu-min}{56.04}
\DefMacro{res-selected-434-nestack2e-no-name-ft-lora-codellama-7b-edit-sim}{79.94}
\DefMacro{res-selected-434-nestack2e-no-name-ft-lora-codellama-7b-edit-sim-avg}{79.94}
\DefMacro{res-selected-434-nestack2e-no-name-ft-lora-codellama-7b-edit-sim-max}{79.94}
\DefMacro{res-selected-434-nestack2e-no-name-ft-lora-codellama-7b-edit-sim-min}{79.94}
\DefMacro{res-selected-434-nestack2e-no-name-ft-lora-codellama-7b-rouge-f}{74.62}
\DefMacro{res-selected-434-nestack2e-no-name-ft-lora-codellama-7b-rouge-f-avg}{74.62}
\DefMacro{res-selected-434-nestack2e-no-name-ft-lora-codellama-7b-rouge-f-max}{74.62}
\DefMacro{res-selected-434-nestack2e-no-name-ft-lora-codellama-7b-rouge-f-min}{74.62}
\DefMacro{res-selected-434-nestack2e-no-name-ft-lora-codellama-7b-rouge-p}{77.07}
\DefMacro{res-selected-434-nestack2e-no-name-ft-lora-codellama-7b-rouge-p-avg}{77.07}
\DefMacro{res-selected-434-nestack2e-no-name-ft-lora-codellama-7b-rouge-p-max}{77.07}
\DefMacro{res-selected-434-nestack2e-no-name-ft-lora-codellama-7b-rouge-p-min}{77.07}
\DefMacro{res-selected-434-nestack2e-no-name-ft-lora-codellama-7b-rouge-r}{74.20}
\DefMacro{res-selected-434-nestack2e-no-name-ft-lora-codellama-7b-rouge-r-avg}{74.20}
\DefMacro{res-selected-434-nestack2e-no-name-ft-lora-codellama-7b-rouge-r-max}{74.20}
\DefMacro{res-selected-434-nestack2e-no-name-ft-lora-codellama-7b-rouge-r-min}{74.20}
\DefMacro{res-selected-434-nestack2e-no-name-ft-lora-codellama-7b-xmatch}{2.30}
\DefMacro{res-selected-434-nestack2e-no-name-ft-lora-codellama-7b-xmatch-top1}{2.30}
\DefMacro{res-selected-434-nestack2e-no-name-ft-lora-codellama-7b-compilable-avg}{82.26}
\DefMacro{res-selected-434-nestack2e-no-name-ft-lora-codellama-7b-compilable-max}{82.26}
\DefMacro{res-selected-434-nestack2e-no-name-ft-lora-codellama-7b-compilable-min}{82.26}
\DefMacro{res-selected-434-nestack2e-no-name-ft-lora-codellama-7b-coverage-avg}{57.14}
\DefMacro{res-selected-434-nestack2e-no-name-ft-lora-codellama-7b-coverage-max}{57.14}
\DefMacro{res-selected-434-nestack2e-no-name-ft-lora-codellama-7b-coverage-min}{57.14}
\DefMacro{res-selected-434-nestack2e-no-name-ft-lora-codellama-7b-match-avg}{100.00}
\DefMacro{res-selected-434-nestack2e-no-name-ft-lora-codellama-7b-match-max}{100.00}
\DefMacro{res-selected-434-nestack2e-no-name-ft-lora-codellama-7b-match-min}{100.00}
\DefMacro{res-selected-434-nestack2e-no-name-ft-lora-codellama-7b-runnable-avg}{80.95}
\DefMacro{res-selected-434-nestack2e-no-name-ft-lora-codellama-7b-runnable-max}{80.95}
\DefMacro{res-selected-434-nestack2e-no-name-ft-lora-codellama-7b-runnable-min}{80.95}
\DefMacro{res-selected-434-nestack2e-no-name-ft-lora-codellama-7b-runnable-overall}{66.59}
\DefMacro{res-selected-434-nestack2e-no-name-ft-lora-codellama-7b-runnable-overall-avg}{66.59}
\DefMacro{res-selected-434-nestack2e-no-name-ft-lora-codellama-7b-runnable-overall-max}{66.59}
\DefMacro{res-selected-434-nestack2e-no-name-ft-lora-codellama-7b-runnable-overall-min}{66.59}
\DefMacro{res-selected-434-nestack2e-no-name-ft-lora-codellama-7b-timeout-avg}{0.28}
\DefMacro{res-selected-434-nestack2e-no-name-ft-lora-codellama-7b-timeout-max}{0.28}
\DefMacro{res-selected-434-nestack2e-no-name-ft-lora-codellama-7b-timeout-min}{0.28}
\DefMacro{res-selected-434-nestack2e-no-name-ft-lora-codellama-13b-bleu}{47.05}
\DefMacro{res-selected-434-nestack2e-no-name-ft-lora-codellama-13b-bleu-avg}{47.05}
\DefMacro{res-selected-434-nestack2e-no-name-ft-lora-codellama-13b-bleu-max}{47.05}
\DefMacro{res-selected-434-nestack2e-no-name-ft-lora-codellama-13b-bleu-min}{47.05}
\DefMacro{res-selected-434-nestack2e-no-name-ft-lora-codellama-13b-code-bleu}{56.30}
\DefMacro{res-selected-434-nestack2e-no-name-ft-lora-codellama-13b-code-bleu-avg}{56.30}
\DefMacro{res-selected-434-nestack2e-no-name-ft-lora-codellama-13b-code-bleu-max}{56.30}
\DefMacro{res-selected-434-nestack2e-no-name-ft-lora-codellama-13b-code-bleu-min}{56.30}
\DefMacro{res-selected-434-nestack2e-no-name-ft-lora-codellama-13b-edit-sim}{79.56}
\DefMacro{res-selected-434-nestack2e-no-name-ft-lora-codellama-13b-edit-sim-avg}{79.56}
\DefMacro{res-selected-434-nestack2e-no-name-ft-lora-codellama-13b-edit-sim-max}{79.56}
\DefMacro{res-selected-434-nestack2e-no-name-ft-lora-codellama-13b-edit-sim-min}{79.56}
\DefMacro{res-selected-434-nestack2e-no-name-ft-lora-codellama-13b-rouge-f}{74.13}
\DefMacro{res-selected-434-nestack2e-no-name-ft-lora-codellama-13b-rouge-f-avg}{74.13}
\DefMacro{res-selected-434-nestack2e-no-name-ft-lora-codellama-13b-rouge-f-max}{74.13}
\DefMacro{res-selected-434-nestack2e-no-name-ft-lora-codellama-13b-rouge-f-min}{74.13}
\DefMacro{res-selected-434-nestack2e-no-name-ft-lora-codellama-13b-rouge-p}{75.87}
\DefMacro{res-selected-434-nestack2e-no-name-ft-lora-codellama-13b-rouge-p-avg}{75.87}
\DefMacro{res-selected-434-nestack2e-no-name-ft-lora-codellama-13b-rouge-p-max}{75.87}
\DefMacro{res-selected-434-nestack2e-no-name-ft-lora-codellama-13b-rouge-p-min}{75.87}
\DefMacro{res-selected-434-nestack2e-no-name-ft-lora-codellama-13b-rouge-r}{74.53}
\DefMacro{res-selected-434-nestack2e-no-name-ft-lora-codellama-13b-rouge-r-avg}{74.53}
\DefMacro{res-selected-434-nestack2e-no-name-ft-lora-codellama-13b-rouge-r-max}{74.53}
\DefMacro{res-selected-434-nestack2e-no-name-ft-lora-codellama-13b-rouge-r-min}{74.53}
\DefMacro{res-selected-434-nestack2e-no-name-ft-lora-codellama-13b-xmatch}{2.07}
\DefMacro{res-selected-434-nestack2e-no-name-ft-lora-codellama-13b-xmatch-avg}{2.07}
\DefMacro{res-selected-434-nestack2e-no-name-ft-lora-codellama-13b-xmatch-max}{2.07}
\DefMacro{res-selected-434-nestack2e-no-name-ft-lora-codellama-13b-xmatch-min}{2.07}
\DefMacro{res-selected-434-nestack2e-no-name-ft-lora-codellama-13b-xmatch-top1}{2.07}
\DefMacro{res-selected-434-nestack2e-no-name-ft-lora-codellama-13b-compilable-avg}{83.87}
\DefMacro{res-selected-434-nestack2e-no-name-ft-lora-codellama-13b-compilable-max}{83.87}
\DefMacro{res-selected-434-nestack2e-no-name-ft-lora-codellama-13b-compilable-min}{83.87}
\DefMacro{res-selected-434-nestack2e-no-name-ft-lora-codellama-13b-coverage-avg}{63.13}
\DefMacro{res-selected-434-nestack2e-no-name-ft-lora-codellama-13b-coverage-max}{63.13}
\DefMacro{res-selected-434-nestack2e-no-name-ft-lora-codellama-13b-coverage-min}{63.13}
\DefMacro{res-selected-434-nestack2e-no-name-ft-lora-codellama-13b-match-avg}{100.00}
\DefMacro{res-selected-434-nestack2e-no-name-ft-lora-codellama-13b-match-max}{100.00}
\DefMacro{res-selected-434-nestack2e-no-name-ft-lora-codellama-13b-match-min}{100.00}
\DefMacro{res-selected-434-nestack2e-no-name-ft-lora-codellama-13b-runnable-avg}{88.19}
\DefMacro{res-selected-434-nestack2e-no-name-ft-lora-codellama-13b-runnable-max}{88.19}
\DefMacro{res-selected-434-nestack2e-no-name-ft-lora-codellama-13b-runnable-min}{88.19}
\DefMacro{res-selected-434-nestack2e-no-name-ft-lora-codellama-13b-runnable-overall}{73.96}
\DefMacro{res-selected-434-nestack2e-no-name-ft-lora-codellama-13b-runnable-overall-avg}{73.96}
\DefMacro{res-selected-434-nestack2e-no-name-ft-lora-codellama-13b-runnable-overall-max}{73.96}
\DefMacro{res-selected-434-nestack2e-no-name-ft-lora-codellama-13b-runnable-overall-min}{73.96}
\DefMacro{res-selected-434-nestack2e-no-name-ft-lora-codellama-13b-timeout-avg}{0.27}
\DefMacro{res-selected-434-nestack2e-no-name-ft-lora-codellama-13b-timeout-max}{0.27}
\DefMacro{res-selected-434-nestack2e-no-name-ft-lora-codellama-13b-timeout-min}{0.27}
\DefMacro{res-selected-434-nestack2e-all-no-name-ft-lora-codellama-7b-bleu}{46.04}
\DefMacro{res-selected-434-nestack2e-all-no-name-ft-lora-codellama-7b-bleu-avg}{46.61}
\DefMacro{res-selected-434-nestack2e-all-no-name-ft-lora-codellama-7b-bleu-max}{51.01}
\DefMacro{res-selected-434-nestack2e-all-no-name-ft-lora-codellama-7b-bleu-min}{42.50}
\DefMacro{res-selected-434-nestack2e-all-no-name-ft-lora-codellama-7b-code-bleu}{55.27}
\DefMacro{res-selected-434-nestack2e-all-no-name-ft-lora-codellama-7b-code-bleu-avg}{55.88}
\DefMacro{res-selected-434-nestack2e-all-no-name-ft-lora-codellama-7b-code-bleu-max}{60.20}
\DefMacro{res-selected-434-nestack2e-all-no-name-ft-lora-codellama-7b-code-bleu-min}{51.79}
\DefMacro{res-selected-434-nestack2e-all-no-name-ft-lora-codellama-7b-edit-sim}{79.73}
\DefMacro{res-selected-434-nestack2e-all-no-name-ft-lora-codellama-7b-edit-sim-avg}{79.83}
\DefMacro{res-selected-434-nestack2e-all-no-name-ft-lora-codellama-7b-edit-sim-max}{83.03}
\DefMacro{res-selected-434-nestack2e-all-no-name-ft-lora-codellama-7b-edit-sim-min}{76.12}
\DefMacro{res-selected-434-nestack2e-all-no-name-ft-lora-codellama-7b-rouge-f}{74.20}
\DefMacro{res-selected-434-nestack2e-all-no-name-ft-lora-codellama-7b-rouge-f-avg}{74.42}
\DefMacro{res-selected-434-nestack2e-all-no-name-ft-lora-codellama-7b-rouge-f-max}{77.76}
\DefMacro{res-selected-434-nestack2e-all-no-name-ft-lora-codellama-7b-rouge-f-min}{70.73}
\DefMacro{res-selected-434-nestack2e-all-no-name-ft-lora-codellama-7b-rouge-p}{76.32}
\DefMacro{res-selected-434-nestack2e-all-no-name-ft-lora-codellama-7b-rouge-p-avg}{76.75}
\DefMacro{res-selected-434-nestack2e-all-no-name-ft-lora-codellama-7b-rouge-p-max}{80.42}
\DefMacro{res-selected-434-nestack2e-all-no-name-ft-lora-codellama-7b-rouge-p-min}{72.38}
\DefMacro{res-selected-434-nestack2e-all-no-name-ft-lora-codellama-7b-rouge-r}{74.11}
\DefMacro{res-selected-434-nestack2e-all-no-name-ft-lora-codellama-7b-rouge-r-avg}{74.15}
\DefMacro{res-selected-434-nestack2e-all-no-name-ft-lora-codellama-7b-rouge-r-max}{77.53}
\DefMacro{res-selected-434-nestack2e-all-no-name-ft-lora-codellama-7b-rouge-r-min}{70.71}
\DefMacro{res-selected-434-nestack2e-all-no-name-ft-lora-codellama-7b-xmatch}{2.07}
\DefMacro{res-selected-434-nestack2e-all-no-name-ft-lora-codellama-7b-xmatch-avg}{2.17}
\DefMacro{res-selected-434-nestack2e-all-no-name-ft-lora-codellama-7b-xmatch-max}{3.23}
\DefMacro{res-selected-434-nestack2e-all-no-name-ft-lora-codellama-7b-xmatch-min}{1.38}
\DefMacro{res-selected-434-nestack2e-all-no-name-ft-lora-codellama-7b-xmatch-top1}{2.07}
\DefMacro{res-selected-434-nestack2e-all-no-name-ft-lora-codellama-7b-compilable-avg}{81.57}
\DefMacro{res-selected-434-nestack2e-all-no-name-ft-lora-codellama-7b-compilable-max}{87.56}
\DefMacro{res-selected-434-nestack2e-all-no-name-ft-lora-codellama-7b-compilable-min}{74.42}
\DefMacro{res-selected-434-nestack2e-all-no-name-ft-lora-codellama-7b-coverage-avg}{59.09}
\DefMacro{res-selected-434-nestack2e-all-no-name-ft-lora-codellama-7b-coverage-max}{68.89}
\DefMacro{res-selected-434-nestack2e-all-no-name-ft-lora-codellama-7b-coverage-min}{48.39}
\DefMacro{res-selected-434-nestack2e-all-no-name-ft-lora-codellama-7b-match-avg}{100.00}
\DefMacro{res-selected-434-nestack2e-all-no-name-ft-lora-codellama-7b-match-max}{100.00}
\DefMacro{res-selected-434-nestack2e-all-no-name-ft-lora-codellama-7b-match-min}{100.00}
\DefMacro{res-selected-434-nestack2e-all-no-name-ft-lora-codellama-7b-runnable-avg}{82.82}
\DefMacro{res-selected-434-nestack2e-all-no-name-ft-lora-codellama-7b-runnable-max}{90.53}
\DefMacro{res-selected-434-nestack2e-all-no-name-ft-lora-codellama-7b-runnable-min}{73.68}
\DefMacro{res-selected-434-nestack2e-all-no-name-ft-lora-codellama-7b-runnable-overall}{79.26}
\DefMacro{res-selected-434-nestack2e-all-no-name-ft-lora-codellama-7b-runnable-overall-avg}{68.49}
\DefMacro{res-selected-434-nestack2e-all-no-name-ft-lora-codellama-7b-runnable-overall-max}{79.26}
\DefMacro{res-selected-434-nestack2e-all-no-name-ft-lora-codellama-7b-runnable-overall-min}{56.68}
\DefMacro{res-selected-434-nestack2e-all-no-name-ft-lora-codellama-7b-timeout-avg}{0.53}
\DefMacro{res-selected-434-nestack2e-all-no-name-ft-lora-codellama-7b-timeout-max}{0.79}
\DefMacro{res-selected-434-nestack2e-all-no-name-ft-lora-codellama-7b-timeout-min}{0.00}
\DefMacro{res-selected-434-conditionne2e-no-name-ft-lora-codellama-7b-bleu}{47.72}
\DefMacro{res-selected-434-conditionne2e-no-name-ft-lora-codellama-7b-bleu-avg}{47.72}
\DefMacro{res-selected-434-conditionne2e-no-name-ft-lora-codellama-7b-bleu-max}{47.72}
\DefMacro{res-selected-434-conditionne2e-no-name-ft-lora-codellama-7b-bleu-min}{47.72}
\DefMacro{res-selected-434-conditionne2e-no-name-ft-lora-codellama-7b-code-bleu}{55.79}
\DefMacro{res-selected-434-conditionne2e-no-name-ft-lora-codellama-7b-code-bleu-avg}{55.79}
\DefMacro{res-selected-434-conditionne2e-no-name-ft-lora-codellama-7b-code-bleu-max}{55.79}
\DefMacro{res-selected-434-conditionne2e-no-name-ft-lora-codellama-7b-code-bleu-min}{55.79}
\DefMacro{res-selected-434-conditionne2e-no-name-ft-lora-codellama-7b-edit-sim}{79.90}
\DefMacro{res-selected-434-conditionne2e-no-name-ft-lora-codellama-7b-edit-sim-avg}{79.90}
\DefMacro{res-selected-434-conditionne2e-no-name-ft-lora-codellama-7b-edit-sim-max}{79.90}
\DefMacro{res-selected-434-conditionne2e-no-name-ft-lora-codellama-7b-edit-sim-min}{79.90}
\DefMacro{res-selected-434-conditionne2e-no-name-ft-lora-codellama-7b-rouge-f}{74.71}
\DefMacro{res-selected-434-conditionne2e-no-name-ft-lora-codellama-7b-rouge-f-avg}{74.71}
\DefMacro{res-selected-434-conditionne2e-no-name-ft-lora-codellama-7b-rouge-f-max}{74.71}
\DefMacro{res-selected-434-conditionne2e-no-name-ft-lora-codellama-7b-rouge-f-min}{74.71}
\DefMacro{res-selected-434-conditionne2e-no-name-ft-lora-codellama-7b-rouge-p}{77.04}
\DefMacro{res-selected-434-conditionne2e-no-name-ft-lora-codellama-7b-rouge-p-avg}{77.04}
\DefMacro{res-selected-434-conditionne2e-no-name-ft-lora-codellama-7b-rouge-p-max}{77.04}
\DefMacro{res-selected-434-conditionne2e-no-name-ft-lora-codellama-7b-rouge-p-min}{77.04}
\DefMacro{res-selected-434-conditionne2e-no-name-ft-lora-codellama-7b-rouge-r}{74.27}
\DefMacro{res-selected-434-conditionne2e-no-name-ft-lora-codellama-7b-rouge-r-avg}{74.27}
\DefMacro{res-selected-434-conditionne2e-no-name-ft-lora-codellama-7b-rouge-r-max}{74.27}
\DefMacro{res-selected-434-conditionne2e-no-name-ft-lora-codellama-7b-rouge-r-min}{74.27}
\DefMacro{res-selected-434-conditionne2e-no-name-ft-lora-codellama-7b-xmatch}{2.07}
\DefMacro{res-selected-434-conditionne2e-no-name-ft-lora-codellama-7b-xmatch-top1}{2.07}
\DefMacro{res-selected-434-conditionne2e-no-name-ft-lora-codellama-7b-compilable-avg}{84.10}
\DefMacro{res-selected-434-conditionne2e-no-name-ft-lora-codellama-7b-compilable-max}{84.10}
\DefMacro{res-selected-434-conditionne2e-no-name-ft-lora-codellama-7b-compilable-min}{84.10}
\DefMacro{res-selected-434-conditionne2e-no-name-ft-lora-codellama-7b-coverage-avg}{55.99}
\DefMacro{res-selected-434-conditionne2e-no-name-ft-lora-codellama-7b-coverage-max}{55.99}
\DefMacro{res-selected-434-conditionne2e-no-name-ft-lora-codellama-7b-coverage-min}{55.99}
\DefMacro{res-selected-434-conditionne2e-no-name-ft-lora-codellama-7b-match-avg}{100.00}
\DefMacro{res-selected-434-conditionne2e-no-name-ft-lora-codellama-7b-match-max}{100.00}
\DefMacro{res-selected-434-conditionne2e-no-name-ft-lora-codellama-7b-match-min}{100.00}
\DefMacro{res-selected-434-conditionne2e-no-name-ft-lora-codellama-7b-runnable-avg}{79.18}
\DefMacro{res-selected-434-conditionne2e-no-name-ft-lora-codellama-7b-runnable-max}{79.18}
\DefMacro{res-selected-434-conditionne2e-no-name-ft-lora-codellama-7b-runnable-min}{79.18}
\DefMacro{res-selected-434-conditionne2e-no-name-ft-lora-codellama-7b-runnable-overall}{66.59}
\DefMacro{res-selected-434-conditionne2e-no-name-ft-lora-codellama-7b-runnable-overall-avg}{66.59}
\DefMacro{res-selected-434-conditionne2e-no-name-ft-lora-codellama-7b-runnable-overall-max}{66.59}
\DefMacro{res-selected-434-conditionne2e-no-name-ft-lora-codellama-7b-runnable-overall-min}{66.59}
\DefMacro{res-selected-434-conditionne2e-no-name-ft-lora-codellama-7b-timeout-avg}{0.00}
\DefMacro{res-selected-434-conditionne2e-no-name-ft-lora-codellama-7b-timeout-max}{0.00}
\DefMacro{res-selected-434-conditionne2e-no-name-ft-lora-codellama-7b-timeout-min}{0.00}
\DefMacro{res-selected-434-conditionnestack2e-no-name-ft-lora-codellama-7b-bleu}{47.84}
\DefMacro{res-selected-434-conditionnestack2e-no-name-ft-lora-codellama-7b-bleu-avg}{47.84}
\DefMacro{res-selected-434-conditionnestack2e-no-name-ft-lora-codellama-7b-bleu-max}{47.84}
\DefMacro{res-selected-434-conditionnestack2e-no-name-ft-lora-codellama-7b-bleu-min}{47.84}
\DefMacro{res-selected-434-conditionnestack2e-no-name-ft-lora-codellama-7b-code-bleu}{55.89}
\DefMacro{res-selected-434-conditionnestack2e-no-name-ft-lora-codellama-7b-code-bleu-avg}{55.89}
\DefMacro{res-selected-434-conditionnestack2e-no-name-ft-lora-codellama-7b-code-bleu-max}{55.89}
\DefMacro{res-selected-434-conditionnestack2e-no-name-ft-lora-codellama-7b-code-bleu-min}{55.89}
\DefMacro{res-selected-434-conditionnestack2e-no-name-ft-lora-codellama-7b-edit-sim}{80.29}
\DefMacro{res-selected-434-conditionnestack2e-no-name-ft-lora-codellama-7b-edit-sim-avg}{80.29}
\DefMacro{res-selected-434-conditionnestack2e-no-name-ft-lora-codellama-7b-edit-sim-max}{80.29}
\DefMacro{res-selected-434-conditionnestack2e-no-name-ft-lora-codellama-7b-edit-sim-min}{80.29}
\DefMacro{res-selected-434-conditionnestack2e-no-name-ft-lora-codellama-7b-rouge-f}{75.07}
\DefMacro{res-selected-434-conditionnestack2e-no-name-ft-lora-codellama-7b-rouge-f-avg}{75.07}
\DefMacro{res-selected-434-conditionnestack2e-no-name-ft-lora-codellama-7b-rouge-f-max}{75.07}
\DefMacro{res-selected-434-conditionnestack2e-no-name-ft-lora-codellama-7b-rouge-f-min}{75.07}
\DefMacro{res-selected-434-conditionnestack2e-no-name-ft-lora-codellama-7b-rouge-p}{78.26}
\DefMacro{res-selected-434-conditionnestack2e-no-name-ft-lora-codellama-7b-rouge-p-avg}{78.26}
\DefMacro{res-selected-434-conditionnestack2e-no-name-ft-lora-codellama-7b-rouge-p-max}{78.26}
\DefMacro{res-selected-434-conditionnestack2e-no-name-ft-lora-codellama-7b-rouge-p-min}{78.26}
\DefMacro{res-selected-434-conditionnestack2e-no-name-ft-lora-codellama-7b-rouge-r}{74.01}
\DefMacro{res-selected-434-conditionnestack2e-no-name-ft-lora-codellama-7b-rouge-r-avg}{74.01}
\DefMacro{res-selected-434-conditionnestack2e-no-name-ft-lora-codellama-7b-rouge-r-max}{74.01}
\DefMacro{res-selected-434-conditionnestack2e-no-name-ft-lora-codellama-7b-rouge-r-min}{74.01}
\DefMacro{res-selected-434-conditionnestack2e-no-name-ft-lora-codellama-7b-xmatch}{2.07}
\DefMacro{res-selected-434-conditionnestack2e-no-name-ft-lora-codellama-7b-xmatch-top1}{2.07}
\DefMacro{res-selected-434-conditionnestack2e-no-name-ft-lora-codellama-7b-compilable-avg}{82.95}
\DefMacro{res-selected-434-conditionnestack2e-no-name-ft-lora-codellama-7b-compilable-max}{82.95}
\DefMacro{res-selected-434-conditionnestack2e-no-name-ft-lora-codellama-7b-compilable-min}{82.95}
\DefMacro{res-selected-434-conditionnestack2e-no-name-ft-lora-codellama-7b-coverage-avg}{58.99}
\DefMacro{res-selected-434-conditionnestack2e-no-name-ft-lora-codellama-7b-coverage-max}{58.99}
\DefMacro{res-selected-434-conditionnestack2e-no-name-ft-lora-codellama-7b-coverage-min}{58.99}
\DefMacro{res-selected-434-conditionnestack2e-no-name-ft-lora-codellama-7b-match-avg}{100.00}
\DefMacro{res-selected-434-conditionnestack2e-no-name-ft-lora-codellama-7b-match-max}{100.00}
\DefMacro{res-selected-434-conditionnestack2e-no-name-ft-lora-codellama-7b-match-min}{100.00}
\DefMacro{res-selected-434-conditionnestack2e-no-name-ft-lora-codellama-7b-runnable-avg}{83.89}
\DefMacro{res-selected-434-conditionnestack2e-no-name-ft-lora-codellama-7b-runnable-max}{83.89}
\DefMacro{res-selected-434-conditionnestack2e-no-name-ft-lora-codellama-7b-runnable-min}{83.89}
\DefMacro{res-selected-434-conditionnestack2e-no-name-ft-lora-codellama-7b-runnable-overall}{69.59}
\DefMacro{res-selected-434-conditionnestack2e-no-name-ft-lora-codellama-7b-runnable-overall-avg}{69.59}
\DefMacro{res-selected-434-conditionnestack2e-no-name-ft-lora-codellama-7b-runnable-overall-max}{69.59}
\DefMacro{res-selected-434-conditionnestack2e-no-name-ft-lora-codellama-7b-runnable-overall-min}{69.59}
\DefMacro{res-selected-434-conditionnestack2e-no-name-ft-lora-codellama-7b-timeout-avg}{0.56}
\DefMacro{res-selected-434-conditionnestack2e-no-name-ft-lora-codellama-7b-timeout-max}{0.56}
\DefMacro{res-selected-434-conditionnestack2e-no-name-ft-lora-codellama-7b-timeout-min}{0.56}
\DefMacro{res-selected-434-conditionnestack2e-all-no-name-ft-lora-codellama-7b-bleu}{47.01}
\DefMacro{res-selected-434-conditionnestack2e-all-no-name-ft-lora-codellama-7b-bleu-avg}{47.54}
\DefMacro{res-selected-434-conditionnestack2e-all-no-name-ft-lora-codellama-7b-bleu-max}{51.58}
\DefMacro{res-selected-434-conditionnestack2e-all-no-name-ft-lora-codellama-7b-bleu-min}{43.84}
\DefMacro{res-selected-434-conditionnestack2e-all-no-name-ft-lora-codellama-7b-code-bleu}{55.58}
\DefMacro{res-selected-434-conditionnestack2e-all-no-name-ft-lora-codellama-7b-code-bleu-avg}{55.87}
\DefMacro{res-selected-434-conditionnestack2e-all-no-name-ft-lora-codellama-7b-code-bleu-max}{59.76}
\DefMacro{res-selected-434-conditionnestack2e-all-no-name-ft-lora-codellama-7b-code-bleu-min}{52.41}
\DefMacro{res-selected-434-conditionnestack2e-all-no-name-ft-lora-codellama-7b-edit-sim}{79.95}
\DefMacro{res-selected-434-conditionnestack2e-all-no-name-ft-lora-codellama-7b-edit-sim-avg}{80.15}
\DefMacro{res-selected-434-conditionnestack2e-all-no-name-ft-lora-codellama-7b-edit-sim-max}{83.23}
\DefMacro{res-selected-434-conditionnestack2e-all-no-name-ft-lora-codellama-7b-edit-sim-min}{76.79}
\DefMacro{res-selected-434-conditionnestack2e-all-no-name-ft-lora-codellama-7b-rouge-f}{74.53}
\DefMacro{res-selected-434-conditionnestack2e-all-no-name-ft-lora-codellama-7b-rouge-f-avg}{74.79}
\DefMacro{res-selected-434-conditionnestack2e-all-no-name-ft-lora-codellama-7b-rouge-f-max}{78.10}
\DefMacro{res-selected-434-conditionnestack2e-all-no-name-ft-lora-codellama-7b-rouge-f-min}{71.59}
\DefMacro{res-selected-434-conditionnestack2e-all-no-name-ft-lora-codellama-7b-rouge-p}{77.19}
\DefMacro{res-selected-434-conditionnestack2e-all-no-name-ft-lora-codellama-7b-rouge-p-avg}{77.53}
\DefMacro{res-selected-434-conditionnestack2e-all-no-name-ft-lora-codellama-7b-rouge-p-max}{81.23}
\DefMacro{res-selected-434-conditionnestack2e-all-no-name-ft-lora-codellama-7b-rouge-p-min}{73.53}
\DefMacro{res-selected-434-conditionnestack2e-all-no-name-ft-lora-codellama-7b-rouge-r}{74.04}
\DefMacro{res-selected-434-conditionnestack2e-all-no-name-ft-lora-codellama-7b-rouge-r-avg}{74.10}
\DefMacro{res-selected-434-conditionnestack2e-all-no-name-ft-lora-codellama-7b-rouge-r-max}{77.30}
\DefMacro{res-selected-434-conditionnestack2e-all-no-name-ft-lora-codellama-7b-rouge-r-min}{71.24}
\DefMacro{res-selected-434-conditionnestack2e-all-no-name-ft-lora-codellama-7b-xmatch}{1.84}
\DefMacro{res-selected-434-conditionnestack2e-all-no-name-ft-lora-codellama-7b-xmatch-avg}{2.17}
\DefMacro{res-selected-434-conditionnestack2e-all-no-name-ft-lora-codellama-7b-xmatch-max}{2.76}
\DefMacro{res-selected-434-conditionnestack2e-all-no-name-ft-lora-codellama-7b-xmatch-min}{1.61}
\DefMacro{res-selected-434-conditionnestack2e-all-no-name-ft-lora-codellama-7b-xmatch-top1}{1.84}
\DefMacro{res-selected-434-conditionnestack2e-all-no-name-ft-lora-codellama-7b-compilable-avg}{84.37}
\DefMacro{res-selected-434-conditionnestack2e-all-no-name-ft-lora-codellama-7b-compilable-max}{89.63}
\DefMacro{res-selected-434-conditionnestack2e-all-no-name-ft-lora-codellama-7b-compilable-min}{77.19}
\DefMacro{res-selected-434-conditionnestack2e-all-no-name-ft-lora-codellama-7b-coverage-avg}{59.88}
\DefMacro{res-selected-434-conditionnestack2e-all-no-name-ft-lora-codellama-7b-coverage-max}{68.66}
\DefMacro{res-selected-434-conditionnestack2e-all-no-name-ft-lora-codellama-7b-coverage-min}{50.23}
\DefMacro{res-selected-434-conditionnestack2e-all-no-name-ft-lora-codellama-7b-match-avg}{100.00}
\DefMacro{res-selected-434-conditionnestack2e-all-no-name-ft-lora-codellama-7b-match-max}{100.00}
\DefMacro{res-selected-434-conditionnestack2e-all-no-name-ft-lora-codellama-7b-match-min}{100.00}
\DefMacro{res-selected-434-conditionnestack2e-all-no-name-ft-lora-codellama-7b-runnable-avg}{81.68}
\DefMacro{res-selected-434-conditionnestack2e-all-no-name-ft-lora-codellama-7b-runnable-max}{88.69}
\DefMacro{res-selected-434-conditionnestack2e-all-no-name-ft-lora-codellama-7b-runnable-min}{72.75}
\DefMacro{res-selected-434-conditionnestack2e-all-no-name-ft-lora-codellama-7b-runnable-overall}{79.49}
\DefMacro{res-selected-434-conditionnestack2e-all-no-name-ft-lora-codellama-7b-runnable-overall-avg}{69.76}
\DefMacro{res-selected-434-conditionnestack2e-all-no-name-ft-lora-codellama-7b-runnable-overall-max}{79.49}
\DefMacro{res-selected-434-conditionnestack2e-all-no-name-ft-lora-codellama-7b-runnable-overall-min}{58.29}
\DefMacro{res-selected-434-conditionnestack2e-all-no-name-ft-lora-codellama-7b-timeout-avg}{0.60}
\DefMacro{res-selected-434-conditionnestack2e-all-no-name-ft-lora-codellama-7b-timeout-max}{0.77}
\DefMacro{res-selected-434-conditionnestack2e-all-no-name-ft-lora-codellama-7b-timeout-min}{0.51}

\DefMacro{res-selected-434-ne2e-few-shot-with-name-gpt-3.5-turbo-16k-bleu}{46.23}
\DefMacro{res-selected-434-ne2e-few-shot-with-name-gpt-3.5-turbo-16k-bleu-avg}{46.23}
\DefMacro{res-selected-434-ne2e-few-shot-with-name-gpt-3.5-turbo-16k-bleu-max}{46.23}
\DefMacro{res-selected-434-ne2e-few-shot-with-name-gpt-3.5-turbo-16k-bleu-min}{46.23}
\DefMacro{res-selected-434-ne2e-few-shot-with-name-gpt-3.5-turbo-16k-code-bleu}{54.97}
\DefMacro{res-selected-434-ne2e-few-shot-with-name-gpt-3.5-turbo-16k-code-bleu-avg}{54.97}
\DefMacro{res-selected-434-ne2e-few-shot-with-name-gpt-3.5-turbo-16k-code-bleu-max}{54.97}
\DefMacro{res-selected-434-ne2e-few-shot-with-name-gpt-3.5-turbo-16k-code-bleu-min}{54.97}
\DefMacro{res-selected-434-ne2e-few-shot-with-name-gpt-3.5-turbo-16k-edit-sim}{72.64}
\DefMacro{res-selected-434-ne2e-few-shot-with-name-gpt-3.5-turbo-16k-edit-sim-avg}{72.64}
\DefMacro{res-selected-434-ne2e-few-shot-with-name-gpt-3.5-turbo-16k-edit-sim-max}{72.64}
\DefMacro{res-selected-434-ne2e-few-shot-with-name-gpt-3.5-turbo-16k-edit-sim-min}{72.64}
\DefMacro{res-selected-434-ne2e-few-shot-with-name-gpt-3.5-turbo-16k-rouge-f}{66.39}
\DefMacro{res-selected-434-ne2e-few-shot-with-name-gpt-3.5-turbo-16k-rouge-f-avg}{66.39}
\DefMacro{res-selected-434-ne2e-few-shot-with-name-gpt-3.5-turbo-16k-rouge-f-max}{66.39}
\DefMacro{res-selected-434-ne2e-few-shot-with-name-gpt-3.5-turbo-16k-rouge-f-min}{66.39}
\DefMacro{res-selected-434-ne2e-few-shot-with-name-gpt-3.5-turbo-16k-rouge-p}{65.24}
\DefMacro{res-selected-434-ne2e-few-shot-with-name-gpt-3.5-turbo-16k-rouge-p-avg}{65.24}
\DefMacro{res-selected-434-ne2e-few-shot-with-name-gpt-3.5-turbo-16k-rouge-p-max}{65.24}
\DefMacro{res-selected-434-ne2e-few-shot-with-name-gpt-3.5-turbo-16k-rouge-p-min}{65.24}
\DefMacro{res-selected-434-ne2e-few-shot-with-name-gpt-3.5-turbo-16k-rouge-r}{71.24}
\DefMacro{res-selected-434-ne2e-few-shot-with-name-gpt-3.5-turbo-16k-rouge-r-avg}{71.24}
\DefMacro{res-selected-434-ne2e-few-shot-with-name-gpt-3.5-turbo-16k-rouge-r-max}{71.24}
\DefMacro{res-selected-434-ne2e-few-shot-with-name-gpt-3.5-turbo-16k-rouge-r-min}{71.24}
\DefMacro{res-selected-434-ne2e-few-shot-with-name-gpt-3.5-turbo-16k-xmatch}{11.52}
\DefMacro{res-selected-434-ne2e-few-shot-with-name-gpt-3.5-turbo-16k-xmatch-top1}{11.52}
\DefMacro{res-selected-434-ne2e-few-shot-with-name-gpt-3.5-turbo-16k-compilable-avg}{56.68}
\DefMacro{res-selected-434-ne2e-few-shot-with-name-gpt-3.5-turbo-16k-compilable-max}{56.68}
\DefMacro{res-selected-434-ne2e-few-shot-with-name-gpt-3.5-turbo-16k-compilable-min}{56.68}
\DefMacro{res-selected-434-ne2e-few-shot-with-name-gpt-3.5-turbo-16k-coverage-avg}{32.26}
\DefMacro{res-selected-434-ne2e-few-shot-with-name-gpt-3.5-turbo-16k-coverage-max}{32.26}
\DefMacro{res-selected-434-ne2e-few-shot-with-name-gpt-3.5-turbo-16k-coverage-min}{32.26}
\DefMacro{res-selected-434-ne2e-few-shot-with-name-gpt-3.5-turbo-16k-match-avg}{98.37}
\DefMacro{res-selected-434-ne2e-few-shot-with-name-gpt-3.5-turbo-16k-match-max}{98.37}
\DefMacro{res-selected-434-ne2e-few-shot-with-name-gpt-3.5-turbo-16k-match-min}{98.37}
\DefMacro{res-selected-434-ne2e-few-shot-with-name-gpt-3.5-turbo-16k-runnable-avg}{75.21}
\DefMacro{res-selected-434-ne2e-few-shot-with-name-gpt-3.5-turbo-16k-runnable-max}{75.21}
\DefMacro{res-selected-434-ne2e-few-shot-with-name-gpt-3.5-turbo-16k-runnable-min}{75.21}
\DefMacro{res-selected-434-ne2e-few-shot-with-name-gpt-3.5-turbo-16k-runnable-overall}{41.94}
\DefMacro{res-selected-434-ne2e-few-shot-with-name-gpt-3.5-turbo-16k-runnable-overall-avg}{41.94}
\DefMacro{res-selected-434-ne2e-few-shot-with-name-gpt-3.5-turbo-16k-runnable-overall-max}{41.94}
\DefMacro{res-selected-434-ne2e-few-shot-with-name-gpt-3.5-turbo-16k-runnable-overall-min}{41.94}
\DefMacro{res-selected-434-ne2e-few-shot-with-name-gpt-3.5-turbo-16k-timeout-avg}{0.00}
\DefMacro{res-selected-434-ne2e-few-shot-with-name-gpt-3.5-turbo-16k-timeout-max}{0.00}
\DefMacro{res-selected-434-ne2e-few-shot-with-name-gpt-3.5-turbo-16k-timeout-min}{0.00}
\DefMacro{res-selected-434-mut2e-with-name-ft-lora-codellama-7b-bleu}{57.17}
\DefMacro{res-selected-434-mut2e-with-name-ft-lora-codellama-7b-bleu-avg}{57.17}
\DefMacro{res-selected-434-mut2e-with-name-ft-lora-codellama-7b-bleu-max}{57.17}
\DefMacro{res-selected-434-mut2e-with-name-ft-lora-codellama-7b-bleu-min}{57.17}
\DefMacro{res-selected-434-mut2e-with-name-ft-lora-codellama-7b-code-bleu}{61.79}
\DefMacro{res-selected-434-mut2e-with-name-ft-lora-codellama-7b-code-bleu-avg}{61.79}
\DefMacro{res-selected-434-mut2e-with-name-ft-lora-codellama-7b-code-bleu-max}{61.79}
\DefMacro{res-selected-434-mut2e-with-name-ft-lora-codellama-7b-code-bleu-min}{61.79}
\DefMacro{res-selected-434-mut2e-with-name-ft-lora-codellama-7b-edit-sim}{82.30}
\DefMacro{res-selected-434-mut2e-with-name-ft-lora-codellama-7b-edit-sim-avg}{82.30}
\DefMacro{res-selected-434-mut2e-with-name-ft-lora-codellama-7b-edit-sim-max}{82.30}
\DefMacro{res-selected-434-mut2e-with-name-ft-lora-codellama-7b-edit-sim-min}{82.30}
\DefMacro{res-selected-434-mut2e-with-name-ft-lora-codellama-7b-rouge-f}{75.47}
\DefMacro{res-selected-434-mut2e-with-name-ft-lora-codellama-7b-rouge-f-avg}{75.47}
\DefMacro{res-selected-434-mut2e-with-name-ft-lora-codellama-7b-rouge-f-max}{75.47}
\DefMacro{res-selected-434-mut2e-with-name-ft-lora-codellama-7b-rouge-f-min}{75.47}
\DefMacro{res-selected-434-mut2e-with-name-ft-lora-codellama-7b-rouge-p}{78.86}
\DefMacro{res-selected-434-mut2e-with-name-ft-lora-codellama-7b-rouge-p-avg}{78.86}
\DefMacro{res-selected-434-mut2e-with-name-ft-lora-codellama-7b-rouge-p-max}{78.86}
\DefMacro{res-selected-434-mut2e-with-name-ft-lora-codellama-7b-rouge-p-min}{78.86}
\DefMacro{res-selected-434-mut2e-with-name-ft-lora-codellama-7b-rouge-r}{74.31}
\DefMacro{res-selected-434-mut2e-with-name-ft-lora-codellama-7b-rouge-r-avg}{74.31}
\DefMacro{res-selected-434-mut2e-with-name-ft-lora-codellama-7b-rouge-r-max}{74.31}
\DefMacro{res-selected-434-mut2e-with-name-ft-lora-codellama-7b-rouge-r-min}{74.31}
\DefMacro{res-selected-434-mut2e-with-name-ft-lora-codellama-7b-xmatch}{13.59}
\DefMacro{res-selected-434-mut2e-with-name-ft-lora-codellama-7b-xmatch-avg}{13.59}
\DefMacro{res-selected-434-mut2e-with-name-ft-lora-codellama-7b-xmatch-max}{13.59}
\DefMacro{res-selected-434-mut2e-with-name-ft-lora-codellama-7b-xmatch-min}{13.59}
\DefMacro{res-selected-434-mut2e-with-name-ft-lora-codellama-7b-xmatch-top1}{13.59}
\DefMacro{res-selected-434-mut2e-with-name-ft-lora-codellama-7b-compilable-avg}{62.21}
\DefMacro{res-selected-434-mut2e-with-name-ft-lora-codellama-7b-compilable-max}{62.21}
\DefMacro{res-selected-434-mut2e-with-name-ft-lora-codellama-7b-compilable-min}{62.21}
\DefMacro{res-selected-434-mut2e-with-name-ft-lora-codellama-7b-coverage-avg}{37.79}
\DefMacro{res-selected-434-mut2e-with-name-ft-lora-codellama-7b-coverage-max}{37.79}
\DefMacro{res-selected-434-mut2e-with-name-ft-lora-codellama-7b-coverage-min}{37.79}
\DefMacro{res-selected-434-mut2e-with-name-ft-lora-codellama-7b-match-avg}{100.00}
\DefMacro{res-selected-434-mut2e-with-name-ft-lora-codellama-7b-match-max}{100.00}
\DefMacro{res-selected-434-mut2e-with-name-ft-lora-codellama-7b-match-min}{100.00}
\DefMacro{res-selected-434-mut2e-with-name-ft-lora-codellama-7b-runnable-avg}{77.04}
\DefMacro{res-selected-434-mut2e-with-name-ft-lora-codellama-7b-runnable-max}{77.04}
\DefMacro{res-selected-434-mut2e-with-name-ft-lora-codellama-7b-runnable-min}{77.04}
\DefMacro{res-selected-434-mut2e-with-name-ft-lora-codellama-7b-runnable-overall}{47.93}
\DefMacro{res-selected-434-mut2e-with-name-ft-lora-codellama-7b-runnable-overall-avg}{47.93}
\DefMacro{res-selected-434-mut2e-with-name-ft-lora-codellama-7b-runnable-overall-max}{47.93}
\DefMacro{res-selected-434-mut2e-with-name-ft-lora-codellama-7b-runnable-overall-min}{47.93}
\DefMacro{res-selected-434-mut2e-with-name-ft-lora-codellama-7b-timeout-avg}{0.37}
\DefMacro{res-selected-434-mut2e-with-name-ft-lora-codellama-7b-timeout-max}{0.37}
\DefMacro{res-selected-434-mut2e-with-name-ft-lora-codellama-7b-timeout-min}{0.37}
\DefMacro{res-selected-434-mut2e-with-name-ft-lora-codellama-13b-bleu}{57.89}
\DefMacro{res-selected-434-mut2e-with-name-ft-lora-codellama-13b-bleu-avg}{57.89}
\DefMacro{res-selected-434-mut2e-with-name-ft-lora-codellama-13b-bleu-max}{57.89}
\DefMacro{res-selected-434-mut2e-with-name-ft-lora-codellama-13b-bleu-min}{57.89}
\DefMacro{res-selected-434-mut2e-with-name-ft-lora-codellama-13b-code-bleu}{63.40}
\DefMacro{res-selected-434-mut2e-with-name-ft-lora-codellama-13b-code-bleu-avg}{63.40}
\DefMacro{res-selected-434-mut2e-with-name-ft-lora-codellama-13b-code-bleu-max}{63.40}
\DefMacro{res-selected-434-mut2e-with-name-ft-lora-codellama-13b-code-bleu-min}{63.40}
\DefMacro{res-selected-434-mut2e-with-name-ft-lora-codellama-13b-edit-sim}{82.52}
\DefMacro{res-selected-434-mut2e-with-name-ft-lora-codellama-13b-edit-sim-avg}{82.52}
\DefMacro{res-selected-434-mut2e-with-name-ft-lora-codellama-13b-edit-sim-max}{82.52}
\DefMacro{res-selected-434-mut2e-with-name-ft-lora-codellama-13b-edit-sim-min}{82.52}
\DefMacro{res-selected-434-mut2e-with-name-ft-lora-codellama-13b-rouge-f}{75.88}
\DefMacro{res-selected-434-mut2e-with-name-ft-lora-codellama-13b-rouge-f-avg}{75.88}
\DefMacro{res-selected-434-mut2e-with-name-ft-lora-codellama-13b-rouge-f-max}{75.88}
\DefMacro{res-selected-434-mut2e-with-name-ft-lora-codellama-13b-rouge-f-min}{75.88}
\DefMacro{res-selected-434-mut2e-with-name-ft-lora-codellama-13b-rouge-p}{78.23}
\DefMacro{res-selected-434-mut2e-with-name-ft-lora-codellama-13b-rouge-p-avg}{78.23}
\DefMacro{res-selected-434-mut2e-with-name-ft-lora-codellama-13b-rouge-p-max}{78.23}
\DefMacro{res-selected-434-mut2e-with-name-ft-lora-codellama-13b-rouge-p-min}{78.23}
\DefMacro{res-selected-434-mut2e-with-name-ft-lora-codellama-13b-rouge-r}{75.43}
\DefMacro{res-selected-434-mut2e-with-name-ft-lora-codellama-13b-rouge-r-avg}{75.43}
\DefMacro{res-selected-434-mut2e-with-name-ft-lora-codellama-13b-rouge-r-max}{75.43}
\DefMacro{res-selected-434-mut2e-with-name-ft-lora-codellama-13b-rouge-r-min}{75.43}
\DefMacro{res-selected-434-mut2e-with-name-ft-lora-codellama-13b-xmatch}{13.59}
\DefMacro{res-selected-434-mut2e-with-name-ft-lora-codellama-13b-xmatch-avg}{13.59}
\DefMacro{res-selected-434-mut2e-with-name-ft-lora-codellama-13b-xmatch-max}{13.59}
\DefMacro{res-selected-434-mut2e-with-name-ft-lora-codellama-13b-xmatch-min}{13.59}
\DefMacro{res-selected-434-mut2e-with-name-ft-lora-codellama-13b-xmatch-top1}{13.59}
\DefMacro{res-selected-434-mut2e-with-name-ft-lora-codellama-13b-compilable-avg}{65.90}
\DefMacro{res-selected-434-mut2e-with-name-ft-lora-codellama-13b-compilable-max}{65.90}
\DefMacro{res-selected-434-mut2e-with-name-ft-lora-codellama-13b-compilable-min}{65.90}
\DefMacro{res-selected-434-mut2e-with-name-ft-lora-codellama-13b-coverage-avg}{43.09}
\DefMacro{res-selected-434-mut2e-with-name-ft-lora-codellama-13b-coverage-max}{43.09}
\DefMacro{res-selected-434-mut2e-with-name-ft-lora-codellama-13b-coverage-min}{43.09}
\DefMacro{res-selected-434-mut2e-with-name-ft-lora-codellama-13b-match-avg}{100.00}
\DefMacro{res-selected-434-mut2e-with-name-ft-lora-codellama-13b-match-max}{100.00}
\DefMacro{res-selected-434-mut2e-with-name-ft-lora-codellama-13b-match-min}{100.00}
\DefMacro{res-selected-434-mut2e-with-name-ft-lora-codellama-13b-runnable-avg}{81.12}
\DefMacro{res-selected-434-mut2e-with-name-ft-lora-codellama-13b-runnable-max}{81.12}
\DefMacro{res-selected-434-mut2e-with-name-ft-lora-codellama-13b-runnable-min}{81.12}
\DefMacro{res-selected-434-mut2e-with-name-ft-lora-codellama-13b-runnable-overall}{53.46}
\DefMacro{res-selected-434-mut2e-with-name-ft-lora-codellama-13b-runnable-overall-avg}{53.46}
\DefMacro{res-selected-434-mut2e-with-name-ft-lora-codellama-13b-runnable-overall-max}{53.46}
\DefMacro{res-selected-434-mut2e-with-name-ft-lora-codellama-13b-runnable-overall-min}{53.46}
\DefMacro{res-selected-434-mut2e-with-name-ft-lora-codellama-13b-timeout-avg}{0.00}
\DefMacro{res-selected-434-mut2e-with-name-ft-lora-codellama-13b-timeout-max}{0.00}
\DefMacro{res-selected-434-mut2e-with-name-ft-lora-codellama-13b-timeout-min}{0.00}
\DefMacro{res-selected-434-catlm-mut2e-with-name-catlm-bleu}{42.02}
\DefMacro{res-selected-434-catlm-mut2e-with-name-catlm-bleu-avg}{42.02}
\DefMacro{res-selected-434-catlm-mut2e-with-name-catlm-bleu-max}{42.02}
\DefMacro{res-selected-434-catlm-mut2e-with-name-catlm-bleu-min}{42.02}
\DefMacro{res-selected-434-catlm-mut2e-with-name-catlm-code-bleu}{51.82}
\DefMacro{res-selected-434-catlm-mut2e-with-name-catlm-code-bleu-avg}{51.82}
\DefMacro{res-selected-434-catlm-mut2e-with-name-catlm-code-bleu-max}{51.82}
\DefMacro{res-selected-434-catlm-mut2e-with-name-catlm-code-bleu-min}{51.82}
\DefMacro{res-selected-434-catlm-mut2e-with-name-catlm-edit-sim}{71.11}
\DefMacro{res-selected-434-catlm-mut2e-with-name-catlm-edit-sim-avg}{71.11}
\DefMacro{res-selected-434-catlm-mut2e-with-name-catlm-edit-sim-max}{71.11}
\DefMacro{res-selected-434-catlm-mut2e-with-name-catlm-edit-sim-min}{71.11}
\DefMacro{res-selected-434-catlm-mut2e-with-name-catlm-rouge-f}{64.89}
\DefMacro{res-selected-434-catlm-mut2e-with-name-catlm-rouge-f-avg}{64.89}
\DefMacro{res-selected-434-catlm-mut2e-with-name-catlm-rouge-f-max}{64.89}
\DefMacro{res-selected-434-catlm-mut2e-with-name-catlm-rouge-f-min}{64.89}
\DefMacro{res-selected-434-catlm-mut2e-with-name-catlm-rouge-p}{64.75}
\DefMacro{res-selected-434-catlm-mut2e-with-name-catlm-rouge-p-avg}{64.75}
\DefMacro{res-selected-434-catlm-mut2e-with-name-catlm-rouge-p-max}{64.75}
\DefMacro{res-selected-434-catlm-mut2e-with-name-catlm-rouge-p-min}{64.75}
\DefMacro{res-selected-434-catlm-mut2e-with-name-catlm-rouge-r}{68.69}
\DefMacro{res-selected-434-catlm-mut2e-with-name-catlm-rouge-r-avg}{68.69}
\DefMacro{res-selected-434-catlm-mut2e-with-name-catlm-rouge-r-max}{68.69}
\DefMacro{res-selected-434-catlm-mut2e-with-name-catlm-rouge-r-min}{68.69}
\DefMacro{res-selected-434-catlm-mut2e-with-name-catlm-xmatch}{5.76}
\DefMacro{res-selected-434-catlm-mut2e-with-name-catlm-xmatch-avg}{5.76}
\DefMacro{res-selected-434-catlm-mut2e-with-name-catlm-xmatch-max}{5.76}
\DefMacro{res-selected-434-catlm-mut2e-with-name-catlm-xmatch-min}{5.76}
\DefMacro{res-selected-434-catlm-mut2e-with-name-catlm-xmatch-top1}{5.76}
\DefMacro{res-selected-434-catlm-mut2e-with-name-catlm-compilable-avg}{41.24}
\DefMacro{res-selected-434-catlm-mut2e-with-name-catlm-compilable-max}{41.24}
\DefMacro{res-selected-434-catlm-mut2e-with-name-catlm-compilable-min}{41.24}
\DefMacro{res-selected-434-catlm-mut2e-with-name-catlm-coverage-avg}{19.59}
\DefMacro{res-selected-434-catlm-mut2e-with-name-catlm-coverage-max}{19.59}
\DefMacro{res-selected-434-catlm-mut2e-with-name-catlm-coverage-min}{19.59}
\DefMacro{res-selected-434-catlm-mut2e-with-name-catlm-match-avg}{100.00}
\DefMacro{res-selected-434-catlm-mut2e-with-name-catlm-match-max}{100.00}
\DefMacro{res-selected-434-catlm-mut2e-with-name-catlm-match-min}{100.00}
\DefMacro{res-selected-434-catlm-mut2e-with-name-catlm-runnable-avg}{59.22}
\DefMacro{res-selected-434-catlm-mut2e-with-name-catlm-runnable-max}{59.22}
\DefMacro{res-selected-434-catlm-mut2e-with-name-catlm-runnable-min}{59.22}
\DefMacro{res-selected-434-catlm-mut2e-with-name-catlm-runnable-overall}{24.42}
\DefMacro{res-selected-434-catlm-mut2e-with-name-catlm-runnable-overall-avg}{24.42}
\DefMacro{res-selected-434-catlm-mut2e-with-name-catlm-runnable-overall-max}{24.42}
\DefMacro{res-selected-434-catlm-mut2e-with-name-catlm-runnable-overall-min}{24.42}
\DefMacro{res-selected-434-catlm-mut2e-with-name-catlm-timeout-avg}{0.00}
\DefMacro{res-selected-434-catlm-mut2e-with-name-catlm-timeout-max}{0.00}
\DefMacro{res-selected-434-catlm-mut2e-with-name-catlm-timeout-min}{0.00}
\DefMacro{res-selected-434-ne2e-with-name-ft-lora-codellama-7b-bleu}{63.61}
\DefMacro{res-selected-434-ne2e-with-name-ft-lora-codellama-7b-bleu-avg}{63.61}
\DefMacro{res-selected-434-ne2e-with-name-ft-lora-codellama-7b-bleu-max}{63.61}
\DefMacro{res-selected-434-ne2e-with-name-ft-lora-codellama-7b-bleu-min}{63.61}
\DefMacro{res-selected-434-ne2e-with-name-ft-lora-codellama-7b-code-bleu}{68.08}
\DefMacro{res-selected-434-ne2e-with-name-ft-lora-codellama-7b-code-bleu-avg}{68.08}
\DefMacro{res-selected-434-ne2e-with-name-ft-lora-codellama-7b-code-bleu-max}{68.08}
\DefMacro{res-selected-434-ne2e-with-name-ft-lora-codellama-7b-code-bleu-min}{68.08}
\DefMacro{res-selected-434-ne2e-with-name-ft-lora-codellama-7b-edit-sim}{85.64}
\DefMacro{res-selected-434-ne2e-with-name-ft-lora-codellama-7b-edit-sim-avg}{85.64}
\DefMacro{res-selected-434-ne2e-with-name-ft-lora-codellama-7b-edit-sim-max}{85.64}
\DefMacro{res-selected-434-ne2e-with-name-ft-lora-codellama-7b-edit-sim-min}{85.64}
\DefMacro{res-selected-434-ne2e-with-name-ft-lora-codellama-7b-rouge-f}{80.54}
\DefMacro{res-selected-434-ne2e-with-name-ft-lora-codellama-7b-rouge-f-avg}{80.54}
\DefMacro{res-selected-434-ne2e-with-name-ft-lora-codellama-7b-rouge-f-max}{80.54}
\DefMacro{res-selected-434-ne2e-with-name-ft-lora-codellama-7b-rouge-f-min}{80.54}
\DefMacro{res-selected-434-ne2e-with-name-ft-lora-codellama-7b-rouge-p}{82.41}
\DefMacro{res-selected-434-ne2e-with-name-ft-lora-codellama-7b-rouge-p-avg}{82.41}
\DefMacro{res-selected-434-ne2e-with-name-ft-lora-codellama-7b-rouge-p-max}{82.41}
\DefMacro{res-selected-434-ne2e-with-name-ft-lora-codellama-7b-rouge-p-min}{82.41}
\DefMacro{res-selected-434-ne2e-with-name-ft-lora-codellama-7b-rouge-r}{80.55}
\DefMacro{res-selected-434-ne2e-with-name-ft-lora-codellama-7b-rouge-r-avg}{80.55}
\DefMacro{res-selected-434-ne2e-with-name-ft-lora-codellama-7b-rouge-r-max}{80.55}
\DefMacro{res-selected-434-ne2e-with-name-ft-lora-codellama-7b-rouge-r-min}{80.55}
\DefMacro{res-selected-434-ne2e-with-name-ft-lora-codellama-7b-xmatch}{18.43}
\DefMacro{res-selected-434-ne2e-with-name-ft-lora-codellama-7b-xmatch-top1}{18.43}
\DefMacro{res-selected-434-ne2e-with-name-ft-lora-codellama-7b-compilable-avg}{80.41}
\DefMacro{res-selected-434-ne2e-with-name-ft-lora-codellama-7b-compilable-max}{80.41}
\DefMacro{res-selected-434-ne2e-with-name-ft-lora-codellama-7b-compilable-min}{80.41}
\DefMacro{res-selected-434-ne2e-with-name-ft-lora-codellama-7b-coverage-avg}{49.31}
\DefMacro{res-selected-434-ne2e-with-name-ft-lora-codellama-7b-coverage-max}{49.31}
\DefMacro{res-selected-434-ne2e-with-name-ft-lora-codellama-7b-coverage-min}{49.31}
\DefMacro{res-selected-434-ne2e-with-name-ft-lora-codellama-7b-match-avg}{100.00}
\DefMacro{res-selected-434-ne2e-with-name-ft-lora-codellama-7b-match-max}{100.00}
\DefMacro{res-selected-434-ne2e-with-name-ft-lora-codellama-7b-match-min}{100.00}
\DefMacro{res-selected-434-ne2e-with-name-ft-lora-codellama-7b-runnable-avg}{75.36}
\DefMacro{res-selected-434-ne2e-with-name-ft-lora-codellama-7b-runnable-max}{75.36}
\DefMacro{res-selected-434-ne2e-with-name-ft-lora-codellama-7b-runnable-min}{75.36}
\DefMacro{res-selected-434-ne2e-with-name-ft-lora-codellama-7b-runnable-overall}{60.60}
\DefMacro{res-selected-434-ne2e-with-name-ft-lora-codellama-7b-runnable-overall-avg}{60.60}
\DefMacro{res-selected-434-ne2e-with-name-ft-lora-codellama-7b-runnable-overall-max}{60.60}
\DefMacro{res-selected-434-ne2e-with-name-ft-lora-codellama-7b-runnable-overall-min}{60.60}
\DefMacro{res-selected-434-ne2e-with-name-ft-lora-codellama-7b-timeout-avg}{0.57}
\DefMacro{res-selected-434-ne2e-with-name-ft-lora-codellama-7b-timeout-max}{0.57}
\DefMacro{res-selected-434-ne2e-with-name-ft-lora-codellama-7b-timeout-min}{0.57}
\DefMacro{res-selected-434-ne2e-all-with-name-ft-lora-codellama-7b-bleu}{63.25}
\DefMacro{res-selected-434-ne2e-all-with-name-ft-lora-codellama-7b-bleu-avg}{63.63}
\DefMacro{res-selected-434-ne2e-all-with-name-ft-lora-codellama-7b-bleu-max}{68.66}
\DefMacro{res-selected-434-ne2e-all-with-name-ft-lora-codellama-7b-bleu-min}{59.08}
\DefMacro{res-selected-434-ne2e-all-with-name-ft-lora-codellama-7b-code-bleu}{67.76}
\DefMacro{res-selected-434-ne2e-all-with-name-ft-lora-codellama-7b-code-bleu-avg}{68.21}
\DefMacro{res-selected-434-ne2e-all-with-name-ft-lora-codellama-7b-code-bleu-max}{72.61}
\DefMacro{res-selected-434-ne2e-all-with-name-ft-lora-codellama-7b-code-bleu-min}{64.49}
\DefMacro{res-selected-434-ne2e-all-with-name-ft-lora-codellama-7b-edit-sim}{85.32}
\DefMacro{res-selected-434-ne2e-all-with-name-ft-lora-codellama-7b-edit-sim-avg}{85.49}
\DefMacro{res-selected-434-ne2e-all-with-name-ft-lora-codellama-7b-edit-sim-max}{88.50}
\DefMacro{res-selected-434-ne2e-all-with-name-ft-lora-codellama-7b-edit-sim-min}{82.54}
\DefMacro{res-selected-434-ne2e-all-with-name-ft-lora-codellama-7b-rouge-f}{80.57}
\DefMacro{res-selected-434-ne2e-all-with-name-ft-lora-codellama-7b-rouge-f-avg}{80.83}
\DefMacro{res-selected-434-ne2e-all-with-name-ft-lora-codellama-7b-rouge-f-max}{84.08}
\DefMacro{res-selected-434-ne2e-all-with-name-ft-lora-codellama-7b-rouge-f-min}{77.61}
\DefMacro{res-selected-434-ne2e-all-with-name-ft-lora-codellama-7b-rouge-p}{82.50}
\DefMacro{res-selected-434-ne2e-all-with-name-ft-lora-codellama-7b-rouge-p-avg}{82.97}
\DefMacro{res-selected-434-ne2e-all-with-name-ft-lora-codellama-7b-rouge-p-max}{86.55}
\DefMacro{res-selected-434-ne2e-all-with-name-ft-lora-codellama-7b-rouge-p-min}{78.70}
\DefMacro{res-selected-434-ne2e-all-with-name-ft-lora-codellama-7b-rouge-r}{80.67}
\DefMacro{res-selected-434-ne2e-all-with-name-ft-lora-codellama-7b-rouge-r-avg}{80.73}
\DefMacro{res-selected-434-ne2e-all-with-name-ft-lora-codellama-7b-rouge-r-max}{83.84}
\DefMacro{res-selected-434-ne2e-all-with-name-ft-lora-codellama-7b-rouge-r-min}{78.05}
\DefMacro{res-selected-434-ne2e-all-with-name-ft-lora-codellama-7b-xmatch}{18.43}
\DefMacro{res-selected-434-ne2e-all-with-name-ft-lora-codellama-7b-xmatch-avg}{19.02}
\DefMacro{res-selected-434-ne2e-all-with-name-ft-lora-codellama-7b-xmatch-max}{22.35}
\DefMacro{res-selected-434-ne2e-all-with-name-ft-lora-codellama-7b-xmatch-min}{16.59}
\DefMacro{res-selected-434-ne2e-all-with-name-ft-lora-codellama-7b-xmatch-top1}{18.43}
\DefMacro{res-selected-434-ne2e-all-with-name-ft-lora-codellama-7b-compilable-avg}{82.78}
\DefMacro{res-selected-434-ne2e-all-with-name-ft-lora-codellama-7b-compilable-max}{88.25}
\DefMacro{res-selected-434-ne2e-all-with-name-ft-lora-codellama-7b-compilable-min}{75.35}
\DefMacro{res-selected-434-ne2e-all-with-name-ft-lora-codellama-7b-coverage-avg}{51.47}
\DefMacro{res-selected-434-ne2e-all-with-name-ft-lora-codellama-7b-coverage-max}{60.37}
\DefMacro{res-selected-434-ne2e-all-with-name-ft-lora-codellama-7b-coverage-min}{43.78}
\DefMacro{res-selected-434-ne2e-all-with-name-ft-lora-codellama-7b-match-avg}{100.00}
\DefMacro{res-selected-434-ne2e-all-with-name-ft-lora-codellama-7b-match-max}{100.00}
\DefMacro{res-selected-434-ne2e-all-with-name-ft-lora-codellama-7b-match-min}{100.00}
\DefMacro{res-selected-434-ne2e-all-with-name-ft-lora-codellama-7b-runnable-avg}{73.73}
\DefMacro{res-selected-434-ne2e-all-with-name-ft-lora-codellama-7b-runnable-max}{79.90}
\DefMacro{res-selected-434-ne2e-all-with-name-ft-lora-codellama-7b-runnable-min}{68.15}
\DefMacro{res-selected-434-ne2e-all-with-name-ft-lora-codellama-7b-runnable-overall}{70.51}
\DefMacro{res-selected-434-ne2e-all-with-name-ft-lora-codellama-7b-runnable-overall-avg}{61.61}
\DefMacro{res-selected-434-ne2e-all-with-name-ft-lora-codellama-7b-runnable-overall-max}{70.51}
\DefMacro{res-selected-434-ne2e-all-with-name-ft-lora-codellama-7b-runnable-overall-min}{52.76}
\DefMacro{res-selected-434-ne2e-all-with-name-ft-lora-codellama-7b-timeout-avg}{0.52}
\DefMacro{res-selected-434-ne2e-all-with-name-ft-lora-codellama-7b-timeout-max}{0.52}
\DefMacro{res-selected-434-ne2e-all-with-name-ft-lora-codellama-7b-timeout-min}{0.52}
\DefMacro{res-selected-434-nestack2e-with-name-ft-lora-codellama-7b-bleu}{64.42}
\DefMacro{res-selected-434-nestack2e-with-name-ft-lora-codellama-7b-bleu-avg}{64.42}
\DefMacro{res-selected-434-nestack2e-with-name-ft-lora-codellama-7b-bleu-max}{64.42}
\DefMacro{res-selected-434-nestack2e-with-name-ft-lora-codellama-7b-bleu-min}{64.42}
\DefMacro{res-selected-434-nestack2e-with-name-ft-lora-codellama-7b-code-bleu}{68.50}
\DefMacro{res-selected-434-nestack2e-with-name-ft-lora-codellama-7b-code-bleu-avg}{68.50}
\DefMacro{res-selected-434-nestack2e-with-name-ft-lora-codellama-7b-code-bleu-max}{68.50}
\DefMacro{res-selected-434-nestack2e-with-name-ft-lora-codellama-7b-code-bleu-min}{68.50}
\DefMacro{res-selected-434-nestack2e-with-name-ft-lora-codellama-7b-edit-sim}{86.05}
\DefMacro{res-selected-434-nestack2e-with-name-ft-lora-codellama-7b-edit-sim-avg}{86.05}
\DefMacro{res-selected-434-nestack2e-with-name-ft-lora-codellama-7b-edit-sim-max}{86.05}
\DefMacro{res-selected-434-nestack2e-with-name-ft-lora-codellama-7b-edit-sim-min}{86.05}
\DefMacro{res-selected-434-nestack2e-with-name-ft-lora-codellama-7b-rouge-f}{81.34}
\DefMacro{res-selected-434-nestack2e-with-name-ft-lora-codellama-7b-rouge-f-avg}{81.34}
\DefMacro{res-selected-434-nestack2e-with-name-ft-lora-codellama-7b-rouge-f-max}{81.34}
\DefMacro{res-selected-434-nestack2e-with-name-ft-lora-codellama-7b-rouge-f-min}{81.34}
\DefMacro{res-selected-434-nestack2e-with-name-ft-lora-codellama-7b-rouge-p}{83.27}
\DefMacro{res-selected-434-nestack2e-with-name-ft-lora-codellama-7b-rouge-p-avg}{83.27}
\DefMacro{res-selected-434-nestack2e-with-name-ft-lora-codellama-7b-rouge-p-max}{83.27}
\DefMacro{res-selected-434-nestack2e-with-name-ft-lora-codellama-7b-rouge-p-min}{83.27}
\DefMacro{res-selected-434-nestack2e-with-name-ft-lora-codellama-7b-rouge-r}{81.15}
\DefMacro{res-selected-434-nestack2e-with-name-ft-lora-codellama-7b-rouge-r-avg}{81.15}
\DefMacro{res-selected-434-nestack2e-with-name-ft-lora-codellama-7b-rouge-r-max}{81.15}
\DefMacro{res-selected-434-nestack2e-with-name-ft-lora-codellama-7b-rouge-r-min}{81.15}
\DefMacro{res-selected-434-nestack2e-with-name-ft-lora-codellama-7b-xmatch}{18.89}
\DefMacro{res-selected-434-nestack2e-with-name-ft-lora-codellama-7b-xmatch-top1}{18.89}
\DefMacro{res-selected-434-nestack2e-with-name-ft-lora-codellama-7b-compilable-avg}{84.10}
\DefMacro{res-selected-434-nestack2e-with-name-ft-lora-codellama-7b-compilable-max}{84.10}
\DefMacro{res-selected-434-nestack2e-with-name-ft-lora-codellama-7b-compilable-min}{84.10}
\DefMacro{res-selected-434-nestack2e-with-name-ft-lora-codellama-7b-coverage-avg}{58.76}
\DefMacro{res-selected-434-nestack2e-with-name-ft-lora-codellama-7b-coverage-max}{58.76}
\DefMacro{res-selected-434-nestack2e-with-name-ft-lora-codellama-7b-coverage-min}{58.76}
\DefMacro{res-selected-434-nestack2e-with-name-ft-lora-codellama-7b-match-avg}{100.00}
\DefMacro{res-selected-434-nestack2e-with-name-ft-lora-codellama-7b-match-max}{100.00}
\DefMacro{res-selected-434-nestack2e-with-name-ft-lora-codellama-7b-match-min}{100.00}
\DefMacro{res-selected-434-nestack2e-with-name-ft-lora-codellama-7b-runnable-avg}{81.10}
\DefMacro{res-selected-434-nestack2e-with-name-ft-lora-codellama-7b-runnable-max}{81.10}
\DefMacro{res-selected-434-nestack2e-with-name-ft-lora-codellama-7b-runnable-min}{81.10}
\DefMacro{res-selected-434-nestack2e-with-name-ft-lora-codellama-7b-runnable-overall}{68.20}
\DefMacro{res-selected-434-nestack2e-with-name-ft-lora-codellama-7b-runnable-overall-avg}{68.20}
\DefMacro{res-selected-434-nestack2e-with-name-ft-lora-codellama-7b-runnable-overall-max}{68.20}
\DefMacro{res-selected-434-nestack2e-with-name-ft-lora-codellama-7b-runnable-overall-min}{68.20}
\DefMacro{res-selected-434-nestack2e-with-name-ft-lora-codellama-7b-timeout-avg}{0.55}
\DefMacro{res-selected-434-nestack2e-with-name-ft-lora-codellama-7b-timeout-max}{0.55}
\DefMacro{res-selected-434-nestack2e-with-name-ft-lora-codellama-7b-timeout-min}{0.55}
\DefMacro{res-selected-434-nestack2e-with-name-ft-lora-codellama-13b-bleu}{65.22}
\DefMacro{res-selected-434-nestack2e-with-name-ft-lora-codellama-13b-bleu-avg}{65.22}
\DefMacro{res-selected-434-nestack2e-with-name-ft-lora-codellama-13b-bleu-max}{65.22}
\DefMacro{res-selected-434-nestack2e-with-name-ft-lora-codellama-13b-bleu-min}{65.22}
\DefMacro{res-selected-434-nestack2e-with-name-ft-lora-codellama-13b-code-bleu}{69.24}
\DefMacro{res-selected-434-nestack2e-with-name-ft-lora-codellama-13b-code-bleu-avg}{69.24}
\DefMacro{res-selected-434-nestack2e-with-name-ft-lora-codellama-13b-code-bleu-max}{69.24}
\DefMacro{res-selected-434-nestack2e-with-name-ft-lora-codellama-13b-code-bleu-min}{69.24}
\DefMacro{res-selected-434-nestack2e-with-name-ft-lora-codellama-13b-edit-sim}{86.71}
\DefMacro{res-selected-434-nestack2e-with-name-ft-lora-codellama-13b-edit-sim-avg}{86.71}
\DefMacro{res-selected-434-nestack2e-with-name-ft-lora-codellama-13b-edit-sim-max}{86.71}
\DefMacro{res-selected-434-nestack2e-with-name-ft-lora-codellama-13b-edit-sim-min}{86.71}
\DefMacro{res-selected-434-nestack2e-with-name-ft-lora-codellama-13b-rouge-f}{82.05}
\DefMacro{res-selected-434-nestack2e-with-name-ft-lora-codellama-13b-rouge-f-avg}{82.05}
\DefMacro{res-selected-434-nestack2e-with-name-ft-lora-codellama-13b-rouge-f-max}{82.05}
\DefMacro{res-selected-434-nestack2e-with-name-ft-lora-codellama-13b-rouge-f-min}{82.05}
\DefMacro{res-selected-434-nestack2e-with-name-ft-lora-codellama-13b-rouge-p}{84.04}
\DefMacro{res-selected-434-nestack2e-with-name-ft-lora-codellama-13b-rouge-p-avg}{84.04}
\DefMacro{res-selected-434-nestack2e-with-name-ft-lora-codellama-13b-rouge-p-max}{84.04}
\DefMacro{res-selected-434-nestack2e-with-name-ft-lora-codellama-13b-rouge-p-min}{84.04}
\DefMacro{res-selected-434-nestack2e-with-name-ft-lora-codellama-13b-rouge-r}{81.89}
\DefMacro{res-selected-434-nestack2e-with-name-ft-lora-codellama-13b-rouge-r-avg}{81.89}
\DefMacro{res-selected-434-nestack2e-with-name-ft-lora-codellama-13b-rouge-r-max}{81.89}
\DefMacro{res-selected-434-nestack2e-with-name-ft-lora-codellama-13b-rouge-r-min}{81.89}
\DefMacro{res-selected-434-nestack2e-with-name-ft-lora-codellama-13b-xmatch}{20.28}
\DefMacro{res-selected-434-nestack2e-with-name-ft-lora-codellama-13b-xmatch-avg}{20.28}
\DefMacro{res-selected-434-nestack2e-with-name-ft-lora-codellama-13b-xmatch-max}{20.28}
\DefMacro{res-selected-434-nestack2e-with-name-ft-lora-codellama-13b-xmatch-min}{20.28}
\DefMacro{res-selected-434-nestack2e-with-name-ft-lora-codellama-13b-xmatch-top1}{20.28}
\DefMacro{res-selected-434-nestack2e-with-name-ft-lora-codellama-13b-compilable-avg}{83.87}
\DefMacro{res-selected-434-nestack2e-with-name-ft-lora-codellama-13b-compilable-max}{83.87}
\DefMacro{res-selected-434-nestack2e-with-name-ft-lora-codellama-13b-compilable-min}{83.87}
\DefMacro{res-selected-434-nestack2e-with-name-ft-lora-codellama-13b-coverage-avg}{65.67}
\DefMacro{res-selected-434-nestack2e-with-name-ft-lora-codellama-13b-coverage-max}{65.67}
\DefMacro{res-selected-434-nestack2e-with-name-ft-lora-codellama-13b-coverage-min}{65.67}
\DefMacro{res-selected-434-nestack2e-with-name-ft-lora-codellama-13b-match-avg}{99.45}
\DefMacro{res-selected-434-nestack2e-with-name-ft-lora-codellama-13b-match-max}{99.45}
\DefMacro{res-selected-434-nestack2e-with-name-ft-lora-codellama-13b-match-min}{99.45}
\DefMacro{res-selected-434-nestack2e-with-name-ft-lora-codellama-13b-runnable-avg}{87.85}
\DefMacro{res-selected-434-nestack2e-with-name-ft-lora-codellama-13b-runnable-max}{87.85}
\DefMacro{res-selected-434-nestack2e-with-name-ft-lora-codellama-13b-runnable-min}{87.85}
\DefMacro{res-selected-434-nestack2e-with-name-ft-lora-codellama-13b-runnable-overall}{73.27}
\DefMacro{res-selected-434-nestack2e-with-name-ft-lora-codellama-13b-runnable-overall-avg}{73.27}
\DefMacro{res-selected-434-nestack2e-with-name-ft-lora-codellama-13b-runnable-overall-max}{73.27}
\DefMacro{res-selected-434-nestack2e-with-name-ft-lora-codellama-13b-runnable-overall-min}{73.27}
\DefMacro{res-selected-434-nestack2e-with-name-ft-lora-codellama-13b-timeout-avg}{0.00}
\DefMacro{res-selected-434-nestack2e-with-name-ft-lora-codellama-13b-timeout-max}{0.00}
\DefMacro{res-selected-434-nestack2e-with-name-ft-lora-codellama-13b-timeout-min}{0.00}
\DefMacro{res-selected-434-nestack2e-all-with-name-ft-lora-codellama-7b-bleu}{62.99}
\DefMacro{res-selected-434-nestack2e-all-with-name-ft-lora-codellama-7b-bleu-avg}{63.42}
\DefMacro{res-selected-434-nestack2e-all-with-name-ft-lora-codellama-7b-bleu-max}{68.21}
\DefMacro{res-selected-434-nestack2e-all-with-name-ft-lora-codellama-7b-bleu-min}{58.86}
\DefMacro{res-selected-434-nestack2e-all-with-name-ft-lora-codellama-7b-code-bleu}{67.58}
\DefMacro{res-selected-434-nestack2e-all-with-name-ft-lora-codellama-7b-code-bleu-avg}{67.84}
\DefMacro{res-selected-434-nestack2e-all-with-name-ft-lora-codellama-7b-code-bleu-max}{72.09}
\DefMacro{res-selected-434-nestack2e-all-with-name-ft-lora-codellama-7b-code-bleu-min}{64.09}
\DefMacro{res-selected-434-nestack2e-all-with-name-ft-lora-codellama-7b-edit-sim}{85.59}
\DefMacro{res-selected-434-nestack2e-all-with-name-ft-lora-codellama-7b-edit-sim-avg}{85.67}
\DefMacro{res-selected-434-nestack2e-all-with-name-ft-lora-codellama-7b-edit-sim-max}{88.38}
\DefMacro{res-selected-434-nestack2e-all-with-name-ft-lora-codellama-7b-edit-sim-min}{82.83}
\DefMacro{res-selected-434-nestack2e-all-with-name-ft-lora-codellama-7b-rouge-f}{80.56}
\DefMacro{res-selected-434-nestack2e-all-with-name-ft-lora-codellama-7b-rouge-f-avg}{80.69}
\DefMacro{res-selected-434-nestack2e-all-with-name-ft-lora-codellama-7b-rouge-f-max}{83.59}
\DefMacro{res-selected-434-nestack2e-all-with-name-ft-lora-codellama-7b-rouge-f-min}{77.55}
\DefMacro{res-selected-434-nestack2e-all-with-name-ft-lora-codellama-7b-rouge-p}{82.37}
\DefMacro{res-selected-434-nestack2e-all-with-name-ft-lora-codellama-7b-rouge-p-avg}{82.73}
\DefMacro{res-selected-434-nestack2e-all-with-name-ft-lora-codellama-7b-rouge-p-max}{86.07}
\DefMacro{res-selected-434-nestack2e-all-with-name-ft-lora-codellama-7b-rouge-p-min}{78.53}
\DefMacro{res-selected-434-nestack2e-all-with-name-ft-lora-codellama-7b-rouge-r}{80.76}
\DefMacro{res-selected-434-nestack2e-all-with-name-ft-lora-codellama-7b-rouge-r-avg}{80.70}
\DefMacro{res-selected-434-nestack2e-all-with-name-ft-lora-codellama-7b-rouge-r-max}{83.34}
\DefMacro{res-selected-434-nestack2e-all-with-name-ft-lora-codellama-7b-rouge-r-min}{78.22}
\DefMacro{res-selected-434-nestack2e-all-with-name-ft-lora-codellama-7b-xmatch}{17.74}
\DefMacro{res-selected-434-nestack2e-all-with-name-ft-lora-codellama-7b-xmatch-avg}{18.29}
\DefMacro{res-selected-434-nestack2e-all-with-name-ft-lora-codellama-7b-xmatch-max}{22.58}
\DefMacro{res-selected-434-nestack2e-all-with-name-ft-lora-codellama-7b-xmatch-min}{15.21}
\DefMacro{res-selected-434-nestack2e-all-with-name-ft-lora-codellama-7b-xmatch-top1}{17.74}
\DefMacro{res-selected-434-nestack2e-all-with-name-ft-lora-codellama-7b-compilable-avg}{84.53}
\DefMacro{res-selected-434-nestack2e-all-with-name-ft-lora-codellama-7b-compilable-max}{88.48}
\DefMacro{res-selected-434-nestack2e-all-with-name-ft-lora-codellama-7b-compilable-min}{79.26}
\DefMacro{res-selected-434-nestack2e-all-with-name-ft-lora-codellama-7b-coverage-avg}{59.35}
\DefMacro{res-selected-434-nestack2e-all-with-name-ft-lora-codellama-7b-coverage-max}{67.97}
\DefMacro{res-selected-434-nestack2e-all-with-name-ft-lora-codellama-7b-coverage-min}{50.92}
\DefMacro{res-selected-434-nestack2e-all-with-name-ft-lora-codellama-7b-match-avg}{100.00}
\DefMacro{res-selected-434-nestack2e-all-with-name-ft-lora-codellama-7b-match-max}{100.00}
\DefMacro{res-selected-434-nestack2e-all-with-name-ft-lora-codellama-7b-match-min}{100.00}
\DefMacro{res-selected-434-nestack2e-all-with-name-ft-lora-codellama-7b-runnable-avg}{80.11}
\DefMacro{res-selected-434-nestack2e-all-with-name-ft-lora-codellama-7b-runnable-max}{87.76}
\DefMacro{res-selected-434-nestack2e-all-with-name-ft-lora-codellama-7b-runnable-min}{73.18}
\DefMacro{res-selected-434-nestack2e-all-with-name-ft-lora-codellama-7b-runnable-overall}{77.65}
\DefMacro{res-selected-434-nestack2e-all-with-name-ft-lora-codellama-7b-runnable-overall-avg}{68.15}
\DefMacro{res-selected-434-nestack2e-all-with-name-ft-lora-codellama-7b-runnable-overall-max}{77.65}
\DefMacro{res-selected-434-nestack2e-all-with-name-ft-lora-codellama-7b-runnable-overall-min}{58.76}
\DefMacro{res-selected-434-nestack2e-all-with-name-ft-lora-codellama-7b-timeout-avg}{0.39}
\DefMacro{res-selected-434-nestack2e-all-with-name-ft-lora-codellama-7b-timeout-max}{0.52}
\DefMacro{res-selected-434-nestack2e-all-with-name-ft-lora-codellama-7b-timeout-min}{0.26}
\DefMacro{res-selected-434-conditionne2e-with-name-ft-lora-codellama-7b-bleu}{64.38}
\DefMacro{res-selected-434-conditionne2e-with-name-ft-lora-codellama-7b-bleu-avg}{64.38}
\DefMacro{res-selected-434-conditionne2e-with-name-ft-lora-codellama-7b-bleu-max}{64.38}
\DefMacro{res-selected-434-conditionne2e-with-name-ft-lora-codellama-7b-bleu-min}{64.38}
\DefMacro{res-selected-434-conditionne2e-with-name-ft-lora-codellama-7b-code-bleu}{69.16}
\DefMacro{res-selected-434-conditionne2e-with-name-ft-lora-codellama-7b-code-bleu-avg}{69.16}
\DefMacro{res-selected-434-conditionne2e-with-name-ft-lora-codellama-7b-code-bleu-max}{69.16}
\DefMacro{res-selected-434-conditionne2e-with-name-ft-lora-codellama-7b-code-bleu-min}{69.16}
\DefMacro{res-selected-434-conditionne2e-with-name-ft-lora-codellama-7b-edit-sim}{86.12}
\DefMacro{res-selected-434-conditionne2e-with-name-ft-lora-codellama-7b-edit-sim-avg}{86.12}
\DefMacro{res-selected-434-conditionne2e-with-name-ft-lora-codellama-7b-edit-sim-max}{86.12}
\DefMacro{res-selected-434-conditionne2e-with-name-ft-lora-codellama-7b-edit-sim-min}{86.12}
\DefMacro{res-selected-434-conditionne2e-with-name-ft-lora-codellama-7b-rouge-f}{81.38}
\DefMacro{res-selected-434-conditionne2e-with-name-ft-lora-codellama-7b-rouge-f-avg}{81.38}
\DefMacro{res-selected-434-conditionne2e-with-name-ft-lora-codellama-7b-rouge-f-max}{81.38}
\DefMacro{res-selected-434-conditionne2e-with-name-ft-lora-codellama-7b-rouge-f-min}{81.38}
\DefMacro{res-selected-434-conditionne2e-with-name-ft-lora-codellama-7b-rouge-p}{83.32}
\DefMacro{res-selected-434-conditionne2e-with-name-ft-lora-codellama-7b-rouge-p-avg}{83.32}
\DefMacro{res-selected-434-conditionne2e-with-name-ft-lora-codellama-7b-rouge-p-max}{83.32}
\DefMacro{res-selected-434-conditionne2e-with-name-ft-lora-codellama-7b-rouge-p-min}{83.32}
\DefMacro{res-selected-434-conditionne2e-with-name-ft-lora-codellama-7b-rouge-r}{81.35}
\DefMacro{res-selected-434-conditionne2e-with-name-ft-lora-codellama-7b-rouge-r-avg}{81.35}
\DefMacro{res-selected-434-conditionne2e-with-name-ft-lora-codellama-7b-rouge-r-max}{81.35}
\DefMacro{res-selected-434-conditionne2e-with-name-ft-lora-codellama-7b-rouge-r-min}{81.35}
\DefMacro{res-selected-434-conditionne2e-with-name-ft-lora-codellama-7b-xmatch}{18.20}
\DefMacro{res-selected-434-conditionne2e-with-name-ft-lora-codellama-7b-xmatch-top1}{18.20}
\DefMacro{res-selected-434-conditionne2e-with-name-ft-lora-codellama-7b-compilable-avg}{82.49}
\DefMacro{res-selected-434-conditionne2e-with-name-ft-lora-codellama-7b-compilable-max}{82.49}
\DefMacro{res-selected-434-conditionne2e-with-name-ft-lora-codellama-7b-compilable-min}{82.49}
\DefMacro{res-selected-434-conditionne2e-with-name-ft-lora-codellama-7b-coverage-avg}{56.91}
\DefMacro{res-selected-434-conditionne2e-with-name-ft-lora-codellama-7b-coverage-max}{56.91}
\DefMacro{res-selected-434-conditionne2e-with-name-ft-lora-codellama-7b-coverage-min}{56.91}
\DefMacro{res-selected-434-conditionne2e-with-name-ft-lora-codellama-7b-match-avg}{100.00}
\DefMacro{res-selected-434-conditionne2e-with-name-ft-lora-codellama-7b-match-max}{100.00}
\DefMacro{res-selected-434-conditionne2e-with-name-ft-lora-codellama-7b-match-min}{100.00}
\DefMacro{res-selected-434-conditionne2e-with-name-ft-lora-codellama-7b-runnable-avg}{81.01}
\DefMacro{res-selected-434-conditionne2e-with-name-ft-lora-codellama-7b-runnable-max}{81.01}
\DefMacro{res-selected-434-conditionne2e-with-name-ft-lora-codellama-7b-runnable-min}{81.01}
\DefMacro{res-selected-434-conditionne2e-with-name-ft-lora-codellama-7b-runnable-overall}{66.82}
\DefMacro{res-selected-434-conditionne2e-with-name-ft-lora-codellama-7b-runnable-overall-avg}{66.82}
\DefMacro{res-selected-434-conditionne2e-with-name-ft-lora-codellama-7b-runnable-overall-max}{66.82}
\DefMacro{res-selected-434-conditionne2e-with-name-ft-lora-codellama-7b-runnable-overall-min}{66.82}
\DefMacro{res-selected-434-conditionne2e-with-name-ft-lora-codellama-7b-timeout-avg}{0.56}
\DefMacro{res-selected-434-conditionne2e-with-name-ft-lora-codellama-7b-timeout-max}{0.56}
\DefMacro{res-selected-434-conditionne2e-with-name-ft-lora-codellama-7b-timeout-min}{0.56}
\DefMacro{res-selected-434-conditionnestack2e-with-name-ft-lora-codellama-7b-bleu}{64.17}
\DefMacro{res-selected-434-conditionnestack2e-with-name-ft-lora-codellama-7b-bleu-avg}{64.17}
\DefMacro{res-selected-434-conditionnestack2e-with-name-ft-lora-codellama-7b-bleu-max}{64.17}
\DefMacro{res-selected-434-conditionnestack2e-with-name-ft-lora-codellama-7b-bleu-min}{64.17}
\DefMacro{res-selected-434-conditionnestack2e-with-name-ft-lora-codellama-7b-code-bleu}{68.60}
\DefMacro{res-selected-434-conditionnestack2e-with-name-ft-lora-codellama-7b-code-bleu-avg}{68.60}
\DefMacro{res-selected-434-conditionnestack2e-with-name-ft-lora-codellama-7b-code-bleu-max}{68.60}
\DefMacro{res-selected-434-conditionnestack2e-with-name-ft-lora-codellama-7b-code-bleu-min}{68.60}
\DefMacro{res-selected-434-conditionnestack2e-with-name-ft-lora-codellama-7b-edit-sim}{85.78}
\DefMacro{res-selected-434-conditionnestack2e-with-name-ft-lora-codellama-7b-edit-sim-avg}{85.78}
\DefMacro{res-selected-434-conditionnestack2e-with-name-ft-lora-codellama-7b-edit-sim-max}{85.78}
\DefMacro{res-selected-434-conditionnestack2e-with-name-ft-lora-codellama-7b-edit-sim-min}{85.78}
\DefMacro{res-selected-434-conditionnestack2e-with-name-ft-lora-codellama-7b-rouge-f}{81.35}
\DefMacro{res-selected-434-conditionnestack2e-with-name-ft-lora-codellama-7b-rouge-f-avg}{81.35}
\DefMacro{res-selected-434-conditionnestack2e-with-name-ft-lora-codellama-7b-rouge-f-max}{81.35}
\DefMacro{res-selected-434-conditionnestack2e-with-name-ft-lora-codellama-7b-rouge-f-min}{81.35}
\DefMacro{res-selected-434-conditionnestack2e-with-name-ft-lora-codellama-7b-rouge-p}{83.18}
\DefMacro{res-selected-434-conditionnestack2e-with-name-ft-lora-codellama-7b-rouge-p-avg}{83.18}
\DefMacro{res-selected-434-conditionnestack2e-with-name-ft-lora-codellama-7b-rouge-p-max}{83.18}
\DefMacro{res-selected-434-conditionnestack2e-with-name-ft-lora-codellama-7b-rouge-p-min}{83.18}
\DefMacro{res-selected-434-conditionnestack2e-with-name-ft-lora-codellama-7b-rouge-r}{81.44}
\DefMacro{res-selected-434-conditionnestack2e-with-name-ft-lora-codellama-7b-rouge-r-avg}{81.44}
\DefMacro{res-selected-434-conditionnestack2e-with-name-ft-lora-codellama-7b-rouge-r-max}{81.44}
\DefMacro{res-selected-434-conditionnestack2e-with-name-ft-lora-codellama-7b-rouge-r-min}{81.44}
\DefMacro{res-selected-434-conditionnestack2e-with-name-ft-lora-codellama-7b-xmatch}{19.82}
\DefMacro{res-selected-434-conditionnestack2e-with-name-ft-lora-codellama-7b-xmatch-top1}{19.82}
\DefMacro{res-selected-434-conditionnestack2e-with-name-ft-lora-codellama-7b-compilable-avg}{82.49}
\DefMacro{res-selected-434-conditionnestack2e-with-name-ft-lora-codellama-7b-compilable-max}{82.49}
\DefMacro{res-selected-434-conditionnestack2e-with-name-ft-lora-codellama-7b-compilable-min}{82.49}
\DefMacro{res-selected-434-conditionnestack2e-with-name-ft-lora-codellama-7b-coverage-avg}{58.53}
\DefMacro{res-selected-434-conditionnestack2e-with-name-ft-lora-codellama-7b-coverage-max}{58.53}
\DefMacro{res-selected-434-conditionnestack2e-with-name-ft-lora-codellama-7b-coverage-min}{58.53}
\DefMacro{res-selected-434-conditionnestack2e-with-name-ft-lora-codellama-7b-match-avg}{100.00}
\DefMacro{res-selected-434-conditionnestack2e-with-name-ft-lora-codellama-7b-match-max}{100.00}
\DefMacro{res-selected-434-conditionnestack2e-with-name-ft-lora-codellama-7b-match-min}{100.00}
\DefMacro{res-selected-434-conditionnestack2e-with-name-ft-lora-codellama-7b-runnable-avg}{82.40}
\DefMacro{res-selected-434-conditionnestack2e-with-name-ft-lora-codellama-7b-runnable-max}{82.40}
\DefMacro{res-selected-434-conditionnestack2e-with-name-ft-lora-codellama-7b-runnable-min}{82.40}
\DefMacro{res-selected-434-conditionnestack2e-with-name-ft-lora-codellama-7b-runnable-overall}{67.97}
\DefMacro{res-selected-434-conditionnestack2e-with-name-ft-lora-codellama-7b-runnable-overall-avg}{67.97}
\DefMacro{res-selected-434-conditionnestack2e-with-name-ft-lora-codellama-7b-runnable-overall-max}{67.97}
\DefMacro{res-selected-434-conditionnestack2e-with-name-ft-lora-codellama-7b-runnable-overall-min}{67.97}
\DefMacro{res-selected-434-conditionnestack2e-with-name-ft-lora-codellama-7b-timeout-avg}{0.56}
\DefMacro{res-selected-434-conditionnestack2e-with-name-ft-lora-codellama-7b-timeout-max}{0.56}
\DefMacro{res-selected-434-conditionnestack2e-with-name-ft-lora-codellama-7b-timeout-min}{0.56}
\DefMacro{res-selected-434-conditionnestack2e-all-with-name-ft-lora-codellama-7b-bleu}{63.44}
\DefMacro{res-selected-434-conditionnestack2e-all-with-name-ft-lora-codellama-7b-bleu-avg}{64.13}
\DefMacro{res-selected-434-conditionnestack2e-all-with-name-ft-lora-codellama-7b-bleu-max}{69.60}
\DefMacro{res-selected-434-conditionnestack2e-all-with-name-ft-lora-codellama-7b-bleu-min}{59.49}
\DefMacro{res-selected-434-conditionnestack2e-all-with-name-ft-lora-codellama-7b-code-bleu}{68.13}
\DefMacro{res-selected-434-conditionnestack2e-all-with-name-ft-lora-codellama-7b-code-bleu-avg}{68.68}
\DefMacro{res-selected-434-conditionnestack2e-all-with-name-ft-lora-codellama-7b-code-bleu-max}{73.40}
\DefMacro{res-selected-434-conditionnestack2e-all-with-name-ft-lora-codellama-7b-code-bleu-min}{64.93}
\DefMacro{res-selected-434-conditionnestack2e-all-with-name-ft-lora-codellama-7b-edit-sim}{85.76}
\DefMacro{res-selected-434-conditionnestack2e-all-with-name-ft-lora-codellama-7b-edit-sim-avg}{85.85}
\DefMacro{res-selected-434-conditionnestack2e-all-with-name-ft-lora-codellama-7b-edit-sim-max}{88.77}
\DefMacro{res-selected-434-conditionnestack2e-all-with-name-ft-lora-codellama-7b-edit-sim-min}{82.99}
\DefMacro{res-selected-434-conditionnestack2e-all-with-name-ft-lora-codellama-7b-rouge-f}{80.98}
\DefMacro{res-selected-434-conditionnestack2e-all-with-name-ft-lora-codellama-7b-rouge-f-avg}{81.21}
\DefMacro{res-selected-434-conditionnestack2e-all-with-name-ft-lora-codellama-7b-rouge-f-max}{84.57}
\DefMacro{res-selected-434-conditionnestack2e-all-with-name-ft-lora-codellama-7b-rouge-f-min}{78.05}
\DefMacro{res-selected-434-conditionnestack2e-all-with-name-ft-lora-codellama-7b-rouge-p}{82.95}
\DefMacro{res-selected-434-conditionnestack2e-all-with-name-ft-lora-codellama-7b-rouge-p-avg}{83.26}
\DefMacro{res-selected-434-conditionnestack2e-all-with-name-ft-lora-codellama-7b-rouge-p-max}{86.97}
\DefMacro{res-selected-434-conditionnestack2e-all-with-name-ft-lora-codellama-7b-rouge-p-min}{79.09}
\DefMacro{res-selected-434-conditionnestack2e-all-with-name-ft-lora-codellama-7b-rouge-r}{81.00}
\DefMacro{res-selected-434-conditionnestack2e-all-with-name-ft-lora-codellama-7b-rouge-r-avg}{81.18}
\DefMacro{res-selected-434-conditionnestack2e-all-with-name-ft-lora-codellama-7b-rouge-r-max}{84.26}
\DefMacro{res-selected-434-conditionnestack2e-all-with-name-ft-lora-codellama-7b-rouge-r-min}{78.52}
\DefMacro{res-selected-434-conditionnestack2e-all-with-name-ft-lora-codellama-7b-xmatch}{18.89}
\DefMacro{res-selected-434-conditionnestack2e-all-with-name-ft-lora-codellama-7b-xmatch-avg}{20.32}
\DefMacro{res-selected-434-conditionnestack2e-all-with-name-ft-lora-codellama-7b-xmatch-max}{25.35}
\DefMacro{res-selected-434-conditionnestack2e-all-with-name-ft-lora-codellama-7b-xmatch-min}{17.74}
\DefMacro{res-selected-434-conditionnestack2e-all-with-name-ft-lora-codellama-7b-xmatch-top1}{18.89}
\DefMacro{res-selected-434-conditionnestack2e-all-with-name-ft-lora-codellama-7b-compilable-avg}{82.26}
\DefMacro{res-selected-434-conditionnestack2e-all-with-name-ft-lora-codellama-7b-compilable-max}{87.10}
\DefMacro{res-selected-434-conditionnestack2e-all-with-name-ft-lora-codellama-7b-compilable-min}{76.50}
\DefMacro{res-selected-434-conditionnestack2e-all-with-name-ft-lora-codellama-7b-coverage-avg}{58.31}
\DefMacro{res-selected-434-conditionnestack2e-all-with-name-ft-lora-codellama-7b-coverage-max}{66.82}
\DefMacro{res-selected-434-conditionnestack2e-all-with-name-ft-lora-codellama-7b-coverage-min}{49.77}
\DefMacro{res-selected-434-conditionnestack2e-all-with-name-ft-lora-codellama-7b-match-avg}{100.00}
\DefMacro{res-selected-434-conditionnestack2e-all-with-name-ft-lora-codellama-7b-match-max}{100.00}
\DefMacro{res-selected-434-conditionnestack2e-all-with-name-ft-lora-codellama-7b-match-min}{100.00}
\DefMacro{res-selected-434-conditionnestack2e-all-with-name-ft-lora-codellama-7b-runnable-avg}{79.73}
\DefMacro{res-selected-434-conditionnestack2e-all-with-name-ft-lora-codellama-7b-runnable-max}{86.24}
\DefMacro{res-selected-434-conditionnestack2e-all-with-name-ft-lora-codellama-7b-runnable-min}{72.22}
\DefMacro{res-selected-434-conditionnestack2e-all-with-name-ft-lora-codellama-7b-runnable-overall}{75.12}
\DefMacro{res-selected-434-conditionnestack2e-all-with-name-ft-lora-codellama-7b-runnable-overall-avg}{66.21}
\DefMacro{res-selected-434-conditionnestack2e-all-with-name-ft-lora-codellama-7b-runnable-overall-max}{75.12}
\DefMacro{res-selected-434-conditionnestack2e-all-with-name-ft-lora-codellama-7b-runnable-overall-min}{56.22}
\DefMacro{res-selected-434-conditionnestack2e-all-with-name-ft-lora-codellama-7b-timeout-avg}{0.60}
\DefMacro{res-selected-434-conditionnestack2e-all-with-name-ft-lora-codellama-7b-timeout-max}{0.79}
\DefMacro{res-selected-434-conditionnestack2e-all-with-name-ft-lora-codellama-7b-timeout-min}{0.53}

\DefMacro{res-exception-coverage-evosuite-coverage}{14.00}
\DefMacro{res-exception-coverage-randoop-coverage}{27.06}

\DefMacro{res-rq2-conditionnestack2e-no-name-ft-lora-codellama-7b-compilable-avg}{62.68}
\DefMacro{res-rq2-conditionnestack2e-no-name-ft-lora-codellama-7b-compilable-max}{62.68}
\DefMacro{res-rq2-conditionnestack2e-no-name-ft-lora-codellama-7b-compilable-min}{62.68}
\DefMacro{res-rq2-conditionnestack2e-no-name-ft-lora-codellama-7b-coverage-avg}{27.88}
\DefMacro{res-rq2-conditionnestack2e-no-name-ft-lora-codellama-7b-coverage-max}{27.88}
\DefMacro{res-rq2-conditionnestack2e-no-name-ft-lora-codellama-7b-coverage-min}{27.88}
\DefMacro{res-rq2-conditionnestack2e-no-name-ft-lora-codellama-7b-match-avg}{100.00}
\DefMacro{res-rq2-conditionnestack2e-no-name-ft-lora-codellama-7b-match-max}{100.00}
\DefMacro{res-rq2-conditionnestack2e-no-name-ft-lora-codellama-7b-match-min}{100.00}
\DefMacro{res-rq2-conditionnestack2e-no-name-ft-lora-codellama-7b-runnable-avg}{66.89}
\DefMacro{res-rq2-conditionnestack2e-no-name-ft-lora-codellama-7b-runnable-max}{66.89}
\DefMacro{res-rq2-conditionnestack2e-no-name-ft-lora-codellama-7b-runnable-min}{66.89}
\DefMacro{res-rq2-conditionnestack2e-no-name-ft-lora-codellama-7b-runnable-overall}{41.93}
\DefMacro{res-rq2-conditionnestack2e-no-name-ft-lora-codellama-7b-runnable-overall-avg}{41.93}
\DefMacro{res-rq2-conditionnestack2e-no-name-ft-lora-codellama-7b-runnable-overall-max}{41.93}
\DefMacro{res-rq2-conditionnestack2e-no-name-ft-lora-codellama-7b-runnable-overall-min}{41.93}
\DefMacro{res-rq2-conditionnestack2e-no-name-ft-lora-codellama-7b-timeout-avg}{0.33}
\DefMacro{res-rq2-conditionnestack2e-no-name-ft-lora-codellama-7b-timeout-max}{0.33}
\DefMacro{res-rq2-conditionnestack2e-no-name-ft-lora-codellama-7b-timeout-min}{0.33}
\DefMacro{res-rq2-conditionnestack2e-all-no-name-ft-lora-codellama-7b-compilable-avg}{66.22}
\DefMacro{res-rq2-conditionnestack2e-all-no-name-ft-lora-codellama-7b-compilable-max}{80.43}
\DefMacro{res-rq2-conditionnestack2e-all-no-name-ft-lora-codellama-7b-compilable-min}{51.62}
\DefMacro{res-rq2-conditionnestack2e-all-no-name-ft-lora-codellama-7b-coverage-avg}{25.28}
\DefMacro{res-rq2-conditionnestack2e-all-no-name-ft-lora-codellama-7b-coverage-max}{28.81}
\DefMacro{res-rq2-conditionnestack2e-all-no-name-ft-lora-codellama-7b-coverage-min}{19.88}
\DefMacro{res-rq2-conditionnestack2e-all-no-name-ft-lora-codellama-7b-match-avg}{99.43}
\DefMacro{res-rq2-conditionnestack2e-all-no-name-ft-lora-codellama-7b-match-max}{99.43}
\DefMacro{res-rq2-conditionnestack2e-all-no-name-ft-lora-codellama-7b-match-min}{99.43}
\DefMacro{res-rq2-conditionnestack2e-all-no-name-ft-lora-codellama-7b-runnable-avg}{56.52}
\DefMacro{res-rq2-conditionnestack2e-all-no-name-ft-lora-codellama-7b-runnable-max}{64.16}
\DefMacro{res-rq2-conditionnestack2e-all-no-name-ft-lora-codellama-7b-runnable-min}{47.78}
\DefMacro{res-rq2-conditionnestack2e-all-no-name-ft-lora-codellama-7b-runnable-overall}{51.31}
\DefMacro{res-rq2-conditionnestack2e-all-no-name-ft-lora-codellama-7b-runnable-overall-avg}{40.21}
\DefMacro{res-rq2-conditionnestack2e-all-no-name-ft-lora-codellama-7b-runnable-overall-max}{51.31}
\DefMacro{res-rq2-conditionnestack2e-all-no-name-ft-lora-codellama-7b-runnable-overall-min}{29.12}
\DefMacro{res-rq2-conditionnestack2e-all-no-name-ft-lora-codellama-7b-timeout-avg}{2.75}
\DefMacro{res-rq2-conditionnestack2e-all-no-name-ft-lora-codellama-7b-timeout-max}{3.08}
\DefMacro{res-rq2-conditionnestack2e-all-no-name-ft-lora-codellama-7b-timeout-min}{2.50}

\DefMacro{Coreoz_Wisp-randoop-num-tests}{2,000}
\DefMacro{Coreoz_Wisp-randoop-time}{2,542}
\DefMacro{Coreoz_Wisp-evosuite-num-tests}{163}
\DefMacro{Coreoz_Wisp-evosuite-time}{271}
\DefMacro{Harium_keel-randoop-num-tests}{8,707}
\DefMacro{Harium_keel-randoop-time}{10,833}
\DefMacro{Harium_keel-evosuite-num-tests}{4,849}
\DefMacro{Harium_keel-evosuite-time}{689}
\DefMacro{JodaOrg_joda-beans-randoop-num-tests}{3,432}
\DefMacro{JodaOrg_joda-beans-randoop-time}{10,816}
\DefMacro{JodaOrg_joda-beans-evosuite-num-tests}{3,331}
\DefMacro{JodaOrg_joda-beans-evosuite-time}{6,607}
\DefMacro{OpenHFT_Chronicle-Map-randoop-num-tests}{0}
\DefMacro{OpenHFT_Chronicle-Map-randoop-time}{-}
\DefMacro{OpenHFT_Chronicle-Map-evosuite-num-tests}{1,287}
\DefMacro{OpenHFT_Chronicle-Map-evosuite-time}{1,904}
\DefMacro{OpenHFT_Chronicle-Network-randoop-num-tests}{5,611}
\DefMacro{OpenHFT_Chronicle-Network-randoop-time}{12,600}
\DefMacro{OpenHFT_Chronicle-Network-evosuite-num-tests}{714}
\DefMacro{OpenHFT_Chronicle-Network-evosuite-time}{398}
\DefMacro{OpenNMS_newts-randoop-num-tests}{0}
\DefMacro{OpenNMS_newts-randoop-time}{-}
\DefMacro{OpenNMS_newts-evosuite-num-tests}{624}
\DefMacro{OpenNMS_newts-evosuite-time}{1,070}
\DefMacro{analogweb_core-randoop-num-tests}{510}
\DefMacro{analogweb_core-randoop-time}{12,002}
\DefMacro{analogweb_core-evosuite-num-tests}{1,686}
\DefMacro{analogweb_core-evosuite-time}{3,193}
\DefMacro{arquillian_arquillian-core-randoop-num-tests}{12,783}
\DefMacro{arquillian_arquillian-core-randoop-time}{10,921}
\DefMacro{arquillian_arquillian-core-evosuite-num-tests}{1,649}
\DefMacro{arquillian_arquillian-core-evosuite-time}{2,612}
\DefMacro{bingoohuang_westcache-randoop-num-tests}{8,038}
\DefMacro{bingoohuang_westcache-randoop-time}{9,098}
\DefMacro{bingoohuang_westcache-evosuite-num-tests}{40}
\DefMacro{bingoohuang_westcache-evosuite-time}{847}
\DefMacro{craftercms_engine-randoop-num-tests}{6,516}
\DefMacro{craftercms_engine-randoop-time}{5,041}
\DefMacro{craftercms_engine-evosuite-num-tests}{582}
\DefMacro{craftercms_engine-evosuite-time}{130}
\DefMacro{davidmoten_ppk-randoop-num-tests}{2,595}
\DefMacro{davidmoten_ppk-randoop-time}{3,300}
\DefMacro{davidmoten_ppk-evosuite-num-tests}{66}
\DefMacro{davidmoten_ppk-evosuite-time}{2,115}
\DefMacro{entrusc_xdata-randoop-num-tests}{30,930}
\DefMacro{entrusc_xdata-randoop-time}{5,023}
\DefMacro{entrusc_xdata-evosuite-num-tests}{161}
\DefMacro{entrusc_xdata-evosuite-time}{557}
\DefMacro{globocom_GloboDNS-Client-randoop-num-tests}{12,622}
\DefMacro{globocom_GloboDNS-Client-randoop-time}{4,432}
\DefMacro{globocom_GloboDNS-Client-evosuite-num-tests}{445}
\DefMacro{globocom_GloboDNS-Client-evosuite-time}{786}
\DefMacro{greenmail-mail-test_greenmail-randoop-num-tests}{8,109}
\DefMacro{greenmail-mail-test_greenmail-randoop-time}{12,600}
\DefMacro{greenmail-mail-test_greenmail-evosuite-num-tests}{1,168}
\DefMacro{greenmail-mail-test_greenmail-evosuite-time}{417}
\DefMacro{jpmml_jpmml-model-randoop-num-tests}{7,002}
\DefMacro{jpmml_jpmml-model-randoop-time}{10,912}
\DefMacro{jpmml_jpmml-model-evosuite-num-tests}{1,382}
\DefMacro{jpmml_jpmml-model-evosuite-time}{2,689}
\DefMacro{kevinsawicki_http-request-randoop-num-tests}{70,079}
\DefMacro{kevinsawicki_http-request-randoop-time}{2,318}
\DefMacro{kevinsawicki_http-request-evosuite-num-tests}{451}
\DefMacro{kevinsawicki_http-request-evosuite-time}{810}
\DefMacro{mguymon_model-citizen-randoop-num-tests}{11,272}
\DefMacro{mguymon_model-citizen-randoop-time}{4,123}
\DefMacro{mguymon_model-citizen-evosuite-num-tests}{264}
\DefMacro{mguymon_model-citizen-evosuite-time}{1,617}
\DefMacro{microfocus-idol_java-content-parameter-api-randoop-num-tests}{3,733}
\DefMacro{microfocus-idol_java-content-parameter-api-randoop-time}{6,626}
\DefMacro{microfocus-idol_java-content-parameter-api-evosuite-num-tests}{767}
\DefMacro{microfocus-idol_java-content-parameter-api-evosuite-time}{17,915}
\DefMacro{mistraltechnologies_smog-randoop-num-tests}{2,906}
\DefMacro{mistraltechnologies_smog-randoop-time}{1,812}
\DefMacro{mistraltechnologies_smog-evosuite-num-tests}{55}
\DefMacro{mistraltechnologies_smog-evosuite-time}{350}
\DefMacro{mp911de_logstash-gelf-randoop-num-tests}{54,616}
\DefMacro{mp911de_logstash-gelf-randoop-time}{12,600}
\DefMacro{mp911de_logstash-gelf-evosuite-num-tests}{189}
\DefMacro{mp911de_logstash-gelf-evosuite-time}{131}
\DefMacro{opencb_java-common-libs-randoop-num-tests}{1,918}
\DefMacro{opencb_java-common-libs-randoop-time}{9,715}
\DefMacro{opencb_java-common-libs-evosuite-num-tests}{1,756}
\DefMacro{opencb_java-common-libs-evosuite-time}{925}
\DefMacro{pinterest_secor-randoop-num-tests}{0}
\DefMacro{pinterest_secor-randoop-time}{-}
\DefMacro{pinterest_secor-evosuite-num-tests}{1,255}
\DefMacro{pinterest_secor-evosuite-time}{700}
\DefMacro{ralscha_wampspring-randoop-num-tests}{19,765}
\DefMacro{ralscha_wampspring-randoop-time}{7,185}
\DefMacro{ralscha_wampspring-evosuite-num-tests}{450}
\DefMacro{ralscha_wampspring-evosuite-time}{426}
\DefMacro{sbtourist_Journal.IO-randoop-num-tests}{4,633}
\DefMacro{sbtourist_Journal.IO-randoop-time}{5,300}
\DefMacro{sbtourist_Journal.IO-evosuite-num-tests}{97}
\DefMacro{sbtourist_Journal.IO-evosuite-time}{128}
\DefMacro{spotify_apollo-randoop-num-tests}{1,584}
\DefMacro{spotify_apollo-randoop-time}{12,600}
\DefMacro{spotify_apollo-evosuite-num-tests}{369}
\DefMacro{spotify_apollo-evosuite-time}{3,054}
\DefMacro{spotify_async-google-pubsub-client-randoop-num-tests}{25,696}
\DefMacro{spotify_async-google-pubsub-client-randoop-time}{8,960}
\DefMacro{spotify_async-google-pubsub-client-evosuite-num-tests}{183}
\DefMacro{spotify_async-google-pubsub-client-evosuite-time}{444}
\DefMacro{stackify_stackify-api-java-randoop-num-tests}{95,275}
\DefMacro{stackify_stackify-api-java-randoop-time}{11,248}
\DefMacro{stackify_stackify-api-java-evosuite-num-tests}{1,166}
\DefMacro{stackify_stackify-api-java-evosuite-time}{1,151}
\DefMacro{statefulj_statefulj-randoop-num-tests}{25,134}
\DefMacro{statefulj_statefulj-randoop-time}{6,769}
\DefMacro{statefulj_statefulj-evosuite-num-tests}{137}
\DefMacro{statefulj_statefulj-evosuite-time}{1,016}
\DefMacro{javadev_moneytostr-russian-randoop-num-tests}{16,471}
\DefMacro{javadev_moneytostr-russian-randoop-time}{4,718}
\DefMacro{javadev_moneytostr-russian-evosuite-num-tests}{114}
\DefMacro{javadev_moneytostr-russian-evosuite-time}{904}
\DefMacro{sharneng_gm4java-randoop-num-tests}{40,276}
\DefMacro{sharneng_gm4java-randoop-time}{3,140}
\DefMacro{sharneng_gm4java-evosuite-num-tests}{236}
\DefMacro{sharneng_gm4java-evosuite-time}{1,287}
\DefMacro{total-randoop-num-tests}{482,213}
\DefMacro{avg-randoop-num-tests}{16,073.8}
\DefMacro{total-randoop-time}{207,234}
\DefMacro{avg-randoop-time}{7,675.3}
\DefMacro{total-evosuite-num-tests}{25,636}
\DefMacro{avg-evosuite-num-tests}{854.5}
\DefMacro{total-evosuite-time}{55,143}
\DefMacro{avg-evosuite-time}{1,838.1}
\DefMacro{num-projs-randoop}{27}
\DefMacro{num-projs-no-tests-randoop}{3}
\DefMacro{num-projs-evosuite}{30}
\DefMacro{num-projs-no-tests-evosuite}{0}

\DefMacro{Coreoz_Wisp-num-public-throw-stmts}{5}
\DefMacro{Harium_keel-num-public-throw-stmts}{196}
\DefMacro{JodaOrg_joda-beans-num-public-throw-stmts}{294}
\DefMacro{OpenHFT_Chronicle-Map-num-public-throw-stmts}{454}
\DefMacro{OpenHFT_Chronicle-Network-num-public-throw-stmts}{92}
\DefMacro{OpenNMS_newts-num-public-throw-stmts}{62}
\DefMacro{analogweb_core-num-public-throw-stmts}{43}
\DefMacro{arquillian_arquillian-core-num-public-throw-stmts}{390}
\DefMacro{bingoohuang_westcache-num-public-throw-stmts}{12}
\DefMacro{craftercms_engine-num-public-throw-stmts}{52}
\DefMacro{davidmoten_ppk-num-public-throw-stmts}{17}
\DefMacro{entrusc_xdata-num-public-throw-stmts}{39}
\DefMacro{globocom_GloboDNS-Client-num-public-throw-stmts}{18}
\DefMacro{greenmail-mail-test_greenmail-num-public-throw-stmts}{96}
\DefMacro{javadev_moneytostr-russian-num-public-throw-stmts}{14}
\DefMacro{jpmml_jpmml-model-num-public-throw-stmts}{108}
\DefMacro{kevinsawicki_http-request-num-public-throw-stmts}{32}
\DefMacro{mguymon_model-citizen-num-public-throw-stmts}{48}
\DefMacro{microfocus-idol_java-content-parameter-api-num-public-throw-stmts}{8}
\DefMacro{mistraltechnologies_smog-num-public-throw-stmts}{5}
\DefMacro{mp911de_logstash-gelf-num-public-throw-stmts}{40}
\DefMacro{opencb_java-common-libs-num-public-throw-stmts}{56}
\DefMacro{pinterest_secor-num-public-throw-stmts}{89}
\DefMacro{ralscha_wampspring-num-public-throw-stmts}{39}
\DefMacro{sbtourist_Journal.IO-num-public-throw-stmts}{46}
\DefMacro{sharneng_gm4java-num-public-throw-stmts}{27}
\DefMacro{spotify_apollo-num-public-throw-stmts}{27}
\DefMacro{spotify_async-google-pubsub-client-num-public-throw-stmts}{4}
\DefMacro{stackify_stackify-api-java-num-public-throw-stmts}{213}
\DefMacro{statefulj_statefulj-num-public-throw-stmts}{54}
\DefMacro{total-num-public-throw-stmts}{2,580}
\DefMacro{avg-num-public-throw-stmts}{86.0}
\DefMacro{Coreoz_Wisp-num-eval-throw-stmts}{2}
\DefMacro{Harium_keel-num-eval-throw-stmts}{12}
\DefMacro{JodaOrg_joda-beans-num-eval-throw-stmts}{32}
\DefMacro{OpenHFT_Chronicle-Map-num-eval-throw-stmts}{51}
\DefMacro{OpenHFT_Chronicle-Network-num-eval-throw-stmts}{23}
\DefMacro{OpenNMS_newts-num-eval-throw-stmts}{28}
\DefMacro{analogweb_core-num-eval-throw-stmts}{33}
\DefMacro{arquillian_arquillian-core-num-eval-throw-stmts}{34}
\DefMacro{bingoohuang_westcache-num-eval-throw-stmts}{9}
\DefMacro{craftercms_engine-num-eval-throw-stmts}{5}
\DefMacro{davidmoten_ppk-num-eval-throw-stmts}{17}
\DefMacro{entrusc_xdata-num-eval-throw-stmts}{18}
\DefMacro{globocom_GloboDNS-Client-num-eval-throw-stmts}{15}
\DefMacro{greenmail-mail-test_greenmail-num-eval-throw-stmts}{37}
\DefMacro{javadev_moneytostr-russian-num-eval-throw-stmts}{14}
\DefMacro{jpmml_jpmml-model-num-eval-throw-stmts}{70}
\DefMacro{kevinsawicki_http-request-num-eval-throw-stmts}{32}
\DefMacro{mguymon_model-citizen-num-eval-throw-stmts}{8}
\DefMacro{microfocus-idol_java-content-parameter-api-num-eval-throw-stmts}{8}
\DefMacro{mistraltechnologies_smog-num-eval-throw-stmts}{4}
\DefMacro{mp911de_logstash-gelf-num-eval-throw-stmts}{2}
\DefMacro{opencb_java-common-libs-num-eval-throw-stmts}{27}
\DefMacro{pinterest_secor-num-eval-throw-stmts}{35}
\DefMacro{ralscha_wampspring-num-eval-throw-stmts}{29}
\DefMacro{sbtourist_Journal.IO-num-eval-throw-stmts}{21}
\DefMacro{sharneng_gm4java-num-eval-throw-stmts}{26}
\DefMacro{spotify_apollo-num-eval-throw-stmts}{8}
\DefMacro{spotify_async-google-pubsub-client-num-eval-throw-stmts}{4}
\DefMacro{stackify_stackify-api-java-num-eval-throw-stmts}{12}
\DefMacro{statefulj_statefulj-num-eval-throw-stmts}{33}
\DefMacro{total-num-eval-throw-stmts}{649}
\DefMacro{avg-num-eval-throw-stmts}{21.6}

\DefMacro{ds-all-avg-stack-trace-len}{2.35}
\DefMacro{ds-all-etest-count}{12,574}
\DefMacro{ds-all-etype-count}{821}
\DefMacro{ds-all-max-stack-trace-len}{16}
\DefMacro{ds-all-median-stack-trace-len}{2.00}
\DefMacro{ds-all-min-stack-trace-len}{1}
\DefMacro{ds-all-module-count}{699}
\DefMacro{ds-all-mut-count}{6,250}
\DefMacro{ds-all-no-stack-trace-count}{0}
\DefMacro{ds-all-project-count}{562}
\DefMacro{ds-all-stack-trace-len>6}{634}
\DefMacro{ds-all-test-count}{111,230}

\DefMacro{res-user-view-with-name-ne2e-few-shot-with-name-gpt-3.5-turbo-16k-bleu}{56.61}
\DefMacro{res-user-view-with-name-ne2e-few-shot-with-name-gpt-3.5-turbo-16k-bleu-avg}{56.61}
\DefMacro{res-user-view-with-name-ne2e-few-shot-with-name-gpt-3.5-turbo-16k-bleu-max}{56.61}
\DefMacro{res-user-view-with-name-ne2e-few-shot-with-name-gpt-3.5-turbo-16k-bleu-min}{56.61}
\DefMacro{res-user-view-with-name-ne2e-few-shot-with-name-gpt-3.5-turbo-16k-code-bleu}{64.28}
\DefMacro{res-user-view-with-name-ne2e-few-shot-with-name-gpt-3.5-turbo-16k-code-bleu-avg}{64.28}
\DefMacro{res-user-view-with-name-ne2e-few-shot-with-name-gpt-3.5-turbo-16k-code-bleu-max}{64.28}
\DefMacro{res-user-view-with-name-ne2e-few-shot-with-name-gpt-3.5-turbo-16k-code-bleu-min}{64.28}
\DefMacro{res-user-view-with-name-ne2e-few-shot-with-name-gpt-3.5-turbo-16k-edit-sim}{82.30}
\DefMacro{res-user-view-with-name-ne2e-few-shot-with-name-gpt-3.5-turbo-16k-edit-sim-avg}{82.30}
\DefMacro{res-user-view-with-name-ne2e-few-shot-with-name-gpt-3.5-turbo-16k-edit-sim-max}{82.30}
\DefMacro{res-user-view-with-name-ne2e-few-shot-with-name-gpt-3.5-turbo-16k-edit-sim-min}{82.30}
\DefMacro{res-user-view-with-name-ne2e-few-shot-with-name-gpt-3.5-turbo-16k-rouge-f}{76.55}
\DefMacro{res-user-view-with-name-ne2e-few-shot-with-name-gpt-3.5-turbo-16k-rouge-f-avg}{76.55}
\DefMacro{res-user-view-with-name-ne2e-few-shot-with-name-gpt-3.5-turbo-16k-rouge-f-max}{76.55}
\DefMacro{res-user-view-with-name-ne2e-few-shot-with-name-gpt-3.5-turbo-16k-rouge-f-min}{76.55}
\DefMacro{res-user-view-with-name-ne2e-few-shot-with-name-gpt-3.5-turbo-16k-rouge-p}{77.39}
\DefMacro{res-user-view-with-name-ne2e-few-shot-with-name-gpt-3.5-turbo-16k-rouge-p-avg}{77.39}
\DefMacro{res-user-view-with-name-ne2e-few-shot-with-name-gpt-3.5-turbo-16k-rouge-p-max}{77.39}
\DefMacro{res-user-view-with-name-ne2e-few-shot-with-name-gpt-3.5-turbo-16k-rouge-p-min}{77.39}
\DefMacro{res-user-view-with-name-ne2e-few-shot-with-name-gpt-3.5-turbo-16k-rouge-r}{78.15}
\DefMacro{res-user-view-with-name-ne2e-few-shot-with-name-gpt-3.5-turbo-16k-rouge-r-avg}{78.15}
\DefMacro{res-user-view-with-name-ne2e-few-shot-with-name-gpt-3.5-turbo-16k-rouge-r-max}{78.15}
\DefMacro{res-user-view-with-name-ne2e-few-shot-with-name-gpt-3.5-turbo-16k-rouge-r-min}{78.15}
\DefMacro{res-user-view-with-name-ne2e-few-shot-with-name-gpt-3.5-turbo-16k-xmatch}{14.98}
\DefMacro{res-user-view-with-name-ne2e-few-shot-with-name-gpt-3.5-turbo-16k-xmatch-avg}{14.98}
\DefMacro{res-user-view-with-name-ne2e-few-shot-with-name-gpt-3.5-turbo-16k-xmatch-max}{14.98}
\DefMacro{res-user-view-with-name-ne2e-few-shot-with-name-gpt-3.5-turbo-16k-xmatch-min}{14.98}
\DefMacro{res-user-view-with-name-ne2e-few-shot-with-name-gpt-3.5-turbo-16k-xmatch-top1}{14.98}
\DefMacro{res-user-view-with-name-ne2e-few-shot-with-name-gpt-3.5-turbo-16k-compilable-avg}{75.12}
\DefMacro{res-user-view-with-name-ne2e-few-shot-with-name-gpt-3.5-turbo-16k-compilable-max}{75.12}
\DefMacro{res-user-view-with-name-ne2e-few-shot-with-name-gpt-3.5-turbo-16k-compilable-min}{75.12}
\DefMacro{res-user-view-with-name-ne2e-few-shot-with-name-gpt-3.5-turbo-16k-coverage-avg}{48.39}
\DefMacro{res-user-view-with-name-ne2e-few-shot-with-name-gpt-3.5-turbo-16k-coverage-max}{48.39}
\DefMacro{res-user-view-with-name-ne2e-few-shot-with-name-gpt-3.5-turbo-16k-coverage-min}{48.39}
\DefMacro{res-user-view-with-name-ne2e-few-shot-with-name-gpt-3.5-turbo-16k-match-avg}{100.00}
\DefMacro{res-user-view-with-name-ne2e-few-shot-with-name-gpt-3.5-turbo-16k-match-max}{100.00}
\DefMacro{res-user-view-with-name-ne2e-few-shot-with-name-gpt-3.5-turbo-16k-match-min}{100.00}
\DefMacro{res-user-view-with-name-ne2e-few-shot-with-name-gpt-3.5-turbo-16k-runnable-avg}{81.60}
\DefMacro{res-user-view-with-name-ne2e-few-shot-with-name-gpt-3.5-turbo-16k-runnable-max}{81.60}
\DefMacro{res-user-view-with-name-ne2e-few-shot-with-name-gpt-3.5-turbo-16k-runnable-min}{81.60}
\DefMacro{res-user-view-with-name-ne2e-few-shot-with-name-gpt-3.5-turbo-16k-runnable-overall}{61.29}
\DefMacro{res-user-view-with-name-ne2e-few-shot-with-name-gpt-3.5-turbo-16k-runnable-overall-avg}{61.29}
\DefMacro{res-user-view-with-name-ne2e-few-shot-with-name-gpt-3.5-turbo-16k-runnable-overall-max}{61.29}
\DefMacro{res-user-view-with-name-ne2e-few-shot-with-name-gpt-3.5-turbo-16k-runnable-overall-min}{61.29}
\DefMacro{res-user-view-with-name-ne2e-few-shot-with-name-gpt-3.5-turbo-16k-timeout-avg}{0.00}
\DefMacro{res-user-view-with-name-ne2e-few-shot-with-name-gpt-3.5-turbo-16k-timeout-max}{0.00}
\DefMacro{res-user-view-with-name-ne2e-few-shot-with-name-gpt-3.5-turbo-16k-timeout-min}{0.00}
\DefMacro{res-user-view-with-name-catlm-ne2e-with-name-catlm-bleu}{53.49}
\DefMacro{res-user-view-with-name-catlm-ne2e-with-name-catlm-bleu-avg}{53.49}
\DefMacro{res-user-view-with-name-catlm-ne2e-with-name-catlm-bleu-max}{53.49}
\DefMacro{res-user-view-with-name-catlm-ne2e-with-name-catlm-bleu-min}{53.49}
\DefMacro{res-user-view-with-name-catlm-ne2e-with-name-catlm-code-bleu}{59.79}
\DefMacro{res-user-view-with-name-catlm-ne2e-with-name-catlm-code-bleu-avg}{59.79}
\DefMacro{res-user-view-with-name-catlm-ne2e-with-name-catlm-code-bleu-max}{59.79}
\DefMacro{res-user-view-with-name-catlm-ne2e-with-name-catlm-code-bleu-min}{59.79}
\DefMacro{res-user-view-with-name-catlm-ne2e-with-name-catlm-edit-sim}{78.91}
\DefMacro{res-user-view-with-name-catlm-ne2e-with-name-catlm-edit-sim-avg}{78.91}
\DefMacro{res-user-view-with-name-catlm-ne2e-with-name-catlm-edit-sim-max}{78.91}
\DefMacro{res-user-view-with-name-catlm-ne2e-with-name-catlm-edit-sim-min}{78.91}
\DefMacro{res-user-view-with-name-catlm-ne2e-with-name-catlm-rouge-f}{74.63}
\DefMacro{res-user-view-with-name-catlm-ne2e-with-name-catlm-rouge-f-avg}{74.63}
\DefMacro{res-user-view-with-name-catlm-ne2e-with-name-catlm-rouge-f-max}{74.63}
\DefMacro{res-user-view-with-name-catlm-ne2e-with-name-catlm-rouge-f-min}{74.63}
\DefMacro{res-user-view-with-name-catlm-ne2e-with-name-catlm-rouge-p}{77.32}
\DefMacro{res-user-view-with-name-catlm-ne2e-with-name-catlm-rouge-p-avg}{77.32}
\DefMacro{res-user-view-with-name-catlm-ne2e-with-name-catlm-rouge-p-max}{77.32}
\DefMacro{res-user-view-with-name-catlm-ne2e-with-name-catlm-rouge-p-min}{77.32}
\DefMacro{res-user-view-with-name-catlm-ne2e-with-name-catlm-rouge-r}{75.08}
\DefMacro{res-user-view-with-name-catlm-ne2e-with-name-catlm-rouge-r-avg}{75.08}
\DefMacro{res-user-view-with-name-catlm-ne2e-with-name-catlm-rouge-r-max}{75.08}
\DefMacro{res-user-view-with-name-catlm-ne2e-with-name-catlm-rouge-r-min}{75.08}
\DefMacro{res-user-view-with-name-catlm-ne2e-with-name-catlm-xmatch}{9.83}
\DefMacro{res-user-view-with-name-catlm-ne2e-with-name-catlm-xmatch-avg}{9.83}
\DefMacro{res-user-view-with-name-catlm-ne2e-with-name-catlm-xmatch-max}{9.83}
\DefMacro{res-user-view-with-name-catlm-ne2e-with-name-catlm-xmatch-min}{9.83}
\DefMacro{res-user-view-with-name-catlm-ne2e-with-name-catlm-xmatch-top1}{9.83}
\DefMacro{res-user-view-with-name-catlm-ne2e-with-name-catlm-compilable-avg}{71.83}
\DefMacro{res-user-view-with-name-catlm-ne2e-with-name-catlm-compilable-max}{71.83}
\DefMacro{res-user-view-with-name-catlm-ne2e-with-name-catlm-compilable-min}{71.83}
\DefMacro{res-user-view-with-name-catlm-ne2e-with-name-catlm-coverage-avg}{30.03}
\DefMacro{res-user-view-with-name-catlm-ne2e-with-name-catlm-coverage-max}{30.03}
\DefMacro{res-user-view-with-name-catlm-ne2e-with-name-catlm-coverage-min}{30.03}
\DefMacro{res-user-view-with-name-catlm-ne2e-with-name-catlm-match-avg}{100.00}
\DefMacro{res-user-view-with-name-catlm-ne2e-with-name-catlm-match-max}{100.00}
\DefMacro{res-user-view-with-name-catlm-ne2e-with-name-catlm-match-min}{100.00}
\DefMacro{res-user-view-with-name-catlm-ne2e-with-name-catlm-runnable-avg}{59.26}
\DefMacro{res-user-view-with-name-catlm-ne2e-with-name-catlm-runnable-max}{59.26}
\DefMacro{res-user-view-with-name-catlm-ne2e-with-name-catlm-runnable-min}{59.26}
\DefMacro{res-user-view-with-name-catlm-ne2e-with-name-catlm-runnable-overall}{36.64}
\DefMacro{res-user-view-with-name-catlm-ne2e-with-name-catlm-runnable-overall-avg}{36.64}
\DefMacro{res-user-view-with-name-catlm-ne2e-with-name-catlm-runnable-overall-max}{36.64}
\DefMacro{res-user-view-with-name-catlm-ne2e-with-name-catlm-runnable-overall-min}{36.64}
\DefMacro{res-user-view-with-name-catlm-ne2e-with-name-catlm-timeout-avg}{0.50}
\DefMacro{res-user-view-with-name-catlm-ne2e-with-name-catlm-timeout-max}{0.50}
\DefMacro{res-user-view-with-name-catlm-ne2e-with-name-catlm-timeout-min}{0.50}
\DefMacro{res-user-view-with-name-selected-434-conditionnestack2e-with-name-zero-shot-lora-codellama-7b-bleu}{38.23}
\DefMacro{res-user-view-with-name-selected-434-conditionnestack2e-with-name-zero-shot-lora-codellama-7b-bleu-avg}{38.23}
\DefMacro{res-user-view-with-name-selected-434-conditionnestack2e-with-name-zero-shot-lora-codellama-7b-bleu-max}{38.23}
\DefMacro{res-user-view-with-name-selected-434-conditionnestack2e-with-name-zero-shot-lora-codellama-7b-bleu-min}{38.23}
\DefMacro{res-user-view-with-name-selected-434-conditionnestack2e-with-name-zero-shot-lora-codellama-7b-code-bleu}{48.63}
\DefMacro{res-user-view-with-name-selected-434-conditionnestack2e-with-name-zero-shot-lora-codellama-7b-code-bleu-avg}{48.63}
\DefMacro{res-user-view-with-name-selected-434-conditionnestack2e-with-name-zero-shot-lora-codellama-7b-code-bleu-max}{48.63}
\DefMacro{res-user-view-with-name-selected-434-conditionnestack2e-with-name-zero-shot-lora-codellama-7b-code-bleu-min}{48.63}
\DefMacro{res-user-view-with-name-selected-434-conditionnestack2e-with-name-zero-shot-lora-codellama-7b-edit-sim}{66.96}
\DefMacro{res-user-view-with-name-selected-434-conditionnestack2e-with-name-zero-shot-lora-codellama-7b-edit-sim-avg}{66.96}
\DefMacro{res-user-view-with-name-selected-434-conditionnestack2e-with-name-zero-shot-lora-codellama-7b-edit-sim-max}{66.96}
\DefMacro{res-user-view-with-name-selected-434-conditionnestack2e-with-name-zero-shot-lora-codellama-7b-edit-sim-min}{66.96}
\DefMacro{res-user-view-with-name-selected-434-conditionnestack2e-with-name-zero-shot-lora-codellama-7b-rouge-f}{61.28}
\DefMacro{res-user-view-with-name-selected-434-conditionnestack2e-with-name-zero-shot-lora-codellama-7b-rouge-f-avg}{61.28}
\DefMacro{res-user-view-with-name-selected-434-conditionnestack2e-with-name-zero-shot-lora-codellama-7b-rouge-f-max}{61.28}
\DefMacro{res-user-view-with-name-selected-434-conditionnestack2e-with-name-zero-shot-lora-codellama-7b-rouge-f-min}{61.28}
\DefMacro{res-user-view-with-name-selected-434-conditionnestack2e-with-name-zero-shot-lora-codellama-7b-rouge-p}{56.39}
\DefMacro{res-user-view-with-name-selected-434-conditionnestack2e-with-name-zero-shot-lora-codellama-7b-rouge-p-avg}{56.39}
\DefMacro{res-user-view-with-name-selected-434-conditionnestack2e-with-name-zero-shot-lora-codellama-7b-rouge-p-max}{56.39}
\DefMacro{res-user-view-with-name-selected-434-conditionnestack2e-with-name-zero-shot-lora-codellama-7b-rouge-p-min}{56.39}
\DefMacro{res-user-view-with-name-selected-434-conditionnestack2e-with-name-zero-shot-lora-codellama-7b-rouge-r}{71.47}
\DefMacro{res-user-view-with-name-selected-434-conditionnestack2e-with-name-zero-shot-lora-codellama-7b-rouge-r-avg}{71.47}
\DefMacro{res-user-view-with-name-selected-434-conditionnestack2e-with-name-zero-shot-lora-codellama-7b-rouge-r-max}{71.47}
\DefMacro{res-user-view-with-name-selected-434-conditionnestack2e-with-name-zero-shot-lora-codellama-7b-rouge-r-min}{71.47}
\DefMacro{res-user-view-with-name-selected-434-conditionnestack2e-with-name-zero-shot-lora-codellama-7b-xmatch}{5.53}
\DefMacro{res-user-view-with-name-selected-434-conditionnestack2e-with-name-zero-shot-lora-codellama-7b-xmatch-avg}{5.53}
\DefMacro{res-user-view-with-name-selected-434-conditionnestack2e-with-name-zero-shot-lora-codellama-7b-xmatch-max}{5.53}
\DefMacro{res-user-view-with-name-selected-434-conditionnestack2e-with-name-zero-shot-lora-codellama-7b-xmatch-min}{5.53}
\DefMacro{res-user-view-with-name-selected-434-conditionnestack2e-with-name-zero-shot-lora-codellama-7b-xmatch-top1}{5.53}
\DefMacro{res-user-view-with-name-selected-434-conditionnestack2e-with-name-zero-shot-lora-codellama-7b-compilable-avg}{57.30}
\DefMacro{res-user-view-with-name-selected-434-conditionnestack2e-with-name-zero-shot-lora-codellama-7b-compilable-max}{57.30}
\DefMacro{res-user-view-with-name-selected-434-conditionnestack2e-with-name-zero-shot-lora-codellama-7b-compilable-min}{57.30}
\DefMacro{res-user-view-with-name-selected-434-conditionnestack2e-with-name-zero-shot-lora-codellama-7b-coverage-avg}{33.87}
\DefMacro{res-user-view-with-name-selected-434-conditionnestack2e-with-name-zero-shot-lora-codellama-7b-coverage-max}{33.87}
\DefMacro{res-user-view-with-name-selected-434-conditionnestack2e-with-name-zero-shot-lora-codellama-7b-coverage-min}{33.87}
\DefMacro{res-user-view-with-name-selected-434-conditionnestack2e-with-name-zero-shot-lora-codellama-7b-match-avg}{96.02}
\DefMacro{res-user-view-with-name-selected-434-conditionnestack2e-with-name-zero-shot-lora-codellama-7b-match-max}{96.02}
\DefMacro{res-user-view-with-name-selected-434-conditionnestack2e-with-name-zero-shot-lora-codellama-7b-match-min}{96.02}
\DefMacro{res-user-view-with-name-selected-434-conditionnestack2e-with-name-zero-shot-lora-codellama-7b-runnable-avg}{73.06}
\DefMacro{res-user-view-with-name-selected-434-conditionnestack2e-with-name-zero-shot-lora-codellama-7b-runnable-max}{73.06}
\DefMacro{res-user-view-with-name-selected-434-conditionnestack2e-with-name-zero-shot-lora-codellama-7b-runnable-min}{73.06}
\DefMacro{res-user-view-with-name-selected-434-conditionnestack2e-with-name-zero-shot-lora-codellama-7b-runnable-overall}{40.17}
\DefMacro{res-user-view-with-name-selected-434-conditionnestack2e-with-name-zero-shot-lora-codellama-7b-runnable-overall-avg}{40.17}
\DefMacro{res-user-view-with-name-selected-434-conditionnestack2e-with-name-zero-shot-lora-codellama-7b-runnable-overall-max}{40.17}
\DefMacro{res-user-view-with-name-selected-434-conditionnestack2e-with-name-zero-shot-lora-codellama-7b-runnable-overall-min}{40.17}
\DefMacro{res-user-view-with-name-selected-434-conditionnestack2e-with-name-zero-shot-lora-codellama-7b-timeout-avg}{0.14}
\DefMacro{res-user-view-with-name-selected-434-conditionnestack2e-with-name-zero-shot-lora-codellama-7b-timeout-max}{0.14}
\DefMacro{res-user-view-with-name-selected-434-conditionnestack2e-with-name-zero-shot-lora-codellama-7b-timeout-min}{0.14}
\DefMacro{res-user-view-with-name-conditionnestack2e-with-name-ft-lora-codellama-7b-bleu}{63.13}
\DefMacro{res-user-view-with-name-conditionnestack2e-with-name-ft-lora-codellama-7b-bleu-avg}{63.13}
\DefMacro{res-user-view-with-name-conditionnestack2e-with-name-ft-lora-codellama-7b-bleu-max}{63.13}
\DefMacro{res-user-view-with-name-conditionnestack2e-with-name-ft-lora-codellama-7b-bleu-min}{63.13}
\DefMacro{res-user-view-with-name-conditionnestack2e-with-name-ft-lora-codellama-7b-code-bleu}{67.49}
\DefMacro{res-user-view-with-name-conditionnestack2e-with-name-ft-lora-codellama-7b-code-bleu-avg}{67.49}
\DefMacro{res-user-view-with-name-conditionnestack2e-with-name-ft-lora-codellama-7b-code-bleu-max}{67.49}
\DefMacro{res-user-view-with-name-conditionnestack2e-with-name-ft-lora-codellama-7b-code-bleu-min}{67.49}
\DefMacro{res-user-view-with-name-conditionnestack2e-with-name-ft-lora-codellama-7b-edit-sim}{85.32}
\DefMacro{res-user-view-with-name-conditionnestack2e-with-name-ft-lora-codellama-7b-edit-sim-avg}{85.32}
\DefMacro{res-user-view-with-name-conditionnestack2e-with-name-ft-lora-codellama-7b-edit-sim-max}{85.32}
\DefMacro{res-user-view-with-name-conditionnestack2e-with-name-ft-lora-codellama-7b-edit-sim-min}{85.32}
\DefMacro{res-user-view-with-name-conditionnestack2e-with-name-ft-lora-codellama-7b-rouge-f}{80.75}
\DefMacro{res-user-view-with-name-conditionnestack2e-with-name-ft-lora-codellama-7b-rouge-f-avg}{80.75}
\DefMacro{res-user-view-with-name-conditionnestack2e-with-name-ft-lora-codellama-7b-rouge-f-max}{80.75}
\DefMacro{res-user-view-with-name-conditionnestack2e-with-name-ft-lora-codellama-7b-rouge-f-min}{80.75}
\DefMacro{res-user-view-with-name-conditionnestack2e-with-name-ft-lora-codellama-7b-rouge-p}{82.85}
\DefMacro{res-user-view-with-name-conditionnestack2e-with-name-ft-lora-codellama-7b-rouge-p-avg}{82.85}
\DefMacro{res-user-view-with-name-conditionnestack2e-with-name-ft-lora-codellama-7b-rouge-p-max}{82.85}
\DefMacro{res-user-view-with-name-conditionnestack2e-with-name-ft-lora-codellama-7b-rouge-p-min}{82.85}
\DefMacro{res-user-view-with-name-conditionnestack2e-with-name-ft-lora-codellama-7b-rouge-r}{80.81}
\DefMacro{res-user-view-with-name-conditionnestack2e-with-name-ft-lora-codellama-7b-rouge-r-avg}{80.81}
\DefMacro{res-user-view-with-name-conditionnestack2e-with-name-ft-lora-codellama-7b-rouge-r-max}{80.81}
\DefMacro{res-user-view-with-name-conditionnestack2e-with-name-ft-lora-codellama-7b-rouge-r-min}{80.81}
\DefMacro{res-user-view-with-name-conditionnestack2e-with-name-ft-lora-codellama-7b-xmatch}{19.05}
\DefMacro{res-user-view-with-name-conditionnestack2e-with-name-ft-lora-codellama-7b-xmatch-avg}{19.05}
\DefMacro{res-user-view-with-name-conditionnestack2e-with-name-ft-lora-codellama-7b-xmatch-max}{19.05}
\DefMacro{res-user-view-with-name-conditionnestack2e-with-name-ft-lora-codellama-7b-xmatch-min}{19.05}
\DefMacro{res-user-view-with-name-conditionnestack2e-with-name-ft-lora-codellama-7b-xmatch-top1}{19.05}
\DefMacro{res-user-view-with-name-conditionnestack2e-with-name-ft-lora-codellama-7b-compilable-avg}{82.10}
\DefMacro{res-user-view-with-name-conditionnestack2e-with-name-ft-lora-codellama-7b-compilable-max}{82.10}
\DefMacro{res-user-view-with-name-conditionnestack2e-with-name-ft-lora-codellama-7b-compilable-min}{82.10}
\DefMacro{res-user-view-with-name-conditionnestack2e-with-name-ft-lora-codellama-7b-coverage-avg}{59.45}
\DefMacro{res-user-view-with-name-conditionnestack2e-with-name-ft-lora-codellama-7b-coverage-max}{59.45}
\DefMacro{res-user-view-with-name-conditionnestack2e-with-name-ft-lora-codellama-7b-coverage-min}{59.45}
\DefMacro{res-user-view-with-name-conditionnestack2e-with-name-ft-lora-codellama-7b-match-avg}{100.00}
\DefMacro{res-user-view-with-name-conditionnestack2e-with-name-ft-lora-codellama-7b-match-max}{100.00}
\DefMacro{res-user-view-with-name-conditionnestack2e-with-name-ft-lora-codellama-7b-match-min}{100.00}
\DefMacro{res-user-view-with-name-conditionnestack2e-with-name-ft-lora-codellama-7b-runnable-avg}{82.05}
\DefMacro{res-user-view-with-name-conditionnestack2e-with-name-ft-lora-codellama-7b-runnable-max}{82.05}
\DefMacro{res-user-view-with-name-conditionnestack2e-with-name-ft-lora-codellama-7b-runnable-min}{82.05}
\DefMacro{res-user-view-with-name-conditionnestack2e-with-name-ft-lora-codellama-7b-runnable-overall}{67.36}
\DefMacro{res-user-view-with-name-conditionnestack2e-with-name-ft-lora-codellama-7b-runnable-overall-avg}{67.36}
\DefMacro{res-user-view-with-name-conditionnestack2e-with-name-ft-lora-codellama-7b-runnable-overall-max}{67.36}
\DefMacro{res-user-view-with-name-conditionnestack2e-with-name-ft-lora-codellama-7b-runnable-overall-min}{67.36}
\DefMacro{res-user-view-with-name-conditionnestack2e-with-name-ft-lora-codellama-7b-timeout-avg}{0.65}
\DefMacro{res-user-view-with-name-conditionnestack2e-with-name-ft-lora-codellama-7b-timeout-max}{0.65}
\DefMacro{res-user-view-with-name-conditionnestack2e-with-name-ft-lora-codellama-7b-timeout-min}{0.65}
\DefMacro{res-user-view-with-name-conditionnestack2e-all-with-name-ft-lora-codellama-7b-bleu}{62.10}
\DefMacro{res-user-view-with-name-conditionnestack2e-all-with-name-ft-lora-codellama-7b-bleu-avg}{61.76}
\DefMacro{res-user-view-with-name-conditionnestack2e-all-with-name-ft-lora-codellama-7b-bleu-max}{70.01}
\DefMacro{res-user-view-with-name-conditionnestack2e-all-with-name-ft-lora-codellama-7b-bleu-min}{53.80}
\DefMacro{res-user-view-with-name-conditionnestack2e-all-with-name-ft-lora-codellama-7b-code-bleu}{66.52}
\DefMacro{res-user-view-with-name-conditionnestack2e-all-with-name-ft-lora-codellama-7b-code-bleu-avg}{66.64}
\DefMacro{res-user-view-with-name-conditionnestack2e-all-with-name-ft-lora-codellama-7b-code-bleu-max}{74.08}
\DefMacro{res-user-view-with-name-conditionnestack2e-all-with-name-ft-lora-codellama-7b-code-bleu-min}{59.92}
\DefMacro{res-user-view-with-name-conditionnestack2e-all-with-name-ft-lora-codellama-7b-edit-sim}{85.20}
\DefMacro{res-user-view-with-name-conditionnestack2e-all-with-name-ft-lora-codellama-7b-edit-sim-avg}{85.26}
\DefMacro{res-user-view-with-name-conditionnestack2e-all-with-name-ft-lora-codellama-7b-edit-sim-max}{90.09}
\DefMacro{res-user-view-with-name-conditionnestack2e-all-with-name-ft-lora-codellama-7b-edit-sim-min}{79.88}
\DefMacro{res-user-view-with-name-conditionnestack2e-all-with-name-ft-lora-codellama-7b-rouge-f}{80.08}
\DefMacro{res-user-view-with-name-conditionnestack2e-all-with-name-ft-lora-codellama-7b-rouge-f-avg}{80.02}
\DefMacro{res-user-view-with-name-conditionnestack2e-all-with-name-ft-lora-codellama-7b-rouge-f-max}{85.50}
\DefMacro{res-user-view-with-name-conditionnestack2e-all-with-name-ft-lora-codellama-7b-rouge-f-min}{74.19}
\DefMacro{res-user-view-with-name-conditionnestack2e-all-with-name-ft-lora-codellama-7b-rouge-p}{82.02}
\DefMacro{res-user-view-with-name-conditionnestack2e-all-with-name-ft-lora-codellama-7b-rouge-p-avg}{81.93}
\DefMacro{res-user-view-with-name-conditionnestack2e-all-with-name-ft-lora-codellama-7b-rouge-p-max}{88.06}
\DefMacro{res-user-view-with-name-conditionnestack2e-all-with-name-ft-lora-codellama-7b-rouge-p-min}{74.57}
\DefMacro{res-user-view-with-name-conditionnestack2e-all-with-name-ft-lora-codellama-7b-rouge-r}{80.14}
\DefMacro{res-user-view-with-name-conditionnestack2e-all-with-name-ft-lora-codellama-7b-rouge-r-avg}{80.15}
\DefMacro{res-user-view-with-name-conditionnestack2e-all-with-name-ft-lora-codellama-7b-rouge-r-max}{85.20}
\DefMacro{res-user-view-with-name-conditionnestack2e-all-with-name-ft-lora-codellama-7b-rouge-r-min}{75.28}
\DefMacro{res-user-view-with-name-conditionnestack2e-all-with-name-ft-lora-codellama-7b-xmatch}{15.28}
\DefMacro{res-user-view-with-name-conditionnestack2e-all-with-name-ft-lora-codellama-7b-xmatch-avg}{15.03}
\DefMacro{res-user-view-with-name-conditionnestack2e-all-with-name-ft-lora-codellama-7b-xmatch-max}{22.92}
\DefMacro{res-user-view-with-name-conditionnestack2e-all-with-name-ft-lora-codellama-7b-xmatch-min}{9.62}
\DefMacro{res-user-view-with-name-conditionnestack2e-all-with-name-ft-lora-codellama-7b-xmatch-top1}{15.28}
\DefMacro{res-user-view-with-name-conditionnestack2e-all-with-name-ft-lora-codellama-7b-compilable-avg}{82.93}
\DefMacro{res-user-view-with-name-conditionnestack2e-all-with-name-ft-lora-codellama-7b-compilable-max}{93.54}
\DefMacro{res-user-view-with-name-conditionnestack2e-all-with-name-ft-lora-codellama-7b-compilable-min}{69.30}
\DefMacro{res-user-view-with-name-conditionnestack2e-all-with-name-ft-lora-codellama-7b-coverage-avg}{55.78}
\DefMacro{res-user-view-with-name-conditionnestack2e-all-with-name-ft-lora-codellama-7b-coverage-max}{71.28}
\DefMacro{res-user-view-with-name-conditionnestack2e-all-with-name-ft-lora-codellama-7b-coverage-min}{39.26}
\DefMacro{res-user-view-with-name-conditionnestack2e-all-with-name-ft-lora-codellama-7b-match-avg}{100.00}
\DefMacro{res-user-view-with-name-conditionnestack2e-all-with-name-ft-lora-codellama-7b-match-max}{100.00}
\DefMacro{res-user-view-with-name-conditionnestack2e-all-with-name-ft-lora-codellama-7b-match-min}{100.00}
\DefMacro{res-user-view-with-name-conditionnestack2e-all-with-name-ft-lora-codellama-7b-runnable-avg}{77.04}
\DefMacro{res-user-view-with-name-conditionnestack2e-all-with-name-ft-lora-codellama-7b-runnable-max}{86.91}
\DefMacro{res-user-view-with-name-conditionnestack2e-all-with-name-ft-lora-codellama-7b-runnable-min}{65.21}
\DefMacro{res-user-view-with-name-conditionnestack2e-all-with-name-ft-lora-codellama-7b-runnable-overall}{81.29}
\DefMacro{res-user-view-with-name-conditionnestack2e-all-with-name-ft-lora-codellama-7b-runnable-overall-avg}{64.94}
\DefMacro{res-user-view-with-name-conditionnestack2e-all-with-name-ft-lora-codellama-7b-runnable-overall-max}{81.29}
\DefMacro{res-user-view-with-name-conditionnestack2e-all-with-name-ft-lora-codellama-7b-runnable-overall-min}{46.11}
\DefMacro{res-user-view-with-name-conditionnestack2e-all-with-name-ft-lora-codellama-7b-timeout-avg}{0.52}
\DefMacro{res-user-view-with-name-conditionnestack2e-all-with-name-ft-lora-codellama-7b-timeout-max}{0.56}
\DefMacro{res-user-view-with-name-conditionnestack2e-all-with-name-ft-lora-codellama-7b-timeout-min}{0.42}

\DefMacro{res-user-view-no-name-ne2e-few-shot-no-name-gpt-3.5-turbo-16k-bleu}{38.67}
\DefMacro{res-user-view-no-name-ne2e-few-shot-no-name-gpt-3.5-turbo-16k-bleu-avg}{38.67}
\DefMacro{res-user-view-no-name-ne2e-few-shot-no-name-gpt-3.5-turbo-16k-bleu-max}{38.67}
\DefMacro{res-user-view-no-name-ne2e-few-shot-no-name-gpt-3.5-turbo-16k-bleu-min}{38.67}
\DefMacro{res-user-view-no-name-ne2e-few-shot-no-name-gpt-3.5-turbo-16k-code-bleu}{49.33}
\DefMacro{res-user-view-no-name-ne2e-few-shot-no-name-gpt-3.5-turbo-16k-code-bleu-avg}{49.33}
\DefMacro{res-user-view-no-name-ne2e-few-shot-no-name-gpt-3.5-turbo-16k-code-bleu-max}{49.33}
\DefMacro{res-user-view-no-name-ne2e-few-shot-no-name-gpt-3.5-turbo-16k-code-bleu-min}{49.33}
\DefMacro{res-user-view-no-name-ne2e-few-shot-no-name-gpt-3.5-turbo-16k-edit-sim}{69.32}
\DefMacro{res-user-view-no-name-ne2e-few-shot-no-name-gpt-3.5-turbo-16k-edit-sim-avg}{69.32}
\DefMacro{res-user-view-no-name-ne2e-few-shot-no-name-gpt-3.5-turbo-16k-edit-sim-max}{69.32}
\DefMacro{res-user-view-no-name-ne2e-few-shot-no-name-gpt-3.5-turbo-16k-edit-sim-min}{69.32}
\DefMacro{res-user-view-no-name-ne2e-few-shot-no-name-gpt-3.5-turbo-16k-rouge-f}{67.87}
\DefMacro{res-user-view-no-name-ne2e-few-shot-no-name-gpt-3.5-turbo-16k-rouge-f-avg}{67.87}
\DefMacro{res-user-view-no-name-ne2e-few-shot-no-name-gpt-3.5-turbo-16k-rouge-f-max}{67.87}
\DefMacro{res-user-view-no-name-ne2e-few-shot-no-name-gpt-3.5-turbo-16k-rouge-f-min}{67.87}
\DefMacro{res-user-view-no-name-ne2e-few-shot-no-name-gpt-3.5-turbo-16k-rouge-p}{68.59}
\DefMacro{res-user-view-no-name-ne2e-few-shot-no-name-gpt-3.5-turbo-16k-rouge-p-avg}{68.59}
\DefMacro{res-user-view-no-name-ne2e-few-shot-no-name-gpt-3.5-turbo-16k-rouge-p-max}{68.59}
\DefMacro{res-user-view-no-name-ne2e-few-shot-no-name-gpt-3.5-turbo-16k-rouge-p-min}{68.59}
\DefMacro{res-user-view-no-name-ne2e-few-shot-no-name-gpt-3.5-turbo-16k-rouge-r}{69.18}
\DefMacro{res-user-view-no-name-ne2e-few-shot-no-name-gpt-3.5-turbo-16k-rouge-r-avg}{69.18}
\DefMacro{res-user-view-no-name-ne2e-few-shot-no-name-gpt-3.5-turbo-16k-rouge-r-max}{69.18}
\DefMacro{res-user-view-no-name-ne2e-few-shot-no-name-gpt-3.5-turbo-16k-rouge-r-min}{69.18}
\DefMacro{res-user-view-no-name-ne2e-few-shot-no-name-gpt-3.5-turbo-16k-xmatch}{0.00}
\DefMacro{res-user-view-no-name-ne2e-few-shot-no-name-gpt-3.5-turbo-16k-xmatch-avg}{0.00}
\DefMacro{res-user-view-no-name-ne2e-few-shot-no-name-gpt-3.5-turbo-16k-xmatch-max}{0.00}
\DefMacro{res-user-view-no-name-ne2e-few-shot-no-name-gpt-3.5-turbo-16k-xmatch-min}{0.00}
\DefMacro{res-user-view-no-name-ne2e-few-shot-no-name-gpt-3.5-turbo-16k-xmatch-top1}{0.00}
\DefMacro{res-user-view-no-name-ne2e-few-shot-no-name-gpt-3.5-turbo-16k-compilable-avg}{82.83}
\DefMacro{res-user-view-no-name-ne2e-few-shot-no-name-gpt-3.5-turbo-16k-compilable-max}{82.83}
\DefMacro{res-user-view-no-name-ne2e-few-shot-no-name-gpt-3.5-turbo-16k-compilable-min}{82.83}
\DefMacro{res-user-view-no-name-ne2e-few-shot-no-name-gpt-3.5-turbo-16k-coverage-avg}{32.32}
\DefMacro{res-user-view-no-name-ne2e-few-shot-no-name-gpt-3.5-turbo-16k-coverage-max}{32.32}
\DefMacro{res-user-view-no-name-ne2e-few-shot-no-name-gpt-3.5-turbo-16k-coverage-min}{32.32}
\DefMacro{res-user-view-no-name-ne2e-few-shot-no-name-gpt-3.5-turbo-16k-match-avg}{100.00}
\DefMacro{res-user-view-no-name-ne2e-few-shot-no-name-gpt-3.5-turbo-16k-match-max}{100.00}
\DefMacro{res-user-view-no-name-ne2e-few-shot-no-name-gpt-3.5-turbo-16k-match-min}{100.00}
\DefMacro{res-user-view-no-name-ne2e-few-shot-no-name-gpt-3.5-turbo-16k-runnable-avg}{69.51}
\DefMacro{res-user-view-no-name-ne2e-few-shot-no-name-gpt-3.5-turbo-16k-runnable-max}{69.51}
\DefMacro{res-user-view-no-name-ne2e-few-shot-no-name-gpt-3.5-turbo-16k-runnable-min}{69.51}
\DefMacro{res-user-view-no-name-ne2e-few-shot-no-name-gpt-3.5-turbo-16k-runnable-overall}{57.58}
\DefMacro{res-user-view-no-name-ne2e-few-shot-no-name-gpt-3.5-turbo-16k-runnable-overall-avg}{57.58}
\DefMacro{res-user-view-no-name-ne2e-few-shot-no-name-gpt-3.5-turbo-16k-runnable-overall-max}{57.58}
\DefMacro{res-user-view-no-name-ne2e-few-shot-no-name-gpt-3.5-turbo-16k-runnable-overall-min}{57.58}
\DefMacro{res-user-view-no-name-ne2e-few-shot-no-name-gpt-3.5-turbo-16k-timeout-avg}{0.00}
\DefMacro{res-user-view-no-name-ne2e-few-shot-no-name-gpt-3.5-turbo-16k-timeout-max}{0.00}
\DefMacro{res-user-view-no-name-ne2e-few-shot-no-name-gpt-3.5-turbo-16k-timeout-min}{0.00}
\DefMacro{res-user-view-no-name-catlm-ne2e-no-name-catlm-bleu}{37.25}
\DefMacro{res-user-view-no-name-catlm-ne2e-no-name-catlm-bleu-avg}{37.25}
\DefMacro{res-user-view-no-name-catlm-ne2e-no-name-catlm-bleu-max}{37.25}
\DefMacro{res-user-view-no-name-catlm-ne2e-no-name-catlm-bleu-min}{37.25}
\DefMacro{res-user-view-no-name-catlm-ne2e-no-name-catlm-code-bleu}{48.47}
\DefMacro{res-user-view-no-name-catlm-ne2e-no-name-catlm-code-bleu-avg}{48.47}
\DefMacro{res-user-view-no-name-catlm-ne2e-no-name-catlm-code-bleu-max}{48.47}
\DefMacro{res-user-view-no-name-catlm-ne2e-no-name-catlm-code-bleu-min}{48.47}
\DefMacro{res-user-view-no-name-catlm-ne2e-no-name-catlm-edit-sim}{68.18}
\DefMacro{res-user-view-no-name-catlm-ne2e-no-name-catlm-edit-sim-avg}{68.18}
\DefMacro{res-user-view-no-name-catlm-ne2e-no-name-catlm-edit-sim-max}{68.18}
\DefMacro{res-user-view-no-name-catlm-ne2e-no-name-catlm-edit-sim-min}{68.18}
\DefMacro{res-user-view-no-name-catlm-ne2e-no-name-catlm-rouge-f}{65.08}
\DefMacro{res-user-view-no-name-catlm-ne2e-no-name-catlm-rouge-f-avg}{65.08}
\DefMacro{res-user-view-no-name-catlm-ne2e-no-name-catlm-rouge-f-max}{65.08}
\DefMacro{res-user-view-no-name-catlm-ne2e-no-name-catlm-rouge-f-min}{65.08}
\DefMacro{res-user-view-no-name-catlm-ne2e-no-name-catlm-rouge-p}{65.03}
\DefMacro{res-user-view-no-name-catlm-ne2e-no-name-catlm-rouge-p-avg}{65.03}
\DefMacro{res-user-view-no-name-catlm-ne2e-no-name-catlm-rouge-p-max}{65.03}
\DefMacro{res-user-view-no-name-catlm-ne2e-no-name-catlm-rouge-p-min}{65.03}
\DefMacro{res-user-view-no-name-catlm-ne2e-no-name-catlm-rouge-r}{69.48}
\DefMacro{res-user-view-no-name-catlm-ne2e-no-name-catlm-rouge-r-avg}{69.48}
\DefMacro{res-user-view-no-name-catlm-ne2e-no-name-catlm-rouge-r-max}{69.48}
\DefMacro{res-user-view-no-name-catlm-ne2e-no-name-catlm-rouge-r-min}{69.48}
\DefMacro{res-user-view-no-name-catlm-ne2e-no-name-catlm-xmatch}{1.23}
\DefMacro{res-user-view-no-name-catlm-ne2e-no-name-catlm-xmatch-avg}{1.23}
\DefMacro{res-user-view-no-name-catlm-ne2e-no-name-catlm-xmatch-max}{1.23}
\DefMacro{res-user-view-no-name-catlm-ne2e-no-name-catlm-xmatch-min}{1.23}
\DefMacro{res-user-view-no-name-catlm-ne2e-no-name-catlm-xmatch-top1}{1.23}
\DefMacro{res-user-view-no-name-catlm-ne2e-no-name-catlm-compilable-avg}{70.83}
\DefMacro{res-user-view-no-name-catlm-ne2e-no-name-catlm-compilable-max}{70.83}
\DefMacro{res-user-view-no-name-catlm-ne2e-no-name-catlm-compilable-min}{70.83}
\DefMacro{res-user-view-no-name-catlm-ne2e-no-name-catlm-coverage-avg}{16.44}
\DefMacro{res-user-view-no-name-catlm-ne2e-no-name-catlm-coverage-max}{16.44}
\DefMacro{res-user-view-no-name-catlm-ne2e-no-name-catlm-coverage-min}{16.44}
\DefMacro{res-user-view-no-name-catlm-ne2e-no-name-catlm-match-avg}{100.00}
\DefMacro{res-user-view-no-name-catlm-ne2e-no-name-catlm-match-max}{100.00}
\DefMacro{res-user-view-no-name-catlm-ne2e-no-name-catlm-match-min}{100.00}
\DefMacro{res-user-view-no-name-catlm-ne2e-no-name-catlm-runnable-avg}{39.38}
\DefMacro{res-user-view-no-name-catlm-ne2e-no-name-catlm-runnable-max}{39.38}
\DefMacro{res-user-view-no-name-catlm-ne2e-no-name-catlm-runnable-min}{39.38}
\DefMacro{res-user-view-no-name-catlm-ne2e-no-name-catlm-runnable-overall}{23.96}
\DefMacro{res-user-view-no-name-catlm-ne2e-no-name-catlm-runnable-overall-avg}{23.96}
\DefMacro{res-user-view-no-name-catlm-ne2e-no-name-catlm-runnable-overall-max}{23.96}
\DefMacro{res-user-view-no-name-catlm-ne2e-no-name-catlm-runnable-overall-min}{23.96}
\DefMacro{res-user-view-no-name-catlm-ne2e-no-name-catlm-timeout-avg}{0.75}
\DefMacro{res-user-view-no-name-catlm-ne2e-no-name-catlm-timeout-max}{0.75}
\DefMacro{res-user-view-no-name-catlm-ne2e-no-name-catlm-timeout-min}{0.75}
\DefMacro{res-user-view-no-name-selected-434-conditionnestack2e-no-name-zero-shot-lora-codellama-7b-bleu}{26.93}
\DefMacro{res-user-view-no-name-selected-434-conditionnestack2e-no-name-zero-shot-lora-codellama-7b-bleu-avg}{26.93}
\DefMacro{res-user-view-no-name-selected-434-conditionnestack2e-no-name-zero-shot-lora-codellama-7b-bleu-max}{26.93}
\DefMacro{res-user-view-no-name-selected-434-conditionnestack2e-no-name-zero-shot-lora-codellama-7b-bleu-min}{26.93}
\DefMacro{res-user-view-no-name-selected-434-conditionnestack2e-no-name-zero-shot-lora-codellama-7b-code-bleu}{37.38}
\DefMacro{res-user-view-no-name-selected-434-conditionnestack2e-no-name-zero-shot-lora-codellama-7b-code-bleu-avg}{37.38}
\DefMacro{res-user-view-no-name-selected-434-conditionnestack2e-no-name-zero-shot-lora-codellama-7b-code-bleu-max}{37.38}
\DefMacro{res-user-view-no-name-selected-434-conditionnestack2e-no-name-zero-shot-lora-codellama-7b-code-bleu-min}{37.38}
\DefMacro{res-user-view-no-name-selected-434-conditionnestack2e-no-name-zero-shot-lora-codellama-7b-edit-sim}{61.88}
\DefMacro{res-user-view-no-name-selected-434-conditionnestack2e-no-name-zero-shot-lora-codellama-7b-edit-sim-avg}{61.88}
\DefMacro{res-user-view-no-name-selected-434-conditionnestack2e-no-name-zero-shot-lora-codellama-7b-edit-sim-max}{61.88}
\DefMacro{res-user-view-no-name-selected-434-conditionnestack2e-no-name-zero-shot-lora-codellama-7b-edit-sim-min}{61.88}
\DefMacro{res-user-view-no-name-selected-434-conditionnestack2e-no-name-zero-shot-lora-codellama-7b-rouge-f}{54.76}
\DefMacro{res-user-view-no-name-selected-434-conditionnestack2e-no-name-zero-shot-lora-codellama-7b-rouge-f-avg}{54.76}
\DefMacro{res-user-view-no-name-selected-434-conditionnestack2e-no-name-zero-shot-lora-codellama-7b-rouge-f-max}{54.76}
\DefMacro{res-user-view-no-name-selected-434-conditionnestack2e-no-name-zero-shot-lora-codellama-7b-rouge-f-min}{54.76}
\DefMacro{res-user-view-no-name-selected-434-conditionnestack2e-no-name-zero-shot-lora-codellama-7b-rouge-p}{50.49}
\DefMacro{res-user-view-no-name-selected-434-conditionnestack2e-no-name-zero-shot-lora-codellama-7b-rouge-p-avg}{50.49}
\DefMacro{res-user-view-no-name-selected-434-conditionnestack2e-no-name-zero-shot-lora-codellama-7b-rouge-p-max}{50.49}
\DefMacro{res-user-view-no-name-selected-434-conditionnestack2e-no-name-zero-shot-lora-codellama-7b-rouge-p-min}{50.49}
\DefMacro{res-user-view-no-name-selected-434-conditionnestack2e-no-name-zero-shot-lora-codellama-7b-rouge-r}{64.40}
\DefMacro{res-user-view-no-name-selected-434-conditionnestack2e-no-name-zero-shot-lora-codellama-7b-rouge-r-avg}{64.40}
\DefMacro{res-user-view-no-name-selected-434-conditionnestack2e-no-name-zero-shot-lora-codellama-7b-rouge-r-max}{64.40}
\DefMacro{res-user-view-no-name-selected-434-conditionnestack2e-no-name-zero-shot-lora-codellama-7b-rouge-r-min}{64.40}
\DefMacro{res-user-view-no-name-selected-434-conditionnestack2e-no-name-zero-shot-lora-codellama-7b-xmatch}{0.15}
\DefMacro{res-user-view-no-name-selected-434-conditionnestack2e-no-name-zero-shot-lora-codellama-7b-xmatch-avg}{0.15}
\DefMacro{res-user-view-no-name-selected-434-conditionnestack2e-no-name-zero-shot-lora-codellama-7b-xmatch-max}{0.15}
\DefMacro{res-user-view-no-name-selected-434-conditionnestack2e-no-name-zero-shot-lora-codellama-7b-xmatch-min}{0.15}
\DefMacro{res-user-view-no-name-selected-434-conditionnestack2e-no-name-zero-shot-lora-codellama-7b-xmatch-top1}{0.15}
\DefMacro{res-user-view-no-name-selected-434-conditionnestack2e-no-name-zero-shot-lora-codellama-7b-compilable-avg}{57.22}
\DefMacro{res-user-view-no-name-selected-434-conditionnestack2e-no-name-zero-shot-lora-codellama-7b-compilable-max}{57.22}
\DefMacro{res-user-view-no-name-selected-434-conditionnestack2e-no-name-zero-shot-lora-codellama-7b-compilable-min}{57.22}
\DefMacro{res-user-view-no-name-selected-434-conditionnestack2e-no-name-zero-shot-lora-codellama-7b-coverage-avg}{33.49}
\DefMacro{res-user-view-no-name-selected-434-conditionnestack2e-no-name-zero-shot-lora-codellama-7b-coverage-max}{33.49}
\DefMacro{res-user-view-no-name-selected-434-conditionnestack2e-no-name-zero-shot-lora-codellama-7b-coverage-min}{33.49}
\DefMacro{res-user-view-no-name-selected-434-conditionnestack2e-no-name-zero-shot-lora-codellama-7b-match-avg}{96.91}
\DefMacro{res-user-view-no-name-selected-434-conditionnestack2e-no-name-zero-shot-lora-codellama-7b-match-max}{96.91}
\DefMacro{res-user-view-no-name-selected-434-conditionnestack2e-no-name-zero-shot-lora-codellama-7b-match-min}{96.91}
\DefMacro{res-user-view-no-name-selected-434-conditionnestack2e-no-name-zero-shot-lora-codellama-7b-runnable-avg}{74.52}
\DefMacro{res-user-view-no-name-selected-434-conditionnestack2e-no-name-zero-shot-lora-codellama-7b-runnable-max}{74.52}
\DefMacro{res-user-view-no-name-selected-434-conditionnestack2e-no-name-zero-shot-lora-codellama-7b-runnable-min}{74.52}
\DefMacro{res-user-view-no-name-selected-434-conditionnestack2e-no-name-zero-shot-lora-codellama-7b-runnable-overall}{41.32}
\DefMacro{res-user-view-no-name-selected-434-conditionnestack2e-no-name-zero-shot-lora-codellama-7b-runnable-overall-avg}{41.32}
\DefMacro{res-user-view-no-name-selected-434-conditionnestack2e-no-name-zero-shot-lora-codellama-7b-runnable-overall-max}{41.32}
\DefMacro{res-user-view-no-name-selected-434-conditionnestack2e-no-name-zero-shot-lora-codellama-7b-runnable-overall-min}{41.32}
\DefMacro{res-user-view-no-name-selected-434-conditionnestack2e-no-name-zero-shot-lora-codellama-7b-timeout-avg}{0.14}
\DefMacro{res-user-view-no-name-selected-434-conditionnestack2e-no-name-zero-shot-lora-codellama-7b-timeout-max}{0.14}
\DefMacro{res-user-view-no-name-selected-434-conditionnestack2e-no-name-zero-shot-lora-codellama-7b-timeout-min}{0.14}
\DefMacro{res-user-view-no-name-conditionnestack2e-no-name-ft-lora-codellama-7b-bleu}{46.66}
\DefMacro{res-user-view-no-name-conditionnestack2e-no-name-ft-lora-codellama-7b-bleu-avg}{46.66}
\DefMacro{res-user-view-no-name-conditionnestack2e-no-name-ft-lora-codellama-7b-bleu-max}{46.66}
\DefMacro{res-user-view-no-name-conditionnestack2e-no-name-ft-lora-codellama-7b-bleu-min}{46.66}
\DefMacro{res-user-view-no-name-conditionnestack2e-no-name-ft-lora-codellama-7b-code-bleu}{55.36}
\DefMacro{res-user-view-no-name-conditionnestack2e-no-name-ft-lora-codellama-7b-code-bleu-avg}{55.36}
\DefMacro{res-user-view-no-name-conditionnestack2e-no-name-ft-lora-codellama-7b-code-bleu-max}{55.36}
\DefMacro{res-user-view-no-name-conditionnestack2e-no-name-ft-lora-codellama-7b-code-bleu-min}{55.36}
\DefMacro{res-user-view-no-name-conditionnestack2e-no-name-ft-lora-codellama-7b-edit-sim}{79.76}
\DefMacro{res-user-view-no-name-conditionnestack2e-no-name-ft-lora-codellama-7b-edit-sim-avg}{79.76}
\DefMacro{res-user-view-no-name-conditionnestack2e-no-name-ft-lora-codellama-7b-edit-sim-max}{79.76}
\DefMacro{res-user-view-no-name-conditionnestack2e-no-name-ft-lora-codellama-7b-edit-sim-min}{79.76}
\DefMacro{res-user-view-no-name-conditionnestack2e-no-name-ft-lora-codellama-7b-rouge-f}{74.21}
\DefMacro{res-user-view-no-name-conditionnestack2e-no-name-ft-lora-codellama-7b-rouge-f-avg}{74.21}
\DefMacro{res-user-view-no-name-conditionnestack2e-no-name-ft-lora-codellama-7b-rouge-f-max}{74.21}
\DefMacro{res-user-view-no-name-conditionnestack2e-no-name-ft-lora-codellama-7b-rouge-f-min}{74.21}
\DefMacro{res-user-view-no-name-conditionnestack2e-no-name-ft-lora-codellama-7b-rouge-p}{76.82}
\DefMacro{res-user-view-no-name-conditionnestack2e-no-name-ft-lora-codellama-7b-rouge-p-avg}{76.82}
\DefMacro{res-user-view-no-name-conditionnestack2e-no-name-ft-lora-codellama-7b-rouge-p-max}{76.82}
\DefMacro{res-user-view-no-name-conditionnestack2e-no-name-ft-lora-codellama-7b-rouge-p-min}{76.82}
\DefMacro{res-user-view-no-name-conditionnestack2e-no-name-ft-lora-codellama-7b-rouge-r}{73.77}
\DefMacro{res-user-view-no-name-conditionnestack2e-no-name-ft-lora-codellama-7b-rouge-r-avg}{73.77}
\DefMacro{res-user-view-no-name-conditionnestack2e-no-name-ft-lora-codellama-7b-rouge-r-max}{73.77}
\DefMacro{res-user-view-no-name-conditionnestack2e-no-name-ft-lora-codellama-7b-rouge-r-min}{73.77}
\DefMacro{res-user-view-no-name-conditionnestack2e-no-name-ft-lora-codellama-7b-xmatch}{2.07}
\DefMacro{res-user-view-no-name-conditionnestack2e-no-name-ft-lora-codellama-7b-xmatch-avg}{2.07}
\DefMacro{res-user-view-no-name-conditionnestack2e-no-name-ft-lora-codellama-7b-xmatch-max}{2.07}
\DefMacro{res-user-view-no-name-conditionnestack2e-no-name-ft-lora-codellama-7b-xmatch-min}{2.07}
\DefMacro{res-user-view-no-name-conditionnestack2e-no-name-ft-lora-codellama-7b-xmatch-top1}{2.07}
\DefMacro{res-user-view-no-name-conditionnestack2e-no-name-ft-lora-codellama-7b-compilable-avg}{82.26}
\DefMacro{res-user-view-no-name-conditionnestack2e-no-name-ft-lora-codellama-7b-compilable-max}{82.26}
\DefMacro{res-user-view-no-name-conditionnestack2e-no-name-ft-lora-codellama-7b-compilable-min}{82.26}
\DefMacro{res-user-view-no-name-conditionnestack2e-no-name-ft-lora-codellama-7b-coverage-avg}{59.83}
\DefMacro{res-user-view-no-name-conditionnestack2e-no-name-ft-lora-codellama-7b-coverage-max}{59.83}
\DefMacro{res-user-view-no-name-conditionnestack2e-no-name-ft-lora-codellama-7b-coverage-min}{59.83}
\DefMacro{res-user-view-no-name-conditionnestack2e-no-name-ft-lora-codellama-7b-match-avg}{100.00}
\DefMacro{res-user-view-no-name-conditionnestack2e-no-name-ft-lora-codellama-7b-match-max}{100.00}
\DefMacro{res-user-view-no-name-conditionnestack2e-no-name-ft-lora-codellama-7b-match-min}{100.00}
\DefMacro{res-user-view-no-name-conditionnestack2e-no-name-ft-lora-codellama-7b-runnable-avg}{84.97}
\DefMacro{res-user-view-no-name-conditionnestack2e-no-name-ft-lora-codellama-7b-runnable-max}{84.97}
\DefMacro{res-user-view-no-name-conditionnestack2e-no-name-ft-lora-codellama-7b-runnable-min}{84.97}
\DefMacro{res-user-view-no-name-conditionnestack2e-no-name-ft-lora-codellama-7b-runnable-overall}{69.89}
\DefMacro{res-user-view-no-name-conditionnestack2e-no-name-ft-lora-codellama-7b-runnable-overall-avg}{69.89}
\DefMacro{res-user-view-no-name-conditionnestack2e-no-name-ft-lora-codellama-7b-runnable-overall-max}{69.89}
\DefMacro{res-user-view-no-name-conditionnestack2e-no-name-ft-lora-codellama-7b-runnable-overall-min}{69.89}
\DefMacro{res-user-view-no-name-conditionnestack2e-no-name-ft-lora-codellama-7b-timeout-avg}{0.56}
\DefMacro{res-user-view-no-name-conditionnestack2e-no-name-ft-lora-codellama-7b-timeout-max}{0.56}
\DefMacro{res-user-view-no-name-conditionnestack2e-no-name-ft-lora-codellama-7b-timeout-min}{0.56}
\DefMacro{res-user-view-no-name-conditionnestack2e-all-no-name-ft-lora-codellama-7b-bleu}{47.00}
\DefMacro{res-user-view-no-name-conditionnestack2e-all-no-name-ft-lora-codellama-7b-bleu-avg}{46.80}
\DefMacro{res-user-view-no-name-conditionnestack2e-all-no-name-ft-lora-codellama-7b-bleu-max}{50.73}
\DefMacro{res-user-view-no-name-conditionnestack2e-all-no-name-ft-lora-codellama-7b-bleu-min}{43.07}
\DefMacro{res-user-view-no-name-conditionnestack2e-all-no-name-ft-lora-codellama-7b-code-bleu}{55.73}
\DefMacro{res-user-view-no-name-conditionnestack2e-all-no-name-ft-lora-codellama-7b-code-bleu-avg}{55.47}
\DefMacro{res-user-view-no-name-conditionnestack2e-all-no-name-ft-lora-codellama-7b-code-bleu-max}{59.38}
\DefMacro{res-user-view-no-name-conditionnestack2e-all-no-name-ft-lora-codellama-7b-code-bleu-min}{51.92}
\DefMacro{res-user-view-no-name-conditionnestack2e-all-no-name-ft-lora-codellama-7b-edit-sim}{79.94}
\DefMacro{res-user-view-no-name-conditionnestack2e-all-no-name-ft-lora-codellama-7b-edit-sim-avg}{79.80}
\DefMacro{res-user-view-no-name-conditionnestack2e-all-no-name-ft-lora-codellama-7b-edit-sim-max}{82.83}
\DefMacro{res-user-view-no-name-conditionnestack2e-all-no-name-ft-lora-codellama-7b-edit-sim-min}{76.40}
\DefMacro{res-user-view-no-name-conditionnestack2e-all-no-name-ft-lora-codellama-7b-rouge-f}{74.58}
\DefMacro{res-user-view-no-name-conditionnestack2e-all-no-name-ft-lora-codellama-7b-rouge-f-avg}{74.23}
\DefMacro{res-user-view-no-name-conditionnestack2e-all-no-name-ft-lora-codellama-7b-rouge-f-max}{77.65}
\DefMacro{res-user-view-no-name-conditionnestack2e-all-no-name-ft-lora-codellama-7b-rouge-f-min}{70.76}
\DefMacro{res-user-view-no-name-conditionnestack2e-all-no-name-ft-lora-codellama-7b-rouge-p}{77.16}
\DefMacro{res-user-view-no-name-conditionnestack2e-all-no-name-ft-lora-codellama-7b-rouge-p-avg}{76.84}
\DefMacro{res-user-view-no-name-conditionnestack2e-all-no-name-ft-lora-codellama-7b-rouge-p-max}{80.61}
\DefMacro{res-user-view-no-name-conditionnestack2e-all-no-name-ft-lora-codellama-7b-rouge-p-min}{72.52}
\DefMacro{res-user-view-no-name-conditionnestack2e-all-no-name-ft-lora-codellama-7b-rouge-r}{74.19}
\DefMacro{res-user-view-no-name-conditionnestack2e-all-no-name-ft-lora-codellama-7b-rouge-r-avg}{73.78}
\DefMacro{res-user-view-no-name-conditionnestack2e-all-no-name-ft-lora-codellama-7b-rouge-r-max}{77.01}
\DefMacro{res-user-view-no-name-conditionnestack2e-all-no-name-ft-lora-codellama-7b-rouge-r-min}{70.80}
\DefMacro{res-user-view-no-name-conditionnestack2e-all-no-name-ft-lora-codellama-7b-xmatch}{1.92}
\DefMacro{res-user-view-no-name-conditionnestack2e-all-no-name-ft-lora-codellama-7b-xmatch-avg}{2.06}
\DefMacro{res-user-view-no-name-conditionnestack2e-all-no-name-ft-lora-codellama-7b-xmatch-max}{2.61}
\DefMacro{res-user-view-no-name-conditionnestack2e-all-no-name-ft-lora-codellama-7b-xmatch-min}{1.54}
\DefMacro{res-user-view-no-name-conditionnestack2e-all-no-name-ft-lora-codellama-7b-xmatch-top1}{1.92}
\DefMacro{res-user-view-no-name-conditionnestack2e-all-no-name-ft-lora-codellama-7b-compilable-avg}{82.36}
\DefMacro{res-user-view-no-name-conditionnestack2e-all-no-name-ft-lora-codellama-7b-compilable-max}{89.40}
\DefMacro{res-user-view-no-name-conditionnestack2e-all-no-name-ft-lora-codellama-7b-compilable-min}{73.12}
\DefMacro{res-user-view-no-name-conditionnestack2e-all-no-name-ft-lora-codellama-7b-coverage-avg}{58.17}
\DefMacro{res-user-view-no-name-conditionnestack2e-all-no-name-ft-lora-codellama-7b-coverage-max}{67.67}
\DefMacro{res-user-view-no-name-conditionnestack2e-all-no-name-ft-lora-codellama-7b-coverage-min}{48.23}
\DefMacro{res-user-view-no-name-conditionnestack2e-all-no-name-ft-lora-codellama-7b-match-avg}{99.91}
\DefMacro{res-user-view-no-name-conditionnestack2e-all-no-name-ft-lora-codellama-7b-match-max}{100.00}
\DefMacro{res-user-view-no-name-conditionnestack2e-all-no-name-ft-lora-codellama-7b-match-min}{99.83}
\DefMacro{res-user-view-no-name-conditionnestack2e-all-no-name-ft-lora-codellama-7b-runnable-avg}{81.65}
\DefMacro{res-user-view-no-name-conditionnestack2e-all-no-name-ft-lora-codellama-7b-runnable-max}{88.15}
\DefMacro{res-user-view-no-name-conditionnestack2e-all-no-name-ft-lora-codellama-7b-runnable-min}{73.45}
\DefMacro{res-user-view-no-name-conditionnestack2e-all-no-name-ft-lora-codellama-7b-runnable-overall}{78.80}
\DefMacro{res-user-view-no-name-conditionnestack2e-all-no-name-ft-lora-codellama-7b-runnable-overall-avg}{68.57}
\DefMacro{res-user-view-no-name-conditionnestack2e-all-no-name-ft-lora-codellama-7b-runnable-overall-max}{78.80}
\DefMacro{res-user-view-no-name-conditionnestack2e-all-no-name-ft-lora-codellama-7b-runnable-overall-min}{56.91}
\DefMacro{res-user-view-no-name-conditionnestack2e-all-no-name-ft-lora-codellama-7b-timeout-avg}{0.49}
\DefMacro{res-user-view-no-name-conditionnestack2e-all-no-name-ft-lora-codellama-7b-timeout-max}{0.77}
\DefMacro{res-user-view-no-name-conditionnestack2e-all-no-name-ft-lora-codellama-7b-timeout-min}{0.26}

\DefMacro{ds-rq2-count-etype}{81}
\DefMacro{ds-rq2-count-module}{55}
\DefMacro{ds-rq2-count-project}{30}
\DefMacro{ds-rq2-count-ts}{649}

\DefMacro{num-methods-invoked}{228}
\DefMacro{num-projects-invoked}{18}
\DefMacro{num-methods-no-context}{1,261}
\DefMacro{num-methods-no-context-subset}{750}
\DefMacro{num-methods-no-context-subset-percentage}{30.4}

\DefMacro{res-exlong-ablation-conditionnestack2e-with-name-ft-lora-codellama-7b-bleu}{63.13}
\DefMacro{res-exlong-ablation-conditionnestack2e-with-name-ft-lora-codellama-7b-bleu-avg}{63.13}
\DefMacro{res-exlong-ablation-conditionnestack2e-with-name-ft-lora-codellama-7b-bleu-max}{63.13}
\DefMacro{res-exlong-ablation-conditionnestack2e-with-name-ft-lora-codellama-7b-bleu-min}{63.13}
\DefMacro{res-exlong-ablation-conditionnestack2e-with-name-ft-lora-codellama-7b-code-bleu}{67.49}
\DefMacro{res-exlong-ablation-conditionnestack2e-with-name-ft-lora-codellama-7b-code-bleu-avg}{67.49}
\DefMacro{res-exlong-ablation-conditionnestack2e-with-name-ft-lora-codellama-7b-code-bleu-max}{67.49}
\DefMacro{res-exlong-ablation-conditionnestack2e-with-name-ft-lora-codellama-7b-code-bleu-min}{67.49}
\DefMacro{res-exlong-ablation-conditionnestack2e-with-name-ft-lora-codellama-7b-edit-sim}{85.32}
\DefMacro{res-exlong-ablation-conditionnestack2e-with-name-ft-lora-codellama-7b-edit-sim-avg}{85.32}
\DefMacro{res-exlong-ablation-conditionnestack2e-with-name-ft-lora-codellama-7b-edit-sim-max}{85.32}
\DefMacro{res-exlong-ablation-conditionnestack2e-with-name-ft-lora-codellama-7b-edit-sim-min}{85.32}
\DefMacro{res-exlong-ablation-conditionnestack2e-with-name-ft-lora-codellama-7b-rouge-f}{80.75}
\DefMacro{res-exlong-ablation-conditionnestack2e-with-name-ft-lora-codellama-7b-rouge-f-avg}{80.75}
\DefMacro{res-exlong-ablation-conditionnestack2e-with-name-ft-lora-codellama-7b-rouge-f-max}{80.75}
\DefMacro{res-exlong-ablation-conditionnestack2e-with-name-ft-lora-codellama-7b-rouge-f-min}{80.75}
\DefMacro{res-exlong-ablation-conditionnestack2e-with-name-ft-lora-codellama-7b-rouge-p}{82.85}
\DefMacro{res-exlong-ablation-conditionnestack2e-with-name-ft-lora-codellama-7b-rouge-p-avg}{82.85}
\DefMacro{res-exlong-ablation-conditionnestack2e-with-name-ft-lora-codellama-7b-rouge-p-max}{82.85}
\DefMacro{res-exlong-ablation-conditionnestack2e-with-name-ft-lora-codellama-7b-rouge-p-min}{82.85}
\DefMacro{res-exlong-ablation-conditionnestack2e-with-name-ft-lora-codellama-7b-rouge-r}{80.81}
\DefMacro{res-exlong-ablation-conditionnestack2e-with-name-ft-lora-codellama-7b-rouge-r-avg}{80.81}
\DefMacro{res-exlong-ablation-conditionnestack2e-with-name-ft-lora-codellama-7b-rouge-r-max}{80.81}
\DefMacro{res-exlong-ablation-conditionnestack2e-with-name-ft-lora-codellama-7b-rouge-r-min}{80.81}
\DefMacro{res-exlong-ablation-conditionnestack2e-with-name-ft-lora-codellama-7b-xmatch}{19.05}
\DefMacro{res-exlong-ablation-conditionnestack2e-with-name-ft-lora-codellama-7b-xmatch-avg}{19.05}
\DefMacro{res-exlong-ablation-conditionnestack2e-with-name-ft-lora-codellama-7b-xmatch-max}{19.05}
\DefMacro{res-exlong-ablation-conditionnestack2e-with-name-ft-lora-codellama-7b-xmatch-min}{19.05}
\DefMacro{res-exlong-ablation-conditionnestack2e-with-name-ft-lora-codellama-7b-xmatch-top1}{19.05}
\DefMacro{res-exlong-ablation-conditionnestack2e-with-name-ft-lora-codellama-7b-compilable-avg}{82.10}
\DefMacro{res-exlong-ablation-conditionnestack2e-with-name-ft-lora-codellama-7b-compilable-max}{82.10}
\DefMacro{res-exlong-ablation-conditionnestack2e-with-name-ft-lora-codellama-7b-compilable-min}{82.10}
\DefMacro{res-exlong-ablation-conditionnestack2e-with-name-ft-lora-codellama-7b-coverage-avg}{59.45}
\DefMacro{res-exlong-ablation-conditionnestack2e-with-name-ft-lora-codellama-7b-coverage-max}{59.45}
\DefMacro{res-exlong-ablation-conditionnestack2e-with-name-ft-lora-codellama-7b-coverage-min}{59.45}
\DefMacro{res-exlong-ablation-conditionnestack2e-with-name-ft-lora-codellama-7b-match-avg}{100.00}
\DefMacro{res-exlong-ablation-conditionnestack2e-with-name-ft-lora-codellama-7b-match-max}{100.00}
\DefMacro{res-exlong-ablation-conditionnestack2e-with-name-ft-lora-codellama-7b-match-min}{100.00}
\DefMacro{res-exlong-ablation-conditionnestack2e-with-name-ft-lora-codellama-7b-runnable-avg}{82.05}
\DefMacro{res-exlong-ablation-conditionnestack2e-with-name-ft-lora-codellama-7b-runnable-max}{82.05}
\DefMacro{res-exlong-ablation-conditionnestack2e-with-name-ft-lora-codellama-7b-runnable-min}{82.05}
\DefMacro{res-exlong-ablation-conditionnestack2e-with-name-ft-lora-codellama-7b-runnable-overall}{67.36}
\DefMacro{res-exlong-ablation-conditionnestack2e-with-name-ft-lora-codellama-7b-runnable-overall-avg}{67.36}
\DefMacro{res-exlong-ablation-conditionnestack2e-with-name-ft-lora-codellama-7b-runnable-overall-max}{67.36}
\DefMacro{res-exlong-ablation-conditionnestack2e-with-name-ft-lora-codellama-7b-runnable-overall-min}{67.36}
\DefMacro{res-exlong-ablation-conditionnestack2e-with-name-ft-lora-codellama-7b-timeout-avg}{0.65}
\DefMacro{res-exlong-ablation-conditionnestack2e-with-name-ft-lora-codellama-7b-timeout-max}{0.65}
\DefMacro{res-exlong-ablation-conditionnestack2e-with-name-ft-lora-codellama-7b-timeout-min}{0.65}
\DefMacro{res-exlong-ablation-conditionne2e-with-name-ft-lora-codellama-7b-bleu}{62.61}
\DefMacro{res-exlong-ablation-conditionne2e-with-name-ft-lora-codellama-7b-bleu-avg}{62.61}
\DefMacro{res-exlong-ablation-conditionne2e-with-name-ft-lora-codellama-7b-bleu-max}{62.61}
\DefMacro{res-exlong-ablation-conditionne2e-with-name-ft-lora-codellama-7b-bleu-min}{62.61}
\DefMacro{res-exlong-ablation-conditionne2e-with-name-ft-lora-codellama-7b-code-bleu}{67.36}
\DefMacro{res-exlong-ablation-conditionne2e-with-name-ft-lora-codellama-7b-code-bleu-avg}{67.36}
\DefMacro{res-exlong-ablation-conditionne2e-with-name-ft-lora-codellama-7b-code-bleu-max}{67.36}
\DefMacro{res-exlong-ablation-conditionne2e-with-name-ft-lora-codellama-7b-code-bleu-min}{67.36}
\DefMacro{res-exlong-ablation-conditionne2e-with-name-ft-lora-codellama-7b-edit-sim}{85.43}
\DefMacro{res-exlong-ablation-conditionne2e-with-name-ft-lora-codellama-7b-edit-sim-avg}{85.43}
\DefMacro{res-exlong-ablation-conditionne2e-with-name-ft-lora-codellama-7b-edit-sim-max}{85.43}
\DefMacro{res-exlong-ablation-conditionne2e-with-name-ft-lora-codellama-7b-edit-sim-min}{85.43}
\DefMacro{res-exlong-ablation-conditionne2e-with-name-ft-lora-codellama-7b-rouge-f}{80.43}
\DefMacro{res-exlong-ablation-conditionne2e-with-name-ft-lora-codellama-7b-rouge-f-avg}{80.43}
\DefMacro{res-exlong-ablation-conditionne2e-with-name-ft-lora-codellama-7b-rouge-f-max}{80.43}
\DefMacro{res-exlong-ablation-conditionne2e-with-name-ft-lora-codellama-7b-rouge-f-min}{80.43}
\DefMacro{res-exlong-ablation-conditionne2e-with-name-ft-lora-codellama-7b-rouge-p}{82.17}
\DefMacro{res-exlong-ablation-conditionne2e-with-name-ft-lora-codellama-7b-rouge-p-avg}{82.17}
\DefMacro{res-exlong-ablation-conditionne2e-with-name-ft-lora-codellama-7b-rouge-p-max}{82.17}
\DefMacro{res-exlong-ablation-conditionne2e-with-name-ft-lora-codellama-7b-rouge-p-min}{82.17}
\DefMacro{res-exlong-ablation-conditionne2e-with-name-ft-lora-codellama-7b-rouge-r}{80.64}
\DefMacro{res-exlong-ablation-conditionne2e-with-name-ft-lora-codellama-7b-rouge-r-avg}{80.64}
\DefMacro{res-exlong-ablation-conditionne2e-with-name-ft-lora-codellama-7b-rouge-r-max}{80.64}
\DefMacro{res-exlong-ablation-conditionne2e-with-name-ft-lora-codellama-7b-rouge-r-min}{80.64}
\DefMacro{res-exlong-ablation-conditionne2e-with-name-ft-lora-codellama-7b-xmatch}{17.74}
\DefMacro{res-exlong-ablation-conditionne2e-with-name-ft-lora-codellama-7b-xmatch-avg}{17.74}
\DefMacro{res-exlong-ablation-conditionne2e-with-name-ft-lora-codellama-7b-xmatch-max}{17.74}
\DefMacro{res-exlong-ablation-conditionne2e-with-name-ft-lora-codellama-7b-xmatch-min}{17.74}
\DefMacro{res-exlong-ablation-conditionne2e-with-name-ft-lora-codellama-7b-xmatch-top1}{17.74}
\DefMacro{res-exlong-ablation-conditionne2e-with-name-ft-lora-codellama-7b-compilable-avg}{81.41}
\DefMacro{res-exlong-ablation-conditionne2e-with-name-ft-lora-codellama-7b-compilable-max}{81.41}
\DefMacro{res-exlong-ablation-conditionne2e-with-name-ft-lora-codellama-7b-compilable-min}{81.41}
\DefMacro{res-exlong-ablation-conditionne2e-with-name-ft-lora-codellama-7b-coverage-avg}{58.14}
\DefMacro{res-exlong-ablation-conditionne2e-with-name-ft-lora-codellama-7b-coverage-max}{58.14}
\DefMacro{res-exlong-ablation-conditionne2e-with-name-ft-lora-codellama-7b-coverage-min}{58.14}
\DefMacro{res-exlong-ablation-conditionne2e-with-name-ft-lora-codellama-7b-match-avg}{100.00}
\DefMacro{res-exlong-ablation-conditionne2e-with-name-ft-lora-codellama-7b-match-max}{100.00}
\DefMacro{res-exlong-ablation-conditionne2e-with-name-ft-lora-codellama-7b-match-min}{100.00}
\DefMacro{res-exlong-ablation-conditionne2e-with-name-ft-lora-codellama-7b-runnable-avg}{82.36}
\DefMacro{res-exlong-ablation-conditionne2e-with-name-ft-lora-codellama-7b-runnable-max}{82.36}
\DefMacro{res-exlong-ablation-conditionne2e-with-name-ft-lora-codellama-7b-runnable-min}{82.36}
\DefMacro{res-exlong-ablation-conditionne2e-with-name-ft-lora-codellama-7b-runnable-overall}{67.05}
\DefMacro{res-exlong-ablation-conditionne2e-with-name-ft-lora-codellama-7b-runnable-overall-avg}{67.05}
\DefMacro{res-exlong-ablation-conditionne2e-with-name-ft-lora-codellama-7b-runnable-overall-max}{67.05}
\DefMacro{res-exlong-ablation-conditionne2e-with-name-ft-lora-codellama-7b-runnable-overall-min}{67.05}
\DefMacro{res-exlong-ablation-conditionne2e-with-name-ft-lora-codellama-7b-timeout-avg}{0.28}
\DefMacro{res-exlong-ablation-conditionne2e-with-name-ft-lora-codellama-7b-timeout-max}{0.28}
\DefMacro{res-exlong-ablation-conditionne2e-with-name-ft-lora-codellama-7b-timeout-min}{0.28}
\DefMacro{res-exlong-ablation-ne2e-with-name-ft-lora-codellama-7b-bleu}{62.53}
\DefMacro{res-exlong-ablation-ne2e-with-name-ft-lora-codellama-7b-bleu-avg}{62.53}
\DefMacro{res-exlong-ablation-ne2e-with-name-ft-lora-codellama-7b-bleu-max}{62.53}
\DefMacro{res-exlong-ablation-ne2e-with-name-ft-lora-codellama-7b-bleu-min}{62.53}
\DefMacro{res-exlong-ablation-ne2e-with-name-ft-lora-codellama-7b-code-bleu}{67.42}
\DefMacro{res-exlong-ablation-ne2e-with-name-ft-lora-codellama-7b-code-bleu-avg}{67.42}
\DefMacro{res-exlong-ablation-ne2e-with-name-ft-lora-codellama-7b-code-bleu-max}{67.42}
\DefMacro{res-exlong-ablation-ne2e-with-name-ft-lora-codellama-7b-code-bleu-min}{67.42}
\DefMacro{res-exlong-ablation-ne2e-with-name-ft-lora-codellama-7b-edit-sim}{84.85}
\DefMacro{res-exlong-ablation-ne2e-with-name-ft-lora-codellama-7b-edit-sim-avg}{84.85}
\DefMacro{res-exlong-ablation-ne2e-with-name-ft-lora-codellama-7b-edit-sim-max}{84.85}
\DefMacro{res-exlong-ablation-ne2e-with-name-ft-lora-codellama-7b-edit-sim-min}{84.85}
\DefMacro{res-exlong-ablation-ne2e-with-name-ft-lora-codellama-7b-rouge-f}{79.68}
\DefMacro{res-exlong-ablation-ne2e-with-name-ft-lora-codellama-7b-rouge-f-avg}{79.68}
\DefMacro{res-exlong-ablation-ne2e-with-name-ft-lora-codellama-7b-rouge-f-max}{79.68}
\DefMacro{res-exlong-ablation-ne2e-with-name-ft-lora-codellama-7b-rouge-f-min}{79.68}
\DefMacro{res-exlong-ablation-ne2e-with-name-ft-lora-codellama-7b-rouge-p}{81.36}
\DefMacro{res-exlong-ablation-ne2e-with-name-ft-lora-codellama-7b-rouge-p-avg}{81.36}
\DefMacro{res-exlong-ablation-ne2e-with-name-ft-lora-codellama-7b-rouge-p-max}{81.36}
\DefMacro{res-exlong-ablation-ne2e-with-name-ft-lora-codellama-7b-rouge-p-min}{81.36}
\DefMacro{res-exlong-ablation-ne2e-with-name-ft-lora-codellama-7b-rouge-r}{80.09}
\DefMacro{res-exlong-ablation-ne2e-with-name-ft-lora-codellama-7b-rouge-r-avg}{80.09}
\DefMacro{res-exlong-ablation-ne2e-with-name-ft-lora-codellama-7b-rouge-r-max}{80.09}
\DefMacro{res-exlong-ablation-ne2e-with-name-ft-lora-codellama-7b-rouge-r-min}{80.09}
\DefMacro{res-exlong-ablation-ne2e-with-name-ft-lora-codellama-7b-xmatch}{17.74}
\DefMacro{res-exlong-ablation-ne2e-with-name-ft-lora-codellama-7b-xmatch-avg}{17.74}
\DefMacro{res-exlong-ablation-ne2e-with-name-ft-lora-codellama-7b-xmatch-max}{17.74}
\DefMacro{res-exlong-ablation-ne2e-with-name-ft-lora-codellama-7b-xmatch-min}{17.74}
\DefMacro{res-exlong-ablation-ne2e-with-name-ft-lora-codellama-7b-xmatch-top1}{17.74}
\DefMacro{res-exlong-ablation-ne2e-with-name-ft-lora-codellama-7b-compilable-avg}{81.41}
\DefMacro{res-exlong-ablation-ne2e-with-name-ft-lora-codellama-7b-compilable-max}{81.41}
\DefMacro{res-exlong-ablation-ne2e-with-name-ft-lora-codellama-7b-compilable-min}{81.41}
\DefMacro{res-exlong-ablation-ne2e-with-name-ft-lora-codellama-7b-coverage-avg}{52.00}
\DefMacro{res-exlong-ablation-ne2e-with-name-ft-lora-codellama-7b-coverage-max}{52.00}
\DefMacro{res-exlong-ablation-ne2e-with-name-ft-lora-codellama-7b-coverage-min}{52.00}
\DefMacro{res-exlong-ablation-ne2e-with-name-ft-lora-codellama-7b-match-avg}{100.00}
\DefMacro{res-exlong-ablation-ne2e-with-name-ft-lora-codellama-7b-match-max}{100.00}
\DefMacro{res-exlong-ablation-ne2e-with-name-ft-lora-codellama-7b-match-min}{100.00}
\DefMacro{res-exlong-ablation-ne2e-with-name-ft-lora-codellama-7b-runnable-avg}{76.13}
\DefMacro{res-exlong-ablation-ne2e-with-name-ft-lora-codellama-7b-runnable-max}{76.13}
\DefMacro{res-exlong-ablation-ne2e-with-name-ft-lora-codellama-7b-runnable-min}{76.13}
\DefMacro{res-exlong-ablation-ne2e-with-name-ft-lora-codellama-7b-runnable-overall}{61.98}
\DefMacro{res-exlong-ablation-ne2e-with-name-ft-lora-codellama-7b-runnable-overall-avg}{61.98}
\DefMacro{res-exlong-ablation-ne2e-with-name-ft-lora-codellama-7b-runnable-overall-max}{61.98}
\DefMacro{res-exlong-ablation-ne2e-with-name-ft-lora-codellama-7b-runnable-overall-min}{61.98}
\DefMacro{res-exlong-ablation-ne2e-with-name-ft-lora-codellama-7b-timeout-avg}{0.38}
\DefMacro{res-exlong-ablation-ne2e-with-name-ft-lora-codellama-7b-timeout-max}{0.38}
\DefMacro{res-exlong-ablation-ne2e-with-name-ft-lora-codellama-7b-timeout-min}{0.38}
\DefMacro{res-exlong-ablation-mut2e-with-name-ft-lora-codellama-7b-bleu}{54.84}
\DefMacro{res-exlong-ablation-mut2e-with-name-ft-lora-codellama-7b-bleu-avg}{54.84}
\DefMacro{res-exlong-ablation-mut2e-with-name-ft-lora-codellama-7b-bleu-max}{54.84}
\DefMacro{res-exlong-ablation-mut2e-with-name-ft-lora-codellama-7b-bleu-min}{54.84}
\DefMacro{res-exlong-ablation-mut2e-with-name-ft-lora-codellama-7b-code-bleu}{60.62}
\DefMacro{res-exlong-ablation-mut2e-with-name-ft-lora-codellama-7b-code-bleu-avg}{60.62}
\DefMacro{res-exlong-ablation-mut2e-with-name-ft-lora-codellama-7b-code-bleu-max}{60.62}
\DefMacro{res-exlong-ablation-mut2e-with-name-ft-lora-codellama-7b-code-bleu-min}{60.62}
\DefMacro{res-exlong-ablation-mut2e-with-name-ft-lora-codellama-7b-edit-sim}{80.70}
\DefMacro{res-exlong-ablation-mut2e-with-name-ft-lora-codellama-7b-edit-sim-avg}{80.70}
\DefMacro{res-exlong-ablation-mut2e-with-name-ft-lora-codellama-7b-edit-sim-max}{80.70}
\DefMacro{res-exlong-ablation-mut2e-with-name-ft-lora-codellama-7b-edit-sim-min}{80.70}
\DefMacro{res-exlong-ablation-mut2e-with-name-ft-lora-codellama-7b-rouge-f}{73.83}
\DefMacro{res-exlong-ablation-mut2e-with-name-ft-lora-codellama-7b-rouge-f-avg}{73.83}
\DefMacro{res-exlong-ablation-mut2e-with-name-ft-lora-codellama-7b-rouge-f-max}{73.83}
\DefMacro{res-exlong-ablation-mut2e-with-name-ft-lora-codellama-7b-rouge-f-min}{73.83}
\DefMacro{res-exlong-ablation-mut2e-with-name-ft-lora-codellama-7b-rouge-p}{76.09}
\DefMacro{res-exlong-ablation-mut2e-with-name-ft-lora-codellama-7b-rouge-p-avg}{76.09}
\DefMacro{res-exlong-ablation-mut2e-with-name-ft-lora-codellama-7b-rouge-p-max}{76.09}
\DefMacro{res-exlong-ablation-mut2e-with-name-ft-lora-codellama-7b-rouge-p-min}{76.09}
\DefMacro{res-exlong-ablation-mut2e-with-name-ft-lora-codellama-7b-rouge-r}{74.03}
\DefMacro{res-exlong-ablation-mut2e-with-name-ft-lora-codellama-7b-rouge-r-avg}{74.03}
\DefMacro{res-exlong-ablation-mut2e-with-name-ft-lora-codellama-7b-rouge-r-max}{74.03}
\DefMacro{res-exlong-ablation-mut2e-with-name-ft-lora-codellama-7b-rouge-r-min}{74.03}
\DefMacro{res-exlong-ablation-mut2e-with-name-ft-lora-codellama-7b-xmatch}{12.06}
\DefMacro{res-exlong-ablation-mut2e-with-name-ft-lora-codellama-7b-xmatch-avg}{12.06}
\DefMacro{res-exlong-ablation-mut2e-with-name-ft-lora-codellama-7b-xmatch-max}{12.06}
\DefMacro{res-exlong-ablation-mut2e-with-name-ft-lora-codellama-7b-xmatch-min}{12.06}
\DefMacro{res-exlong-ablation-mut2e-with-name-ft-lora-codellama-7b-xmatch-top1}{12.06}
\DefMacro{res-exlong-ablation-mut2e-with-name-ft-lora-codellama-7b-compilable-avg}{61.52}
\DefMacro{res-exlong-ablation-mut2e-with-name-ft-lora-codellama-7b-compilable-max}{61.52}
\DefMacro{res-exlong-ablation-mut2e-with-name-ft-lora-codellama-7b-compilable-min}{61.52}
\DefMacro{res-exlong-ablation-mut2e-with-name-ft-lora-codellama-7b-coverage-avg}{38.25}
\DefMacro{res-exlong-ablation-mut2e-with-name-ft-lora-codellama-7b-coverage-max}{38.25}
\DefMacro{res-exlong-ablation-mut2e-with-name-ft-lora-codellama-7b-coverage-min}{38.25}
\DefMacro{res-exlong-ablation-mut2e-with-name-ft-lora-codellama-7b-match-avg}{100.00}
\DefMacro{res-exlong-ablation-mut2e-with-name-ft-lora-codellama-7b-match-max}{100.00}
\DefMacro{res-exlong-ablation-mut2e-with-name-ft-lora-codellama-7b-match-min}{100.00}
\DefMacro{res-exlong-ablation-mut2e-with-name-ft-lora-codellama-7b-runnable-avg}{75.92}
\DefMacro{res-exlong-ablation-mut2e-with-name-ft-lora-codellama-7b-runnable-max}{75.92}
\DefMacro{res-exlong-ablation-mut2e-with-name-ft-lora-codellama-7b-runnable-min}{75.92}
\DefMacro{res-exlong-ablation-mut2e-with-name-ft-lora-codellama-7b-runnable-overall}{46.70}
\DefMacro{res-exlong-ablation-mut2e-with-name-ft-lora-codellama-7b-runnable-overall-avg}{46.70}
\DefMacro{res-exlong-ablation-mut2e-with-name-ft-lora-codellama-7b-runnable-overall-max}{46.70}
\DefMacro{res-exlong-ablation-mut2e-with-name-ft-lora-codellama-7b-runnable-overall-min}{46.70}
\DefMacro{res-exlong-ablation-mut2e-with-name-ft-lora-codellama-7b-timeout-avg}{0.12}
\DefMacro{res-exlong-ablation-mut2e-with-name-ft-lora-codellama-7b-timeout-max}{0.12}
\DefMacro{res-exlong-ablation-mut2e-with-name-ft-lora-codellama-7b-timeout-min}{0.12}

\DefMacro{res-netest-diversity-selected-253-conditionnestack2e-with-name-ft-lora-codellama-7b-bleu}{61.44}
\DefMacro{res-netest-diversity-selected-253-conditionnestack2e-with-name-ft-lora-codellama-7b-bleu-avg}{61.44}
\DefMacro{res-netest-diversity-selected-253-conditionnestack2e-with-name-ft-lora-codellama-7b-bleu-max}{61.44}
\DefMacro{res-netest-diversity-selected-253-conditionnestack2e-with-name-ft-lora-codellama-7b-bleu-min}{61.44}
\DefMacro{res-netest-diversity-selected-253-conditionnestack2e-with-name-ft-lora-codellama-7b-code-bleu}{66.21}
\DefMacro{res-netest-diversity-selected-253-conditionnestack2e-with-name-ft-lora-codellama-7b-code-bleu-avg}{66.21}
\DefMacro{res-netest-diversity-selected-253-conditionnestack2e-with-name-ft-lora-codellama-7b-code-bleu-max}{66.21}
\DefMacro{res-netest-diversity-selected-253-conditionnestack2e-with-name-ft-lora-codellama-7b-code-bleu-min}{66.21}
\DefMacro{res-netest-diversity-selected-253-conditionnestack2e-with-name-ft-lora-codellama-7b-edit-sim}{85.34}
\DefMacro{res-netest-diversity-selected-253-conditionnestack2e-with-name-ft-lora-codellama-7b-edit-sim-avg}{85.34}
\DefMacro{res-netest-diversity-selected-253-conditionnestack2e-with-name-ft-lora-codellama-7b-edit-sim-max}{85.34}
\DefMacro{res-netest-diversity-selected-253-conditionnestack2e-with-name-ft-lora-codellama-7b-edit-sim-min}{85.34}
\DefMacro{res-netest-diversity-selected-253-conditionnestack2e-with-name-ft-lora-codellama-7b-rouge-f}{79.77}
\DefMacro{res-netest-diversity-selected-253-conditionnestack2e-with-name-ft-lora-codellama-7b-rouge-f-avg}{79.77}
\DefMacro{res-netest-diversity-selected-253-conditionnestack2e-with-name-ft-lora-codellama-7b-rouge-f-max}{79.77}
\DefMacro{res-netest-diversity-selected-253-conditionnestack2e-with-name-ft-lora-codellama-7b-rouge-f-min}{79.77}
\DefMacro{res-netest-diversity-selected-253-conditionnestack2e-with-name-ft-lora-codellama-7b-rouge-p}{81.49}
\DefMacro{res-netest-diversity-selected-253-conditionnestack2e-with-name-ft-lora-codellama-7b-rouge-p-avg}{81.49}
\DefMacro{res-netest-diversity-selected-253-conditionnestack2e-with-name-ft-lora-codellama-7b-rouge-p-max}{81.49}
\DefMacro{res-netest-diversity-selected-253-conditionnestack2e-with-name-ft-lora-codellama-7b-rouge-p-min}{81.49}
\DefMacro{res-netest-diversity-selected-253-conditionnestack2e-with-name-ft-lora-codellama-7b-rouge-r}{80.01}
\DefMacro{res-netest-diversity-selected-253-conditionnestack2e-with-name-ft-lora-codellama-7b-rouge-r-avg}{80.01}
\DefMacro{res-netest-diversity-selected-253-conditionnestack2e-with-name-ft-lora-codellama-7b-rouge-r-max}{80.01}
\DefMacro{res-netest-diversity-selected-253-conditionnestack2e-with-name-ft-lora-codellama-7b-rouge-r-min}{80.01}
\DefMacro{res-netest-diversity-selected-253-conditionnestack2e-with-name-ft-lora-codellama-7b-xmatch}{13.44}
\DefMacro{res-netest-diversity-selected-253-conditionnestack2e-with-name-ft-lora-codellama-7b-xmatch-avg}{13.44}
\DefMacro{res-netest-diversity-selected-253-conditionnestack2e-with-name-ft-lora-codellama-7b-xmatch-max}{13.44}
\DefMacro{res-netest-diversity-selected-253-conditionnestack2e-with-name-ft-lora-codellama-7b-xmatch-min}{13.44}
\DefMacro{res-netest-diversity-selected-253-conditionnestack2e-with-name-ft-lora-codellama-7b-xmatch-top1}{13.44}
\DefMacro{res-netest-diversity-selected-253-conditionnestack2e-with-name-ft-lora-codellama-7b-compilable-avg}{83.40}
\DefMacro{res-netest-diversity-selected-253-conditionnestack2e-with-name-ft-lora-codellama-7b-compilable-max}{83.40}
\DefMacro{res-netest-diversity-selected-253-conditionnestack2e-with-name-ft-lora-codellama-7b-compilable-min}{83.40}
\DefMacro{res-netest-diversity-selected-253-conditionnestack2e-with-name-ft-lora-codellama-7b-coverage-avg}{56.79}
\DefMacro{res-netest-diversity-selected-253-conditionnestack2e-with-name-ft-lora-codellama-7b-coverage-max}{56.79}
\DefMacro{res-netest-diversity-selected-253-conditionnestack2e-with-name-ft-lora-codellama-7b-coverage-min}{56.79}
\DefMacro{res-netest-diversity-selected-253-conditionnestack2e-with-name-ft-lora-codellama-7b-match-avg}{100.00}
\DefMacro{res-netest-diversity-selected-253-conditionnestack2e-with-name-ft-lora-codellama-7b-match-max}{100.00}
\DefMacro{res-netest-diversity-selected-253-conditionnestack2e-with-name-ft-lora-codellama-7b-match-min}{100.00}
\DefMacro{res-netest-diversity-selected-253-conditionnestack2e-with-name-ft-lora-codellama-7b-runnable-avg}{79.31}
\DefMacro{res-netest-diversity-selected-253-conditionnestack2e-with-name-ft-lora-codellama-7b-runnable-max}{79.31}
\DefMacro{res-netest-diversity-selected-253-conditionnestack2e-with-name-ft-lora-codellama-7b-runnable-min}{79.31}
\DefMacro{res-netest-diversity-selected-253-conditionnestack2e-with-name-ft-lora-codellama-7b-runnable-overall}{66.14}
\DefMacro{res-netest-diversity-selected-253-conditionnestack2e-with-name-ft-lora-codellama-7b-runnable-overall-avg}{66.14}
\DefMacro{res-netest-diversity-selected-253-conditionnestack2e-with-name-ft-lora-codellama-7b-runnable-overall-max}{66.14}
\DefMacro{res-netest-diversity-selected-253-conditionnestack2e-with-name-ft-lora-codellama-7b-runnable-overall-min}{66.14}
\DefMacro{res-netest-diversity-selected-253-conditionnestack2e-with-name-ft-lora-codellama-7b-timeout-avg}{0.47}
\DefMacro{res-netest-diversity-selected-253-conditionnestack2e-with-name-ft-lora-codellama-7b-timeout-max}{0.47}
\DefMacro{res-netest-diversity-selected-253-conditionnestack2e-with-name-ft-lora-codellama-7b-timeout-min}{0.47}
\DefMacro{res-netest-diversity-conditionnestack2e-sample-with-name-ft-lora-codellama-7b-bleu}{60.91}
\DefMacro{res-netest-diversity-conditionnestack2e-sample-with-name-ft-lora-codellama-7b-bleu-avg}{60.91}
\DefMacro{res-netest-diversity-conditionnestack2e-sample-with-name-ft-lora-codellama-7b-bleu-max}{63.64}
\DefMacro{res-netest-diversity-conditionnestack2e-sample-with-name-ft-lora-codellama-7b-bleu-min}{58.32}
\DefMacro{res-netest-diversity-conditionnestack2e-sample-with-name-ft-lora-codellama-7b-code-bleu}{65.75}
\DefMacro{res-netest-diversity-conditionnestack2e-sample-with-name-ft-lora-codellama-7b-code-bleu-avg}{65.67}
\DefMacro{res-netest-diversity-conditionnestack2e-sample-with-name-ft-lora-codellama-7b-code-bleu-max}{68.28}
\DefMacro{res-netest-diversity-conditionnestack2e-sample-with-name-ft-lora-codellama-7b-code-bleu-min}{63.30}
\DefMacro{res-netest-diversity-conditionnestack2e-sample-with-name-ft-lora-codellama-7b-edit-sim}{84.79}
\DefMacro{res-netest-diversity-conditionnestack2e-sample-with-name-ft-lora-codellama-7b-edit-sim-avg}{84.78}
\DefMacro{res-netest-diversity-conditionnestack2e-sample-with-name-ft-lora-codellama-7b-edit-sim-max}{86.70}
\DefMacro{res-netest-diversity-conditionnestack2e-sample-with-name-ft-lora-codellama-7b-edit-sim-min}{82.82}
\DefMacro{res-netest-diversity-conditionnestack2e-sample-with-name-ft-lora-codellama-7b-rouge-f}{79.06}
\DefMacro{res-netest-diversity-conditionnestack2e-sample-with-name-ft-lora-codellama-7b-rouge-f-avg}{79.17}
\DefMacro{res-netest-diversity-conditionnestack2e-sample-with-name-ft-lora-codellama-7b-rouge-f-max}{81.18}
\DefMacro{res-netest-diversity-conditionnestack2e-sample-with-name-ft-lora-codellama-7b-rouge-f-min}{77.18}
\DefMacro{res-netest-diversity-conditionnestack2e-sample-with-name-ft-lora-codellama-7b-rouge-p}{81.06}
\DefMacro{res-netest-diversity-conditionnestack2e-sample-with-name-ft-lora-codellama-7b-rouge-p-avg}{81.21}
\DefMacro{res-netest-diversity-conditionnestack2e-sample-with-name-ft-lora-codellama-7b-rouge-p-max}{83.67}
\DefMacro{res-netest-diversity-conditionnestack2e-sample-with-name-ft-lora-codellama-7b-rouge-p-min}{78.60}
\DefMacro{res-netest-diversity-conditionnestack2e-sample-with-name-ft-lora-codellama-7b-rouge-r}{79.29}
\DefMacro{res-netest-diversity-conditionnestack2e-sample-with-name-ft-lora-codellama-7b-rouge-r-avg}{79.35}
\DefMacro{res-netest-diversity-conditionnestack2e-sample-with-name-ft-lora-codellama-7b-rouge-r-max}{81.04}
\DefMacro{res-netest-diversity-conditionnestack2e-sample-with-name-ft-lora-codellama-7b-rouge-r-min}{77.75}
\DefMacro{res-netest-diversity-conditionnestack2e-sample-with-name-ft-lora-codellama-7b-xmatch}{13.18}
\DefMacro{res-netest-diversity-conditionnestack2e-sample-with-name-ft-lora-codellama-7b-xmatch-avg}{13.07}
\DefMacro{res-netest-diversity-conditionnestack2e-sample-with-name-ft-lora-codellama-7b-xmatch-max}{15.02}
\DefMacro{res-netest-diversity-conditionnestack2e-sample-with-name-ft-lora-codellama-7b-xmatch-min}{11.59}
\DefMacro{res-netest-diversity-conditionnestack2e-sample-with-name-ft-lora-codellama-7b-xmatch-top1}{13.18}
\DefMacro{res-netest-diversity-conditionnestack2e-sample-with-name-ft-lora-codellama-7b-compilable-avg}{83.19}
\DefMacro{res-netest-diversity-conditionnestack2e-sample-with-name-ft-lora-codellama-7b-compilable-max}{87.35}
\DefMacro{res-netest-diversity-conditionnestack2e-sample-with-name-ft-lora-codellama-7b-compilable-min}{79.18}
\DefMacro{res-netest-diversity-conditionnestack2e-sample-with-name-ft-lora-codellama-7b-coverage-avg}{54.99}
\DefMacro{res-netest-diversity-conditionnestack2e-sample-with-name-ft-lora-codellama-7b-coverage-max}{61.53}
\DefMacro{res-netest-diversity-conditionnestack2e-sample-with-name-ft-lora-codellama-7b-coverage-min}{48.62}
\DefMacro{res-netest-diversity-conditionnestack2e-sample-with-name-ft-lora-codellama-7b-match-avg}{100.00}
\DefMacro{res-netest-diversity-conditionnestack2e-sample-with-name-ft-lora-codellama-7b-match-max}{100.00}
\DefMacro{res-netest-diversity-conditionnestack2e-sample-with-name-ft-lora-codellama-7b-match-min}{100.00}
\DefMacro{res-netest-diversity-conditionnestack2e-sample-with-name-ft-lora-codellama-7b-runnable-avg}{77.34}
\DefMacro{res-netest-diversity-conditionnestack2e-sample-with-name-ft-lora-codellama-7b-runnable-max}{82.96}
\DefMacro{res-netest-diversity-conditionnestack2e-sample-with-name-ft-lora-codellama-7b-runnable-min}{71.35}
\DefMacro{res-netest-diversity-conditionnestack2e-sample-with-name-ft-lora-codellama-7b-runnable-overall}{72.46}
\DefMacro{res-netest-diversity-conditionnestack2e-sample-with-name-ft-lora-codellama-7b-runnable-overall-avg}{64.69}
\DefMacro{res-netest-diversity-conditionnestack2e-sample-with-name-ft-lora-codellama-7b-runnable-overall-max}{72.46}
\DefMacro{res-netest-diversity-conditionnestack2e-sample-with-name-ft-lora-codellama-7b-runnable-overall-min}{56.92}
\DefMacro{res-netest-diversity-conditionnestack2e-sample-with-name-ft-lora-codellama-7b-timeout-avg}{0.45}
\DefMacro{res-netest-diversity-conditionnestack2e-sample-with-name-ft-lora-codellama-7b-timeout-max}{0.45}
\DefMacro{res-netest-diversity-conditionnestack2e-sample-with-name-ft-lora-codellama-7b-timeout-min}{0.45}
\DefMacro{res-netest-diversity-diversity-conditionnestack2e-all-with-name-ft-lora-codellama-7b-bleu}{64.12}
\DefMacro{res-netest-diversity-diversity-conditionnestack2e-all-with-name-ft-lora-codellama-7b-bleu-avg}{63.08}
\DefMacro{res-netest-diversity-diversity-conditionnestack2e-all-with-name-ft-lora-codellama-7b-bleu-max}{71.66}
\DefMacro{res-netest-diversity-diversity-conditionnestack2e-all-with-name-ft-lora-codellama-7b-bleu-min}{55.15}
\DefMacro{res-netest-diversity-diversity-conditionnestack2e-all-with-name-ft-lora-codellama-7b-code-bleu}{68.37}
\DefMacro{res-netest-diversity-diversity-conditionnestack2e-all-with-name-ft-lora-codellama-7b-code-bleu-avg}{67.55}
\DefMacro{res-netest-diversity-diversity-conditionnestack2e-all-with-name-ft-lora-codellama-7b-code-bleu-max}{75.18}
\DefMacro{res-netest-diversity-diversity-conditionnestack2e-all-with-name-ft-lora-codellama-7b-code-bleu-min}{61.01}
\DefMacro{res-netest-diversity-diversity-conditionnestack2e-all-with-name-ft-lora-codellama-7b-edit-sim}{86.15}
\DefMacro{res-netest-diversity-diversity-conditionnestack2e-all-with-name-ft-lora-codellama-7b-edit-sim-avg}{85.86}
\DefMacro{res-netest-diversity-diversity-conditionnestack2e-all-with-name-ft-lora-codellama-7b-edit-sim-max}{90.62}
\DefMacro{res-netest-diversity-diversity-conditionnestack2e-all-with-name-ft-lora-codellama-7b-edit-sim-min}{80.93}
\DefMacro{res-netest-diversity-diversity-conditionnestack2e-all-with-name-ft-lora-codellama-7b-rouge-f}{81.71}
\DefMacro{res-netest-diversity-diversity-conditionnestack2e-all-with-name-ft-lora-codellama-7b-rouge-f-avg}{80.91}
\DefMacro{res-netest-diversity-diversity-conditionnestack2e-all-with-name-ft-lora-codellama-7b-rouge-f-max}{86.17}
\DefMacro{res-netest-diversity-diversity-conditionnestack2e-all-with-name-ft-lora-codellama-7b-rouge-f-min}{75.49}
\DefMacro{res-netest-diversity-diversity-conditionnestack2e-all-with-name-ft-lora-codellama-7b-rouge-p}{83.57}
\DefMacro{res-netest-diversity-diversity-conditionnestack2e-all-with-name-ft-lora-codellama-7b-rouge-p-avg}{83.02}
\DefMacro{res-netest-diversity-diversity-conditionnestack2e-all-with-name-ft-lora-codellama-7b-rouge-p-max}{89.01}
\DefMacro{res-netest-diversity-diversity-conditionnestack2e-all-with-name-ft-lora-codellama-7b-rouge-p-min}{76.08}
\DefMacro{res-netest-diversity-diversity-conditionnestack2e-all-with-name-ft-lora-codellama-7b-rouge-r}{81.52}
\DefMacro{res-netest-diversity-diversity-conditionnestack2e-all-with-name-ft-lora-codellama-7b-rouge-r-avg}{80.71}
\DefMacro{res-netest-diversity-diversity-conditionnestack2e-all-with-name-ft-lora-codellama-7b-rouge-r-max}{85.63}
\DefMacro{res-netest-diversity-diversity-conditionnestack2e-all-with-name-ft-lora-codellama-7b-rouge-r-min}{75.99}
\DefMacro{res-netest-diversity-diversity-conditionnestack2e-all-with-name-ft-lora-codellama-7b-xmatch}{17.79}
\DefMacro{res-netest-diversity-diversity-conditionnestack2e-all-with-name-ft-lora-codellama-7b-xmatch-avg}{15.78}
\DefMacro{res-netest-diversity-diversity-conditionnestack2e-all-with-name-ft-lora-codellama-7b-xmatch-max}{24.51}
\DefMacro{res-netest-diversity-diversity-conditionnestack2e-all-with-name-ft-lora-codellama-7b-xmatch-min}{10.67}
\DefMacro{res-netest-diversity-diversity-conditionnestack2e-all-with-name-ft-lora-codellama-7b-xmatch-top1}{17.79}
\DefMacro{res-netest-diversity-diversity-conditionnestack2e-all-with-name-ft-lora-codellama-7b-compilable-avg}{85.89}
\DefMacro{res-netest-diversity-diversity-conditionnestack2e-all-with-name-ft-lora-codellama-7b-compilable-max}{95.65}
\DefMacro{res-netest-diversity-diversity-conditionnestack2e-all-with-name-ft-lora-codellama-7b-compilable-min}{74.70}
\DefMacro{res-netest-diversity-diversity-conditionnestack2e-all-with-name-ft-lora-codellama-7b-coverage-avg}{57.29}
\DefMacro{res-netest-diversity-diversity-conditionnestack2e-all-with-name-ft-lora-codellama-7b-coverage-max}{71.94}
\DefMacro{res-netest-diversity-diversity-conditionnestack2e-all-with-name-ft-lora-codellama-7b-coverage-min}{42.29}
\DefMacro{res-netest-diversity-diversity-conditionnestack2e-all-with-name-ft-lora-codellama-7b-match-avg}{99.79}
\DefMacro{res-netest-diversity-diversity-conditionnestack2e-all-with-name-ft-lora-codellama-7b-match-max}{100.00}
\DefMacro{res-netest-diversity-diversity-conditionnestack2e-all-with-name-ft-lora-codellama-7b-match-min}{99.59}
\DefMacro{res-netest-diversity-diversity-conditionnestack2e-all-with-name-ft-lora-codellama-7b-runnable-avg}{75.10}
\DefMacro{res-netest-diversity-diversity-conditionnestack2e-all-with-name-ft-lora-codellama-7b-runnable-max}{86.78}
\DefMacro{res-netest-diversity-diversity-conditionnestack2e-all-with-name-ft-lora-codellama-7b-runnable-min}{62.81}
\DefMacro{res-netest-diversity-diversity-conditionnestack2e-all-with-name-ft-lora-codellama-7b-runnable-overall}{83.00}
\DefMacro{res-netest-diversity-diversity-conditionnestack2e-all-with-name-ft-lora-codellama-7b-runnable-overall-avg}{65.63}
\DefMacro{res-netest-diversity-diversity-conditionnestack2e-all-with-name-ft-lora-codellama-7b-runnable-overall-max}{83.00}
\DefMacro{res-netest-diversity-diversity-conditionnestack2e-all-with-name-ft-lora-codellama-7b-runnable-overall-min}{49.01}
\DefMacro{res-netest-diversity-diversity-conditionnestack2e-all-with-name-ft-lora-codellama-7b-timeout-avg}{0.41}
\DefMacro{res-netest-diversity-diversity-conditionnestack2e-all-with-name-ft-lora-codellama-7b-timeout-max}{0.41}
\DefMacro{res-netest-diversity-diversity-conditionnestack2e-all-with-name-ft-lora-codellama-7b-timeout-min}{0.41}

\DefMacro{res-model-size-mut2e-with-name-ft-lora-codellama-13b-bleu}{57.23}
\DefMacro{res-model-size-mut2e-with-name-ft-lora-codellama-13b-bleu-avg}{57.23}
\DefMacro{res-model-size-mut2e-with-name-ft-lora-codellama-13b-bleu-max}{57.23}
\DefMacro{res-model-size-mut2e-with-name-ft-lora-codellama-13b-bleu-min}{57.23}
\DefMacro{res-model-size-mut2e-with-name-ft-lora-codellama-13b-code-bleu}{61.99}
\DefMacro{res-model-size-mut2e-with-name-ft-lora-codellama-13b-code-bleu-avg}{61.99}
\DefMacro{res-model-size-mut2e-with-name-ft-lora-codellama-13b-code-bleu-max}{61.99}
\DefMacro{res-model-size-mut2e-with-name-ft-lora-codellama-13b-code-bleu-min}{61.99}
\DefMacro{res-model-size-mut2e-with-name-ft-lora-codellama-13b-edit-sim}{82.61}
\DefMacro{res-model-size-mut2e-with-name-ft-lora-codellama-13b-edit-sim-avg}{82.61}
\DefMacro{res-model-size-mut2e-with-name-ft-lora-codellama-13b-edit-sim-max}{82.61}
\DefMacro{res-model-size-mut2e-with-name-ft-lora-codellama-13b-edit-sim-min}{82.61}
\DefMacro{res-model-size-mut2e-with-name-ft-lora-codellama-13b-rouge-f}{75.50}
\DefMacro{res-model-size-mut2e-with-name-ft-lora-codellama-13b-rouge-f-avg}{75.50}
\DefMacro{res-model-size-mut2e-with-name-ft-lora-codellama-13b-rouge-f-max}{75.50}
\DefMacro{res-model-size-mut2e-with-name-ft-lora-codellama-13b-rouge-f-min}{75.50}
\DefMacro{res-model-size-mut2e-with-name-ft-lora-codellama-13b-rouge-p}{78.37}
\DefMacro{res-model-size-mut2e-with-name-ft-lora-codellama-13b-rouge-p-avg}{78.37}
\DefMacro{res-model-size-mut2e-with-name-ft-lora-codellama-13b-rouge-p-max}{78.37}
\DefMacro{res-model-size-mut2e-with-name-ft-lora-codellama-13b-rouge-p-min}{78.37}
\DefMacro{res-model-size-mut2e-with-name-ft-lora-codellama-13b-rouge-r}{74.65}
\DefMacro{res-model-size-mut2e-with-name-ft-lora-codellama-13b-rouge-r-avg}{74.65}
\DefMacro{res-model-size-mut2e-with-name-ft-lora-codellama-13b-rouge-r-max}{74.65}
\DefMacro{res-model-size-mut2e-with-name-ft-lora-codellama-13b-rouge-r-min}{74.65}
\DefMacro{res-model-size-mut2e-with-name-ft-lora-codellama-13b-xmatch}{12.78}
\DefMacro{res-model-size-mut2e-with-name-ft-lora-codellama-13b-xmatch-avg}{12.78}
\DefMacro{res-model-size-mut2e-with-name-ft-lora-codellama-13b-xmatch-max}{12.78}
\DefMacro{res-model-size-mut2e-with-name-ft-lora-codellama-13b-xmatch-min}{12.78}
\DefMacro{res-model-size-mut2e-with-name-ft-lora-codellama-13b-xmatch-top1}{12.78}
\DefMacro{res-model-size-mut2e-with-name-ft-lora-codellama-13b-compilable-avg}{61.23}
\DefMacro{res-model-size-mut2e-with-name-ft-lora-codellama-13b-compilable-max}{61.23}
\DefMacro{res-model-size-mut2e-with-name-ft-lora-codellama-13b-compilable-min}{61.23}
\DefMacro{res-model-size-mut2e-with-name-ft-lora-codellama-13b-coverage-avg}{41.92}
\DefMacro{res-model-size-mut2e-with-name-ft-lora-codellama-13b-coverage-max}{41.92}
\DefMacro{res-model-size-mut2e-with-name-ft-lora-codellama-13b-coverage-min}{41.92}
\DefMacro{res-model-size-mut2e-with-name-ft-lora-codellama-13b-match-avg}{100.00}
\DefMacro{res-model-size-mut2e-with-name-ft-lora-codellama-13b-match-max}{100.00}
\DefMacro{res-model-size-mut2e-with-name-ft-lora-codellama-13b-match-min}{100.00}
\DefMacro{res-model-size-mut2e-with-name-ft-lora-codellama-13b-runnable-avg}{83.66}
\DefMacro{res-model-size-mut2e-with-name-ft-lora-codellama-13b-runnable-max}{83.66}
\DefMacro{res-model-size-mut2e-with-name-ft-lora-codellama-13b-runnable-min}{83.66}
\DefMacro{res-model-size-mut2e-with-name-ft-lora-codellama-13b-runnable-overall}{51.23}
\DefMacro{res-model-size-mut2e-with-name-ft-lora-codellama-13b-runnable-overall-avg}{51.23}
\DefMacro{res-model-size-mut2e-with-name-ft-lora-codellama-13b-runnable-overall-max}{51.23}
\DefMacro{res-model-size-mut2e-with-name-ft-lora-codellama-13b-runnable-overall-min}{51.23}
\DefMacro{res-model-size-mut2e-with-name-ft-lora-codellama-13b-timeout-avg}{0.13}
\DefMacro{res-model-size-mut2e-with-name-ft-lora-codellama-13b-timeout-max}{0.13}
\DefMacro{res-model-size-mut2e-with-name-ft-lora-codellama-13b-timeout-min}{0.13}
\DefMacro{res-model-size-conditionnestack2e-with-name-ft-lora-codellama-13b-bleu}{63.68}
\DefMacro{res-model-size-conditionnestack2e-with-name-ft-lora-codellama-13b-bleu-avg}{63.68}
\DefMacro{res-model-size-conditionnestack2e-with-name-ft-lora-codellama-13b-bleu-max}{63.68}
\DefMacro{res-model-size-conditionnestack2e-with-name-ft-lora-codellama-13b-bleu-min}{63.68}
\DefMacro{res-model-size-conditionnestack2e-with-name-ft-lora-codellama-13b-code-bleu}{67.57}
\DefMacro{res-model-size-conditionnestack2e-with-name-ft-lora-codellama-13b-code-bleu-avg}{67.57}
\DefMacro{res-model-size-conditionnestack2e-with-name-ft-lora-codellama-13b-code-bleu-max}{67.57}
\DefMacro{res-model-size-conditionnestack2e-with-name-ft-lora-codellama-13b-code-bleu-min}{67.57}
\DefMacro{res-model-size-conditionnestack2e-with-name-ft-lora-codellama-13b-edit-sim}{86.07}
\DefMacro{res-model-size-conditionnestack2e-with-name-ft-lora-codellama-13b-edit-sim-avg}{86.07}
\DefMacro{res-model-size-conditionnestack2e-with-name-ft-lora-codellama-13b-edit-sim-max}{86.07}
\DefMacro{res-model-size-conditionnestack2e-with-name-ft-lora-codellama-13b-edit-sim-min}{86.07}
\DefMacro{res-model-size-conditionnestack2e-with-name-ft-lora-codellama-13b-rouge-f}{81.07}
\DefMacro{res-model-size-conditionnestack2e-with-name-ft-lora-codellama-13b-rouge-f-avg}{81.07}
\DefMacro{res-model-size-conditionnestack2e-with-name-ft-lora-codellama-13b-rouge-f-max}{81.07}
\DefMacro{res-model-size-conditionnestack2e-with-name-ft-lora-codellama-13b-rouge-f-min}{81.07}
\DefMacro{res-model-size-conditionnestack2e-with-name-ft-lora-codellama-13b-rouge-p}{83.74}
\DefMacro{res-model-size-conditionnestack2e-with-name-ft-lora-codellama-13b-rouge-p-avg}{83.74}
\DefMacro{res-model-size-conditionnestack2e-with-name-ft-lora-codellama-13b-rouge-p-max}{83.74}
\DefMacro{res-model-size-conditionnestack2e-with-name-ft-lora-codellama-13b-rouge-p-min}{83.74}
\DefMacro{res-model-size-conditionnestack2e-with-name-ft-lora-codellama-13b-rouge-r}{80.35}
\DefMacro{res-model-size-conditionnestack2e-with-name-ft-lora-codellama-13b-rouge-r-avg}{80.35}
\DefMacro{res-model-size-conditionnestack2e-with-name-ft-lora-codellama-13b-rouge-r-max}{80.35}
\DefMacro{res-model-size-conditionnestack2e-with-name-ft-lora-codellama-13b-rouge-r-min}{80.35}
\DefMacro{res-model-size-conditionnestack2e-with-name-ft-lora-codellama-13b-xmatch}{18.94}
\DefMacro{res-model-size-conditionnestack2e-with-name-ft-lora-codellama-13b-xmatch-avg}{18.94}
\DefMacro{res-model-size-conditionnestack2e-with-name-ft-lora-codellama-13b-xmatch-max}{18.94}
\DefMacro{res-model-size-conditionnestack2e-with-name-ft-lora-codellama-13b-xmatch-min}{18.94}
\DefMacro{res-model-size-conditionnestack2e-with-name-ft-lora-codellama-13b-xmatch-top1}{18.94}
\DefMacro{res-model-size-conditionnestack2e-with-name-ft-lora-codellama-13b-compilable-avg}{82.72}
\DefMacro{res-model-size-conditionnestack2e-with-name-ft-lora-codellama-13b-compilable-max}{82.72}
\DefMacro{res-model-size-conditionnestack2e-with-name-ft-lora-codellama-13b-compilable-min}{82.72}
\DefMacro{res-model-size-conditionnestack2e-with-name-ft-lora-codellama-13b-coverage-avg}{61.90}
\DefMacro{res-model-size-conditionnestack2e-with-name-ft-lora-codellama-13b-coverage-max}{61.90}
\DefMacro{res-model-size-conditionnestack2e-with-name-ft-lora-codellama-13b-coverage-min}{61.90}
\DefMacro{res-model-size-conditionnestack2e-with-name-ft-lora-codellama-13b-match-avg}{100.00}
\DefMacro{res-model-size-conditionnestack2e-with-name-ft-lora-codellama-13b-match-max}{100.00}
\DefMacro{res-model-size-conditionnestack2e-with-name-ft-lora-codellama-13b-match-min}{100.00}
\DefMacro{res-model-size-conditionnestack2e-with-name-ft-lora-codellama-13b-runnable-avg}{84.95}
\DefMacro{res-model-size-conditionnestack2e-with-name-ft-lora-codellama-13b-runnable-max}{84.95}
\DefMacro{res-model-size-conditionnestack2e-with-name-ft-lora-codellama-13b-runnable-min}{84.95}
\DefMacro{res-model-size-conditionnestack2e-with-name-ft-lora-codellama-13b-runnable-overall}{70.28}
\DefMacro{res-model-size-conditionnestack2e-with-name-ft-lora-codellama-13b-runnable-overall-avg}{70.28}
\DefMacro{res-model-size-conditionnestack2e-with-name-ft-lora-codellama-13b-runnable-overall-max}{70.28}
\DefMacro{res-model-size-conditionnestack2e-with-name-ft-lora-codellama-13b-runnable-overall-min}{70.28}
\DefMacro{res-model-size-conditionnestack2e-with-name-ft-lora-codellama-13b-timeout-avg}{0.19}
\DefMacro{res-model-size-conditionnestack2e-with-name-ft-lora-codellama-13b-timeout-max}{0.19}
\DefMacro{res-model-size-conditionnestack2e-with-name-ft-lora-codellama-13b-timeout-min}{0.19}

\DefMacro{res-gpt4o-with-name-ne2e-few-shot-with-name-gpt-4o-bleu}{60.07}
\DefMacro{res-gpt4o-with-name-ne2e-few-shot-with-name-gpt-4o-bleu-avg}{60.07}
\DefMacro{res-gpt4o-with-name-ne2e-few-shot-with-name-gpt-4o-bleu-max}{60.07}
\DefMacro{res-gpt4o-with-name-ne2e-few-shot-with-name-gpt-4o-bleu-min}{60.07}
\DefMacro{res-gpt4o-with-name-ne2e-few-shot-with-name-gpt-4o-code-bleu}{65.56}
\DefMacro{res-gpt4o-with-name-ne2e-few-shot-with-name-gpt-4o-code-bleu-avg}{65.56}
\DefMacro{res-gpt4o-with-name-ne2e-few-shot-with-name-gpt-4o-code-bleu-max}{65.56}
\DefMacro{res-gpt4o-with-name-ne2e-few-shot-with-name-gpt-4o-code-bleu-min}{65.56}
\DefMacro{res-gpt4o-with-name-ne2e-few-shot-with-name-gpt-4o-edit-sim}{84.00}
\DefMacro{res-gpt4o-with-name-ne2e-few-shot-with-name-gpt-4o-edit-sim-avg}{84.00}
\DefMacro{res-gpt4o-with-name-ne2e-few-shot-with-name-gpt-4o-edit-sim-max}{84.00}
\DefMacro{res-gpt4o-with-name-ne2e-few-shot-with-name-gpt-4o-edit-sim-min}{84.00}
\DefMacro{res-gpt4o-with-name-ne2e-few-shot-with-name-gpt-4o-rouge-f}{78.30}
\DefMacro{res-gpt4o-with-name-ne2e-few-shot-with-name-gpt-4o-rouge-f-avg}{78.30}
\DefMacro{res-gpt4o-with-name-ne2e-few-shot-with-name-gpt-4o-rouge-f-max}{78.30}
\DefMacro{res-gpt4o-with-name-ne2e-few-shot-with-name-gpt-4o-rouge-f-min}{78.30}
\DefMacro{res-gpt4o-with-name-ne2e-few-shot-with-name-gpt-4o-rouge-p}{80.36}
\DefMacro{res-gpt4o-with-name-ne2e-few-shot-with-name-gpt-4o-rouge-p-avg}{80.36}
\DefMacro{res-gpt4o-with-name-ne2e-few-shot-with-name-gpt-4o-rouge-p-max}{80.36}
\DefMacro{res-gpt4o-with-name-ne2e-few-shot-with-name-gpt-4o-rouge-p-min}{80.36}
\DefMacro{res-gpt4o-with-name-ne2e-few-shot-with-name-gpt-4o-rouge-r}{78.60}
\DefMacro{res-gpt4o-with-name-ne2e-few-shot-with-name-gpt-4o-rouge-r-avg}{78.60}
\DefMacro{res-gpt4o-with-name-ne2e-few-shot-with-name-gpt-4o-rouge-r-max}{78.60}
\DefMacro{res-gpt4o-with-name-ne2e-few-shot-with-name-gpt-4o-rouge-r-min}{78.60}
\DefMacro{res-gpt4o-with-name-ne2e-few-shot-with-name-gpt-4o-xmatch}{16.82}
\DefMacro{res-gpt4o-with-name-ne2e-few-shot-with-name-gpt-4o-xmatch-avg}{16.82}
\DefMacro{res-gpt4o-with-name-ne2e-few-shot-with-name-gpt-4o-xmatch-max}{16.82}
\DefMacro{res-gpt4o-with-name-ne2e-few-shot-with-name-gpt-4o-xmatch-min}{16.82}
\DefMacro{res-gpt4o-with-name-ne2e-few-shot-with-name-gpt-4o-xmatch-top1}{16.82}
\DefMacro{res-gpt4o-with-name-ne2e-few-shot-with-name-gpt-4o-compilable-avg}{81.87}
\DefMacro{res-gpt4o-with-name-ne2e-few-shot-with-name-gpt-4o-compilable-max}{81.87}
\DefMacro{res-gpt4o-with-name-ne2e-few-shot-with-name-gpt-4o-compilable-min}{81.87}
\DefMacro{res-gpt4o-with-name-ne2e-few-shot-with-name-gpt-4o-coverage-avg}{55.53}
\DefMacro{res-gpt4o-with-name-ne2e-few-shot-with-name-gpt-4o-coverage-max}{55.53}
\DefMacro{res-gpt4o-with-name-ne2e-few-shot-with-name-gpt-4o-coverage-min}{55.53}
\DefMacro{res-gpt4o-with-name-ne2e-few-shot-with-name-gpt-4o-match-avg}{99.73}
\DefMacro{res-gpt4o-with-name-ne2e-few-shot-with-name-gpt-4o-match-max}{99.73}
\DefMacro{res-gpt4o-with-name-ne2e-few-shot-with-name-gpt-4o-match-min}{99.73}
\DefMacro{res-gpt4o-with-name-ne2e-few-shot-with-name-gpt-4o-runnable-avg}{87.60}
\DefMacro{res-gpt4o-with-name-ne2e-few-shot-with-name-gpt-4o-runnable-max}{87.60}
\DefMacro{res-gpt4o-with-name-ne2e-few-shot-with-name-gpt-4o-runnable-min}{87.60}
\DefMacro{res-gpt4o-with-name-ne2e-few-shot-with-name-gpt-4o-runnable-overall}{71.52}
\DefMacro{res-gpt4o-with-name-ne2e-few-shot-with-name-gpt-4o-runnable-overall-avg}{71.52}
\DefMacro{res-gpt4o-with-name-ne2e-few-shot-with-name-gpt-4o-runnable-overall-max}{71.52}
\DefMacro{res-gpt4o-with-name-ne2e-few-shot-with-name-gpt-4o-runnable-overall-min}{71.52}
\DefMacro{res-gpt4o-with-name-ne2e-few-shot-with-name-gpt-4o-timeout-avg}{0.28}
\DefMacro{res-gpt4o-with-name-ne2e-few-shot-with-name-gpt-4o-timeout-max}{0.28}
\DefMacro{res-gpt4o-with-name-ne2e-few-shot-with-name-gpt-4o-timeout-min}{0.28}
\DefMacro{res-gpt4o-with-name-conditionnestack2e-few-shot-with-name-gpt-4o-bleu}{60.48}
\DefMacro{res-gpt4o-with-name-conditionnestack2e-few-shot-with-name-gpt-4o-bleu-avg}{60.48}
\DefMacro{res-gpt4o-with-name-conditionnestack2e-few-shot-with-name-gpt-4o-bleu-max}{60.48}
\DefMacro{res-gpt4o-with-name-conditionnestack2e-few-shot-with-name-gpt-4o-bleu-min}{60.48}
\DefMacro{res-gpt4o-with-name-conditionnestack2e-few-shot-with-name-gpt-4o-code-bleu}{66.77}
\DefMacro{res-gpt4o-with-name-conditionnestack2e-few-shot-with-name-gpt-4o-code-bleu-avg}{66.77}
\DefMacro{res-gpt4o-with-name-conditionnestack2e-few-shot-with-name-gpt-4o-code-bleu-max}{66.77}
\DefMacro{res-gpt4o-with-name-conditionnestack2e-few-shot-with-name-gpt-4o-code-bleu-min}{66.77}
\DefMacro{res-gpt4o-with-name-conditionnestack2e-few-shot-with-name-gpt-4o-edit-sim}{84.77}
\DefMacro{res-gpt4o-with-name-conditionnestack2e-few-shot-with-name-gpt-4o-edit-sim-avg}{84.77}
\DefMacro{res-gpt4o-with-name-conditionnestack2e-few-shot-with-name-gpt-4o-edit-sim-max}{84.77}
\DefMacro{res-gpt4o-with-name-conditionnestack2e-few-shot-with-name-gpt-4o-edit-sim-min}{84.77}
\DefMacro{res-gpt4o-with-name-conditionnestack2e-few-shot-with-name-gpt-4o-rouge-f}{79.24}
\DefMacro{res-gpt4o-with-name-conditionnestack2e-few-shot-with-name-gpt-4o-rouge-f-avg}{79.24}
\DefMacro{res-gpt4o-with-name-conditionnestack2e-few-shot-with-name-gpt-4o-rouge-f-max}{79.24}
\DefMacro{res-gpt4o-with-name-conditionnestack2e-few-shot-with-name-gpt-4o-rouge-f-min}{79.24}
\DefMacro{res-gpt4o-with-name-conditionnestack2e-few-shot-with-name-gpt-4o-rouge-p}{80.33}
\DefMacro{res-gpt4o-with-name-conditionnestack2e-few-shot-with-name-gpt-4o-rouge-p-avg}{80.33}
\DefMacro{res-gpt4o-with-name-conditionnestack2e-few-shot-with-name-gpt-4o-rouge-p-max}{80.33}
\DefMacro{res-gpt4o-with-name-conditionnestack2e-few-shot-with-name-gpt-4o-rouge-p-min}{80.33}
\DefMacro{res-gpt4o-with-name-conditionnestack2e-few-shot-with-name-gpt-4o-rouge-r}{80.30}
\DefMacro{res-gpt4o-with-name-conditionnestack2e-few-shot-with-name-gpt-4o-rouge-r-avg}{80.30}
\DefMacro{res-gpt4o-with-name-conditionnestack2e-few-shot-with-name-gpt-4o-rouge-r-max}{80.30}
\DefMacro{res-gpt4o-with-name-conditionnestack2e-few-shot-with-name-gpt-4o-rouge-r-min}{80.30}
\DefMacro{res-gpt4o-with-name-conditionnestack2e-few-shot-with-name-gpt-4o-xmatch}{17.74}
\DefMacro{res-gpt4o-with-name-conditionnestack2e-few-shot-with-name-gpt-4o-xmatch-avg}{17.74}
\DefMacro{res-gpt4o-with-name-conditionnestack2e-few-shot-with-name-gpt-4o-xmatch-max}{17.74}
\DefMacro{res-gpt4o-with-name-conditionnestack2e-few-shot-with-name-gpt-4o-xmatch-min}{17.74}
\DefMacro{res-gpt4o-with-name-conditionnestack2e-few-shot-with-name-gpt-4o-xmatch-top1}{17.74}
\DefMacro{res-gpt4o-with-name-conditionnestack2e-few-shot-with-name-gpt-4o-compilable-avg}{82.49}
\DefMacro{res-gpt4o-with-name-conditionnestack2e-few-shot-with-name-gpt-4o-compilable-max}{82.49}
\DefMacro{res-gpt4o-with-name-conditionnestack2e-few-shot-with-name-gpt-4o-compilable-min}{82.49}
\DefMacro{res-gpt4o-with-name-conditionnestack2e-few-shot-with-name-gpt-4o-coverage-avg}{64.75}
\DefMacro{res-gpt4o-with-name-conditionnestack2e-few-shot-with-name-gpt-4o-coverage-max}{64.75}
\DefMacro{res-gpt4o-with-name-conditionnestack2e-few-shot-with-name-gpt-4o-coverage-min}{64.75}
\DefMacro{res-gpt4o-with-name-conditionnestack2e-few-shot-with-name-gpt-4o-match-avg}{100.00}
\DefMacro{res-gpt4o-with-name-conditionnestack2e-few-shot-with-name-gpt-4o-match-max}{100.00}
\DefMacro{res-gpt4o-with-name-conditionnestack2e-few-shot-with-name-gpt-4o-match-min}{100.00}
\DefMacro{res-gpt4o-with-name-conditionnestack2e-few-shot-with-name-gpt-4o-runnable-avg}{91.34}
\DefMacro{res-gpt4o-with-name-conditionnestack2e-few-shot-with-name-gpt-4o-runnable-max}{91.34}
\DefMacro{res-gpt4o-with-name-conditionnestack2e-few-shot-with-name-gpt-4o-runnable-min}{91.34}
\DefMacro{res-gpt4o-with-name-conditionnestack2e-few-shot-with-name-gpt-4o-runnable-overall}{75.35}
\DefMacro{res-gpt4o-with-name-conditionnestack2e-few-shot-with-name-gpt-4o-runnable-overall-avg}{75.35}
\DefMacro{res-gpt4o-with-name-conditionnestack2e-few-shot-with-name-gpt-4o-runnable-overall-max}{75.35}
\DefMacro{res-gpt4o-with-name-conditionnestack2e-few-shot-with-name-gpt-4o-runnable-overall-min}{75.35}
\DefMacro{res-gpt4o-with-name-conditionnestack2e-few-shot-with-name-gpt-4o-timeout-avg}{0.00}
\DefMacro{res-gpt4o-with-name-conditionnestack2e-few-shot-with-name-gpt-4o-timeout-max}{0.00}
\DefMacro{res-gpt4o-with-name-conditionnestack2e-few-shot-with-name-gpt-4o-timeout-min}{0.00}

\newcommand{\NumOfGoodEBTs}{187}
\newcommand{\NumOfMergedTests}{23}

\newcommand{\NumOfPRprojects}{9}
\newcommand{\NumSubmtEBTS}{35}
\newcommand{\NumOfNPRprojects}{2}  %
\newcommand{\NumOfPRs}{7}
\newcommand{\NumOfAcPRs}{4}
\newcommand{\NumPendPRs}{3}
\newcommand{\NumPendEBTs}{12}
\newcommand{\TotalCollectedTests}{111,230}
\newcommand{\TotalCollectedETests}{12,574}
\newcommand{\TotalProjects}{562}
\newcommand{\overlapEvalProjects}{27}
\newcommand{\allEvalProjects}{30}

\newcommand{\improvedPctCoverageThanCAT}{83.8\%}

\newcommand{\algoCollectNodes}{CollectNodes\xspace}

\newcommand{\algoMerge}{Merge\xspace}

\newcommand{\algoCollectStackTraceCorpus}{CollectStackTraceSet\xspace}
\newcommand{\algoAssemblePrompt}{AssemblePrompt\xspace}

\newcommand{\exampleRepo}{\texttt{greenmail-mail-test/greenmail}\xspace}
\newcommand{\TSCoverage}{ThrowCov\%\xspace}
\newcommand{\runnable}{Runnable\%\xspace}
\newcommand{\compile}{Compilable\%\xspace}
\newcommand{\SampleSize}{5}
\newcommand{\ToolMoreNebt}{\Tool-sample\xspace}

\newcommand{\OverGPTRun}{9.9\%}
\newcommand{\OverGPTCov}{22.8\%}
\newcommand{\OverCATRun}{83.8\%}
\newcommand{\OverCATCov}{97.5\%}
\newcommand{\OverGPTFour}{16.6\%}

\DefMacro{TCap-models-results}{Models Results. MUT+FILE2E: model inputs: MUT, etype, etest\_name, etest\_file; model output: etest. NE2E: model inputs: MUT, etype, etest\_name, netest, etest\_file; model output: etest. NESTACK2E: model inputs: MUT, etype, etest\_name, netest, stack\_trace, etest\_file; model output: etest.}

\DefMacro{TCap-models-with-name-results}{Model Results providing the \etest name in the prompt.}
\DefMacro{TCap-models-no-name-results}{Model Results not providing the \etest name in the prompt.}
\DefMacro{TCap-models-ablations}{Models Ablations. Comparison between the same model with stack traces in the input with not providing.}
\DefMacro{TCap-mut2e-dataset-split}{Dataset Split.}
\DefMacro{TCap-models-gpt}{GPT Models results.}
\DefMacro{TCap-models-gt-stacktrace}{Models' results with ground-truth stack trace}
\DefMacro{TCap-models-gt-stacktrace-with-name}{Models results with ground-truth stack trace and providing the \etest name in the prompt.}
\DefMacro{TCap-models-gt-stacktrace-no-name}{Models results with ground-truth stack trace and not providing the \etest name in the prompt.}
\DefMacro{TCap-models-ts-no-name-results}{RQ1 models' results without \etest name.}
\DefMacro{TCap-models-ts-coverage-with-name-results}{RQ1 models' throw statement coverage with \etest name. Pct. of generated runnable \etests correctly cover the target throw statement.}
\DefMacro{TCap-models-ts-coverage-no-name-results}{RQ1 models' throw statement coverage without \etest name.}
\DefMacro{TCap-models-ts-with-name-results}{RQ1 models' results with \etest name.}
\DefMacro{TCap-models-test-coverage}{Tools' coverage compared to data used in RQ1}
\DefMacro{TCap-models-rq2-results}{Models's results on RQ2}

\begin{document}

\title{\Title}

\makeatletter
\newcommand{\linebreakand}{%
\end{@IEEEauthorhalign}
\hfill\mbox{}\par
\mbox{}\hfill\begin{@IEEEauthorhalign}
}
\makeatother
\author{
    \IEEEauthorblockN{Jiyang Zhang}
\IEEEauthorblockA{\textit{The University of Texas at Austin, USA} \\
jiyang.zhang@utexas.edu}
\and
\IEEEauthorblockN{Yu Liu}
\IEEEauthorblockA{\textit{The University of Texas at Austin, USA}\\
yuki.liu@utexas.edu}
\and
\IEEEauthorblockN{Pengyu Nie}
\IEEEauthorblockA{\textit{University of Waterloo, Canada} \\
pynie@uwaterloo.ca
}
\linebreakand
\IEEEauthorblockN{Junyi Jessy Li}
\IEEEauthorblockA{\textit{The University of Texas at Austin, USA}\\
jessy@austin.utexas.edu}
\and
\IEEEauthorblockN{Milos Gligoric}
\IEEEauthorblockA{\textit{The University of Texas at Austin, USA}\\
gligoric@utexas.edu}
}
\IEEEaftertitletext{\vspace{-2\baselineskip}}

\maketitle

\begin{abstract}

Many popular programming languages, including C\#, Java, and Python,
support exceptions.  Exceptions are thrown during program execution if
an unwanted
event happens, e.g., a method is invoked with an illegal argument
value.
Software developers write exceptional behavior tests (\EBTs) to check
that their code detects unwanted events and throws appropriate
exceptions.
Prior research studies have shown the importance of \EBTs, but those
studies also highlighted that developers put most of their efforts on
``happy paths'', e.g., paths without unwanted events.
To help developers fill the gap, we present the first framework,
dubbed \Tool, that automatically generates \EBTs.
\Tool is a large language model instruction fine-tuned from 
CodeLlama and embeds reasoning about traces that
lead to throw statements,
conditional expressions that guard throw statements,
and non-exceptional behavior tests
that execute similar traces.
We compare \Tool with the state-of-the-art models for test generation
(\CAT) and one of the strongest foundation models (\GPTFour), as well
as with analysis-based tools for test generation (\Randoop and
\EvoSuite).  Our results show that \Tool outperforms existing models
and tools.  Furthermore, we contributed several pull requests to
open-source projects and \NumOfMergedTests{} \EBTs generated by \Tool
were already accepted.

\end{abstract}


\section{Introduction}

\begin{figure}[t]
  \begin{subfigure}{\columnwidth}
\begin{lstlisting}[language=java-pretty]
class SearchCommandParser extends CommandParser {
  static final char CHR_SPACE = ' ';
  static final char CHR_CR = '\r';

  public SearchTerm searchTerm(ImapRequestLineReader request) throws ProtocolException {
    ...
    char next;
    while ((next = request.nextChar()) != '\n' &&
	   next != CHR_CR) {
      next = request.consumeAll(CHR_SPACE); 
      if (isAtomSpecial(next)) {  (*@\label{example:is-atom}@*)
        if (next == '(') {  (*@\label{example:opening-parenthesis}@*)
	        ...
        } else if (next == ')') { (*@\label{example:closing-parenthesis}@*)
          ...
        } else {
        (*@\ColorBackX{blue!20}{throw new IllegalStateException("Unsupported atom}@*) (*@\ColorBackX{blue!20}{special char <" + next + ">");}@*) (*@\label{example:exception}@*)
        }
      }
      ...
    }
    return handleOperators();
  }
}
\end{lstlisting}
    \begin{tikzpicture}[remember picture, overlay, every node/.style={circle, inner sep=0, radius=0}]
      \draw[annotatecolor, thick] (1.3, 3.1) rectangle (8.7, 2.5);
      \node[draw=none, annotatecolor, thick, fill=none, text width=3cm, font=\small] at (5, 3.3) {\ts};
    \end{tikzpicture}
    \vspace{-15pt}
    \caption{Method under test: \CodeIn{searchTerm}.}
    \label{fig:example-mut}
  \end{subfigure}
  \begin{subfigure}{\columnwidth}
\begin{lstlisting}[language=java-pretty]
public class SearchCommandParserTest {
  private SearchTerm parse(String line)
      throws ProtocolException {
    final byte[] bytes = (line.endsWith("\n")?line:(line + '\n')).getBytes(); (*@\label{example:create-object-1}@*)
    ByteArrayInputStream ins = new ByteArrayInputStream(bytes); (*@\label{example:create-object-2}@*)
    return new SearchCommandParser().searchTerm(new ImapRequestLineReader(ins, null)); (*@\label{example:invoke-mut}@*)
  }
    
  @Test(expected = IllegalStateException.class) (*@\label{example:expected-exception}@*)
  public void testUnsupportedAtomSpecialChar()
     throws ProtocolException {
    parse("*");
  }
}
\end{lstlisting}
    \begin{tikzpicture}[remember picture, overlay, every node/.style={circle, inner sep=0, radius=0}]
      \draw[annotatecolor, thick] (5.8, 3.4) rectangle (7.3, 3.1);
      \node[draw=none, annotatecolor, thick, fill=none, text width=3cm, font=\small] at (7.8, 3.55) {method under test};
      \draw[annotatecolor, thick] (2.4, 2.0) rectangle (7.3, 1.7);
      \node[draw=none, annotatecolor, thick, fill=none, text width=5cm, font=\small] at (7.7, 1.5) {exceptional behavior test};
    \end{tikzpicture}
    \vspace{-10pt}
    \caption{Exceptional behavior test \MR{written using JUnit 4} that covers the highlighted statement above.}
    \label{fig:example-test}
  \end{subfigure}
  \caption{An \EBT (`testUnsupportedAtomSpecialChar') from \exampleRepo
    and the target throw statement.\label{fig:example}}
  \vspace{-15pt}
\end{figure}

Many popular programming languages, including C\#, Java, and Python,
support exceptions~\cite{hejlsberg2003c, gosling2000java,
vanrossum2010python}. Exceptions are thrown during program execution
if an unwanted event happens, e.g., a method is invoked with an
illegal argument value.
To throw an exception, a developer writes a \emph{throw statement} in
their code.  These throw statements are commonly guarded with
conditional statements (e.g., \CodeIn{if}),
as exceptions should be thrown only under exceptional circumstances.
Figure~\ref{fig:example-mut} shows a code snippet, in Java, that
throws an \ExampleException (line~\ref{example:exception}) when the
\CodeIn{next} character, parsed from argument \CodeIn{request}, is
identified as a special atom (line~\ref{example:is-atom}) but is
neither an opening (line~\ref{example:opening-parenthesis}) nor a
closing parenthesis (line~\ref{example:closing-parenthesis}).

Software developers write \emph{exceptional behavior tests} (\EBTs) to
check that their code properly detects unwanted events and throws
desired exceptions.
Figure~\ref{fig:example-test} shows an example \EBT.
An \EBT, similar to a \emph{non-exceptional behavior test} (\nEBT),
first performs necessary setup of the system under test, e.g., creates
objects
(lines~\ref{example:create-object-1}-\ref{example:create-object-2}),
then invokes a method under test (line~\ref{example:invoke-mut}), and
finally checks the expected behavior
(line~\ref{example:expected-exception}). For an \EBT, the expected
behavior is that an exception was thrown and the type of the exception
matches the one specified by a developer.

Prior research has studied \EBTs in
practice~\cite{DaltonETAL20ExceptionalBehaviorTesting,
  BernardoETAL11AgileTestingOfExceptionalBehavior,
  GoffiETAL16AutomaticGenerationOfOraclesForExceptionalBehaviors,
  MarcilioFuria21HowJavaProgrammersTestExceptionalBehavior,
  LimaETAL21AssessingExceptionHandlingTesting} and observed that most
projects already have some \EBTs, but that the number of \EBTs is not
as high as the number of \nEBTs.  Simply put, developers focus on
``happy paths'' and have limited time to test exceptional behaviors.
Furthermore, through interviews and
surveys~\cite{DaltonETAL20ExceptionalBehaviorTesting,
  MarcilioFuria21HowJavaProgrammersTestExceptionalBehavior}, prior
studies confirmed the importance of \EBTs and developers' desire to
improve the testing of exceptional behaviors.

Sadly, tool support for automatically generating \EBTs is limited.
Most existing analysis-based test generation tools (e.g.,
\Randoop~\cite{PachecoETAL07Randoop, RobinsonETAL11ScalingTestGen} and
\EvoSuite~\cite{FraserAndArcuri11EvoSuite}) and learning-based test
generation tools (e.g., \CAT~\cite{RaoETAL23CAT} and
\TeCo~\cite{NieETAL23TeCo}) have no special settings for targeting
\EBTs and are primarily evaluated on \nEBTs.
Random test generation tools can be guided by reinforcement learning
to target exceptional
behaviors~\cite{AlmullaETAL20GeneratingExceptionTriggeringTests}, but
the generation works only on the entire codebase, and not for a
specific throw statement that a develop might select.
Additionally, tests produced by analysis-based tools lack
readability~\cite{daka2017generating, panichella2022test,
daka2018improving}.

We present the first framework, dubbed \Tool, an instruction fine-tuned
large language model (\LLM) that automatically generates \EBTs.
\LLMs are shown to be effective in code generation, including test
generation~\cite{TufanoETAL20TestGeneration, RaoETAL23CAT,
NieETAL23TeCo, LemieuxETAL23CodaMosa, yuan2023no}. Such strong prior
provides a good foundation yet is not enough.
\EBTs contribute to only a
very small percentage in existing codebases, \ie, they are not
well-represented in \LLM training data. The special conditions
that trigger an \EBT during execution are not included in the training phase of standard code \LLMs, thus they do not perform well on the task of generating \EBTs.

Using \CodeLlama~\cite{roziere2023code} as its base, \Tool is
\finetuned~\cite{selfInstruct,wei2021finetuned,sanh2021multitask} with a novel task instruction and \finetuning 
data, designed specifically to embed 
the reasoning about a context that includes:
(a)~traces that lead to target throw statements, (b)~\guardexps (\ie,
conditional expressions that guard those throw statements), and
(c)~\nEBTs that execute similar traces.
This context is used as the input
to generate an \EBT that triggers the target throw statement.
The \EBT that we already showed in Figure~\ref{fig:example-test} was
generated by \Tool.

We assess the power of \Tool using two use cases.
In the first use case, which we call \emph{\userView}, a developer
selects a method under test and a target throw statement, as well as a
\testfile.
\Tool takes these inputs and automatically generates an \EBT that
executes the target throw statement.

In this use case,
we compare \Tool with the state-of-the-art models for test generation
(\CAT~\cite{RaoETAL23CAT}) and strongest foundation models
(\GPTThree~\cite{gpt3.5-turbo} and \GPTFour~\cite{gpt4o}).  We use a number
of standard metrics (BLEU~\cite{papineni2002bleu},
CodeBLEU~\cite{ren2020codebleu}, edit
similarity~\cite{yujian2007normalized, svyatkovskiy2020intellicode}
and exact match), as well as metrics specific to code, including
percentage of compilable tests, executable tests, and executable tests
that cover the \ts.  Our results show that \Tool generates
\improvedPctCoverageThanCAT{} and \OverGPTRun{} more executable \EBTs
than \CAT and \GPTThree, respectively.

In the second use case, which we call \emph{machine-oriented use
case}, a developer uses \Tool to automatically generate \EBTs for the
entire codebase with the goal to cover all existing throw statements
(with one \EBT per statement). \Tool takes the entire codebase as
input, finds throw statements that are in public methods already
covered by at least one \nEBT and generates one \EBT for each of the
throw statements.
This use case is similar to the traditional test generation setup
targeted by analysis-based generation tools.

In this use case,
we compare \Tool with popular analysis-based test generation tools:
\Randoop~\cite{PachecoETAL07Randoop, RobinsonETAL11ScalingTestGen} and
\EvoSuite~\cite{FraserAndArcuri11EvoSuite}.  Although
tools complement each other (\ie, each tool can generate \EBTs for
some target throw statements that other tools cannot), our findings
show that \Tool outperforms \Randoop and \EvoSuite.

\MR{Additionally, we built \Tool on \GPTFour (a state-of-the-art 
  language model) without fine-tuning and evaluated it in \userView.
  Our results show that \Tool--\GPTFour outperforms \GPTFour by up to
  \OverGPTFour. This emphasizes that our technique is generalizable to 
  the most advanced proprietary LLMs.}

Finally, we selected a subset of \EBTs generated by \Tool and created
pull requests for several open-source projects.  By the time of this
writing, \NumOfMergedTests{} tests generated by \Tool have already been
accepted by developers of those projects.

\vspace{5pt}
\noindent
The key contributions of this paper include:

\begin{itemize}[topsep=0pt,itemsep=3pt,partopsep=0ex,parsep=0ex,leftmargin=*]
\item \textbf{Task}. We define a novel task for \LLMs: generating
  exceptional behavior tests (\EBTs).
\item \textbf{Model}. We designed and implemented \Tool, an instruction fine-tuned
LLM built
on \CodeLlama, 
which reasons about traces to methods that contain
  throw statements, \guardexps, and \nEBTs that cover similar traces.
\item \textbf{Use cases}.  We recognized two use cases for \Tool:
  \umViews.
\item \textbf{Evaluation}.  We assess the power of \Tool in both use
  cases.  In \userView, we compare \Tool with existing models for code
  and test generation.  In machine-oriented use case, we compare \Tool
  with analysis-based testing tools: \Randoop and \EvoSuite.  We find
  that \Tool outperforms existing state-of-the-art models and tools.
\item \textbf{Dataset}.  We developed a novel dataset for the
  presented task and this dataset is publicly available.  
\end{itemize}

\noindent
\Tool is available on GitHub at \url{https://github.com/EngineeringSoftware/exLong}.

\section{\MR{Use Cases}}
\label{sec:task}

\begin{figure*}[t]
  \centering \includegraphics[width=\textwidth]{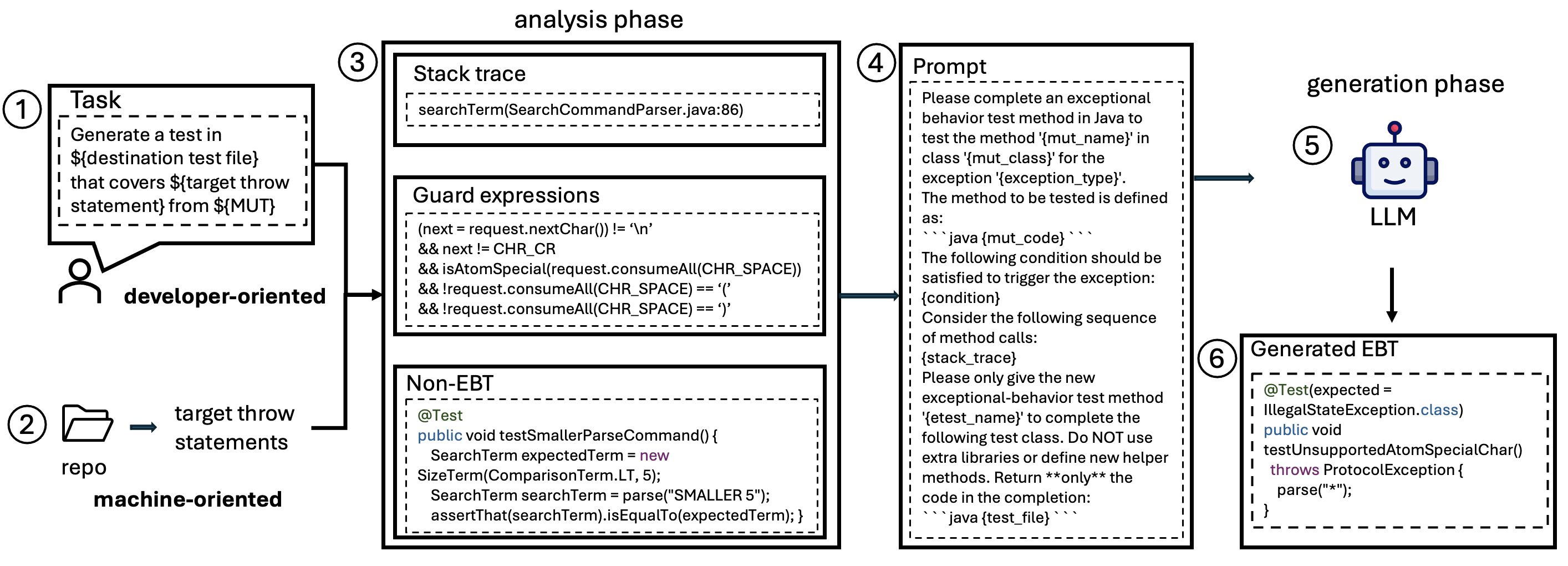}
  \caption{Overview of \Tool. Two use cases for \Tool: (1)~\userView and (2) \machineView. In the \userView, a developer specify the method under test, a \ts and a \testfile and ask \Tool to generate an \EBT that cover the \ts. In the \machineView, a developer gives an entire \repo to \Tool.\label{fig:overview}}
\vspace{-15pt}
\end{figure*}

At a high level, \Tool is designed to help software developers write
\EBTs that cover the throw statements within the given \repo.  We
propose two use cases for \Tool: \emph{\userView}
(Section~\ref{sec:userview}) and \emph{\machineView}
(Section~\ref{sec:machineview}).
Note that in this work we do \emph{not} consider generating \EBTs that
cover throw statements in the dependency libraries of the given
\repo, e.g., \CodeIn{ArithmeticException} thrown from the
\CodeIn{java.lang.Math}, as we assume that those throw statements are
already covered by \EBTs available in dependency libraries; developers
can configure \Tool to include such throw statements if necessary.

\subsection{\UserView}\label{sec:userview}

In the \userView (box \circled{1} in Figure~\ref{fig:overview}), a
developer will specify the \emph{\methodundertest} (\MUT), a
\emph{target throw statement}, and a \emph{\testfile}, then ask \Tool
to generate an \EBT that invokes \MUT and triggers the target throw
statement.
When the developer specifies a target throw statement that is not
within the \MUT, \Tool will first use a static program analysis
technique to find all possible throw statements reachable through a
sequence of method calls starting from the \MUT, and then prompt the
developer to select the target statement from the list.

\subsection{\MachineView}\label{sec:machineview}

In the \machineView (box \circled{2} in Figure~\ref{fig:overview}), a
developer gives an entire repository to \Tool.
\Tool will first find all public methods and throw statements in those
methods.  Furthermore, \Tool will heuristically select a
  \testfile. Then, \Tool will create (public method, throw
statement, \testfile) triples.  Finally, it
will generate one \EBT to test each triple,
without requiring any developer intervention.
\Tool uses
static+dynamic program analyses to obtain necessary context such as
traces leading to throw statements and \guardexps.

\section{\Tool}
\label{sec:tech}

Figure~\ref{fig:overview} shows the workflow of \Tool.  Given an \MUT
(\aMUT), a \ts (\aThrowStmt), and a \testfile (\aTestFile), \Tool
reasons about the following context, collected using static and
dynamic program analyses (\circled{3}):
\begin{itemize}
\item \stacktrace (\aStackTrace): the sequence of method invocations
that start from the \MUT and lead to the \ts.
\item \guardexp (\aGuardExp): the logical formula representing the
  constraints on the symbolic variables that must be true to follow
  that particular trace.
\item relevant \nEBTs (\aNETestSetRel): \MR{a subset of \nEBTs that
  invoke the given \MUT or are in the given \testfile;
  their snippets and coding styles help with generating desired \EBTs.}

\end{itemize}

\noindent
The above constitutes the prompt $\aPrompt = (\aMUT, \aThrowStmt,
\aTestFile, \aStackTrace, \aGuardExp, \aNETestSetRel)$ that
encompasses the task inputs and the task context (\circled{4}). During
training, \Tool is instruction fine-tuned over a base \LLM to produce
\EBT (\aETest) conditioned on the input \aPrompt.
During inference, \Tool generates the \EBT (\circled{6})
given the prompt.

While the format of the \LLM's input (\aPrompt) and output (\aETest)
is static, the steps for preparing the task inputs and reasoning about
the task context differ for training and inference, and
differ slightly between the \umViews.  We describe the methodology for
training \Tool in Section~\ref{sec:tech:train} and inference in
Section~\ref{sec:tech:eval}.

\subsection{Training}
\label{sec:tech:train}

\begin{algorithm}
\caption{Collecting training corpus. \label{algo:training-data}}
\small
\begin{algorithmic}[1]

\State \textbf{inputs:} \aETestSet, \aNETestSet\ - existing \EBTs and \nEBTs\label{line:training-data:input}
\State \qquad\quad\ \aSUT\ - the system under test
\State \textbf{outputs:} \aCorpus\ - the training corpus of \{$(\aMUT, \aThrowStmt, \aTestFile, \aStackTrace, \aGuardExp, \aNETestSetRel, \aETest)$\}

\Procedure{CollectTrainingCorpus}{\aETestSet, \aNETestSet, \aSUT}
  \State \aCorpus $\gets$ $\emptyset$
  \For{\aETest $\in$ \aETestSet}
    \State \aTestFile $\gets$ GetFile(\aETest)\label{line:training-data:tf}
    \State \aStackTrace $\gets$ Execute(InstrumentPrintException(\aETest))\label{line:training-data:stacktrace}
    \State \AlgoComment{instrumenting and executing \EBT to get \stacktrace}
    \State \aStackTrace $\gets$ ExcludeTestAndUtilMethods(\aStackTrace, \aTestFile)\label{line:training-data:stacktrace-postproc}
    \State \aMUT $\gets$ \aStackTrace{[}0].method \AlgoComment{\MUT comes first in \stacktrace}\label{line:training-data:mut}
    \State \aThrowStmt $\gets$ GetSourceCode(\aStackTrace{[}-1])\label{line:training-data:ts}
    \State \AlgoComment{last \stacktrace item points to \ts}
    \State \aGuardExp $\gets$ ComputeGuardExp(\aStackTrace)
    \State \aNETestSetRel $\gets$ $\emptyset$ \AlgoComment{initialize set of relevant \nEBTs}
    \State \aCorpus $\gets$ \aCorpus $\cup$ \{(\aMUT, \aThrowStmt, \aTestFile, \aStackTrace, \aGuardExp, \aNETestSetRel, \aETest)\}
  \EndFor

  \State \aSUT $\gets$ InstrumentPrintMethod(\aSUT)\label{line:training-data:nebt-instrumentation}
  \State \AlgoComment{instrumentation for getting methods covered by \nEBTs}
  \For{\aNETest $\in$ \aNETestSet}
    \State $\aMUT'$ $\gets$ Execute(\aNETest)[0]  
    \State \AlgoComment{get \MUT directly invoked by \nEBT}
    \For{\aCorpusItem $\in$ \aCorpus}
      \If{\aCorpusItem.\aMUT == $\aMUT' \lor$ \aCorpusItem.\aTestFile == GetFile(\aNETest)}\label{line:training-data:nebt-condition}
        \State $\aCorpusItem.\aNETestSetRel \gets \aCorpusItem.\aNETestSetRel \cup \{\aNETest\}$\label{line:training-data:nebt-add}
      \EndIf
    \EndFor
  \EndFor

  \State \Return \aCorpus
\EndProcedure
\end{algorithmic}
\vspace{-2pt}
\end{algorithm}

To perform supervised fine-tuning (SFT) on
\Tool's foundation \LLM, we collect a corpus (\aCorpus) of
\instdata $\aCorpusItem = (\aPrompt, \aETest) = (\aMUT, \aThrowStmt,
\aTestFile, \aStackTrace, \aGuardExp, \aNETestSetRel, \aETest)$ from a
set of \repos with \developerwritten \EBTs and \nEBTs.
Algorithm~\ref{algo:training-data} shows the process of collecting the
training corpus for \Tool.

\subsubsection{Identifying \EBTs and \nEBTs}
\label{sec:tech:train:ebt-nebt}

For each \repo in the training set, we first parse the source code in
the repository
to identify test methods and categorize them into \EBTs and \nEBTs.
A test method is categorized as an \EBT if it conforms to one of the
four patterns that are widely used by
developers~\cite{MarcilioFuria21HowJavaProgrammersTestExceptionalBehavior}: 
\UseMacro{THead-expect},
\UseMacro{THead-assertThrows},
\UseMacro{THead-expectedRule},
\UseMacro{THead-try{fail}catch}.
All the other test methods are categorized as \nEBTs.  The set of
\EBTs and \nEBTs are used as the inputs to the training corpus
collection algorithm (line~\ref{line:training-data:input} in
Algorithm~\ref{algo:training-data}).

\subsubsection{Executing \EBT and collecting \stacktrace}
\label{sec:tech:train:stacktrace}

Each \EBT will be expanded to one SFT example (\aCorpusItem) in the
training corpus (\aCorpus).  Naturally, the file that contains the
\EBT is the \testfile (line~\ref{line:training-data:tf} in
Algorithm~\ref{algo:training-data}).  To avoid data leakage problems,
we remove all test methods in the test file (\aTestFile) and only keep
the test class structure and utility methods.

For training, \stacktrace is the sequence of method invocations from
\EBT (non inclusive) that lead to the \ts under test (inclusive).
Line~\ref{line:training-data:stacktrace} in
Algorithm~\ref{algo:training-data} shows how to collect the
\stacktrace: first, instrument the \EBT by adding
``\CodeIn{print(exception.getCause())}''
to the code location after the exception is thrown and caught by the
\EBT; then, execute the instrumented \EBT to get the printed
\stacktrace. To avoid duplicate information, we exclude \EBT itself
and any utility methods in the \testfile from the \stacktrace
(line~\ref{line:training-data:stacktrace-postproc} in
Algorithm~\ref{algo:training-data}).

The first method invoked in the \stacktrace is the \MUT by definition
(line~\ref{line:training-data:mut} in
Algorithm~\ref{algo:training-data}).  The last method invocation and
line number in the \stacktrace point to the \ts
(line~\ref{line:training-data:ts} in
Algorithm~\ref{algo:training-data}).

\subsubsection{Computing the \guardexp}
\label{sec:tech:train:guardexp}

\begin{algorithm}
\caption{Collect AST nodes along the \stacktrace.}\label{algo:collect-nodes-cut}
\small
\begin{algorithmic}[1]
\State \textbf{inputs:} \aStackTrace\ - the \stacktrace for a \ts
\State \textbf{outputs:} \aNodes\ - the collected nodes
\Procedure{\algoCollectNodes}{\aStackTrace}\label{line:collect-nodes:collect-nodes}
\State \aNodes $\gets \emptyset$
\For{(\aMethod, $\mathtt{lineno}$) $\in$ Reversed(\aStackTrace)}\label{line:collect-nodes:method-loop}
\State \aCurrent $\gets$ \textsc{GetSourceCode}(\aMethod, $\mathtt{lineno}$)
\State $\aParent \gets \aCurrent$
\State \aNodes $\gets$ \aNodes $\cup$ \{\aCurrent{}\} \label{line:collect-nodes:add-start-node}
\While{$\aCurrent \neq \aMethod$}
\If{$\aParent$ is ForStmt}\label{line:collect-nodes:for-beg}
\State \aNodes $\gets$ \aNodes $\cup$ \{\aParent.CompareExpression\}\label{line:collect-nodes:for-end}
\EndIf
\If{$\aParent$ is IfStmt}\label{line:collect-nodes:if-beg}
\If{\aParent.ThenStmt == \aCurrent}
\State \aNodes $\gets$ \aNodes $\cup$ \{\aParent.ConditionExpression\}
\EndIf
\If{\aParent.ElseStmt == \aCurrent}
\State \aNodes $\gets$ \aNodes $\cup$ \{$\lnot$ \aParent.ConditionExpression\}\label{line:collect-nodes:if-end}
\EndIf
\EndIf

\State \AlgoComment{Other cases (while, switch, block and assignment statements) are in supplementary material.}

\State $\aCurrent = \aParent$
\State $\aParent = \aParent$.getParent()
\EndWhile

\EndFor
\State \Return \aNodes
\EndProcedure

\end{algorithmic}

\end{algorithm}

\begin{algorithm}
\caption{Compute \guardexp based on \stacktrace.} \label{algo:compute-guardexp}
\small
\begin{algorithmic}[1]
\State \textbf{inputs:} \aStackTrace\ - the \stacktrace for a \ts
\State \textbf{outputs:} \aGuardExp\ - the \guardexp for the \ts
\Procedure{ComputeGuardExp}{\aStackTrace}
\State \aNodes $\gets$ \textsc{CollectNodes}(\aStackTrace)\label{line:compute-guardexp:collect-nodes}
\State \AlgoComment{collect all condition (if, while, etc.) and assignment (assign, method call) nodes along the \stacktrace}
\State $\aCondSet \gets \emptyset$  \AlgoComment{the set of conditions in the \guardexp}
\For{\aNode $\in$ \aNodes}\label{line:compute-guardexp:compute-beg}
\If{\aNode is ConditionalExpr}
\State \aCondSet $\gets$ \aCondSet $\cup$ \{\aNode{}\}\label{line:compute-guardexp:condition}
\EndIf
\If{\aNode is AssignStmt}
\State \aCondSet $\gets$ \textsc{Merge}(\aCondSet, \{\aNode.lhs $\mapsto$ \aNode.rhs\})\label{line:compute-guardexp:assignstmt}
\EndIf
\If{\aNode is MethodDeclaration}
\State $\aNode' \gets \aNode$ \AlgoComment{process with next node (method call)}
\EndIf
\If{\aNode is MethodCallExpr}
\State \aArgMap $\gets \emptyset$\label{line:method-map-0}\label{line:compute-guardexp:methodcall-beg}
\For{$\mathtt{arg}, \mathtt{argname} \in$ \aNode.getArgs(), $\aNode'$.getParams()}
\State \aArgMap $\gets$ \aArgMap $\cup \{\mathtt{argname} \mapsto \mathtt{arg}\}$
\EndFor
\State \aCondSet $\gets$ \textsc{Merge}(\aCondSet, \aArgMap)\label{line:compute-guardexp:methodcall-end}
\EndIf
\EndFor
\State \Return $\lAnd{}\aCondSet$\label{line:compute-guardexp:compute-end}
\EndProcedure
\\
\Procedure{Merge}{\aCondSet, \aArgMap}\label{line:compute-guardexp:merge}
\State \aCondSet' $\gets \emptyset$
\For{\aCond $\in$ \aCondSet}
\For{$\mathtt{name} \mapsto \mathtt{expr} \in \aArgMap$}
\If{\aCond.contains($\mathtt{name}$)}
\State \aCond $\gets$ \aCond.replace($\mathtt{name}$, $\mathtt{expr}$) \label{line:replace}
\EndIf
\EndFor
\State \aCondSet' $\gets$ \aCondSet' $\cup$ \{\aCond{}\}
\EndFor
\State \Return \aCondSet'
\EndProcedure
\end{algorithmic}
\vspace{-1pt}
\end{algorithm}

\Stacktrace provides the sequence of method invocations
that lead to the \ts, but knowing only the names of the methods is
insufficient for generating \EBTs.  To aid the reasoning about the
setup of the system under test, which lead to exceptional behaviors, we
propose \emph{\guardexp}: a logical formula representing the
constraints on the symbolic variables that must be true to follow the
particular code trace.
Specifically, we use conjunctions of expressions extracted from the
invoked methods in the \stacktrace to form the \guardexps.
For instance, the \guardexp for the throw statement highlighted in
Figure~\ref{fig:example-mut} is present in the
Figure~\ref{fig:overview} (box \circled{3}).

The first step in the computation is to collect the list of
guard-related AST nodes along the \stacktrace starting from the \ts to
the \MUT, as described in Algorithm~\ref{algo:collect-nodes-cut}.  We
traverse each method in the \stacktrace in reversed order
(line~\ref{line:collect-nodes:method-loop}). Inside each method, we
start from the AST node specified by the line number in the
\stacktrace.
We always include this node into the list of collected nodes
(line~\ref{line:collect-nodes:add-start-node}) because 
the variable names in that statement may be used by the next step for
replacing method call arguments.  Starting from the statement, we
traverse the AST by maintaining the pointer \aCurrent that constantly
moves from the child AST node to its parent node.  AST nodes
entailing condition expressions within the `for' loop
(lines~\ref{line:collect-nodes:for-beg}~to~\ref{line:collect-nodes:for-end}),
`if' statement
(lines~\ref{line:collect-nodes:if-beg}~to~\ref{line:collect-nodes:if-end}),
`while' loop and `switch' statement will be added to \aNodes.
We also collect the assignment statements, method call expressions,
and method declarations.
Some of these are omitted from Algorithm~\ref{algo:collect-nodes-cut} to keep it simple.

The second step of computing \guardexps is to process the collected
nodes by propagating the symbolic variables through the \stacktrace,
as described in Algorithm~\ref{algo:compute-guardexp}.
The nodes are visited in the order being collected, and
conditional expressions are directly added to the \guardexp
(line~\ref{line:compute-guardexp:condition}).  Assignment statements,
method declaration, and method call expressions will trigger a
\algoMerge operation (line~\ref{line:compute-guardexp:merge}).
The goal of the \algoMerge operation is to review the current
\guardexp and replace the symbolic variables (e.g., variables
appearing in the \ts) with their corresponding values (e.g., constant
values or \MUT's arguments).
Therefore, the \guardexp reflects
the \MUT's arguments and public fields that are usable by \EBT,
rather than local variables that \EBT may not have access to.
For assignment statement
(line~\ref{line:compute-guardexp:assignstmt}), we replace the left
hand side variable in \aCondSet with the right hand side expression.
For method declaration and method call expression
(lines~\ref{line:compute-guardexp:methodcall-beg}~to~\ref{line:compute-guardexp:methodcall-end};
these two nodes always appear in pairs in the collected nodes), we
replace the method declaration's argument names in \aCondSet with the
actual arguments in the method call expression.

\subsubsection{Connecting \EBTs to relevant \nEBTs}
\label{sec:tec:train:nbt}

We add the \emph{relevant} \nEBTs to the prompt
(lines~\ref{line:training-data:nebt-instrumentation}~to~\ref{line:training-data:nebt-add} in
Algorithm~\ref{algo:training-data}), to encourage the \LLM to reason
about the procedures to set up the object under test and the condition
under which the exception will be triggered grounding the existing
\nEBTs.
Additionally, we believe the \nEBTs in the same repository will
promote the consistency between the generated code and the existing
code in terms of the both format and coding conventions.
Given an \MUT, we use two approaches to retrieve the relevant \nEBTs:
\nEBTs that directly invoke the same \MUT (used first when the context
window of the \LLM cannot fit all relevant \nEBTs) and \nEBTs that are
already present in the \testfile
(line~\ref{line:training-data:nebt-condition} in
Algorithm~\ref{algo:training-data}).  If there is no relevant \nEBT
retrieved, this part of the prompt is left empty.

\subsubsection{Instruction \finetuning}
\label{sec:tech:train:model}

We use \CodeLlama-Instruct-7B~\cite{roziere2023code}, an open-source
foundation model designed for code generation and instruction following, as the
foundation model of \Tool. 
\CodeLlama is \pretrained on auto-regressive and infilling
objectives, enabling tasks like code completion and document
generation.  
We \finetune \CodeLlama on our collected training corpus
of \instdata $\aCorpus = \{(\aPrompt, \aETest)\}$, with the novel
instructions shown in Figure~\ref{fig:overview} (\circled{4}).
Given the instruction that includes the collected context, the model
is expected to produce the \EBT: $\mathtt{LLM}(\aPrompt) = \aETest$.

We \finetune the \CodeLlama model using the
parameter-efficient Low-Rank Adaptation (LoRA)
technique~\cite{hu2021lora}.
Rather than updating the entire set of parameters in the
LLM, LoRA injects trainable low-rank matrices into each layer of the
model.  This approach dramatically reduces the number of trainable
parameters and the amount of required computational resources.

\subsection{Inference}
\label{sec:tech:eval}

\begin{algorithm}
\caption{Prepare the pool \stacktraces from \nEBTs and assemble the prompt for evaluating \Tool.  \label{algo:eval-data}}
\small
\begin{algorithmic}[1]

\State \textbf{inputs:} \aNETestSet\ - existing \nEBTs\label{line:eval-data:input}
\State \qquad\quad\ \aSUT\ - the system under test
\State \textbf{outputs:} \aStackTraceTupleSet\ - \stacktraces to reach \tss

\Procedure{\algoCollectStackTraceCorpus}{\aNETestSet, \aSUT}\label{line:eval-data:stacktrace-corpus-beg}
    \State \aSUT $\gets$ InstrumentPrintTrace(\aSUT)\label{line:eval-data:instrument}
    \State \aStackTraceTupleSet $\gets \emptyset$

    \For{\aNETest $\in$ \aNETestSet}\label{line:eval-data:nebt-beg}
        \For{\aStackTrace $\in$ Execute(\aNETest)}
            \State \aStackTrace $\gets$ ExcludeTestAndUtilMethods(\aStackTrace, GetFile(\aNETest))
            \For{\aThrowStmtTarget $\in$ GetThrowStmts(\aStackTrace)}
                \State \aStackTraceTupleSet $\gets$ \aStackTraceTupleSet $\cup$ (\aStackTrace, \aNETest, \aThrowStmtTarget)\label{line:eval-data:nebt-end}
            \EndFor
        \EndFor
    \EndFor
    \State \Return \aStackTraceTupleSet
\EndProcedure\label{line:eval-data:stacktrace-corpus-end}
\\
\State \textbf{global var:} \aStackTraceTupleSet = \textsc{\algoCollectStackTraceCorpus}(\aNETestSet, \aSUT)
\State \textbf{inputs:} \aMUT, \aThrowStmtTarget, \aTestFile\ - task inputs selected by developers or inferred\label{sec:eval-data:assemble-prompt-inputs}
\State \qquad\quad\ \aNETestSet\ - existing \nEBTs
\State \textbf{outputs:} \aPrompt\ - the prompt to give to the LLM
\Procedure{\algoAssemblePrompt}{\aMUT, \aThrowStmtTarget, \aTestFile}
\State $\mathtt{R} \gets \emptyset$
\State \aNETestSetRel $\gets \emptyset$
\For{\aStackTraceTuple $\in$ \aStackTraceTupleSet}
    \If{\aMUT $\in$ \aStackTraceTuple.\aStackTrace $\land$ \aStackTraceTuple.\aThrowStmt == \aThrowStmtTarget}\label{sec:eval-data:match}
        \State $\mathtt{R} \gets \mathtt{R}\ \cup$ \{\aStackTraceTuple.\aStackTrace{}\}
        \If{\aMUT == \aStackTraceTuple.\aStackTrace{}[0].method}
            \State \aNETestSetRel $\gets$ \aNETestSetRel $\cup$ \{\aStackTraceTuple.\aNETest{}\}\label{sec:eval-data:nebt-rel-mut}
        \EndIf
    \EndIf
\EndFor
\If {$\mathtt{R} \neq \emptyset$}
    \State \aStackTrace $\gets$ RandomSelect($\mathtt{R}$)\label{sec:eval-data:random}
    \State \aGuardExp = \textsc{ExtractGuardExp}(\aStackTrace)\label{sec:eval-data:guardexp}
    \State \aNETestSetRel $\gets$ \aNETestSetRel $\cup$ $\{\aNETest | \text{GetFile}(\aNETest) \text{==} d\}$\label{sec:eval-data:nebt-rel-testfile}
    \State \Return (\aMUT, \aThrowStmtTarget, \aTestFile, \aStackTrace, \aGuardExp, \aNETestSetRel)
\EndIf
\EndProcedure

\end{algorithmic}
\end{algorithm}

The inference workflow of \Tool is different from its training
workflow in that we cannot rely on executing \EBTs to collect the
context (e.g., \stacktrace), as our goal is to generate those \EBTs.
Instead, \Tool reasons about the context based on task inputs (\MUT,
\ts, and \testfile) and leveraging existing \nEBTs for the system
under test.
Algorithm~\ref{algo:eval-data} describes the key steps in preparing
the inference prompt of \Tool.

\subsubsection{Collecting \nEBTs' \stacktraces to reach potential \tss}
\label{sec:tech:eval:stacktrace}

We first prepare a set of \stacktraces, from the execution of \nEBTs,
that can reach potential \tss in the \repo (lines~\ref{line:eval-data:stacktrace-corpus-beg}~to~\ref{line:eval-data:stacktrace-corpus-end}).
This only needs
to be done once per \repo.
Specifically, we first instrument all methods that have throw
statements to log the 
current \stacktrace (using
\CodeIn{Thread.currentThread().getStackTrace();}) upon invoking those
methods (line~\ref{line:eval-data:instrument}).  Then, we execute all
\nEBTs and collect the logged \stacktraces
(lines~\ref{line:eval-data:nebt-beg}~to~\ref{line:eval-data:nebt-end}).
The execution of each \nEBT may generate multiple \stacktraces, as it
may cover multiple methods with throw statements.

\subsubsection{Selecting task inputs}
\label{sec:tech:eval:usecases}

Next, we select the task inputs (\MUT, \ts, and \testfile), which can
be specified by the developer or inferred by heuristics depending on
the use case that \Tool is targeting:

\begin{itemize}
\item \UserView: developer specify the \MUT and \ts to generate the \EBT
for, and the \testfile where the \EBT should be placed.
\item \MachineView: given a \repo, \Tool locates all throw statements
and generates one \EBT for each of them. For a \ts, the \MUT is the
method containing the throw statement, and the \testfile is selected
based on (a)~file name matching, and (b)~test coverage analysis.
Specifically, similar to prior work~\cite{RaoETAL23CAT}, given a code
file named \texttt{FNM}, we search for test file named
\texttt{FNMTest} or \texttt{TestFNM}.  If there is no result based on
file name matching, we run the existing \nEBTs to find any existing
test class that cover the \MUT or the class of \MUT. If there is again
no result based on test coverage analysis, \Tool will not generate an
\EBT for this \ts (and it will move to the next one).
\end{itemize}

\noindent
The selected task inputs are then used to assemble the prompt for
\Tool (line~\ref{sec:eval-data:assemble-prompt-inputs}).

\subsubsection{Assembling the prompt}
\label{sec:tech:eval:prompt}

We first need to find \stacktraces from the set of \nEBTs'
\stacktraces that match the given \MUT and \ts
(line~\ref{sec:eval-data:match}).  If multiple matching \stacktraces
are found, we randomly select one (line~\ref{sec:eval-data:random}).
Given the \stacktrace, we use the same algorithm in
Section~\ref{sec:tech:train:guardexp} to compute the corresponding
\guardexp (line~\ref{sec:eval-data:guardexp}).  
The relevant \nEBTs are selected using the similar criteria as
Section~\ref{sec:tec:train:nbt}, i.e., having the same \MUT
(line~\ref{sec:eval-data:nebt-rel-mut}) or in the same \testfile
(line~\ref{sec:eval-data:nebt-rel-testfile}).  However, if no matching
\stacktrace is found, \Tool will not generate an \EBT for the given
inputs.

\section{Dataset}
\label{sec:dataset}

In this section, we describe details on collecting the dataset
(Section~\ref{sec:data:collection}), as well as the statistics of our
dataset used for training and evaluation
(Section~\ref{sec:data:stats}).

\begin{table}[t]
  \begin{small}
  \begin{center}
\caption{\UseMacro{TCap-dataset}\label{tab:dataset}}
\scalebox{0.95}{
\begin{tabular}{ c | r  r  r  r  r r }
\toprule
 
 & \textbf{\UseMacro{THead-project-count}}
 & \textbf{\UseMacro{THead-test-count}}
 & \textbf{\UseMacro{THead-etest-count}}
& \textbf{\UseMacro{THead-mut-count}}
& \makecell{\textbf{\#Exception} \\ \textbf{Types}}
\\
\midrule
All
 & \UseMacro{ds-all-project-count}
 & \UseMacro{ds-all-test-count}
 & \UseMacro{ds-all-etest-count}
 & \UseMacro{ds-all-mut-count}
 & \UseMacro{ds-all-etype-count}
\\
\UseMacro{THead-train}
 & \UseMacro{ds-train-project-count}
 & \UseMacro{ds-train-test-count}
 & \UseMacro{ds-train-etest-count}
 & \UseMacro{ds-train-mut-count}
 & \UseMacro{ds-train-etype-count}
\\
\UseMacro{THead-valid}
 & \UseMacro{ds-valid-project-count}
 & \UseMacro{ds-valid-test-count}
 & \UseMacro{ds-valid-etest-count}
 & \UseMacro{ds-valid-mut-count}
 & \UseMacro{ds-valid-etype-count}
\\
\UseMacro{THead-test}
 & \UseMacro{ds-test-project-count}
& \UseMacro{ds-test-test-count}
& \UseMacro{ds-test-etest-count}
& -
& -
\\
\bottomrule
\end{tabular}
}
  \end{center}
  \end{small}
  \vspace{-15px}
\end{table}

\begin{table}
  \begin{small}
\begin{center}
  \caption{\UseMacro{TCap-rq1-eval-dataset-stats}\label{tab:userview}}
    \setlength{\tabcolsep}{2pt} 
\begin{tabular}{l | r  r  r  r }
  \toprule
 & \textbf{\UseMacro{THead-ebts}}
 & \textbf{\UseMacro{THead-count-mut}}
 & \makecell{\textbf{\#Exception}\\\textbf{Types}}
 & \makecell{\textbf{\#Throw}\\\textbf{Statements}}
\\
\midrule
\UseMacro{THead-rq1-eval}
 & \UseMacro{ds-rq1-eval-ebts}
 & \UseMacro{ds-rq1-eval-count-mut}
 & \UseMacro{ds-rq1-eval-count-etype}
 & \UseMacro{ds-rq1-eval-count-ts}
\\
\bottomrule
\end{tabular}
\end{center}
  \end{small}
  \vspace{-10pt}
\end{table}

\begin{table}[t]
  \begin{small}
\begin{center}
\caption{\UseMacro{TCap-eval-dataset-stats}\label{tab:machineview}}
\begin{tabular}{l | r  r  r r }
  \toprule
  & \makecell{\textbf{\#Throw} \\ \textbf{Statements}}
    & \makecell{\textbf{\#Exception} \\ \textbf{Types}}
\\
\midrule
Machine-Oriented
& \UseMacro{ds-rq2-count-ts}
 & \UseMacro{ds-rq2-count-etype}
\\
\bottomrule
\end{tabular}
\end{center}
\end{small}
\vspace{-15px}
\end{table}

\subsection{Dataset Collection}
\label{sec:data:collection}
Following prior work~\cite{NieETAL23TeCo}, we collect data from Java
projects from CodeSearchNet~\cite{CodeSearchNet}, which are available
on GitHub and satisfy the following: (1)~use the Maven build system;
(2)~compile successfully; (3)~do not have test failures; (4)~have at
least one \EBT that follows one of the four
patterns~\cite{MarcilioFuria21HowJavaProgrammersTestExceptionalBehavior}
(Section~\ref{sec:tech:train:ebt-nebt}), and (5)~have a license that
permits the use of its data.  Requirements 1-4 simplify the automation
steps and ensure that we can run existing tests to collect dynamic
data (\eg, \stacktraces), as well as run \EBTs that we generate.

\subsection{Dataset Statistics}
\label{sec:data:stats}

The statistics for the collected dataset are presented in
Table~\ref{tab:dataset}.  In total, we collected
\TotalCollectedTests{} tests from \TotalProjects{} projects, where
\TotalCollectedETests{} of these tests are \EBTs. Collected \EBTs
cover a range of \UseMacro{ds-all-etype-count} unique exception types
(e.g., \CodeIn{RuntimeException}, \CodeIn{IllegalArgumentException}).

The dataset is randomly split by projects into \train
(\UseMacro{THead-train}), \val (\UseMacro{THead-valid}), and \eval
(\UseMacro{THead-test}) sets, where the \train set is the SFT data used to
instruction \finetune \Tool, the \val set is used for early-stopping the training
process and guiding our design decision of \Tool, and the \eval set is
used for evaluating the performance of \Tool and baselines.

Table~\ref{tab:userview} presents the statistics of the evaluation
data for \userView.  Note that this is a subset of the last row from
Table~\ref{tab:dataset}.
Under \userView, we benchmark \Tool on the subset of
\UseMacro{tsEtests} examples for
which we are able to extract stack traces.
\MR{In this paper, we focus on cases where accurate \stacktraces can be
  extracted by executing existing \nEBTs.
  When such \nEBTs are not available, namely the \stacktraces cannot be
  obtained, developers can first write or
generate \nEBTs for the \MUT with the help of other test generation
tools, and then use \Tool to generate \EBTs.}

Table~\ref{tab:machineview} presents the statistics of the evaluation
data for \machineView.  Note that this is a subset of the last row
from Table~\ref{tab:dataset}.
For \machineView, we evaluate on \UseMacro{machineViewSize} examples 
as we filter the data for which we were not able to locate the
\testfile with our designed heuristics (Section~\ref{sec:tech:eval:usecases}).

\section{Evaluation Design}
\label{sec:exp}

We assess the performance of \Tool by answering the
following research questions:

\DefMacro{rq-userView}{RQ1\xspace}
\DefMacro{rq-stackTraceCondition}{RQ2\xspace}
\DefMacro{rq-nEBTs}{RQ3\xspace}
\DefMacro{rq-GPT4}{RQ4\xspace}
\DefMacro{rq-machineView}{RQ5\xspace}

\vspace{2pt}
\MyParaOnly{\UseMacro{rq-userView}}: How does \Tool perform under the
\userView compared with the state-of-the-art models?

\vspace{2pt}
\MyParaOnly{\UseMacro{rq-stackTraceCondition}}: How much do
\stacktraces and \guardexps help \Tool in generating \EBTs?

\vspace{2pt}
\MyParaOnly{\UseMacro{rq-nEBTs}}: How much does the selection of
\nEBTs help \Tool in generating \EBTs?

\vspace{2pt}
\MyParaOnly{\UseMacro{rq-GPT4}}: \MR{How does \Tool perform with different
underlying \LLM model?}

\vspace{2pt}
\MyParaOnly{\UseMacro{rq-machineView}}: How does \Tool perform under
the \machineView compared with analysis-based test generation tools?

\vspace{2pt}
We next describe metrics used to compare models and tools
(Section~\ref{sec:exp:metrics}) and then describe the baselines used
in our comparison (Section~\ref{sec:exp:baseline}). We answer all
research questions in Section~\ref{sec:eval}.

\subsection{Evaluation Metrics}
\label{sec:exp:metrics}

\subsubsection{\UserView}

For \userView, we compare (using data shown in Table~\ref{tab:userview}) the
generated \EBTs against the \developerwritten \EBTs by benchmarking on
similarity-based and functional-correctness metrics.

Following prior work on learning-based test
generation~\cite{TufanoETAL20TestGeneration, RaoETAL23CAT,
  NieETAL23TeCo}, we use the following
similarity-based metrics to compare generated \EBTs and \gt (i.e.,
\developerwritten ones):

\MyParaOnly{Exact-match accuracy (xMatch):} the
percentage of the predictions that are exactly the same as
the \gt.

\MyParaOnly{BLEU}~\cite{papineni2002bleu}: the number of n-grams in
the prediction that also appear in the \gt.

\MyParaOnly{CodeBLEU}~\cite{ren2020codebleu}: adapted version of BLEU
score for code. In addition to n-grams overlapping, it also computes
the overlap of AST nodes, nodes in the data-flow graph between the
prediction and \gt.

\MyParaOnly{Edit similarity}~\cite{yujian2007normalized,
svyatkovskiy2020intellicode}: calculates 1-Levenshtein distance
which is the minimum number of character-level edits (insertions,
deletions, or substitutions) required to change the prediction into
the \gt.

The similarity metrics only capture the surface-level similarity
between the prediction against an existing \EBT; \MR{among them,
xMatch is the most strict one as it requires perfect matches, while
the others account for partial matches.}
However, such surface metrics do not adequately capture the functional
validity of the generated \EBT (e.g., whether the code can be compiled
or executed), especially since
the \developerwritten \EBTs may not be the only correct implementation
to cover a specific \ts.
Thus, we additionally include the following functional-correctness
metrics:

\MyParaOnly{\compile}: percentage of the generated \EBTs that can
be compiled. \MR{Being compilable is a basic functional requirement for the generated tests.}

\MyParaOnly{Matched-E\%}: percentage of \EBTs that check the specified
exception type.  Namely, whether the exception class following
`\CodeIn{@Test(expected=}' is the same as user specified one.
\MR{This metric checks if the model hallucinates the exception type.}

\MyParaOnly{Runnable\%}: percentage of \EBTs that check the specified
exception type, and can be compiled and executed without any
error.
\MR{This metric, unlike others, requires the generated \EBTs to be
  semantically valid.}

\MyParaOnly{\TSCoverage}: 
out of all developer-specified \tss (Table~\ref{tab:userview}), the
percentage of \tss with successfully generated \EBTs, i.e.,
compilable, runnable, and checking the specified exception type.
\MR{This is the strictest metric, ensuring that the generated \EBTs
  are semantically valid and are targeting the throw statement
  specified by developers.}

\subsubsection{\MachineView}

For \machineView, we benchmark tools ability to cover the
throw statements within a given repository:

\MyParaOnly{\TSCoverage}: out of all \tss selected in \repos
(Table~\ref{tab:machineview}), the percentage of the \tss with
successfully generated \EBTs, i.e., compilable, runnable, and checking
the correct exception type.

\subsection{Baselines}
\label{sec:exp:baseline}

\subsubsection{Learning-based tools}
We compare \Tool with one of the strongest foundation models and one
\LLM that is specifically pretrained to generate tests.

\MyParaOnly{\GPTThree}: We instruct \GPTThree~\cite{gpt3.5-turbo}
to write \EBTs by first providing one random example
from the training data. Namely, one prompt and the corresponding \gt \EBT.
The prompt
we use to query \GPTThree includes the \MUT, the target %
exception type to test, the method containing the \ts, one relevant
\nEBT, and the \testfile.
We sample a single \EBT from the output.

\MyParaOnly{\CAT}: \CAT{}~\cite{RaoETAL23CAT} is an \LLM 
pretrained on \Java and \Python repositories. It is pretrained with
a novel objective that considers the mapping between source code and
the corresponding test files. 
\MR{\CAT{} is pretrained to
generate the remaining test methods given a code under test and
the beginning of the test file.}
It has shown strong performance on
several test generation tasks. To be consistent with its pretraining
objective and \MR{intended use case}, 
we prompt \CAT with the \MUT followed by the \testfile, one
randomly-selected relevant \nEBT, and the test annotation
(`\CodeIn{@Test(expected=}'), encouraging the model to complete the
\EBT.  Just like in other cases, we sample a single \EBT.

\subsubsection{Automatic test generation tools}
In \machineView, we compare \Tool with two widely-used analysis-based
test generation tools.

\MyParaOnly{\Randoop}:
\Randoop~\cite{PachecoETAL07Randoop,RobinsonETAL11ScalingTestGen} is a
random test generation tool that creates tests by randomly generating
inputs and recording the sequences of method calls. We run \Randoop
with a time limit of 100 seconds per class to generate unit tests for
each project (per the \Randoop user manual~\cite{RandoopManual}), we
set \CodeIn{seed} to 42, \CodeIn{usethreads} to true, and other
options to their default values.

\MyParaOnly{\EvoSuite}: \EvoSuite~\cite{FraserAndArcuri11EvoSuite} is
a search based test generation tool that randomly generates inputs and
employs a genetic algorithm to evolve these inputs, aiming to maximize
code coverage.  We run \EvoSuite for 120 seconds per class (as
suggested in a recent SBST
competition~\cite{SebastianETAL22SBSTEvosuite}).  We also set the
\CodeIn{seed} to 42.
Unlike \Randoop, which generates tests for the entire project, to
generate more \EBTs for the \tss within the time limit, we generate
tests on a subset of classes when running \EvoSuite. Starting from
classes that contain throw statements, we use jdeps~\cite{JDeps} to
retrieve all classes that transitively depend on these initial
classes, thereby creating a targeted subset for evaluation.

\subsection{Hardware}
\label{sec:exp:setup}

\MR{We run \Tool's program analyses part, \Randoop, and \EvoSuite on a
machine with Intel Core i7-11700K @ 3.60GHz (8 cores, 16 threads) CPU,
64 GB RAM, Ubuntu 20.04, Java 8, and Maven 3.8.6.  We perform \Tool's \LLM
\finetuning and generation, as well as \CAT on a server
with 4 Nvidia A100 GPUs, 2 AMD Milan 7413 @ 2.65 GHz.}
We run \finetuning and generation, for \Tool and baselines, three
times with different random seeds and report the average numbers
across three runs.

\section{Results}
\label{sec:eval}

In this section, we present the evaluation results and answer each
research question.

\begin{table*}[t]
\begin{small}
\begin{center}
\caption{\UseMacro{TCap-models-user-view-with-name}\label{tab:user-view-with-name}}
\begin{tabular}{l | c  c  c  c | c  c  c  c }
\toprule
\textbf{\UseMacro{THead-models}}
 & \textbf{\UseMacro{THead-bleu-max}}
 & \textbf{\UseMacro{THead-code-bleu-max}}
 & \textbf{\UseMacro{THead-edit-sim-max}}
 & \textbf{\UseMacro{THead-xmatch-top1}}
 & \textbf{\UseMacro{THead-compilable-max}}
 & \textbf{\UseMacro{THead-match-max}}
 & \textbf{\UseMacro{THead-runnable-overall-max}}
 & \textbf{\UseMacro{THead-coverage-max}}
\\
\midrule
\UseMacro{THead-ne2e-few-shot-with-name-gpt-3.5-turbo-16k}
 & \UseMacro{res-user-view-with-name-ne2e-few-shot-with-name-gpt-3.5-turbo-16k-bleu-max}
 & \UseMacro{res-user-view-with-name-ne2e-few-shot-with-name-gpt-3.5-turbo-16k-code-bleu-max}
 & \UseMacro{res-user-view-with-name-ne2e-few-shot-with-name-gpt-3.5-turbo-16k-edit-sim-max}
 & \UseMacro{res-user-view-with-name-ne2e-few-shot-with-name-gpt-3.5-turbo-16k-xmatch-top1}
 & \UseMacro{res-user-view-with-name-ne2e-few-shot-with-name-gpt-3.5-turbo-16k-compilable-max}
 & \textbf{\UseMacro{res-user-view-with-name-ne2e-few-shot-with-name-gpt-3.5-turbo-16k-match-max}}
 & \UseMacro{res-user-view-with-name-ne2e-few-shot-with-name-gpt-3.5-turbo-16k-runnable-overall-max}
 & \UseMacro{res-user-view-with-name-ne2e-few-shot-with-name-gpt-3.5-turbo-16k-coverage-max}
\\
\UseMacro{THead-catlm-ne2e-with-name-catlm}
 & \UseMacro{res-user-view-with-name-catlm-ne2e-with-name-catlm-bleu-max}
 & \UseMacro{res-user-view-with-name-catlm-ne2e-with-name-catlm-code-bleu-max}
 & \UseMacro{res-user-view-with-name-catlm-ne2e-with-name-catlm-edit-sim-max}
 & \UseMacro{res-user-view-with-name-catlm-ne2e-with-name-catlm-xmatch-top1}
 & \UseMacro{res-user-view-with-name-catlm-ne2e-with-name-catlm-compilable-max}
 & \textbf{\UseMacro{res-user-view-with-name-catlm-ne2e-with-name-catlm-match-max}}
 & \UseMacro{res-user-view-with-name-catlm-ne2e-with-name-catlm-runnable-overall-max}
 & \UseMacro{res-user-view-with-name-catlm-ne2e-with-name-catlm-coverage-max}
\\
\UseMacro{THead-selected-434-conditionnestack2e-with-name-zero-shot-lora-codellama-7b}
 & \UseMacro{res-user-view-with-name-selected-434-conditionnestack2e-with-name-zero-shot-lora-codellama-7b-bleu-max}
 & \UseMacro{res-user-view-with-name-selected-434-conditionnestack2e-with-name-zero-shot-lora-codellama-7b-code-bleu-max}
 & \UseMacro{res-user-view-with-name-selected-434-conditionnestack2e-with-name-zero-shot-lora-codellama-7b-edit-sim-max}
 & \UseMacro{res-user-view-with-name-selected-434-conditionnestack2e-with-name-zero-shot-lora-codellama-7b-xmatch-top1}
 & \UseMacro{res-user-view-with-name-selected-434-conditionnestack2e-with-name-zero-shot-lora-codellama-7b-compilable-max}
 & \UseMacro{res-user-view-with-name-selected-434-conditionnestack2e-with-name-zero-shot-lora-codellama-7b-match-max}
 & \UseMacro{res-user-view-with-name-selected-434-conditionnestack2e-with-name-zero-shot-lora-codellama-7b-runnable-overall-max}
 & \UseMacro{res-user-view-with-name-selected-434-conditionnestack2e-with-name-zero-shot-lora-codellama-7b-coverage-max}
\\
\UseMacro{THead-conditionnestack2e-with-name-ft-lora-codellama-7b}
 & \UseMacro{res-user-view-with-name-conditionnestack2e-with-name-ft-lora-codellama-7b-bleu-max}
 & \UseMacro{res-user-view-with-name-conditionnestack2e-with-name-ft-lora-codellama-7b-code-bleu-max}
 & \UseMacro{res-user-view-with-name-conditionnestack2e-with-name-ft-lora-codellama-7b-edit-sim-max}
 & \textbf{\UseMacro{res-user-view-with-name-conditionnestack2e-with-name-ft-lora-codellama-7b-xmatch-top1}}
 & \UseMacro{res-user-view-with-name-conditionnestack2e-with-name-ft-lora-codellama-7b-compilable-max}
 & \textbf{\UseMacro{res-user-view-with-name-conditionnestack2e-with-name-ft-lora-codellama-7b-match-max}}
 & \UseMacro{res-user-view-with-name-conditionnestack2e-with-name-ft-lora-codellama-7b-runnable-overall-max}
 & \UseMacro{res-user-view-with-name-conditionnestack2e-with-name-ft-lora-codellama-7b-coverage-max}
\\
\midrule
\UseMacro{THead-conditionnestack2e-all-with-name-ft-lora-codellama-7b}
 & \textbf{\UseMacro{res-user-view-with-name-conditionnestack2e-all-with-name-ft-lora-codellama-7b-bleu-max}}
 & \textbf{\UseMacro{res-user-view-with-name-conditionnestack2e-all-with-name-ft-lora-codellama-7b-code-bleu-max}}
 & \textbf{\UseMacro{res-user-view-with-name-conditionnestack2e-all-with-name-ft-lora-codellama-7b-edit-sim-max}}
 & \UseMacro{res-user-view-with-name-conditionnestack2e-all-with-name-ft-lora-codellama-7b-xmatch-top1}
 & \textbf{\UseMacro{res-user-view-with-name-conditionnestack2e-all-with-name-ft-lora-codellama-7b-compilable-max}}
 & \textbf{\UseMacro{res-user-view-with-name-conditionnestack2e-all-with-name-ft-lora-codellama-7b-match-max}}
 & \textbf{\UseMacro{res-user-view-with-name-conditionnestack2e-all-with-name-ft-lora-codellama-7b-runnable-overall-max}}
 & \textbf{\UseMacro{res-user-view-with-name-conditionnestack2e-all-with-name-ft-lora-codellama-7b-coverage-max}}
\\
\bottomrule
\end{tabular}
\end{center}
\end{small}
\end{table*}

\begin{table*}[t]
\begin{small}
\begin{center}
\caption{\UseMacro{TCap-models-user-view-no-name}\label{tab:user-view-no-name}}
\begin{tabular}{l | c  c  c  c | c  c  c  c }
\toprule
\textbf{\UseMacro{THead-models}}
 & \textbf{\UseMacro{THead-bleu-max}}
 & \textbf{\UseMacro{THead-code-bleu-max}}
 & \textbf{\UseMacro{THead-edit-sim-max}}
 & \textbf{\UseMacro{THead-xmatch-top1}}
 & \textbf{\UseMacro{THead-compilable-max}}
 & \textbf{\UseMacro{THead-match-max}}
 & \textbf{\UseMacro{THead-runnable-overall-max}}
 & \textbf{\UseMacro{THead-coverage-max}}
\\
\midrule
\UseMacro{THead-ne2e-few-shot-no-name-gpt-3.5-turbo-16k}
 & \UseMacro{res-user-view-no-name-ne2e-few-shot-no-name-gpt-3.5-turbo-16k-bleu-max}
 & \UseMacro{res-user-view-no-name-ne2e-few-shot-no-name-gpt-3.5-turbo-16k-code-bleu-max}
 & \UseMacro{res-user-view-no-name-ne2e-few-shot-no-name-gpt-3.5-turbo-16k-edit-sim-max}
 & \UseMacro{res-user-view-no-name-ne2e-few-shot-no-name-gpt-3.5-turbo-16k-xmatch-top1}
 & \UseMacro{res-user-view-no-name-ne2e-few-shot-no-name-gpt-3.5-turbo-16k-compilable-max}
 & \textbf{\UseMacro{res-user-view-no-name-ne2e-few-shot-no-name-gpt-3.5-turbo-16k-match-max}}
 & \UseMacro{res-user-view-no-name-ne2e-few-shot-no-name-gpt-3.5-turbo-16k-runnable-overall-max}
 & \UseMacro{res-user-view-no-name-ne2e-few-shot-no-name-gpt-3.5-turbo-16k-coverage-max}
\\
\UseMacro{THead-catlm-ne2e-no-name-catlm}
 & \UseMacro{res-user-view-no-name-catlm-ne2e-no-name-catlm-bleu-max}
 & \UseMacro{res-user-view-no-name-catlm-ne2e-no-name-catlm-code-bleu-max}
 & \UseMacro{res-user-view-no-name-catlm-ne2e-no-name-catlm-edit-sim-max}
 & \UseMacro{res-user-view-no-name-catlm-ne2e-no-name-catlm-xmatch-top1}
 & \UseMacro{res-user-view-no-name-catlm-ne2e-no-name-catlm-compilable-max}
 & \textbf{\UseMacro{res-user-view-no-name-catlm-ne2e-no-name-catlm-match-max}}
 & \UseMacro{res-user-view-no-name-catlm-ne2e-no-name-catlm-runnable-overall-max}
 & \UseMacro{res-user-view-no-name-catlm-ne2e-no-name-catlm-coverage-max}
\\
\UseMacro{THead-selected-434-conditionnestack2e-no-name-zero-shot-lora-codellama-7b}
 & \UseMacro{res-user-view-no-name-selected-434-conditionnestack2e-no-name-zero-shot-lora-codellama-7b-bleu-max}
 & \UseMacro{res-user-view-no-name-selected-434-conditionnestack2e-no-name-zero-shot-lora-codellama-7b-code-bleu-max}
 & \UseMacro{res-user-view-no-name-selected-434-conditionnestack2e-no-name-zero-shot-lora-codellama-7b-edit-sim-max}
 & \UseMacro{res-user-view-no-name-selected-434-conditionnestack2e-no-name-zero-shot-lora-codellama-7b-xmatch-top1}
 & \UseMacro{res-user-view-no-name-selected-434-conditionnestack2e-no-name-zero-shot-lora-codellama-7b-compilable-max}
 & \UseMacro{res-user-view-no-name-selected-434-conditionnestack2e-no-name-zero-shot-lora-codellama-7b-match-max}
 & \UseMacro{res-user-view-no-name-selected-434-conditionnestack2e-no-name-zero-shot-lora-codellama-7b-runnable-overall-max}
 & \UseMacro{res-user-view-no-name-selected-434-conditionnestack2e-no-name-zero-shot-lora-codellama-7b-coverage-max}
\\
\UseMacro{THead-conditionnestack2e-no-name-ft-lora-codellama-7b}
 & \UseMacro{res-user-view-no-name-conditionnestack2e-no-name-ft-lora-codellama-7b-bleu-max}
 & \UseMacro{res-user-view-no-name-conditionnestack2e-no-name-ft-lora-codellama-7b-code-bleu-max}
 & \UseMacro{res-user-view-no-name-conditionnestack2e-no-name-ft-lora-codellama-7b-edit-sim-max}
 & \textbf{\UseMacro{res-user-view-no-name-conditionnestack2e-no-name-ft-lora-codellama-7b-xmatch-top1}}
 & \UseMacro{res-user-view-no-name-conditionnestack2e-no-name-ft-lora-codellama-7b-compilable-max}
 & \textbf{\UseMacro{res-user-view-no-name-conditionnestack2e-no-name-ft-lora-codellama-7b-match-max}}
 & \UseMacro{res-user-view-no-name-conditionnestack2e-no-name-ft-lora-codellama-7b-runnable-overall-max}
 & \UseMacro{res-user-view-no-name-conditionnestack2e-no-name-ft-lora-codellama-7b-coverage-max}
\\
\midrule
\UseMacro{THead-conditionnestack2e-all-no-name-ft-lora-codellama-7b}
 & \textbf{\UseMacro{res-user-view-no-name-conditionnestack2e-all-no-name-ft-lora-codellama-7b-bleu-max}}
 & \textbf{\UseMacro{res-user-view-no-name-conditionnestack2e-all-no-name-ft-lora-codellama-7b-code-bleu-max}}
 & \textbf{\UseMacro{res-user-view-no-name-conditionnestack2e-all-no-name-ft-lora-codellama-7b-edit-sim-max}}
 & \UseMacro{res-user-view-no-name-conditionnestack2e-all-no-name-ft-lora-codellama-7b-xmatch-top1}
 & \textbf{\UseMacro{res-user-view-no-name-conditionnestack2e-all-no-name-ft-lora-codellama-7b-compilable-max}}
 & \textbf{\UseMacro{res-user-view-no-name-conditionnestack2e-all-no-name-ft-lora-codellama-7b-match-max}}
 & \textbf{\UseMacro{res-user-view-no-name-conditionnestack2e-all-no-name-ft-lora-codellama-7b-runnable-overall-max}}
 & \textbf{\UseMacro{res-user-view-no-name-conditionnestack2e-all-no-name-ft-lora-codellama-7b-coverage-max}}
\\
\bottomrule
\end{tabular}
\end{center}
\end{small}
  \vspace{-20pt}
\end{table*}

\subsection{\UseMacro{rq-userView}: \XUserView}

To answer \UseMacro{rq-userView}, we compare the \EBTs generated by
\Tool with \developerwritten tests.  The results of \Tool and
baselines are shown in tables~\ref{tab:user-view-with-name} and
\ref{tab:user-view-no-name}.
Table~\ref{tab:user-view-with-name} presents the results when we
inform \LLMs the method name of the target \EBT while in
Table~\ref{tab:user-view-no-name} we do not.  (We describe the last
row in these tables in a later subsection.)

\Tool outperforms all the baselines on both similarity-based metrics
(left side in tables) and functional-correctness metrics (right side
in tables).
\Tool achieves higher performance than baselines for both generating
executable \EBTs (Runnable\%) and \EBTs that cover the \tss
(\TSCoverage).
\MR{This highlights that \Tool can generate more \EBTs that can be directly adopted by developers.}
In Table~\ref{tab:user-view-with-name}, we can see that \Tool
outperforms \GPTThree by \OverGPTRun{} and \OverGPTCov{} on
\runnable and \TSCoverage, respectively.
Similarly, we can see that \Tool outperforms CAT-LM by \OverCATRun{}
and \OverCATCov{} on \runnable and \TSCoverage, respectively.
This further underlines the benefit of the \stacktraces and \guardexps
extracted via program analysis, and \Tool's capability of reasoning
about them.

\MR{To further understand the performance difference, we inspect the
  \EBTs generated by \GPTThree and \Tool.
  Although \GPTThree generates comparable number of compilable \EBTs
as \Tool, it struggles to cover the correct \tss especially when they
are not in the \MUT (but could be reached through a sequence of method
calls).
For example, in Figure~\ref{fig:expression}, we show the \guardexp extracted by \Tool
to trigger the \CodeIn{IllegalAccessException} with regard to the first argument (\CodeIn{className}) of the \MUT
(\Code{createFileWriter}) in Figure~\ref{fig:mut-example}.
The \EBT generated by \GPTThree can be compiled but fails to check the \CodeIn{IllegalAccessException} (Figure~\ref{fig:gpt3-ebt}).
\Tool uses the correct class \CodeIn{``java.lang.String"} that satisfies the condition and successfully covers the
  target throw statement (Figure~\ref{fig:exlong-ebt}).}

\begin{figure}[t]
  \begin{subfigure}{\columnwidth}
\begin{lstlisting}[language=java-pretty, numbers=none, xleftmargin=0pt]
public static FileWriter createFileWriter(String className, LogFilePath logFilePath, CompressionCodec codec, SecorConfig config)
    throws Exception {
    return createFileReaderWriterFactory(className, config).BuildFileWriter(logFilePath, codec);
}
\end{lstlisting}
  \vspace{-5pt}
  \caption{The \MUT to be tested.}\label{fig:mut-example}
\end{subfigure}
  \begin{subfigure}{\columnwidth}
\begin{lstlisting}[language=java-pretty, numbers=none, xleftmargin=0pt]
!FileReaderWriterFactory.class.isAssignableFrom(Class.forName(className))
\end{lstlisting}
  \vspace{-5pt}
  \caption{The \guardexp for the target throw statement.}\label{fig:expression}
\end{subfigure}
\begin{subfigure}{\columnwidth}
\begin{lstlisting}[language=java-pretty, numbers=none, xleftmargin=0pt]
@Test(expected = IllegalArgumentException.class)
public void testFileWriterConstructorMissing() throws Exception {
    ReflectionUtil.createFileWriter((*@\ColorBack{"MissingClass"}@*), mLogFilePath, null, mSecorConfig);
}
\end{lstlisting}
  \vspace{-5pt}
  \caption{Compilable but failing \EBT generated by \GPTThree}\label{fig:gpt3-ebt}
\end{subfigure}
\begin{subfigure}{\columnwidth}
\begin{lstlisting}[language=java-pretty, numbers=none, xleftmargin=0pt]
@Test(expected = IllegalArgumentException.class)
public void testFileWriterConstructorMissing() throws Exception {
    ReflectionUtil.createFileWriter("java.lang.String", mLogFilePath, null, mSecorConfig);
}
\end{lstlisting}
  \vspace{-5pt}
  \caption{Compilable and runnable \EBT generated by \Tool}\label{fig:exlong-ebt}
\end{subfigure}
\caption{\MR{\EBT (\CodeIn{testFileWriterConstructorMissing})
    generated by \GPTThree and \Tool. The \EBT generated by \Tool
 covers the target throw statement satisfying the correct condition.}\label{fig:compile-error-example}}
\vspace{-10pt}
\end{figure}

Comparing functional-correctness metrics in
Table~\ref{tab:user-view-with-name} and
Table~\ref{tab:user-view-no-name}, performance of both \GPTThree and
CAT-LM declines %
significantly when the \EBT method name is omitted.
This result aligns with expectations, as the method name frequently
implies the conditions under which the exception is supposed to be
thrown, \eg, ``\CodeIn{should\_fail\_if\_time\_provider\_is\_null}''.
In contrast, \Tool demonstrates robustness regarding the inclusion or
exclusion of the test method name, maintaining consistent performance
on functional-correctness metrics.
This further emphasizes the reasoning ability of \Tool on the
\stacktraces and \guardexps.

\subsection{\UseMacro{rq-stackTraceCondition}: Ablation Study of Stack Traces and Conditions}
\label{sec:results:ablation}
To evaluate the contribution of the components in \Tool, we perform an
ablation study.  In Table~\ref{tab:exlong-ablation}, we show the results
while including \EBT's name in the prompt.  We find that ablating each
component
deteriorates performance especially across functional correctness
metrics.  Removing \stacktraces slightly hurts the performance of
\Tool \MR{in terms of functional correctness} which is expected, because a \guardexp is, in a way, the summary
of a \stacktrace.

We observe a small drop in \compile but a rather larger drop in
\TSCoverage if removing both \stacktrace and \guardexp from the context.
This underlines the importance of the condition information present by
these
two components on which \Tool reasons about when generating \EBTs that
cover the \tss.  Relevant \nEBTs mostly contribute to the
\compile because the relevant \nEBT covers the same
\MUT which gives model a starting point to construct the \EBTs.
\MR{The ablation study underscores the importance of each
  component in \Tool.}

\subsection{\UseMacro{rq-nEBTs}: Selection of \nEBTs}
\label{sec:results:nebt}

In addition to randomly choosing one relevant \nEBT to add to the
prompt for \Tool, we try running the inference of \Tool for no more than \SampleSize{} times,
each time with a different relevant \nEBT, and reporting the best performance,
dubbed \ToolMoreNebt in
tables~\ref{tab:user-view-with-name} and \ref{tab:user-view-no-name}.
We see that that performance differences can be substantial.

To study how the diversity of \nEBTs used in the prompt affect the
\Tool performance, our ablation study on the number of different
relevant \nEBTs 
is shown in Table~\ref{tab:netest-diversity}. We present results for 
1)~\Tool generating only one \EBT for each \ts{}; 
2)~\ToolMoreNebt but using the same \nEBT as 1) in multiple inference runs;
3)~\ToolMoreNebt using different \nEBTs in multiple inference runs.
For a fair comparison, we keep the number of
generated \EBTs for 2) and 3) same and we report the best-of-$k$ metrics on a
subset of \UseMacro{NumDiversitySubset} examples on which we extracted more than one relevant \nEBT.

We find that increasing the number of sampled \EBTs improves the
performance when compared to only generating one \EBT, which can be
attributed to the randomness.
Moreover, sampling with a diverse set of \nEBTs further boost the performance
across most of the metrics under the same sample size.  This motivates
a new research question on how to choose better \nEBTs to the prompt
for \EBT generation, which we leave for future work.

\subsection{\MR{\UseMacro{rq-GPT4}: \Tool Built with \GPTFour}}

\begin{table*}[t]
\begin{small}
\begin{center}
\caption{\UseMacro{TCap-models-exlong-ablation}\label{tab:exlong-ablation}}
\setlength{\tabcolsep}{1.5pt}
\begin{tabular}{l | c  c  c  c | c  c  c  c }
\toprule
\textbf{\UseMacro{THead-models}}
 & \textbf{\UseMacro{THead-bleu-max}}
 & \textbf{\UseMacro{THead-code-bleu-max}}
 & \textbf{\UseMacro{THead-edit-sim-max}}
 & \textbf{\UseMacro{THead-xmatch-top1}}
 & \textbf{\UseMacro{THead-compilable-max}}
 & \textbf{\UseMacro{THead-match-max}}
 & \textbf{\UseMacro{THead-runnable-overall-max}}
 & \textbf{\UseMacro{THead-coverage-max}}
\\
\midrule
\UseMacro{THead-conditionnestack2e-with-name-ft-lora-codellama-7b}
 & \textbf{\UseMacro{res-exlong-ablation-conditionnestack2e-with-name-ft-lora-codellama-7b-bleu-max}}
 & \textbf{\UseMacro{res-exlong-ablation-conditionnestack2e-with-name-ft-lora-codellama-7b-code-bleu-max}}
 & \UseMacro{res-exlong-ablation-conditionnestack2e-with-name-ft-lora-codellama-7b-edit-sim-max}
 & \textbf{\UseMacro{res-exlong-ablation-conditionnestack2e-with-name-ft-lora-codellama-7b-xmatch-top1}}
 & \textbf{\UseMacro{res-exlong-ablation-conditionnestack2e-with-name-ft-lora-codellama-7b-compilable-max}}
 & \textbf{\UseMacro{res-exlong-ablation-conditionnestack2e-with-name-ft-lora-codellama-7b-match-max}}
 & \textbf{\UseMacro{res-exlong-ablation-conditionnestack2e-with-name-ft-lora-codellama-7b-runnable-overall-max}}
 & \textbf{\UseMacro{res-exlong-ablation-conditionnestack2e-with-name-ft-lora-codellama-7b-coverage-max}}
\\
\midrule
\UseMacro{THead-conditionne2e-with-name-ft-lora-codellama-7b}
 & \UseMacro{res-exlong-ablation-conditionne2e-with-name-ft-lora-codellama-7b-bleu-max}
 & \UseMacro{res-exlong-ablation-conditionne2e-with-name-ft-lora-codellama-7b-code-bleu-max}
 & \textbf{\UseMacro{res-exlong-ablation-conditionne2e-with-name-ft-lora-codellama-7b-edit-sim-max}}
 & \UseMacro{res-exlong-ablation-conditionne2e-with-name-ft-lora-codellama-7b-xmatch-top1}
 & \UseMacro{res-exlong-ablation-conditionne2e-with-name-ft-lora-codellama-7b-compilable-max}
 & \textbf{\UseMacro{res-exlong-ablation-conditionne2e-with-name-ft-lora-codellama-7b-match-max}}
 & \UseMacro{res-exlong-ablation-conditionne2e-with-name-ft-lora-codellama-7b-runnable-overall-max}
 & \UseMacro{res-exlong-ablation-conditionne2e-with-name-ft-lora-codellama-7b-coverage-max}
\\
\UseMacro{THead-ne2e-with-name-ft-lora-codellama-7b}
 & \UseMacro{res-exlong-ablation-ne2e-with-name-ft-lora-codellama-7b-bleu-max}
 & \UseMacro{res-exlong-ablation-ne2e-with-name-ft-lora-codellama-7b-code-bleu-max}
 & \UseMacro{res-exlong-ablation-ne2e-with-name-ft-lora-codellama-7b-edit-sim-max}
 & \UseMacro{res-exlong-ablation-ne2e-with-name-ft-lora-codellama-7b-xmatch-top1}
 & \UseMacro{res-exlong-ablation-ne2e-with-name-ft-lora-codellama-7b-compilable-max}
 & \textbf{\UseMacro{res-exlong-ablation-ne2e-with-name-ft-lora-codellama-7b-match-max}}
 & \UseMacro{res-exlong-ablation-ne2e-with-name-ft-lora-codellama-7b-runnable-overall-max}
 & \UseMacro{res-exlong-ablation-ne2e-with-name-ft-lora-codellama-7b-coverage-max}
\\
\UseMacro{THead-mut2e-with-name-ft-lora-codellama-7b}
 & \UseMacro{res-exlong-ablation-mut2e-with-name-ft-lora-codellama-7b-bleu-max}
 & \UseMacro{res-exlong-ablation-mut2e-with-name-ft-lora-codellama-7b-code-bleu-max}
 & \UseMacro{res-exlong-ablation-mut2e-with-name-ft-lora-codellama-7b-edit-sim-max}
 & \UseMacro{res-exlong-ablation-mut2e-with-name-ft-lora-codellama-7b-xmatch-top1}
 & \UseMacro{res-exlong-ablation-mut2e-with-name-ft-lora-codellama-7b-compilable-max}
 & \textbf{\UseMacro{res-exlong-ablation-mut2e-with-name-ft-lora-codellama-7b-match-max}}
 & \UseMacro{res-exlong-ablation-mut2e-with-name-ft-lora-codellama-7b-runnable-overall-max}
 & \UseMacro{res-exlong-ablation-mut2e-with-name-ft-lora-codellama-7b-coverage-max}
\\
\bottomrule
\end{tabular}
\end{center}
\end{small}
\vspace{-10pt}
\end{table*}

\begin{table*}[t]
\begin{small}
\begin{center}
\caption{\UseMacro{TCap-models-netest-diversity}\label{tab:netest-diversity}}
\setlength{\tabcolsep}{1.5pt}
\begin{tabular}{l | c  c  c  c | c  c  c  c }
\toprule
\textbf{\UseMacro{THead-models}}
 & \textbf{\UseMacro{THead-bleu-max}}
 & \textbf{\UseMacro{THead-code-bleu-max}}
 & \textbf{\UseMacro{THead-edit-sim-max}}
 & \textbf{\UseMacro{THead-xmatch-top1}}
 & \textbf{\UseMacro{THead-compilable-max}}
 & \textbf{\UseMacro{THead-match-max}}
 & \textbf{\UseMacro{THead-runnable-overall-max}}
 & \textbf{\UseMacro{THead-coverage-max}}
\\
\midrule
\UseMacro{THead-selected-253-conditionnestack2e-with-name-ft-lora-codellama-7b}
 & \UseMacro{res-netest-diversity-selected-253-conditionnestack2e-with-name-ft-lora-codellama-7b-bleu-max}
 & \UseMacro{res-netest-diversity-selected-253-conditionnestack2e-with-name-ft-lora-codellama-7b-code-bleu-max}
 & \UseMacro{res-netest-diversity-selected-253-conditionnestack2e-with-name-ft-lora-codellama-7b-edit-sim-max}
 & \UseMacro{res-netest-diversity-selected-253-conditionnestack2e-with-name-ft-lora-codellama-7b-xmatch-top1}
 & \UseMacro{res-netest-diversity-selected-253-conditionnestack2e-with-name-ft-lora-codellama-7b-compilable-max}
 & \textbf{\UseMacro{res-netest-diversity-selected-253-conditionnestack2e-with-name-ft-lora-codellama-7b-match-max}}
 & \UseMacro{res-netest-diversity-selected-253-conditionnestack2e-with-name-ft-lora-codellama-7b-runnable-overall-max}
 & \UseMacro{res-netest-diversity-selected-253-conditionnestack2e-with-name-ft-lora-codellama-7b-coverage-max}
\\
\UseMacro{THead-conditionnestack2e-sample-with-name-ft-lora-codellama-7b}
 & \UseMacro{res-netest-diversity-conditionnestack2e-sample-with-name-ft-lora-codellama-7b-bleu-max}
 & \UseMacro{res-netest-diversity-conditionnestack2e-sample-with-name-ft-lora-codellama-7b-code-bleu-max}
 & \UseMacro{res-netest-diversity-conditionnestack2e-sample-with-name-ft-lora-codellama-7b-edit-sim-max}
 & \UseMacro{res-netest-diversity-conditionnestack2e-sample-with-name-ft-lora-codellama-7b-xmatch-top1}
 & \UseMacro{res-netest-diversity-conditionnestack2e-sample-with-name-ft-lora-codellama-7b-compilable-max}
 & \textbf{\UseMacro{res-netest-diversity-conditionnestack2e-sample-with-name-ft-lora-codellama-7b-match-max}}
 & \UseMacro{res-netest-diversity-conditionnestack2e-sample-with-name-ft-lora-codellama-7b-runnable-overall-max}
 & \UseMacro{res-netest-diversity-conditionnestack2e-sample-with-name-ft-lora-codellama-7b-coverage-max}
\\
\UseMacro{THead-diversity-conditionnestack2e-all-with-name-ft-lora-codellama-7b}
 & \textbf{\UseMacro{res-netest-diversity-diversity-conditionnestack2e-all-with-name-ft-lora-codellama-7b-bleu-max}}
 & \textbf{\UseMacro{res-netest-diversity-diversity-conditionnestack2e-all-with-name-ft-lora-codellama-7b-code-bleu-max}}
 & \textbf{\UseMacro{res-netest-diversity-diversity-conditionnestack2e-all-with-name-ft-lora-codellama-7b-edit-sim-max}}
 & \textbf{\UseMacro{res-netest-diversity-diversity-conditionnestack2e-all-with-name-ft-lora-codellama-7b-xmatch-top1}}
 & \textbf{\UseMacro{res-netest-diversity-diversity-conditionnestack2e-all-with-name-ft-lora-codellama-7b-compilable-max}}
 & \textbf{\UseMacro{res-netest-diversity-diversity-conditionnestack2e-all-with-name-ft-lora-codellama-7b-match-max}}
 & \textbf{\UseMacro{res-netest-diversity-diversity-conditionnestack2e-all-with-name-ft-lora-codellama-7b-runnable-overall-max}}
 & \textbf{\UseMacro{res-netest-diversity-diversity-conditionnestack2e-all-with-name-ft-lora-codellama-7b-coverage-max}}
\\
\bottomrule
\end{tabular}
\end{center}
\end{small}
\vspace{-10pt}
\end{table*}

\begin{table*}[t]
\begin{small}
\begin{center}
\caption{\UseMacro{TCap-models-gpt4o-with-name}\label{tab:gpt4o-with-name}}
\begin{tabular}{l | c  c  c  c | c  c  c  c }
\toprule
\textbf{\UseMacro{THead-models}}
 & \textbf{\UseMacro{THead-bleu-max}}
 & \textbf{\UseMacro{THead-code-bleu-max}}
 & \textbf{\UseMacro{THead-edit-sim-max}}
 & \textbf{\UseMacro{THead-xmatch-top1}}
 & \textbf{\UseMacro{THead-compilable-max}}
 & \textbf{\UseMacro{THead-match-max}}
 & \textbf{\UseMacro{THead-runnable-overall-max}}
 & \textbf{\UseMacro{THead-coverage-max}}
\\
\midrule
\UseMacro{THead-ne2e-few-shot-with-name-gpt-4o}
 & \UseMacro{res-gpt4o-with-name-ne2e-few-shot-with-name-gpt-4o-bleu-max}
 & \UseMacro{res-gpt4o-with-name-ne2e-few-shot-with-name-gpt-4o-code-bleu-max}
 & \UseMacro{res-gpt4o-with-name-ne2e-few-shot-with-name-gpt-4o-edit-sim-max}
 & \UseMacro{res-gpt4o-with-name-ne2e-few-shot-with-name-gpt-4o-xmatch-top1}
 & \UseMacro{res-gpt4o-with-name-ne2e-few-shot-with-name-gpt-4o-compilable-max}
 & \UseMacro{res-gpt4o-with-name-ne2e-few-shot-with-name-gpt-4o-match-max}
 & \UseMacro{res-gpt4o-with-name-ne2e-few-shot-with-name-gpt-4o-runnable-overall-max}
 & \UseMacro{res-gpt4o-with-name-ne2e-few-shot-with-name-gpt-4o-coverage-max}
\\
\UseMacro{THead-conditionnestack2e-few-shot-with-name-gpt-4o}
 & \textbf{\UseMacro{res-gpt4o-with-name-conditionnestack2e-few-shot-with-name-gpt-4o-bleu-max}}
 & \textbf{\UseMacro{res-gpt4o-with-name-conditionnestack2e-few-shot-with-name-gpt-4o-code-bleu-max}}
 & \textbf{\UseMacro{res-gpt4o-with-name-conditionnestack2e-few-shot-with-name-gpt-4o-edit-sim-max}}
 & \textbf{\UseMacro{res-gpt4o-with-name-conditionnestack2e-few-shot-with-name-gpt-4o-xmatch-top1}}
 & \textbf{\UseMacro{res-gpt4o-with-name-conditionnestack2e-few-shot-with-name-gpt-4o-compilable-max}}
 & \textbf{\UseMacro{res-gpt4o-with-name-conditionnestack2e-few-shot-with-name-gpt-4o-match-max}}
 & \textbf{\UseMacro{res-gpt4o-with-name-conditionnestack2e-few-shot-with-name-gpt-4o-runnable-overall-max}}
 & \textbf{\UseMacro{res-gpt4o-with-name-conditionnestack2e-few-shot-with-name-gpt-4o-coverage-max}}
\\
\bottomrule
\end{tabular}
\end{center}
\end{small}
\vspace{-10pt}
\end{table*}

\MR{
Our technique, which assists \LLMs in generating \EBTs
can be transferred to other \LLMs with significant benefits.
In addition to results with open-sourced \LLMs in prior sections, we further
present the results of \Tool built on \GPTFour~\cite{gpt4o}, currently one of the most powerful proprietary \LLMs.
Table~\ref{tab:gpt4o-with-name} compares the performance of \exlongGPTFour with baseline \GPTFour under the \userView.
The inputs to \GPTFour includes the \MUT, the target
exception type, the method containing the \ts, one relevant
\nEBT, and the \testfile.
We instruct \exlongGPTFour with the
prompt containing \MUT, \stacktrace, \guardexp, relevant \nEBTs and the \testfile.
One example pair of input and \gt \EBT
is provided for both models.
Results show that \exlongGPTFour outperforms
\GPTFour across both similarity metrics and functional-correctness
metrics.
}

\subsection{\UseMacro{rq-machineView}: \XMachineView}

\begin{table}[t]
\begin{small}
\begin{center}
\caption{\UseMacro{TCap-Tool-cmp-table}\label{tab:coverage}}
\begin{tabular}{l | c  c }
\toprule
\multirow{2}{*}{\textbf{Tools}}
 & \multicolumn{2}{c}{\textbf{\TSCoverage}}
\\
& \UseMacro{THead-subset-projects}
& \UseMacro{THead-all-projects}
\\
\midrule
\Tool
 & \textbf{\UseMacro{res-tools-cmp-conditionnestack2e-all-no-name-ft-lora-codellama-7b-eval-rq2-intersect-coverage-max}}
 & \textbf{\UseMacro{res-tools-cmp-conditionnestack2e-all-no-name-ft-lora-codellama-7b-eval-rq2-coverage-max}}
\\
EvoSuite
 & \UseMacro{Evosuite-subset-projects-coverage}
 & \UseMacro{Evosuite-all-projects-coverage}
\\
Randoop
 & \UseMacro{Randoop-subset-projects-coverage}
 & \UseMacro{Randoop-all-projects-coverage}
\\
\bottomrule
\end{tabular}
\end{center}
\end{small}
\vspace{-20px}
\end{table}

\begin{figure}[t!]
  \centering
  \begin{subfigure}[t]{0.23\textwidth}
    \centering
    \includegraphics[width=\textwidth]{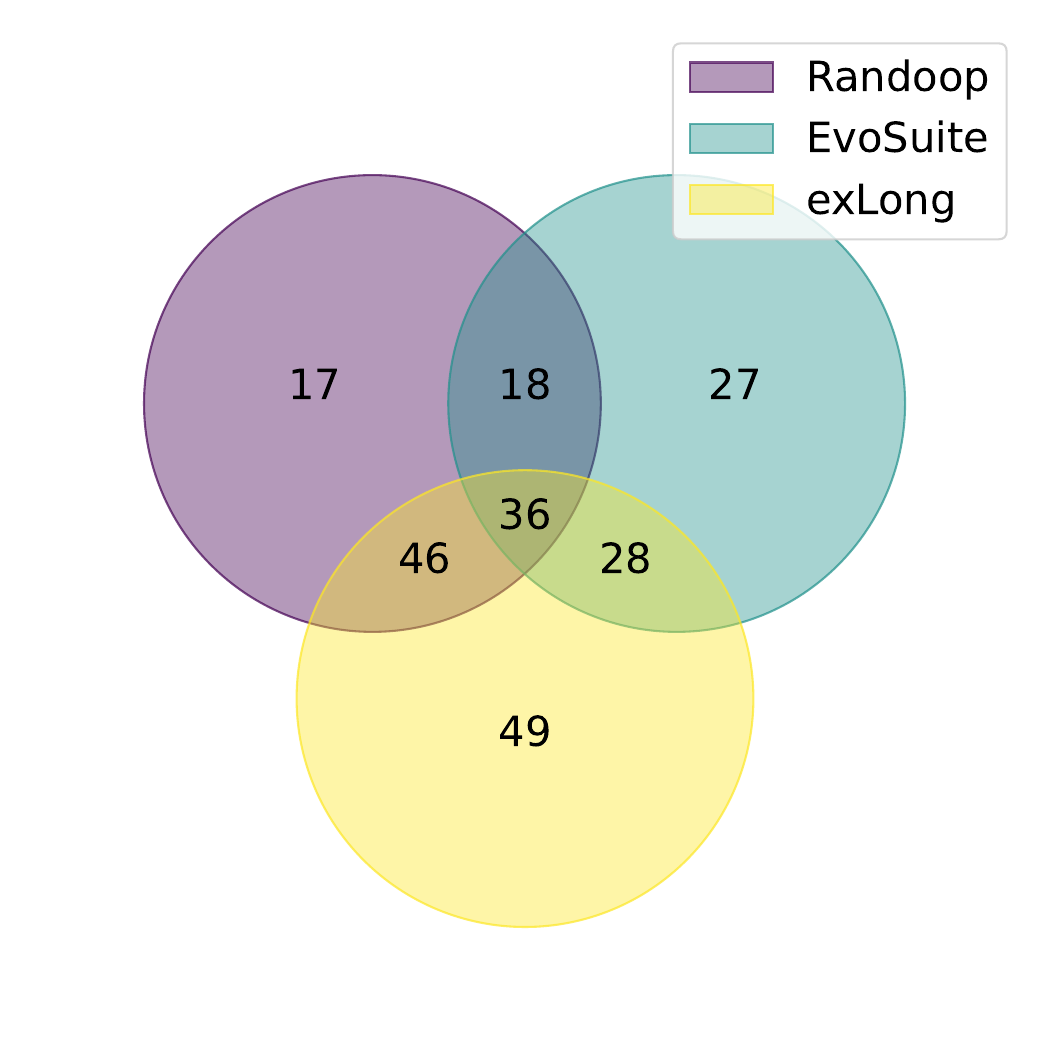}
    \vspace{-28pt}
    \caption{Covered throw statements on a subset of \overlapEvalProjects{} projects.}
  \end{subfigure}
  \hfill
  \begin{subfigure}[t]{0.23\textwidth}
    \centering
    \includegraphics[width=\textwidth]{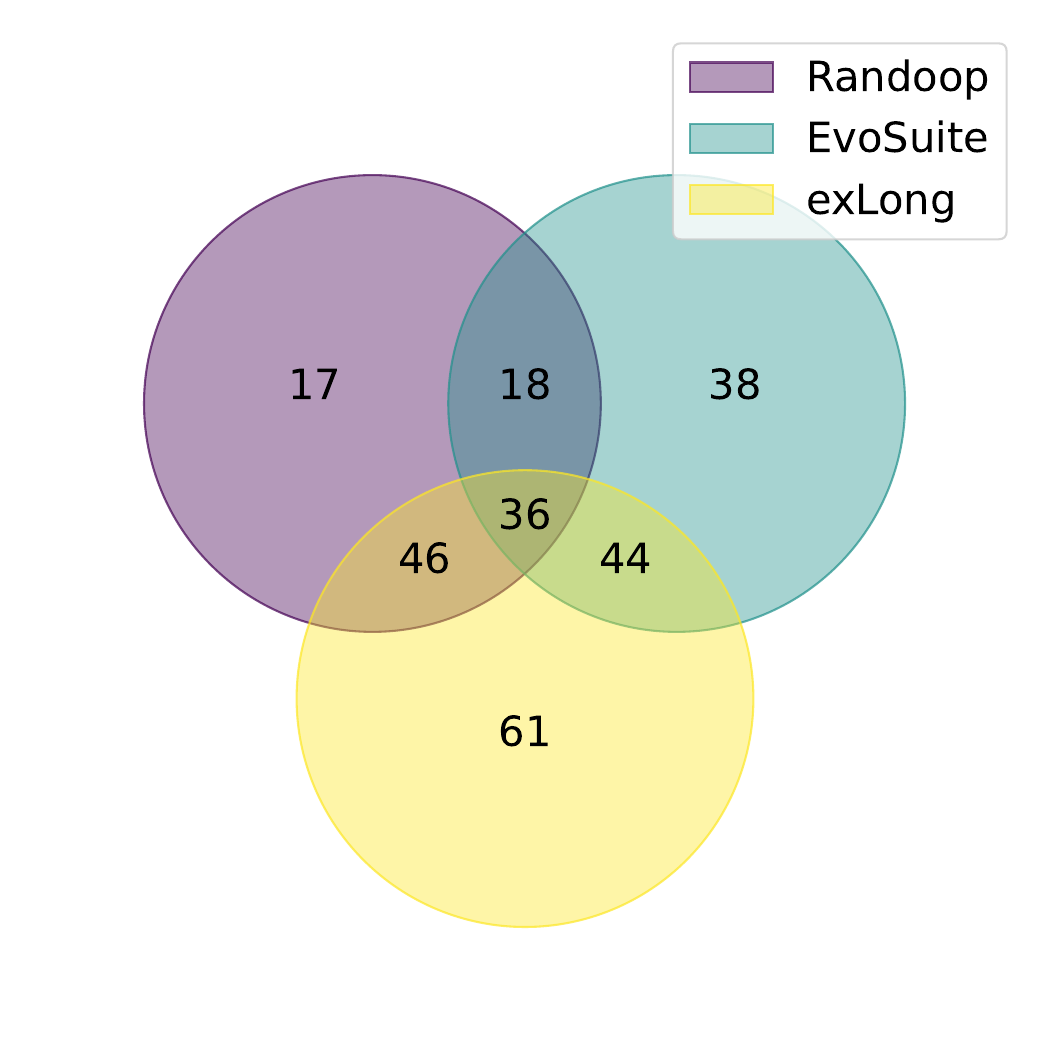}
    \vspace{-28pt}
    \caption{Covered throw statements on all \allEvalProjects{} projects.}
  \end{subfigure}
  \caption{Venn diagram that shows \tss coverage by \Tool, \Randoop, and \EvoSuite.\label{fig:Venn}}
  \vspace{-15pt}
\end{figure}

In Table~\ref{tab:coverage}, we present the throw statement coverage
rate of \ToolMoreNebt and two analysis-based test generation tools.
\EvoSuite could generate tests for all projects, while \Randoop could
not generate tests for \UseMacro{RandoopFailProjs} projects. We
inspected the issues and found that: (1)~\Randoop crashed on
\CodeIn{OpenNMS/newts} because this project kept throwing runtime
exceptions (all related to \CodeIn{com.codahale.metrics.ScheduledReporter});
(2)~\Randoop crashed on \CodeIn{pinterest/secor} because this project
requires the configuration of Kafka; (3)~\Randoop could not load a
class and crashed on \CodeIn{OpenHFT/Chronicle\text{-}Map}.

We report results on the subset of \overlapEvalProjects{} projects
where all the tools can be run successfully (Subset Projects) and
results on all \allEvalProjects{} projects (All Projects).
Among the given \tss, \Tool achieves higher throw statement coverage
rate than analysis-based tools.  Figure~\ref{fig:Venn} illustrates the
overlap and difference among the sets of \tss covered by \Tool,
\EvoSuite, and \Randoop. All three tools cover different sets of throw
statements.  \Tool covers the most \tss that other two cannot.

\section{\MR{Case Study}}
\label{sec:case-study}

\MR{We performed a case study where we submitted the \EBTs generated 
  by \Tool to the open-source projects (where the data was extracted 
  from) to collect developers' feedback.
  Among the evaluation set for
  \machineView, \Tool generated \NumOfGoodEBTs{} \EBTs across \allEvalProjects{} projects
  that are runnable and cover the correct throw statements.
  We selected a subset of \NumOfPRprojects{} projects that are actively maintained, i.e.,
  they had at least one commit, accepted pull request
  (PR) or responded issue within the past six months (at the time of the paper submission).  We found that
  the generated \EBTs of \NumOfNPRprojects{} projects were the same as
  those added by developers on later commits (commits after the ones
  we used during evaluation), thus refrained from submitting PRs to
  them.  In total, we submitted \NumOfPRs{} PRs which include \NumSubmtEBTS{} \EBTs (one PR per project).
  Among them, \NumOfAcPRs{} PRs (\NumOfMergedTests{}
  \EBTs) have been accepted, and \NumPendPRs{} PRs (\NumPendEBTs{}
  \EBTs) are still pending.  No PR was rejected.  In one
  instance, a developer responded and merged our PR only 30 minutes
  after we create the PR.  This was encouraging, and future tool
  development should integrate \Tool into an IDE, such that \Tool
  continuously provide \EBTs for code that a developer is editing.}

\section{Limitations}
\label{sec:limitations}

We discuss several limitations and potential future work.

\MyPara{Programming language}
In this work, we focused on supporting the Java programming language,
which is among the most popular languages nowadays.  We expect no
substantial differences in our approach for similar programming
languages, \eg, C\#.  Future work could evaluate and tune our model
for dynamically typed languages, \eg, Python.

\MyPara{Project boundaries}
In our evaluation, we generate \EBTs for throw statements within a
single project, and we ignore throw statements that are in libraries
used by the project.
We could not come up with a use case that targets throw statements in
libraries, so we left it out of our work.  If we were to target such a
case, we would need to collect context (throw statement and
conditions) from those libraries.  One could take several directions,
e.g., finding code of those libraries, building a model on bytecode
level, or decompiling code and then extracting the context.

\MyPara{\Testfile}
Not every \MUT has a \testfile. We leave the problem on finding or
generating \testfile for any given method under test as future work.

\MyPara{\LLMs}
We built \Tool around \CodeLlama~\cite{roziere2023code}, a recent
open-source model.
We believe that building on \CodeLlama provides reproducibility
guarantees that will help us and others to build on this work.
Our contributions are the task, definition of a context for the task,
tools for extracting the context, instruction fine-tuned model, and
extensive evaluation.
We expect that building \Tool on other open-source \LLMs
would lead to similar results.

\section{Related Work}
\label{sec:related:work}

There has been significant work on
test generation~\cite{RaoETAL23CAT,
PachecoETAL07Randoop, FraserAndArcuri11EvoSuite, ernst2007daikon,
wang2024software, siddiq2024using}
and code
generation~\cite{zhang2023multilingual,zhang2022coditt5,
  PanthaplackelETAL20CommentUpdate,muennighoff2023octopack,lu2023llama,li2023starcoder,nijkamp2022conversational}.
We cover related work
on
(1)~\LLM-based test generation, (2)~generating tests for exceptional
behavior, and (3)~other test generation techniques.

\MyPara{LLM-based test generation}
Transformer models have been used to generate
tests~\cite{wang2024software, el2024using, SchaferETAL23TestPilot,
TufanoETAL20TestGeneration, RaoETAL23CAT, NieETAL23TeCo, Nie23Thesis,
LemieuxETAL23CodaMosa, yuan2023no} and test
oracles~\cite{WatsonETAL20ATLAS, DinellaETAL22TOGA,
tufano2022generating}. \CAT~\cite{RaoETAL23CAT} is a 2.7B model that
is \pretrained on a large dataset of \Java and \Python projects. It
outperforms existing test generation tools
StarCoder~\cite{li2023starcoder} and CodeGen
16B~\cite{nijkamp2022conversational} in terms of the number of valid
tests and test completion tool TeCo~\cite{NieETAL23TeCo, Nie23Thesis}.
So we compared our work with \CAT in this paper.

Conditions are useful for guiding the generation of tests and finding
bugs~\cite{bouzenia2023say, blasi2022call, ryan2024code}.
SymPrompt~\cite{ryan2024code} introduced path constraint prompting to
guide \LLMs to generate high-coverage tests without additional
training.
They collect constraints from
each possible execution path in the target method and prompt
the LLM to generate tests that cover those
paths. We extract guard expressions by analyzing multiple methods along the stack trace starting from the throw statement to the target method for
generating EBTs.

Existing test cases (including the setup and teardown methods) serve
as a useful context to guide the generation~\cite{RaoETAL23CAT,
TufanoETAL20TestGeneration, NieETAL23TeCo}. Haji et
al.~\cite{el2024using} empirically studies the effectiveness of
generating tests using GitHub Copilot and discovers that using
existing test cases as context can increase the passing rate of
generated tests by 37.73\%. Our work uses existing \nEBTs as context
and collects \stacktraces from those tests to guide the generation of
\EBTs.

\MyPara{Generating tests for exceptional behavior}
Exception handling~\cite{ZhangETAL21LearningToHandleExceptions,
ZhangETAL23DetectingExceptionHandlingBugsInCpp,
MarcilioFuria21HowJavaProgrammersTestExceptionalBehavior,
zhong2022exception} is an important aspect of software development.
There are several
techniques~\cite{TraceyETAL00TestDataGenerationForExceptionConditions,
GoffiETAL16AutomaticGenerationOfOraclesForExceptionalBehaviors,
BernardoETAL11AgileTestingOfExceptionalBehavior,
AlmullaETAL20GeneratingExceptionTriggeringTests} to generate \EBTs.
However, they either generate \EBTs from specifications or use
random-based or search-based strategies to generate \EBTs. Our work is
the first to use \LLMs to generate \EBTs. Also, prior work generates tests for
the whole program, while our \Tool allows users to specify which throw
statements to cover. 

Guo et al.~\cite{guo2024optimal} introduces boundary coverage distance
(BCD) to evaluate the quality of test inputs, which can be used to
guide the random generation of test inputs by minimizing BCD.
Goffi~\cite{GoffiETAL16AutomaticGenerationOfOraclesForExceptionalBehaviors}
proposed throw statement coverage to measure the effectiveness of test
inputs in triggering exceptions. We also use throw statement coverage
in our evaluation.

\MyPara{Other test generation techniques} 
Other techniques incorporate random-based~\cite{PachecoETAL07Randoop,
RobinsonETAL11ScalingTestGen},
search-based~\cite{FraserAndArcuri11EvoSuite, harman2009theoretical, LiuISSTA23EXLI, LiuFSE24EXLI}
and constraint-based~\cite{ernst2007daikon,
HolmesETAL20TestGeneration,godefroid2012test} strategies to
automatically generate tests. Tests can be derived from multiple
sources, including the code under test, error
messages~\cite{han2018perflearner}, and
specifications~\cite{MotwaniETAL19OraclesFromNLSpec} like
comments~\cite{TanETAL12TComment, tan2007icomment, tan2011acomment}.

\Randoop~\cite{PachecoETAL07Randoop, RobinsonETAL11ScalingTestGen}
generates tests by randomly generating inputs and saving the sequences
of method calls. \EvoSuite~\cite{FraserAndArcuri11EvoSuite} is a
search based test generation tool that randomly generates inputs and
uses a genetic algorithm to evolve the inputs to maximize code
coverage.  During the test generation process, both \Randoop and
\EvoSuite create tests that cover normal as well as exceptional
behaviors. However, since they generate inputs randomly, they do not
guarantee to generate tests with high coverage or meaningful and
readable inputs. Moreover, there is no assurance that these inputs
will successfully trigger certain exceptional behaviors.
EvoSuiteFIT~\cite{AlmullaETAL20GeneratingExceptionTriggeringTests}
adapted \EvoSuite's search algorithm with reinforcement learning to
generate exceptional tests. Unfortunately, we cannot directly use
EvoSuiteFIT (has error ``Invalid or corrupt jarfile'').

\section{Conclusion}

We presented the first work on generating tests for exceptional
behavior (\EBTs) using large language models.  We introduced \Tool
that builds on top of \CodeLlama to embed reasoning about traces that
lead to throw statements, conditional expressions along those traces,
and non-exceptional tests that cover similar traces.
We evaluated \Tool in two use cases: \userView (\ie, generate \EBT for
a given method under test and a target throw statement) and
\machineView (\ie, automatically generate tests for all throw
statements available in a repository).  Our results show that \Tool
outperforms existing test generation models and analysis-based test
generation tools.  We contributed a number of tests generated by \Tool
to open-source projects, and \NumOfMergedTests{} \EBTs are already
accepted.  We believe that \Tool targets an important task, has good
performance, and helps developers increase code quality assurance by
automatically providing high quality \EBTs.

\section*{Acknowledgments}
We thank Nader Al Awar, Jayanth Srinivasa, Aditya Thimmaiah, Zijian
Yi, Samuel Yuan, Zhiqiang Zang, Linghan Zhong, and the anonymous
reviewers for their comments and feedback.
This work is partially supported by the US National Science Foundation
under Grant Nos.~CCF-2107291, CCF-2217696, CCF-2313027, 
CCF-2403036; as well as AST-2421782 and Simons Foundation MPS-AI-00010515 
(NSF-Simons AI Institute for Cosmic Origins -- CosmicAI).  
This work was in part supported by Cisco Research.  Any
opinions, findings and conclusions, or recommendations expressed in
this material are those of the authors and do not necessarily reflect
the views of Cisco Research.
\balance
\bibliography{bib}

\end{document}